\documentclass[a4paper,11pt]{article}
\pdfoutput=1 

\usepackage{jheppub} 

\usepackage[T1]{fontenc} 

\title{\boldmath Estimating QCD uncertainties in Monte Carlo event generators for gamma-ray dark matter searches}

\author[a]{Simone Amoroso,}
\author[b,c]{Sascha Caron,}
\author[d]{Adil Jueid,}
\author[e]{Roberto Ruiz de Austri}
\author[f]{and Peter Skands}
\affiliation[a]{Deutsches Elektronen-Synchrotron (DESY), Notkestrasse 85, D-22607 Hamburg, Germany}
\affiliation[b]{Institute for Mathematics, Astrophysics and Particle Physics, Faculty of Science, Mailbox 79,
Radboud University Nijmegen, P.O. Box 9010, NL-6500 GL Nijmegen, The Netherlands}
\affiliation[c]{Nikhef, Science Park, Amsterdam, The Netherlands}
\affiliation[d]{INPAC, Shanghai Key Laboratory for Particle Physics and Cosmology,
Department of Physics and Astronomy, Shanghai Jiao Tong University, Shanghai 200240, China}
\affiliation[e]{Instituto de F\'isica Corpuscular, IFIC-UV/CSIC, Valencia, Spain}
\affiliation[f]{School of Physics and Astronomy, Monash University, Australia}

\emailAdd{simone.amoroso@desy.de}
\emailAdd{scaron@cern.ch}
\emailAdd{adil.jueid@sjtu.edu.cn}
\emailAdd{rruiz@ific.uv.es}
\emailAdd{peter.skands@monash.edu}

\abstract{Motivated by the recent galactic center gamma-ray excess identified in the Fermi-LAT data, we perform a detailed study of QCD fragmentation uncertainties in the modeling of the energy spectra of gamma-rays from Dark-Matter (DM) annihilation. When Dark-Matter particles annihilate to coloured final states, either directly or via decays such as $W^{(*)}\to q\bar{q}'$, photons are produced from a complex sequence of shower, hadronisation and hadron decays. In phenomenological studies their energy spectra are typically computed using Monte Carlo event generators. These results have however intrinsic uncertainties due to the specific model used and the choice of model parameters, which are difficult to asses and which are typically neglected. We derive a new set of hadronisation parameters (tunes) for the \textsc{Pythia~8.2} Monte Carlo generator from a fit to LEP and SLD data at the $Z$ peak. For the first time we also derive a conservative set of uncertainties on the shower and hadronisation model parameters. Their impact on the gamma-ray energy spectra is evaluated and discussed for a range of DM masses and annihilation channels. 
The spectra and their uncertainties are also provided in tabulated form for future use. The fragmentation-parameter uncertainties may be useful for collider studies as well.}

\begin{document} 
\maketitle
\flushbottom 

\section{Introduction}
\label{sec:introduction}

The existence of Dark Matter (DM), making up $\sim 85\%$ of the matter of the universe, is today an accepted part of the Standard Cosmological Model. Observations of cosmological large scale structure moreover favour that the DM was not relativistic when galaxies formed, the so called cold DM (CDM) scenario. 

In particle physics, the CDM scenario is most straightforwardly realised by extending the Standard Model (SM) with weakly interacting massive particles (WIMPs). Unlike neutrinos, WIMPs can be non-relativistic and hence compatible with the CDM scenario. They can also provide a simple and compelling explanation of the density of DM we observe today in the Universe through a thermal mechanism \cite{Bertone:2004pz}. Namely, WIMPs were in thermal equilibrium with the thermal bath when the temperature of the plasma was larger than their mass. As long the temperature was dropping their number density was decreasing due to the cooling of the Universe. Eventually they froze out due to the expansion of the Universe and the comoving number density was fixed. What is interesting is that weak interactions and WIMP masses $\sim 100$ GeV give rise to the relic abundance of DM observed by the Planck satellite ($\Omega_{DM} h^2 = 0.1188 \pm 0.0010$) \cite{Ade:2015xua}. This is the so-called WIMP miracle.

WIMPs could be detected indirectly through annihilation to gamma-rays, positrons, antiprotons, neutrinos and other particles that can be observed in experiments such as the Fermi Large Area Telescope (LAT), AMS-02 or IceCube. Those annihilation products might leave footprints in the fluxes of cosmic rays and any excess could be interpreted as a WIMP signature. This is the case of the excess detected in the gamma-ray data collected by the Fermi-LAT from the inner Galaxy \cite{TheFermi-LAT:2015kwa}, the so-called Galactic Center Excess (GCE). Its spectral energy distribution and morphology are consistent with predictions from DM annihilation \cite{Goodenough:2009gk, Vitale:2009hr, Hooper:2010mq, Gordon:2013vta, Hooper:2011ti, Daylan:2014rsa, Calore:2014xka, Abazajian:2014fta,Zhou:2014lva}.

There have been many attempts to explain the GCE in the particle physics framework, particularly in the supersymmetry context \cite{Caron:2015wda, 2015arXiv150707008B, Butter:2016tjc, Achterberg:2017emt}. In its minimal phenomenological realization, called the Minimal Supersymmetric Standard Model, it has been shown that the precision in the determination of the gamma-ray spectrum from DM annihilation plays a fundamental role in the quality of fitting the model to the data \cite{Caron:2015wda}. 

For dark-matter annihilation processes that produce hadronic final states --- either directly or via the hadronic decays of intermediate resonances like $W$, $Z$, or $H$ bosons --- the dominant source of particle production is QCD jet fragmentation (for dark-matter masses above a few GeV). 
Long-lived particles such as photons are then produced as the final result of a complex sequence of physical processes which include bremsstrahlung, hadronisation, and hadron decays. 
There is no first-principles solution to the problem of hadronisation. However, by exploiting that it is a long-distance effect, as compared to the scales involved in typical high-energy (sub-femtometre)
production processes, it can be formally factorised off in a universal (process-independent) way and represented either by parametric fits (called Fragmentation Functions, or FFs for short; see~\cite{Metz:2016swz}) or by explicit dynamical models, such as the string~\cite{Artru:1974hr,Andersson:1983ia} or cluster~\cite{Webber:1983if,Winter:2003tt} models which are embedded in Monte Carlo (MC) event generators~\cite{Buckley:2011ms}. Note that the former (FFs) typically only parametrise the spectra of one specific type of particle at a time, with all other degrees of freedom inclusively summed over. This allows to reach a higher formal accuracy, albeit typically only over a limited range in the
energy fraction of the produced hadrons. In contrast, the MC models provide fully exclusive simulated ``events'' from which in principle any desired observable can be  constructed, with formally less accuracy but typically a broader range of applicability. In both formalisms, the essential point is that the parameters governing the long-distance physics are to a good approximation independent of the detailed nature of the short-distance process; they can therefore be constrained by fits to data (such as $e^+e^- \to \mbox{hadrons}$) and applied universally to make predictions for other processes (such as dark-matter annihilation). 

In the context of such predictions, a crucial question is what is the uncertainty on the predicted spectra? 
A recent comprehensive study~\cite{Cembranos:2013cfa} highlighted that different MC models make different default assumptions for which physics effects to include and over which dynamical ranges; this can result in large differences in particular in the tails of distributions, e.g.\ for very low photon energies (for which it becomes important whether and how photon radiation off hadrons is treated, and/or how soft radiation off leptons is regulated) or for very high photon energies (for which prompt QED bremsstrahlung, $q \to q\gamma$, dominates). However the study also showed that in the bulk of the distributions, well inside these limits, there was a high level of agreement between the codes. This agrees with what one would expect from the default parameter sets for the MC models being chosen to essentially provide ``central'' fits to roughly the same set of constraining data, comprised mostly of LEP measurements; see~\cite{Buckley:2009bj,Buckley:2010ar,Skands:2010ak,Platzer:2011bc,Karneyeu:2013aha,Skands:2014pea,Fischer:2014bja,Fischer:2016vfv,Reichelt:2017hts,Kile:2017ryy}. This degeneracy of fitting different models to the same data, however, also implies that the envelope spanned by them is  not a particularly systematic or exhaustive way of exploring the true region of allowed uncertainty. Thus, while there can be huge differences in the tails caused by intrinsically different modeling assumptions, we believe that the uncertainty in the bulk of the distributions is not well represented by the envelope of different MC models, or is at least not \emph{guaranteed} to be well represented by it. 

The question we wish to address in this paper is therefore: can we provide meaningful and exhaustive sets of alternative model parameters that are able to faithfully represent the uncertainty with respect to a given set of constraining measurement data, within a given modeling paradigm? To answer this question, we take the default Monash 2013 tune~\cite{Skands:2014pea} of the \textsc{Pythia}~8 event generator~\cite{Sjostrand:2014zea} as our baseline\footnote{Specifically, we have used version 8.2.35 in this work.} and --- using a selection of fragmentation constraints from $e^+e^-$ colliders encoded in the \textsc{Rivet}~\cite{Buckley:2010ar} analysis preservation package combined with the \textsc{Professor}~\cite{Buckley:2009bj} parameter optimisation tool --- define a small set of systematic parameter variations which we argue explores the uncertainty envelopes relevant for estimating QCD fragmentation uncertainties on dark-matter annihilation processes in a meaningful way, and could be useful for estimating at least the flavour-insensitive component of fragmentation uncertainties on collider observables as well. As will be discussed in the main part of the paper, a straightforward $\chi^2$ minimisation with uncertainty envelopes defined by parameter variations along the eigenvectors corresponding to $\Delta \chi^2 = 1$ variations (called ``eigentunes''~\cite{Buckley:2009bj}) does not immediately result in what we could call a faithful representation of the true uncertainty envelope, but with minor and well-motivated modifications can be adapted to produce such representations. 

The paper is organized as follows. In Sec. \ref{sec:physics-modeling} we describe how the photon spectra from DM annihilation is modeled in MC event generators. In Sec. \ref{sec:measurements} a detailed study of the origin of the gamma spectrum is presented while in Sec. \ref{sec:tune} the tunning of the MC to data is performed and in Sec. \ref{sec:results} we
present the results of the tuning and the QCD uncertainties with emphasis on the impact of those 
uncertainties on two benchmark points of the MSSM. We conclude in Sec.~\ref{sec:conclusions}.

\section{Physics Modeling}
\label{sec:physics-modeling}
Consider a generic dark-matter annihilation process, $\chi\chi \to X$. Typically, we have a lowest-order picture in mind for what $X$ can be, in which additional physics (like sequential resonance decays, or fragmentation of coloured particles) are implicitly summed over. These aspects must be dealt with explicitly before final-state observables like photon spectra can be estimated. 

If $X$  includes short-lived resonances such as $Z/W$ or $H$ bosons, then the narrow-width approximation allows us to factorise the complete physics process into a production part, $\chi\chi \to X_1 \ldots X_n$, and a decay part, $X_i \to Y_{i1} \ldots Y_{in}$. This factorisation is reliable up to corrections of order $\Gamma_i/M_i$, and is hence a good approximation for states with $\Gamma \ll M$ such as the SM gauge and Higgs bosons. Note that at wavelengths above $(\hbar c)/\Gamma_i$, we would still expect interference effects between the decay products of different resonances, suppressed by boost effects if the resonances have non-zero relative velocities. 

\begin{figure}[t]
\centering
\begin{tabular}{c|cc}
QED bremsstrahlung & \multicolumn{2}{c}{QCD fragmentation and hadron decays}\\
\includegraphics[scale=0.65]{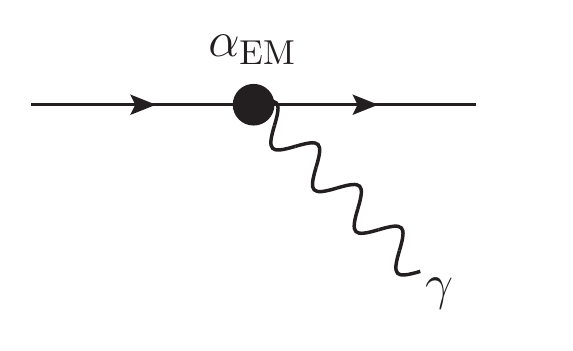} &
\includegraphics[scale=0.65]
{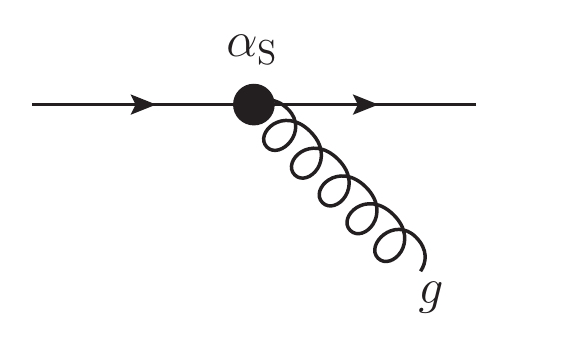} &
\raisebox{-2.5mm}{\includegraphics[scale=0.65]{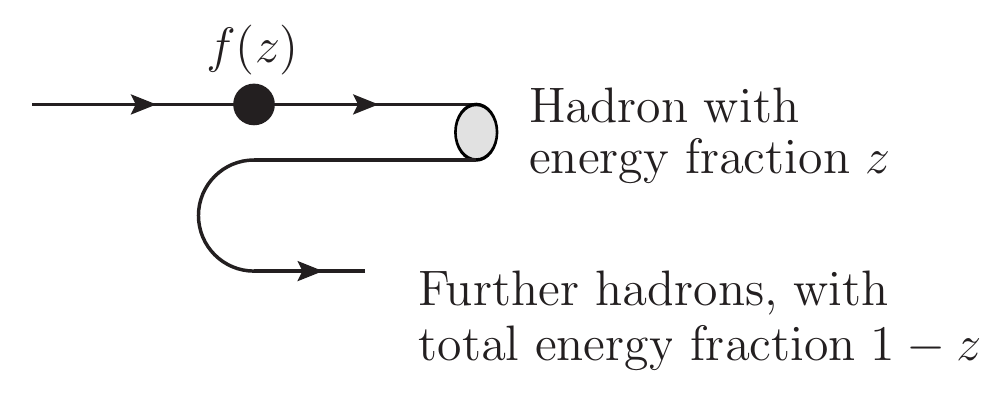}}\\
Dominates at high $x_\gamma$ & \multicolumn{2}{c}{Photons from $\pi^0\to\gamma\gamma$ dominate bulk (and peak) of spectra}
\end{tabular}
\caption{
Illustration of the main parameters that affect the $x_\gamma$ distributions in jets. From left to right: the electromagnetic coupling $\alpha_\mathrm{EM}$, the strong coupling $\alpha_S$, and the nonperturbative fragmentation function $f(z)$. For coloured final states, all three processes are active. QED bremsstrahlung is then only dominant at very high $x_\gamma$ where the total rate is low, while the QCD fragmentation parameters govern the bulk of the photon spectrum, via $\pi^0\to\gamma\gamma$ decays. (We note that $\tau\tau$ final states will also have some contributions from $\pi^0\to\gamma\gamma$ decays but these are controlled by the $\tau$ decay branching fractions and phase-space distributions and are independent of the $\alpha_S$ and $f(z)$ parameters.)
\label{fig:MCparams}}
\end{figure}

If $X$ (or decay products $Y$) includes  photons or electrically charged particles, then those will undergo QED bremsstrahlung showers. Additional photons are produced via $X_i^\pm \to X_i^\pm\gamma$ branchings, which are enhanced for both soft (low $x_\gamma = E_\gamma/m_\chi$) and (quasi)collinear photons. Note that the latter type of photons can have high energies (the only requirement for the enhancement being a small angle between the photon and its parent particle) and tend to dominate the ultra-hard tail of the final-state photon spectra towards $x_\gamma\to 1$.
Charged fermion-antifermion pairs can also be produced, at a subleading level, via $\gamma\to f\bar{f}$ branchings, which are enhanced at very low values of $Q^2/m_\chi^2 = (p_f + p_{\bar{f}})^2/m_\chi^2$. The main modeling parameter that governs the rate of both types of QED processes is the effective value assumed for the QED fine-structure constant, $\alpha_\mathrm{EM}$, illustrated by fig.~\ref{fig:MCparams}a. Nominally, this parameter is of course extremely well constrained by measurements, but it may still be useful to subject its effective value to variations, as a poor man's way to estimate the possible effects of missing higher-order or non-universal (process-dependent) contributions to the spectra. Since this work focuses on DM annihilation to coloured particles, however, variations of $\alpha_\mathrm{EM}$ are quite subleading with respect to the larger variations in the QCD sector we shall discuss below, and are not considered further. 

Obviously, there are also cases in which resonance decays and QED showers occur together, such as in $\chi\chi\to W^+W^-$; in an MC model like \textsc{Pythia}, this will be treated by first allowing the $W$ bosons to undergo QED showers, i.e.\ allowing for $W\to W\gamma$ branchings, with a phase space that is vanishing at the $WW$ threshold and proportional to $\Delta m_{\chi _W} = m_\chi - m_W$ above it. The $W$ bosons are then decayed, and their decay products undergo further showering (and hadronisation). Note that, as $m_\chi$ is increased, the QED shower off the $WW$ system is the only component that will become more active, due to the increasing phase space. The decays of each of the $W$ systems themselves are only affected by an overall boost (in the narrow-width limit). By Lorentz invariance, they therefore remain well constrained by measurements of decays of $W$ (or $Z$) bosons at rest as long as the range of well-measured rest-frame final-state energies still covers the region of interest in the boosted ($\chi\chi$ annihilation) frame. 

If $X$ (or decay products $Y$) includes coloured particles, then those will undergo QCD showers + hadronisation. The QCD shower stage is modeled similarly to the QED one, reflecting the enhancement of soft and collinear emissions and of $g\to q\bar{q}$ splittings at low virtualities. The default treatment in \textsc{Pythia} is based on a combination of DGLAP splitting kernels for QED+QCD radiation with dipole ($2\to 3$) kinematics~\cite{Sjostrand:2004ef}. The main parameter governing the rate of QCD shower branchings is the effective value assumed for the strong coupling constant, $\alpha_S$, at each branching vertex, cf.\ fig.~\ref{fig:MCparams}b. There are strong arguments in the literature that the renormalised coupling for shower branching processes should be evaluated at a scale proportional to the $p_\perp$ of each branching, and that a further set of universal corrections in the soft limit can be absorbed by using the so-called Monte Carlo, or CMW~\cite{Catani:1990rr}, scheme to define the running coupling, rather than the conventional $\overline{\mathrm{MS}}$ scheme. This has the net effect of increasing the effective value of $\alpha_s(M_Z)$ by about 10\%. In \textsc{Pythia} tunes, the effective value of the strong coupling is typically further increased by about 10\%, to reach agreement with measured rates for $e^+e^- \to 3~\mathrm{jets}$; see e.g.\ \cite{Skands:2010ak,Skands:2014pea}. The standard recommendation for perturbative uncertainty estimates is to perform a variation of the renormalisation scale by a factor of 2 in each direction, but since this would actually destroy some of the universal corrections obtained from the CMW scheme, the framework for automated scale variations that was recently  implemented in \textsc{Pythia}~\cite{Mrenna:2016sih} allows for a second-order compensation term to be imposed, which reduces the effect of the variations somewhat and reestablishes agreement with the CMW scheme at second order. This is the prescription we advocate for a realistic and still reasonably conservative uncertainty estimate on the perturbative part of the QCD fragmentation process. If desired, variations of the splitting functions by non-enhanced terms can also be included, as described in~\cite{Mrenna:2016sih}.

An example of a physical process that combines all of the elements of sequential resonance decays, QED showers,  and QCD showers, is $\chi\chi \to t\bar{t}$. Here, the $t\bar{t}$ system will first undergo a QCD+QED shower, with a phase space proportional to how far above $t\bar{t}$ threshold the $\chi\chi$ annihilation process is. The top quarks will then decay, and the resulting $bW$ systems showered, upon which the $W$ bosons will decay, etc. 

\begin{figure}[t]
\centering
\includegraphics[scale=0.65]{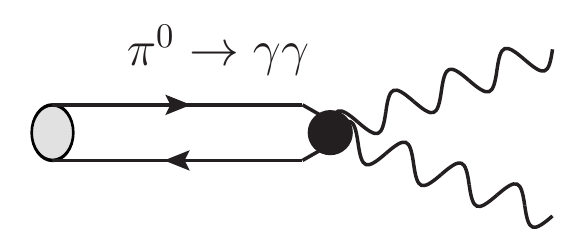}~~
\caption{In hadronic jets, $\pi^0$ decays are the main source of photons. 
\label{fig:pi0decay}}
\end{figure}

\begin{figure}[t]
\centering
\includegraphics*[scale=0.35]{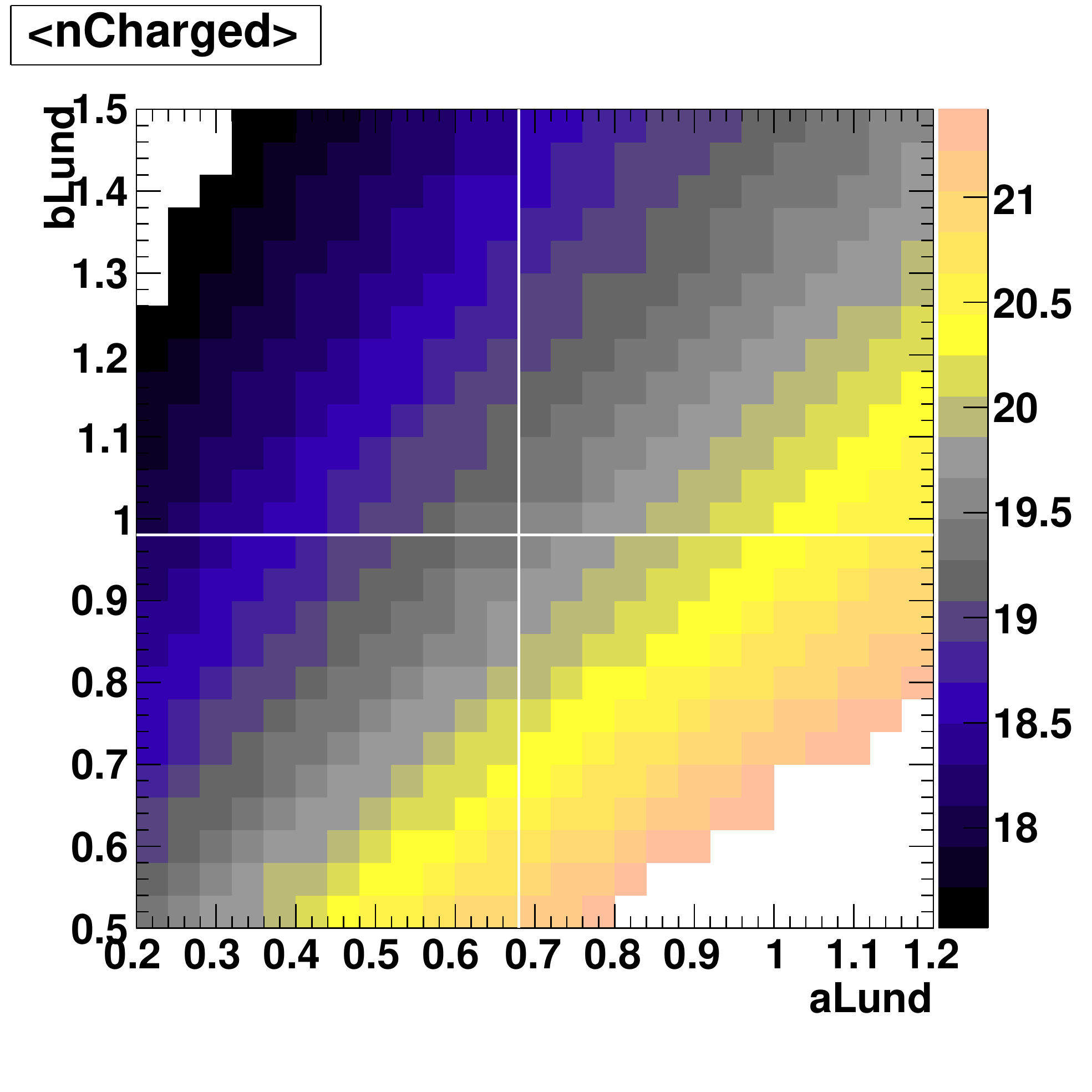}~~
\includegraphics*[scale=0.35]{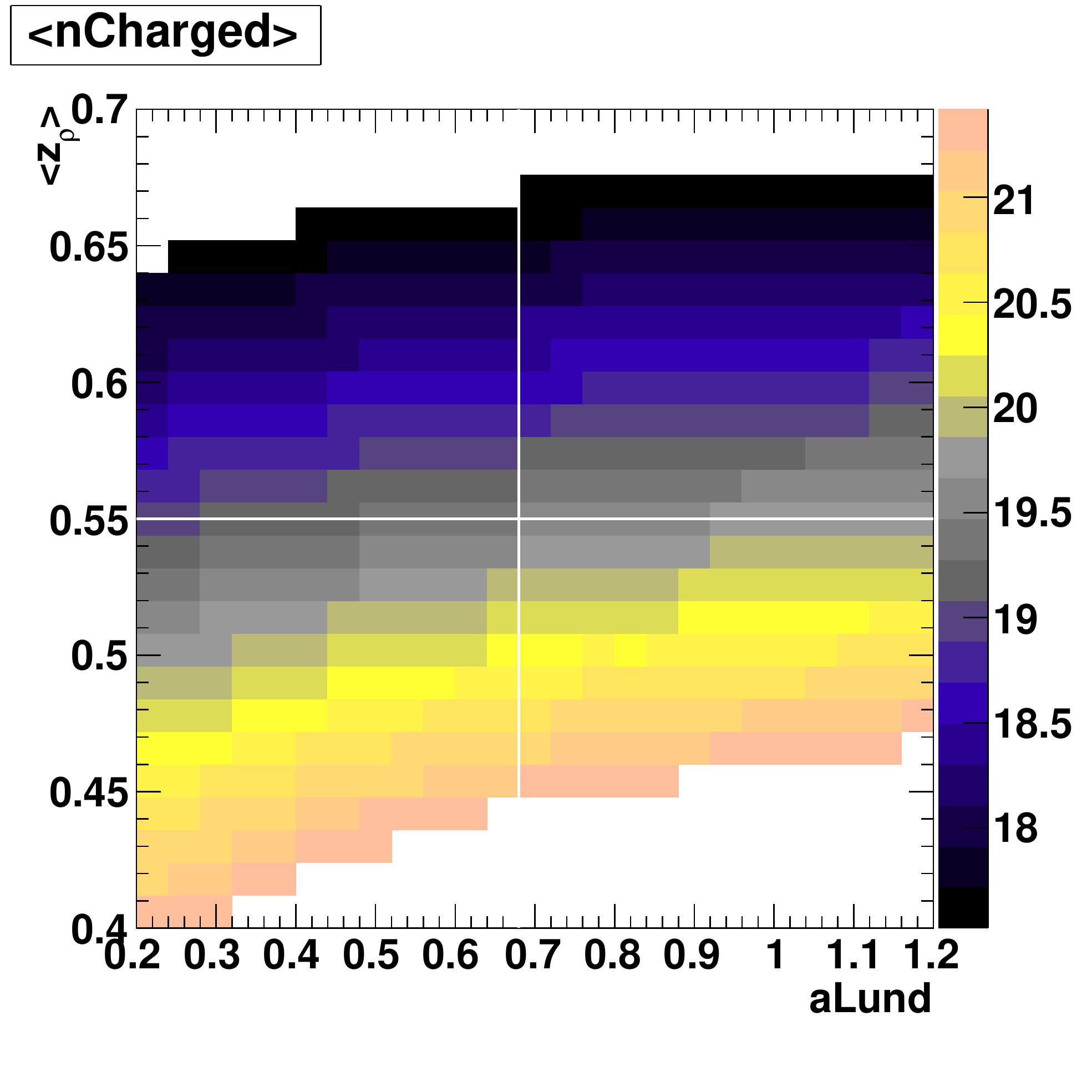}
\caption{The average number of charged particles produced in $Z\to d\bar{d}$ decays at $E_\mathrm{cm}=M_Z=91.2$ GeV. {\sl Left:} as a function of $a$ and $b$. {\sl Right:} as a function of $a$ and $\left<z_\rho\right>$.\label{fig:corrab}}
\end{figure}

Finally, any produced coloured particles must be confined inside colourless hadrons. This process --- \emph{hadronisation} --- takes place at a distance scale of order the proton size $\sim 10^{-15}$m and in \textsc{Pythia} is modelled by the Lund string model; see~\cite{Andersson:1983ia} for details. Since pions are the most copiously produced particles in jets and the branching fraction for $\pi^0\to\gamma \gamma$ is $\sim 99\%$~\cite{Patrignani:2016xqp}, the vast majority of photons in jets are produced from decays of neutral pions, illustrated in fig.~\ref{fig:pi0decay}.
The number and hardness of produced photons therefore correlates very strongly with the predicted pion spectra, and the uncertainties in turn are dominated by the quality of the available constraints on pion spectra, as well as the model's ability to reproduce them.  
A crucial component of this description is the \emph{fragmentation function}, $f(z)$, which parametrises the probability for a hadron to take a fraction $z \in [0,1]$ of the remaining energy at each step of the (iterative) string fragmentation process, cf.~fig.~\ref{fig:MCparams}c. While $f(z)$ cannot be calculated from first principles by current methods, its functional form is strongly constrained by self-consistency requirements (essentially, causality) within the string-fragmentation framework, so that its general form can be cast in terms of just two effective parameters, called $a$ and $b$:
\begin{equation}
f(z,m_{\perp h}) = N \frac{(1-z)^a}{z}\exp\left(\frac{-b m_{\perp h}^2}{z}\right)~,
\label{eq:fz}
\end{equation}
where $N$ is a normalisation constant that ensures the distribution is normalised to unit integral, and $m_{\perp h}= \sqrt{m_h^2 + p_{\perp h}^2}$ is called the ``transverse mass'', with $m_h$ the mass of the produced hadron and $p_{\perp h}$ its momentum transverse to the string direction. 
For non-experts: if $f(z)$ is peaked near 1 (low $a$ and/or high $b$), then QCD jets will tend to consist of only a few hadrons,  each taking a rather large fraction of the energy of the jet. Conversely, if $f(z)$ is peaked near zero (low $b$ and/or high $a$), then  the prediction is for jets which consist of very many hadrons, each taking only a small fraction of the total available energy. An illustration of this is given in the left-hand pane of fig.~\ref{fig:corrab}, which shows the average number of charged particles produced in $Z\to d\bar{d}$ decays, after hadronisation, as a function of the $a$ and $b$ parameters, with all other parameters fixed to their Monash 2013 tune values~\cite{Skands:2014pea}. (The Monash values \texttt{StringZ:aLund         =  0.68} and \texttt{StringZ:bLund         =  0.98} are indicated by the white cross hair in the centre of the plot.)

Since the average number of charged particles in jets is one of the most salient constraining observables, this plot also illustrates an oft-encountered problem; in tuning contexts, the $a$ and $b$ parameters are extremely highly correlated. This makes it meaningless to assign independent $\pm$ uncertainties on them; likewise sensitivity estimates and the like cannot be interpreted without taking the correlation into account carefully. Therefore, in the context of the current work, we have implemented an alternative parametrization of $f(z)$, with  $b$ replaced by a parameter representing the average $z$ fraction taken by a typical hadron (specifically, primary $\rho$ mesons), 
\begin{equation}
\left<z_\rho\right> = \int_0^1 \mathrm{d}z \ z f(z,\left<m_{\perp\rho}\right>)~,
\label{eq:zrho}
\end{equation}
which we solve (numerically) for $b$ at initialisation when the option \texttt{StringZ:deriveBLund = on} is selected in \textsc{Pythia} 8.235, using the following parameters:
\begin{eqnarray}
\left<m_{\perp\rho}\right>^2 & = & 
m^2_\rho + 2( \mbox{\texttt{StringPT:sigma}})^2~,
\\
\left<z_\rho\right> & = &\mbox{\texttt{StringZ:avgZLund}}~.
\end{eqnarray}

As illustrated by the right-hand pane in fig.~\ref{fig:corrab}, a measurement of the mean charged multiplicity puts a rather tight bound on the value of $\left< z_\rho\right>$,  and there is a much smaller degree of correlation with the $a$ parameter, the latter of which is essentially unconstrained by this observable. Below, this observation of reduced correlations and hence more meaningful independent $\pm$ uncertainty ranges for the new parameter set will be explored further and quantified in the context of full-fledged parameter optimisations ('tunes'). 
\section{Photon Origins and Measurements}
\label{sec:measurements}
\subsection{Photon Origins}

\begin{figure}[!t]
\centering
\includegraphics[width=7.5cm, height=6.5cm]{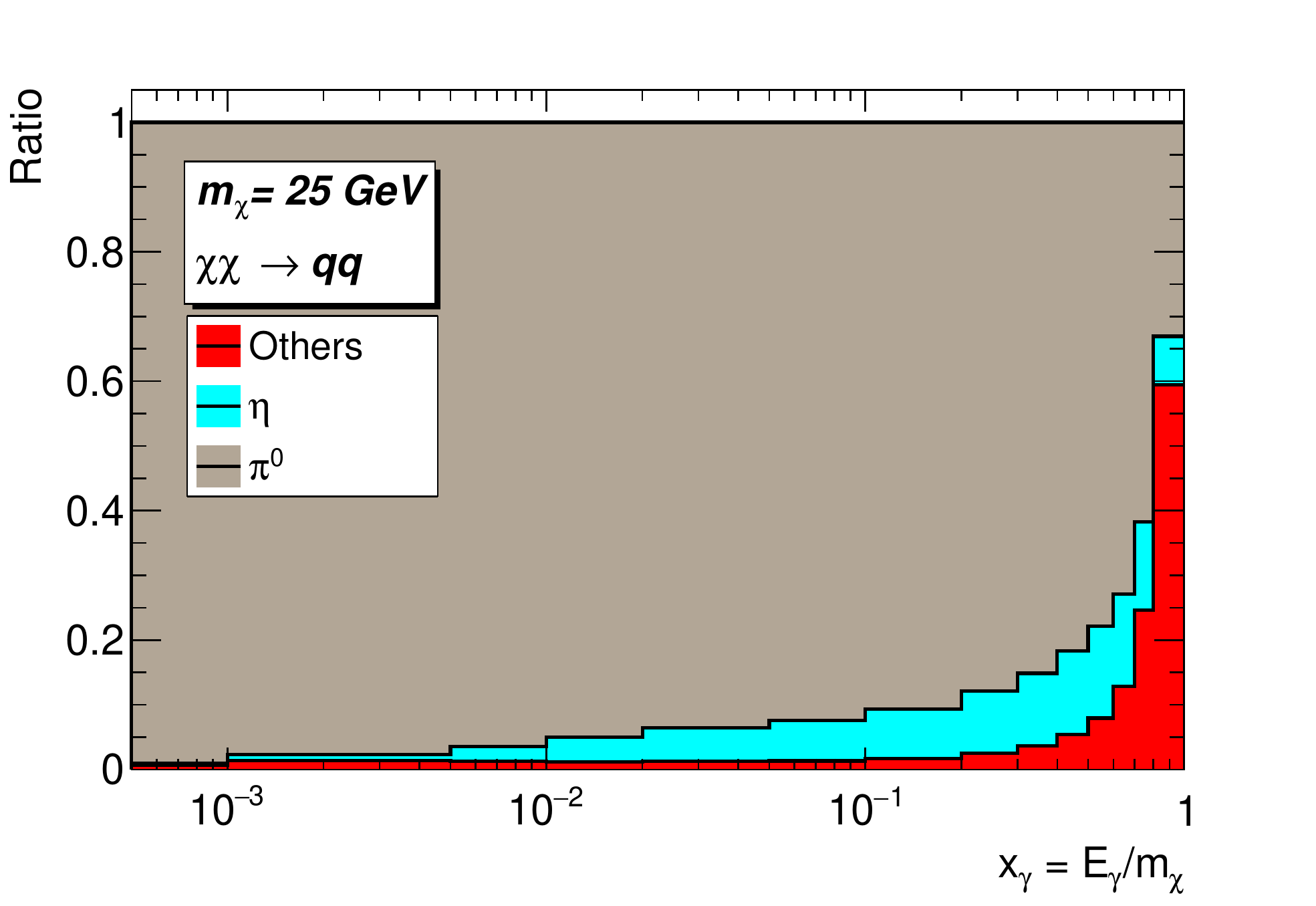}
\hfill
\includegraphics[width=7.5cm, height=6.5cm]{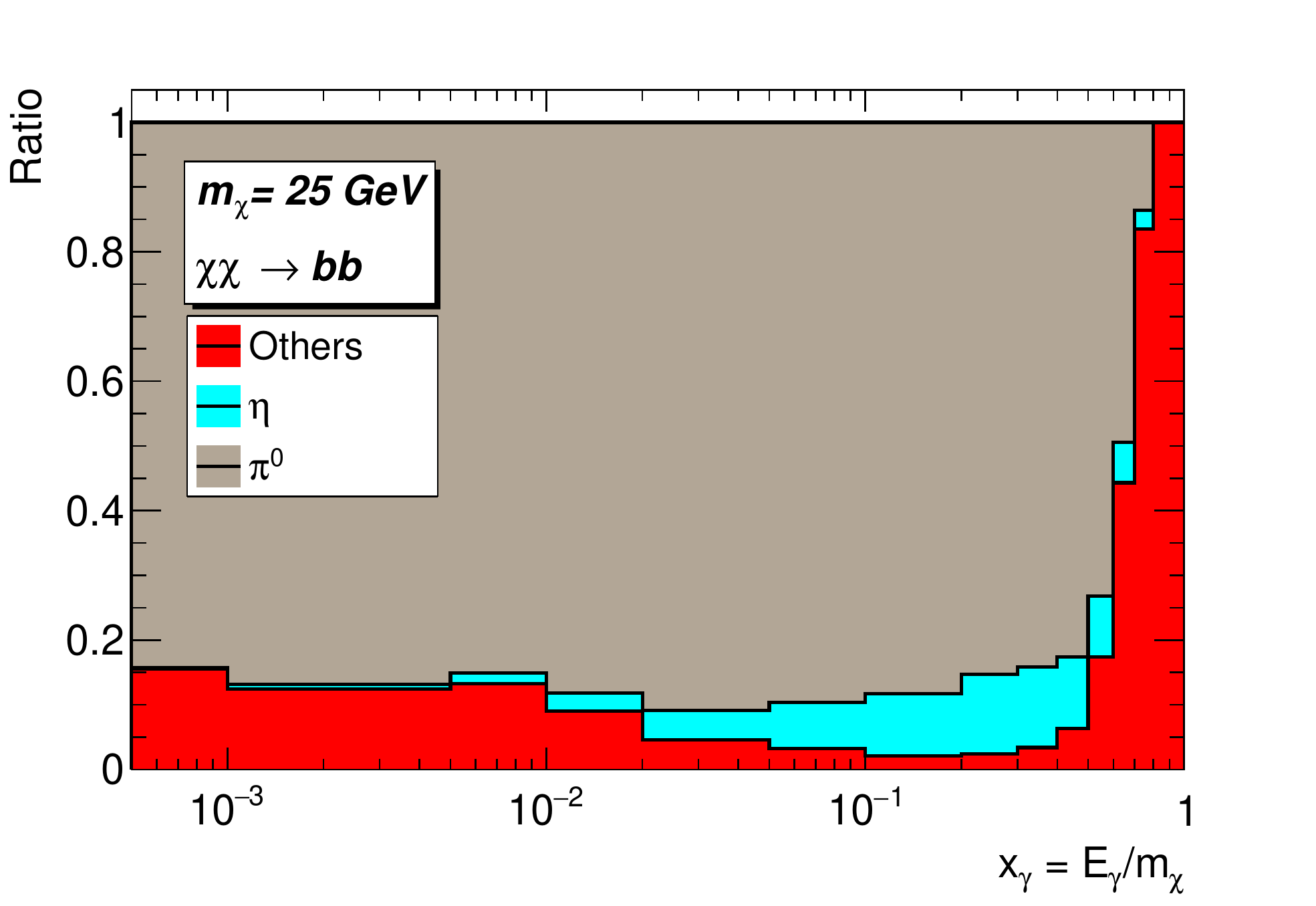}
\caption{Ratio plots for photon origins in $\chi\chi$ annihilation
into $q\bar{q}, q=u,d,s$ (left panel) and $b\bar{b}$ (right panel)
for $m_\chi=25$ GeV.}
\label{PhotonOrigins-quarks}
\end{figure}

In the previous section, it was highlighted that the spectra of 
photons in QCD jets is dominated by $\pi^0\to \gamma \gamma$ decays. This will be studied in more details in this section. To do so, we identify the origin of photons in generic dark-matter annihilation processes with $m_\chi=25$ GeV and 
$m_\chi=250$ GeV. This allows the dark-matter candidate to be annihilated
into quarks and gluons only for the former case and into all the SM particles in the latter case. Given that we
are interested specifically in the modeling of QCD uncertainties here, 
final states such as $\ell^+ \ell^-$ and $\gamma\gamma$ are not considered.
We show in Fig. \ref{PhotonOrigins-quarks}, the relative composition of the photon spectrum in terms of photons coming from $\pi^0$ decays, photons coming from $\eta^0$ decays, and ``others'' (for all other photon sources), resulting from a generic dark matter annihilation into $q\bar{q}$ and $b\bar{b}$ for $m_\chi=25~\mathrm{ GeV}$.
We see that the vast majority of photons indeed come from pion decays, 
about $95\%$ in the $q\bar{q}$ final state and $88\%$ in the $b\bar{b}$ one.
The fraction is somewhat lower in the $b\bar{b}$ final state since most of these events will have one or two photons that come from $B^*$ decays (which are here lumped into the ``others'' category)\footnote{The radiative decays $B^*\to B \gamma$ have a $100\%$ branching ratio, and the probability for a $b$ quark to fragment into $B_{0,\pm}^*$ is higher than 50\%, phase space permitting.}. 
However, as one goes to higher DM masses, far above the $b\bar{b}$ threshold -- see Fig. \ref{photons-sources-250-light-ratio} -- the contribution from neutral pions eventually dominates again and one approaches the same distribution as in the $q\bar{q}$ case. Very subleading contributions come from photon bremsstrahlung off charged 
quarks; these typically dominate in the high $x_\gamma$ region but the total integrated rate is small. (We note however that the rate of these photons is proportional to the charge squared of the emitting particle, so there will be more such photons in $c\bar{c}$ final states, for example, than in $b\bar{b}$ ones.)
On the other hand, the contribution from decays 
of $\eta$ mesons is about $4\%$ ($3\%$ in our $b\bar{b}$ final state example). 

Photon origins in
other final states and for the case of $m_\chi=250$ GeV are shown in the appendix 
specifically in 
Figs. \ref{photons-sources-25}, \ref{photons-sources-250-light-ratio}-\ref{photons-sources-250}. For higher dark matter masses, the relative contribution
of pions to the photon spectrum becomes similar in almost all final states. 
For other final states such as 
the di-Higgs, the ``others'' contributions occur from many sources since the full set of Higgs decay channels include not only $b\bar{b}$ but also $\tau\tau$, $c\bar{c}$, and $gg$ final states, the former three of which radiate photons before 
they hadronize or decay (for the case of $\tau$-leptons). Nevertheless, since this contribution is important in regions with low number of photons (see e.g Fig. \ref{photons-sources-250}) and, furthermore, is of QED nature, it will not affect QCD uncertainties that we estimate in what follows.

Given that most of the photons produced in jets come from pion decays, a study of QCD uncertainties on photon spectra ultimately boils down to a study of  uncertainties on pion spectra, with a very small additional component coming from $\eta^0$ decays. 
The main experimental constraints on such spectra come from $e^+ e^- \to \mathrm{hadrons}$.
At CM energies of $m_Z = 91.2$ GeV, experimental collaborations at \textsc{Lep} (and \textsc{Sld}) measured the mean pion multiplicities and found $\langle n_{\pi^0} \rangle = 9.5$ \cite{Barate:1996fi} and
$\langle n_{\pi^\pm} \rangle = 17.1$ \cite{Abreu:1996na}
compared to the mean charged  multiplicity $\langle n_{\mathrm{ch}} \rangle = 21$ \cite{Barate:1996fi}. 
However, most of these pions are in fact themselves the decay products of more massive particles. One therefore roughly distinguishes between ``primary'' pions, produced directly in the quark/gluon fragmentation process, and ``secondary'' ones, produced from the decay of heavier hadrons and $\tau$-leptons; see e.g.~\cite{Skands:2012ts,Buckley:2011ms}. 
To access these details, we studied the contributions to the 
spectra of $\pi^0$ in different final states and for $m_\chi=25$ GeV and $m_\chi=250$ GeV. To avoid clutter, the corresponding distributions and ratio plots are collected in the appendix (Figs. \ref{pions-sources-25}-\ref{pions-sources-25-ratio-hadrons} 
for $m_\chi=25$ GeV and in Figs. \ref{pions-sources-250-light}-\ref{pions-sources-250-ratio-hadrons} for $m_\chi=250$ GeV). 

In all cases, the number of secondary pions is larger than the number of primary ones, with secondaries accounting for a fraction of $70\%$--$87\%$ of the total. The highest fraction of secondaries occurs in $b\bar{b}$ production for $m_\chi = 25$ GeV (bottom left pane of Fig. \ref{pions-sources-25-ratio}); this is not surprising since a significant chunk of the energy is here tied up in the $B$ hadron masses, and any pions produced in the decay of those hadrons are secondaries by definition. As soon as we go far from the $b\bar{b}$ threshold, cf.~Fig. \ref{pions-sources-250-light-ratio}, the fraction of $\pi^0$ coming from hadron decays becomes similar to that in e.g.\ $q\bar{q}$. Another observation is that in $gg$ final states, there is a possibility that no $g\to q\bar{q}$ branchings are produced in the parton shower, in which case the hadronising string system is a closed ``gluon loop''; this is indicated in the plots by using the label ``g'' instead of ``q'' for the mother and happens about $10\%$ of the time for $m_\chi=25$ GeV, decreasing to about $2\%$ for $m_\chi=250$ GeV. 
The contribution from $\tau$ leptons only accounts for about $1.5\%$ ($0.5\%$) of the total number of pions for $hh$ ($b\bar{b}$) final states. 

For completeness, we note that the secondary pions mainly come from five sources: $\rho^\pm, \eta, \omega, D^{0,\pm}$ and $K_{S,L}$ (see e.g Fig. \ref{pions-sources-25-ratio-hadrons}). In final states such as $b\bar{b}$, $B$ hadrons will naturally also contribute, with a fraction of about $7\%$ for $m_\chi = 25$ GeV, dropping to about $3.5\%$ for $m_\chi=250$ GeV (Fig. \ref{pions-sources-250-light-ratio-hadrons}). 

In principle, one could follow the chain of secondaries further up, but the main point we wish to make here is simply that, in addition to the direct measurements of $\pi^0$ spectra, we are able to use information as well from a wide range of other measurements, in particular $\pi^\pm$ spectra (via isospin, see below), but also the spectra of the dominant immediate ancestors of the pions can provide relevant constraints.

\subsection{Measurements}
In the previous subsection, we reviewed the origins of photons and neutral pions in different dark-matter annihilation channels. The immediate conclusion is that, in addition to direct measurements of the photon spectrum itself, the modeling of photon spectra in QCD jets can be constrained by the following observables:
\begin{itemize}
\item $\pi^0$ spectrum: Since $\pi^0$ decays are the dominant source of photons in QCD jets, a correct modeling of $\pi^0$ is the sine qua non for modeling the $\gamma$ spectrum. There are several measurements carried out by LEP which can be used in our fit. We note however that, since the $\pi^0$ are reconstructed from their two-photon decays (see e.g.~\cite{Adeva:1991it,Adam:1995rf,Barate:1996uh,Ackerstaff:1998ap}), we do not expect these measurements to add much genuinely new information to our fits (i.e., we expect them to be highly correlated with the $\gamma$ spectrum measurements).
\item $\pi^\pm$ spectrum: Since they are members of the same $SU(2)$ multiplet, $\pi^\pm$ are related to $\pi^0$ by isospin symmetry, i.e.\ one expects $\langle n_{\pi^\pm} \rangle \sim 2 \langle n_{\pi^0} \rangle$. Slight breakings of this relation due to isospin-violating effects are accounted for by the event-generator modeling, and we may therefore include the constraints on the fragmentation-function parameters obtained for charged pions, which are typically far better measured in collider experiments. In particular, whereas the photon and $\pi^0$ measurements typically do not cover the peak region of the spectra well, the $\pi^\pm$ spectra are well measured down to much lower momenta, with small uncertainties on both sides of the peak; this is illustrated in Fig.~\ref{comparison-1} below. The ability to include these constraints therefore adds significantly to the overall constraining power, especially in the peak region, and is statistically independent of the $\pi^0$ and $\gamma$ constraints. 
\item $\eta$ spectrum: These are the second-most important source of photons in QCD jets; they contribute both directly through $\eta\to \gamma+X$ or via cascade decays $\eta \to \pi^0 +X \to \gamma \gamma + X$. At LEP, the multiplicity of $\eta$ mesons was about 10\% of the $\pi^0$ one. Again, we expect there to be a significant correlation with the $\gamma$, $\pi^0$, and $\pi^\pm$ measurements, since the $\eta$ mesons are reconstructed from their $\eta\to\gamma\gamma$ or $\eta \to \pi^+\pi^-\pi^0$ decays (see e.g.~\cite{Buskulic:1992hn,Adriani:1992hd,Ackerstaff:1998ap,Heister:2001kp}). 
\end{itemize}
\begin{figure}[!t]
\includegraphics[width=0.48\linewidth]{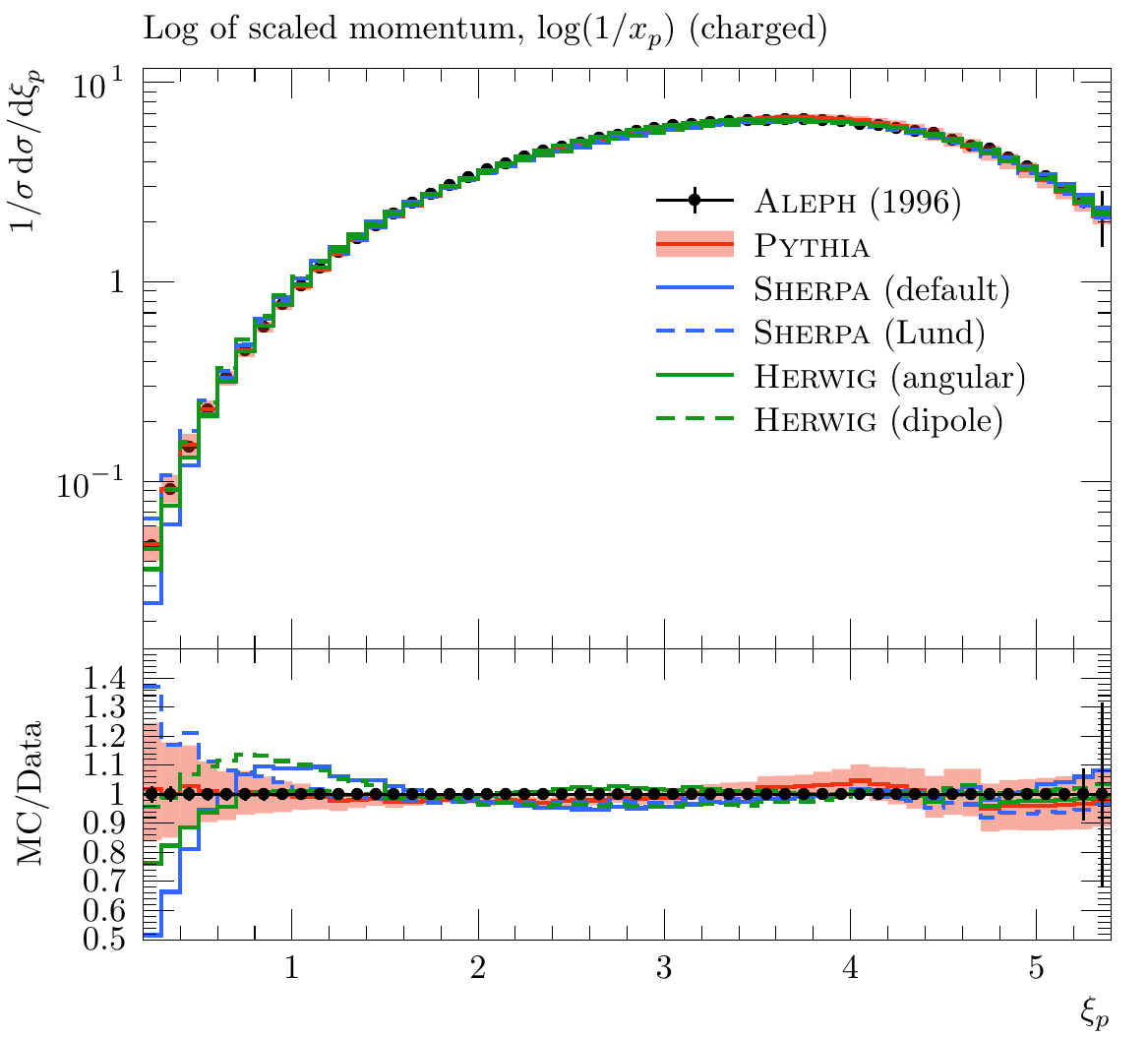}
\hfill
\includegraphics[width=0.48\linewidth]{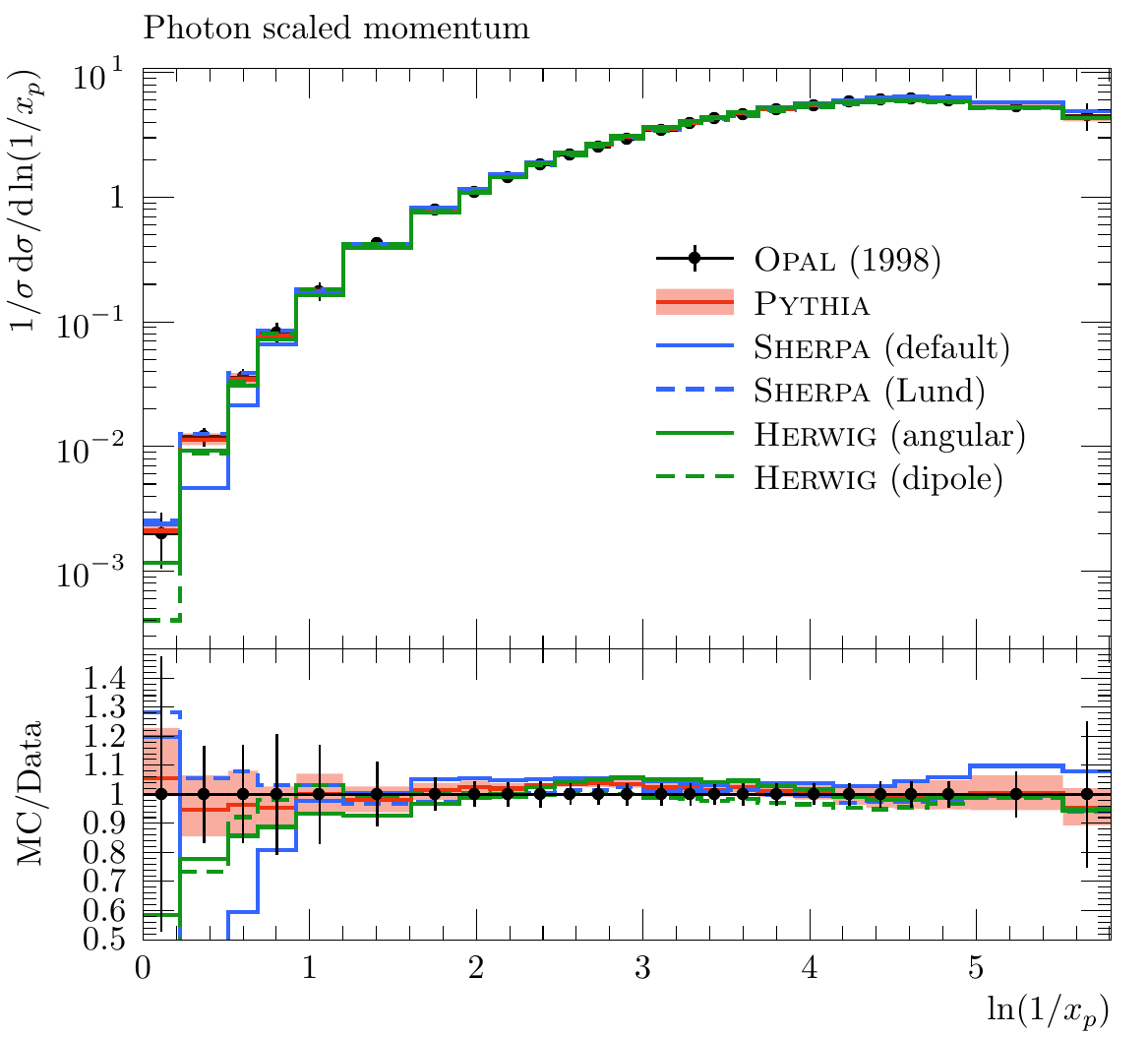}\\
\includegraphics[width=0.48\linewidth]{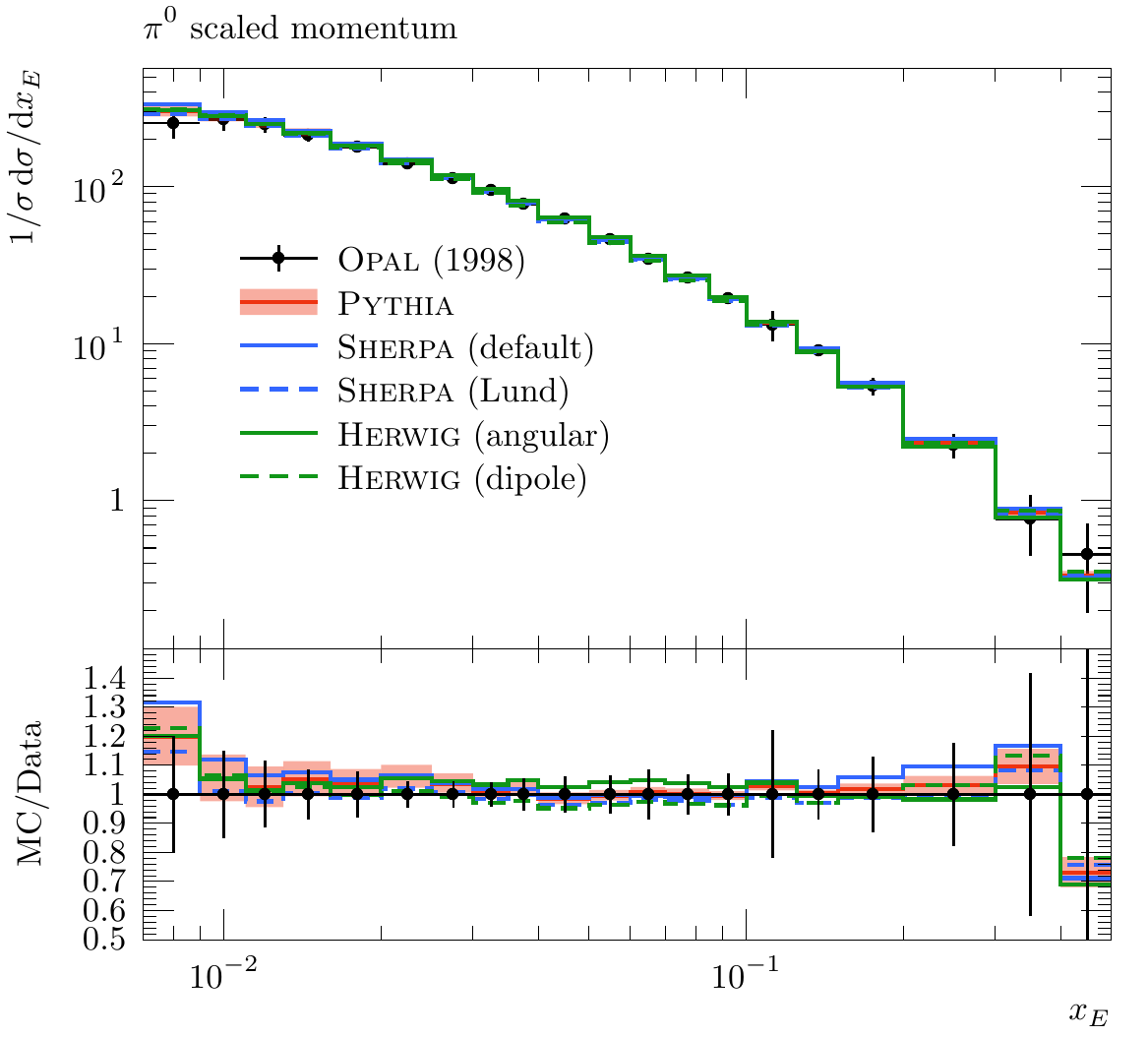}
\hfill
\includegraphics[width=0.48\linewidth]{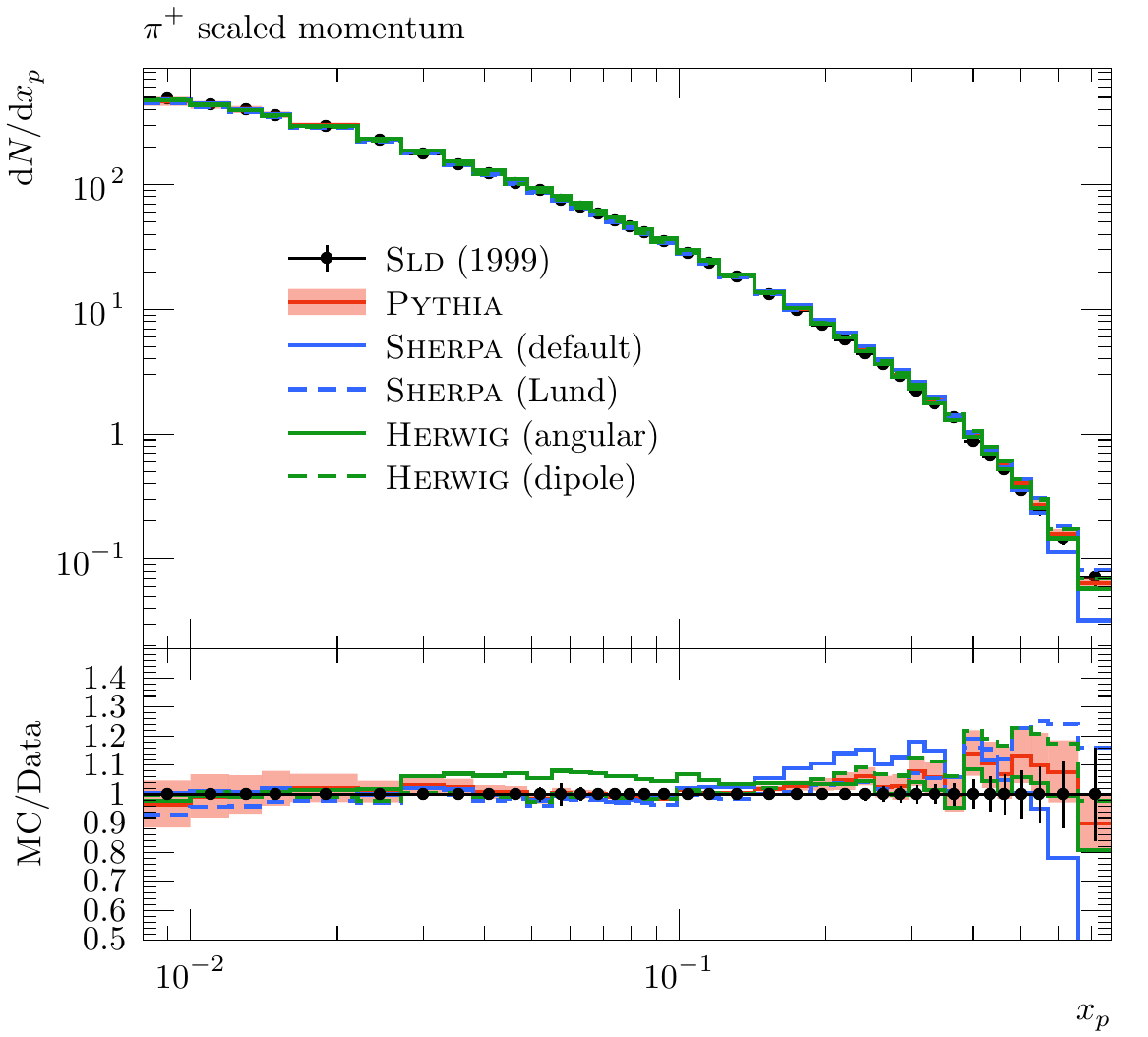}
\caption{Comparison between MC event generators and LEP and SLD measurements for the log of scaled momentum (top left pane),
$\gamma$ spectrum (top right pane), $\pi^0$ spectrum (bottom left pane) and $\pi^\pm$ spectrum (bottom right pane). 
Data is taken from \cite{Barate:1996fi, Ackerstaff:1998ap, Abe:1998zs}}
\label{comparison-1}
\end{figure}

In addition to the particle spectra, it is important to ensure that a nonperturbative tuning  does not produce ``too large'' corrections to infrared and collinear safe observables. In the following, we focus on the Thrust and $C$-parameter event shapes as IRC safe controls, which the baseline Monash 2013 tune is known to describe reasonably well~\cite{Skands:2014pea}. (The full set of observables is described in appendix \ref{app:observables}.)
Note that we include the full range of these observables, including also the back-to-back region near $\tau = 1-T \to 0$ and $C\to 0$, where the nonperturbative corrections are not power suppressed; this provides a complementary sensitivity to the fragmentation function parameters. Specifically, whereas the particle  spectra are only sensitive to the total magnitude of the momentum of the produced hadrons, the nonperturbative corrections to the event shapes are mostly sensitive to the transverse components (the property of IRC safety implies that the correction vanishes for a purely longitudinal breakup), hence they provide important additional sensitivity to the \texttt{StringPT:sigma} parameter in particular. 

\begin{figure}[!h]
\includegraphics[width=0.48\linewidth]{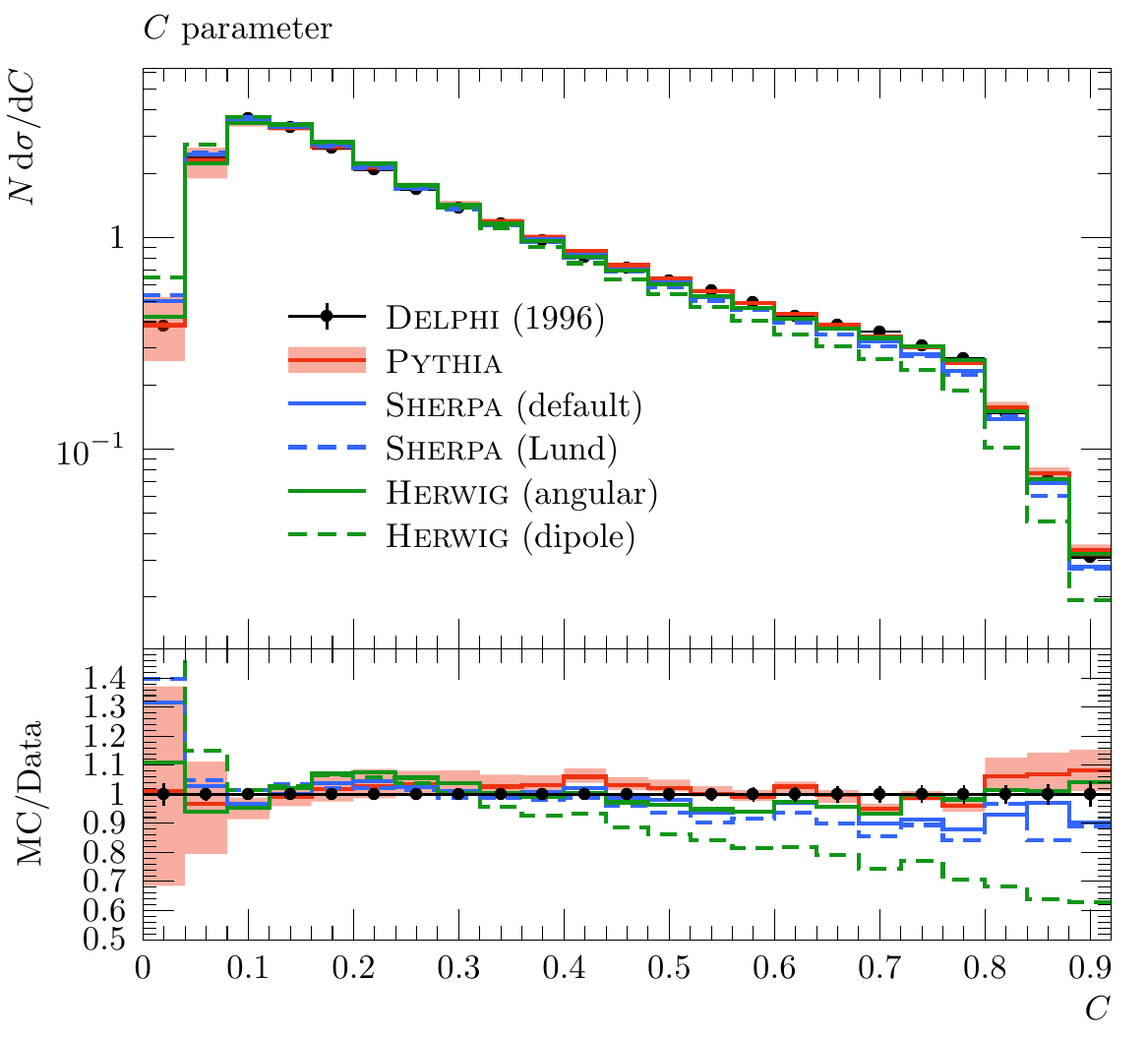}
\hfill
\includegraphics[width=0.48\linewidth]{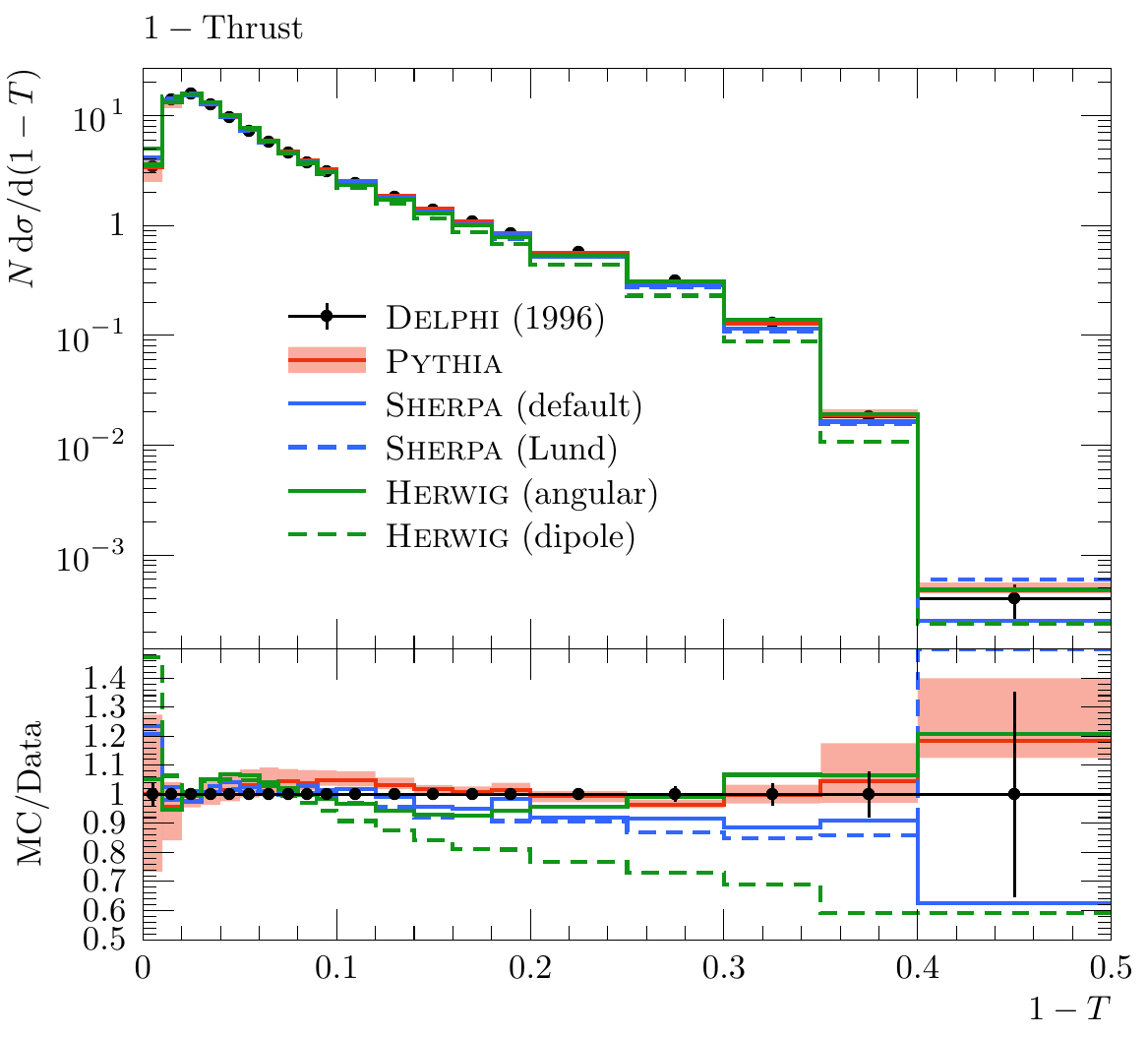}\\
\includegraphics[width=0.48\linewidth]{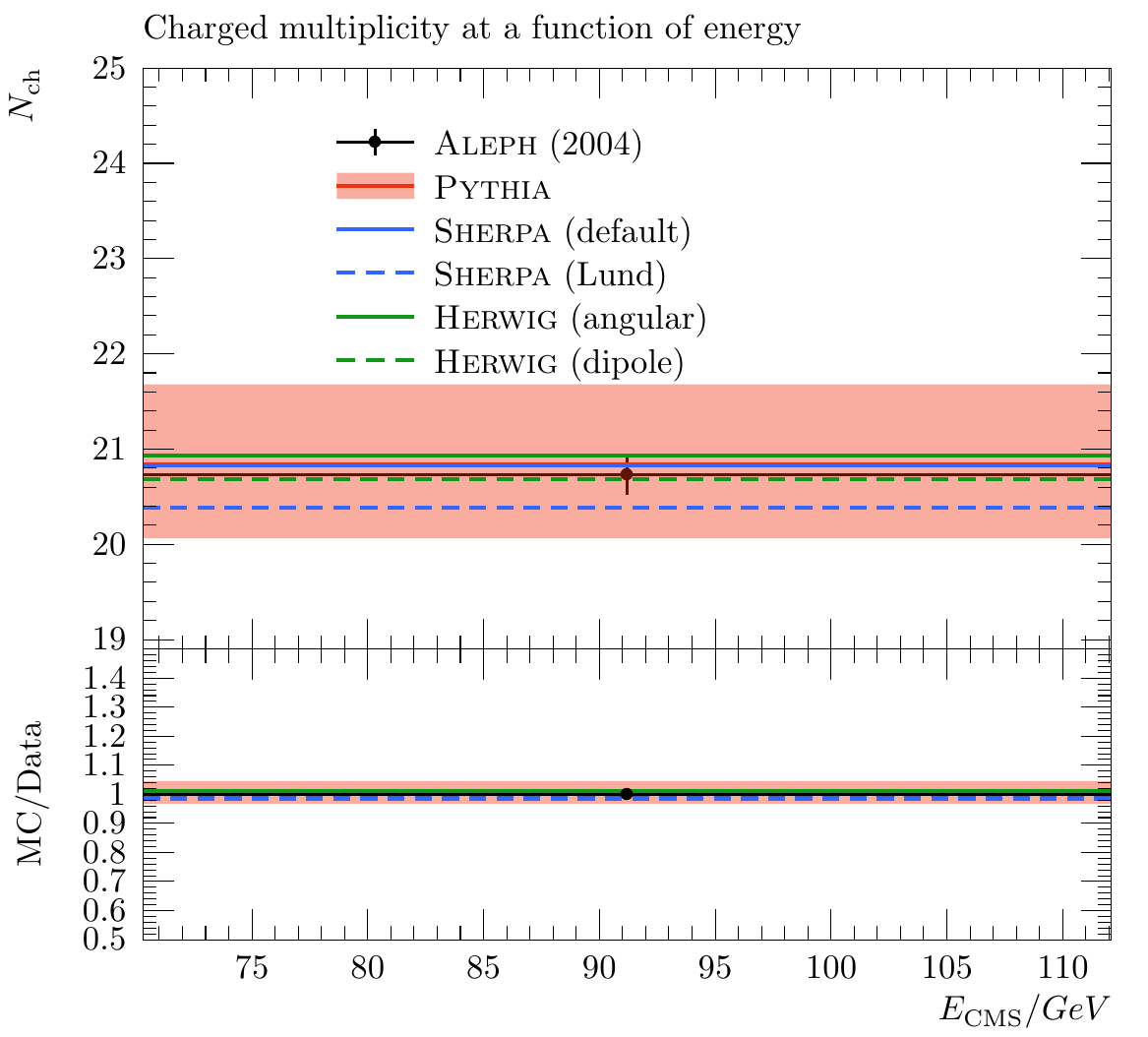}
\hfill
\includegraphics[width=0.48\linewidth]{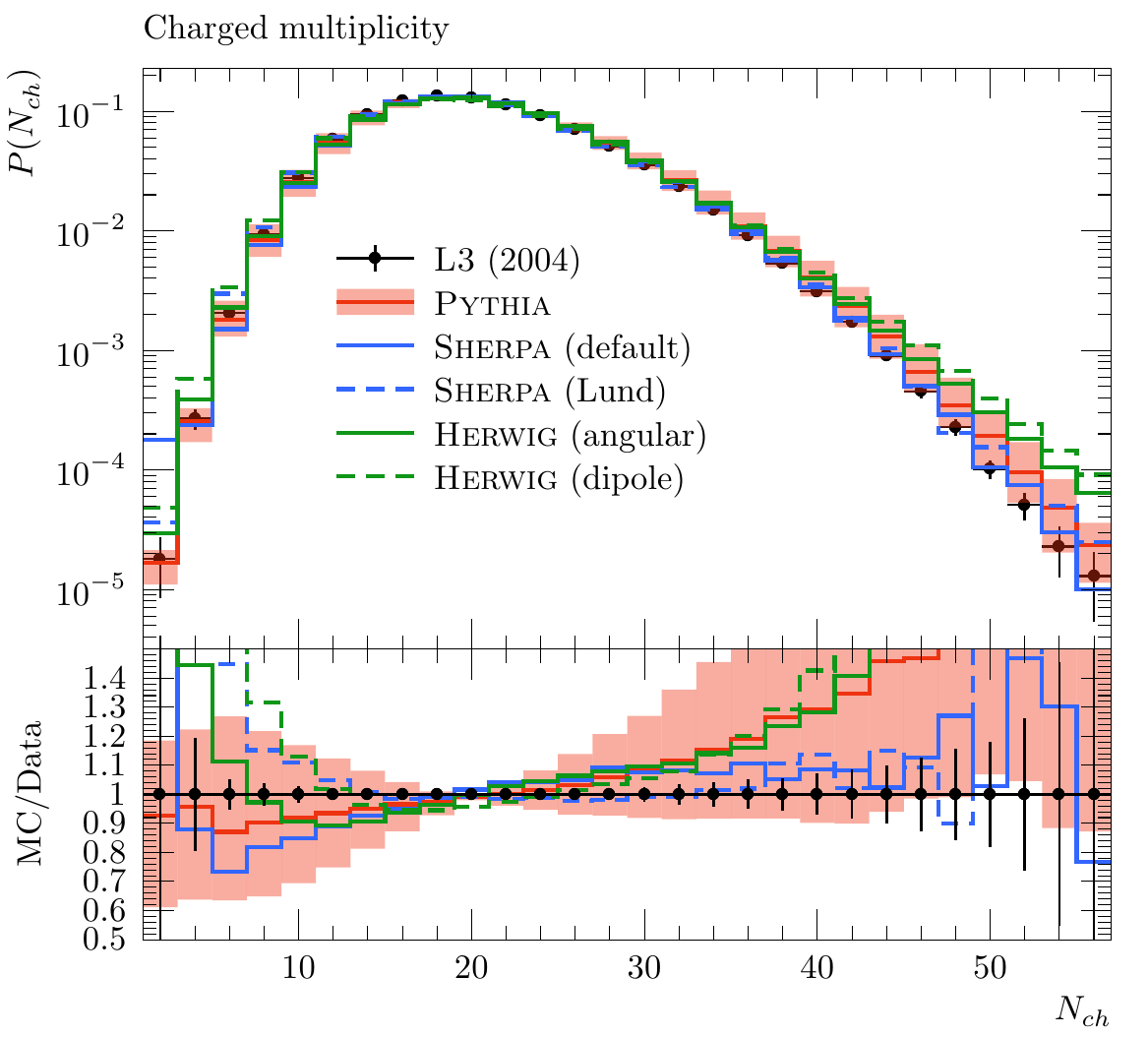}
\caption{Same as in Fig. \ref{comparison-1} but for $C$-parameter (top left pane), $1-T$ (top right pane), mean of charged multiplicity (bottom left pane) and Total charged multiplicity (bottom right pane). Data is taken from \cite{Abreu:1996na, 
Heister:2003aj, Achard:2004sv}}
\label{comparison-2}
\end{figure}

For completeness and as a cross check on the validity of the often used method of estimating QCD uncertainties by comparing the predictions of different MC generators, we compare several different multi-purpose MC event generators to measurements of the most important observables that we included in the tunes in Figs.~\ref{comparison-1} and \ref{comparison-2}. Three event generators are considered in these comparisons; \textsc{Herwig} 7.1.3 \cite{Bellm:2015jjp} using both the angular-ordered \cite{Gieseke:2003rz} and dipole \cite{Platzer:2009jq, Platzer:2011bc} shower algorithms and a cluster based hadronisation model~\cite{Webber:1983if}, \textsc{Pythia} 8.2.35 with the default
model of hadronisation (and using our central tune parameters) \cite{Sjostrand:2014zea} and \textsc{Sherpa} 2.2.5 \cite{Gleisberg:2008ta} with the CSS parton shower \cite{Schumann:2007mg} using both the Ahadic~\cite{Winter:2003tt} (based on the cluster model) and the \textsc{Pythia} 6.4 Lund hadronisation~\cite{Sjostrand:2006za} models. The curve corresponding to \textsc{Pythia} is shown with an  uncertainty band (red) obtained using the results of our new tune, based on the recent \textsc{Monash} tune but refitting the three main hadronisation parameters (see below). We can see from 
Fig. \ref{comparison-1} that the multi-purpose event generators agree pretty well
except in a few regions; 
\begin{itemize}
\item In the tails towards hard high-energy fragmentation $\xi=\log(1/x_\gamma) < 1$ (corresponding to $x_\gamma \simeq 0.35$ in the bottom-row plots), where a substantial fraction of the energy of the jet is carried by a single particle.
\item For very soft photons ($\log(1/x_p) > 4.5$) in the top right-hand plot, the \textsc{Sherpa} curve (corresponding to the 
Cluster hadronization model) is above all the other predictions (differences are within $5\%$-$10\%$)
\item In the momentum of charged pions, \textsc{Herwig} (with the angular shower algorithm) disagrees with the other generators by less than $10\%$ in 
$0.03 < x_p < 0.1$.
\end{itemize}

For the event shapes shown in Fig. \ref{comparison-2}, the \textsc{Pythia} prediction agrees fairly well with the experimental measurements while there are some tensions with some of the other multi-purpose event generators; e.g. the \textsc{Herwig} dipole shower in the 3-jet regions at large values of the $C$ and $T$ parameters. We conclude that the relative differences between multi-purpose event generators do not faithfully map out the allowed range of uncertainty allowed by data (at least not in the whole spectrum). This finding 
agrees with the results shown in \cite{Cembranos:2013cfa} where they find that differences between event generators are more important 
in the edges and not in the peak region of the photon spectrum. Furthermore, in the context of dark matter searches through $\gamma$-rays, the high $x_\gamma$ region of the photon spectrum usually has low statistics and therefore, relative differences between MC event generators, in that region, won't have a significant impact on the interpretation of the results.

\section{Tuning}
\label{sec:tune}
\textsc{Pythia8} version 8.235 is used throughout this study.
The most recent Monash~\cite{Skands:2014pea} tune is used as baseline for the  parameter optimisation (``tuning'').
The tuning is performed using \textsc{Professor} v2.2~\cite{Buckley:2009bj} for the fit to the data,
and \textsc{Rivet} v2.5.4~\cite{Buckley:2010ar} for the implementation of the measurements.
The method implemented in \textsc{Professor} permits the simultaneous optimisation of several parameters
by using an analytic approximation for the dependence of the physical observables on the model parameters,
an idea first introduced in Ref.~\cite{Abreu:1996na}.
Polynomials of fourth-order are used to parametrise the response of the observables to the generator parameter.
The coefficients in the polynomials are obtained by fitting MC predictions generated at a set of randomly selected parameter points, called anchor points.
The optimal values of the model parameters are then obtained with a standard $\chi^2$ minimisation
of the analytic approximation to the corresponding data using \textsc{Minuit}~\cite{James:1975dr}. \\

These are the $a$ and $b$ parameters of the Lund fragmentation
function ($a$ and $<z_\rho>$ in the new parametrization), which govern
the longitudinal momentum fractions of produced primary hadrons
relative to the jet direction, and the $\sigma$ parameter which
governs the transverse components (see e.g. \cite{Skands:2012ts}). The default values of 
the parameters and their allowed range in \textsc{Pythia8} are shown in Table \ref{tab:ranges}. \\

To protect against over-fitting effects\footnote{Overfitting is a situation where a theoretical model fits perfectly a given data. However, in this case, there is a possibility that other data cannot be fit well by the same model and, furthermore, predictions for other measurements are not reliable.}
 and as a baseline sanity limit for the achievable accuracy 
in both the perturbative and nonperturbative modeling regimes, we introduce an additional $5\%$ uncertainty 
on each bin and for each observable. This also substantially reduces
the value of the goodness-of-fit measure so that the resulting
$\chi^2/\textrm{ndf}$ is consistent with unity (see Table \ref{tab:T2tune}). The absence of such an additional uncertainty in the tune 
leads to an overconstrained fit with a large central $\chi^2$ value
and artificially small parameter variations which cannot be interpreted as conservative 
estimate of the uncertainty. The MC statistical uncertainties are treated as uncorrelated and included in the definition 
of the $\chi^2$ function, which thus becomes:
\begin{equation}
 \frac{\chi^2}{N_\textrm{DoF}} = \frac{1}{\sum_{\mathcal{O}} \omega_\mathcal{O} |b \in \mathcal{O}|}\frac{\sum_{\mathcal{O}}
 \omega_\mathcal{O} \sum_{b\in \mathcal{O}} (f_{(b)}(\textbf{p}) - \mathcal{R}_b)^2}
 {(\Delta_b^2+ (0.05 f_{(b)}(\textbf{p}))^2)}~.
\label{Gof-Ndf}
 \end{equation}
 
Here $\omega_\mathcal{O}$ represents the weight per observable and per bin, $f_{(b)}(\textbf{p})$
is the interpolation function per bin $b$, $\mathcal{R}_b$ is the experimental value of the observable
$\mathcal{O}$ and $\Delta_b$ is the experimental error per bin. A fourth-order polynomial was used as 
interpolating function $f_{(b)}$. 
We have checked the robustness of the interpolations by comparing the response functions to
the real generated runs at the minimum and found excellent agreement. To get a good tune, we have to use a 
large number of MC runs. For this study, 120 random combinations of 1000 independent runs, with 2M events for each, have been used.

\begin{table}[t!]
  \begin{center}
    \begin{tabular}{llcl}
      \hline
      parameter & \textsc{Pythia8} setting & Variation range & \textsc{Monash}\\
      \hline
      $\sigma_{\perp}$~[GeV] & \verb|StringPT:Sigma|   & 0.0 -- 1.0 & 0.335\\
      $a$              & \verb|StringZ:aLund|    & 0.0 -- 2.0 & 0.68 \\
      $b$              & \verb|StringZ:bLund|    & 0.2 -- 2.0 & 0.98 \\
      $\left<z_\rho\right>$       & \verb|StringZ:avgZLund| & 0.3 -- 0.7 & (0.55) \\
      \hline
    \end{tabular}
  \end{center}
  \caption{\label{tab:ranges} Parameter ranges used for the \textsc{Pythia} 8 tuning,
    and their corresponding value in the Monash tune. The parenthesis around the Monash value of the $\left<z_\rho\right>$ parameter 
    indicates that this is a derived quantity, not an independent parameter.}
\end{table}

As a first step in the  tuning to the $e^+e^-$ measurements,
a study of the sensitivity of the various observables to the MC parameters is performed.
The results of the sensitivity study is used to guide the selection of the observables to use for tuning. The sensitivity of each observable bin to a set of parameters $p_i$,
is estimated from the interpolated response of the observables to the parameters,
with the following formula:
\begin{eqnarray}
 \mathcal{S}_i = \frac{\partial \mathrm{MC}(p)}{|\mathrm{MC}(p_0)|+\epsilon w_{\rm MC}} 
 \frac{|p_{0,i}|+\epsilon w_{p_i}}{\partial p_i},
 \label{sens-eq}
\end{eqnarray}

where $p_0$ is the centre of the parameters range, MC$(p_0)$ is the interpolated
MC prediction at $p_0$ and the $\epsilon$ terms,
set to 1\% of the parameter range, are introduced to avoid the ill defined case MC$(p_0)=0$, $\partial p_i = 0$.
$w_{p_i}$ corresponds to 80\% of the original sampling range and is used to construct $w_{\rm MC}$. \\

\begin{figure}[!h]
  \centering
  \includegraphics[width=0.44\textwidth]{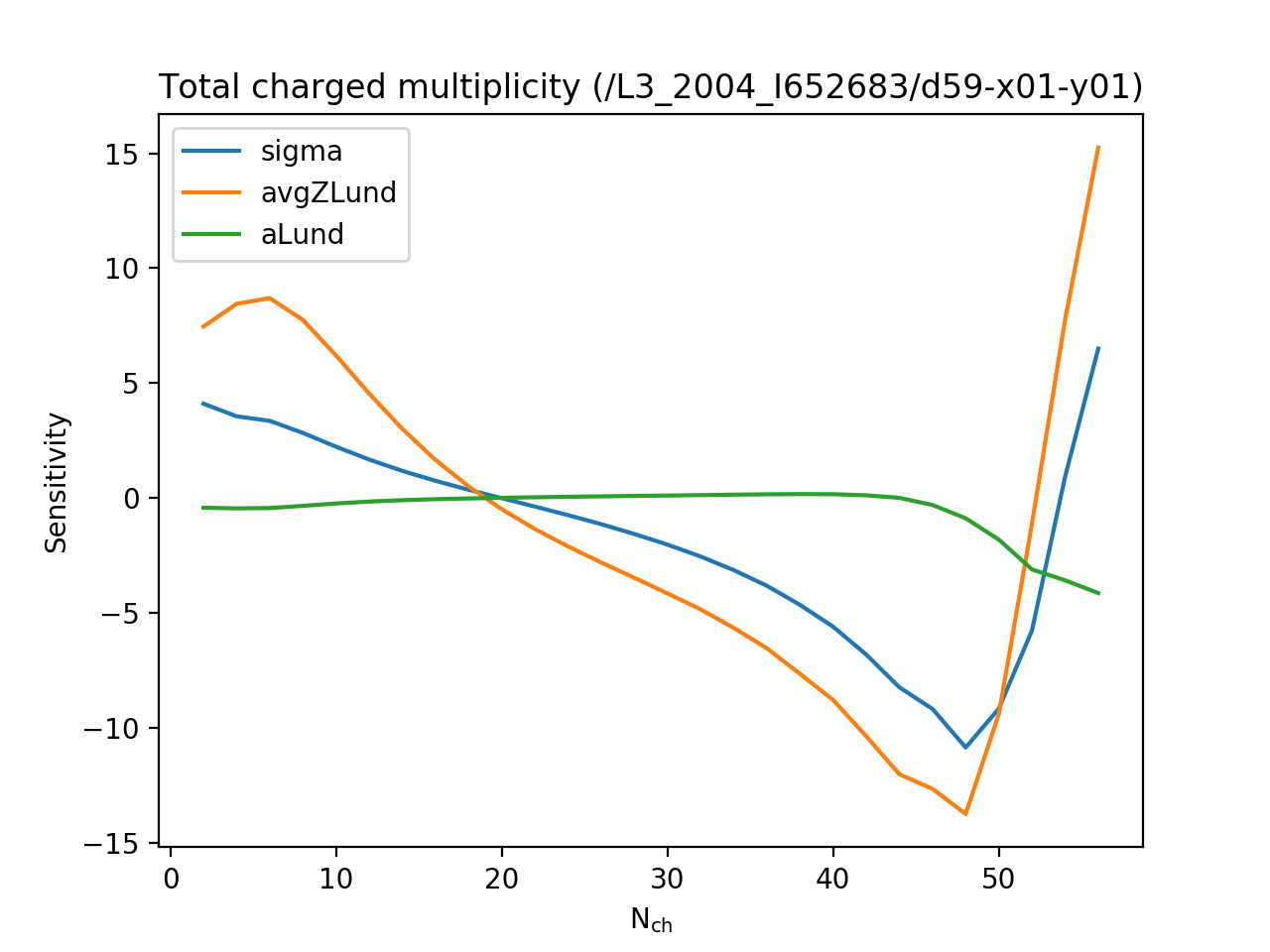}
  \includegraphics[width=0.44\textwidth]{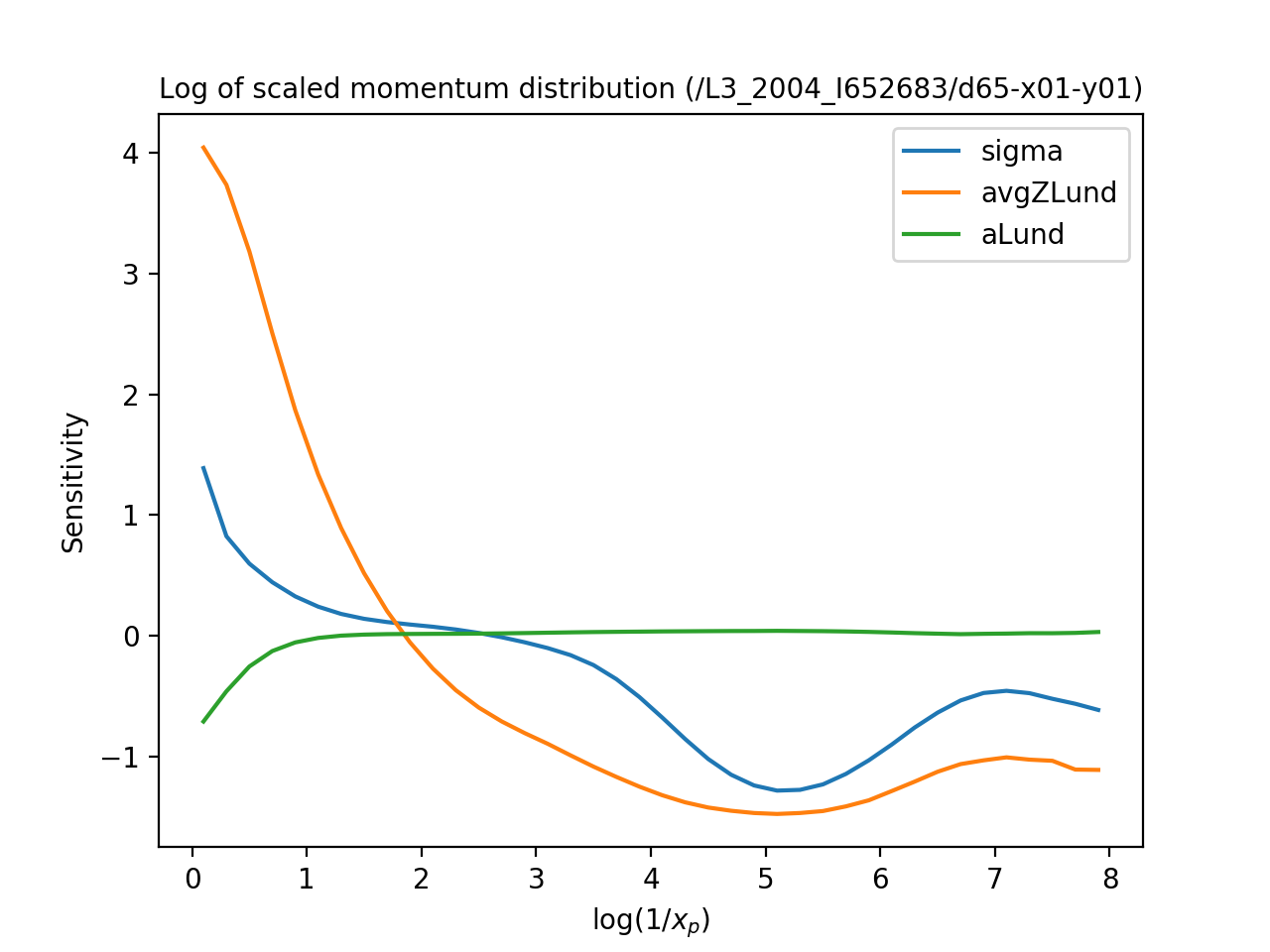}\\
  \includegraphics[width=0.44\textwidth]{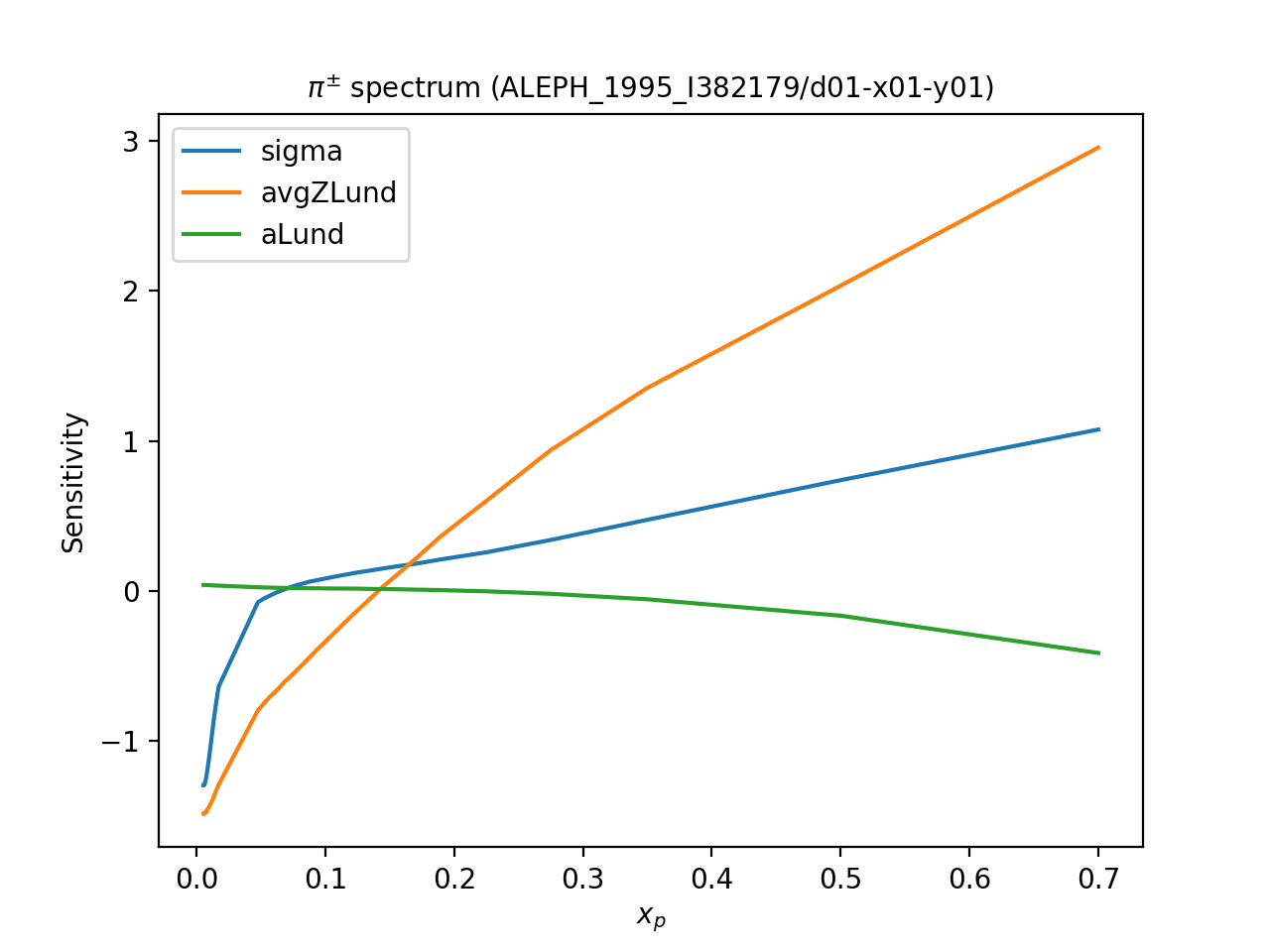}
  \includegraphics[width=0.44\textwidth]{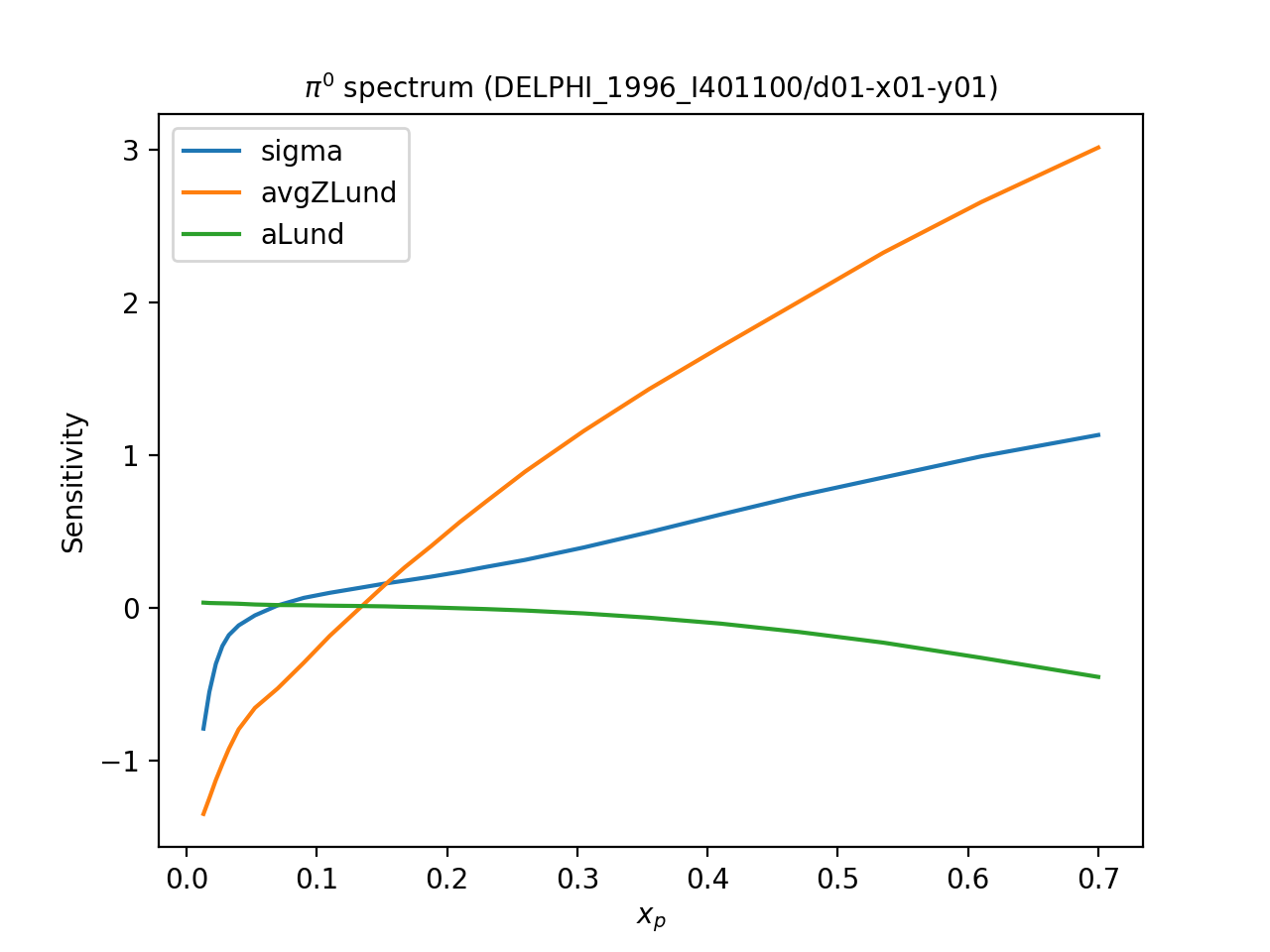}\\
  \caption{\label{fig:sensitivity1} Sensitivity of the different observables to the \textsc{Pythia} 8 
  fragmentation parameters
    $\sigma_{\perp}$ (blue), $\left<z_\rho\right>$ (orange),  $a$ (green).
    The sensitivities are shown as functions of the charged particle multiplicity $N_{\rm ch}$ (a),
    the total scaled momentum $x_p$ (b), the $\pi^{\pm}$ scaled momentum (c), and
    the $\pi^{0}$ scaled momentum (d). }
\end{figure}

\begin{figure}[!t]
\centering
  \includegraphics[width=0.44\textwidth]{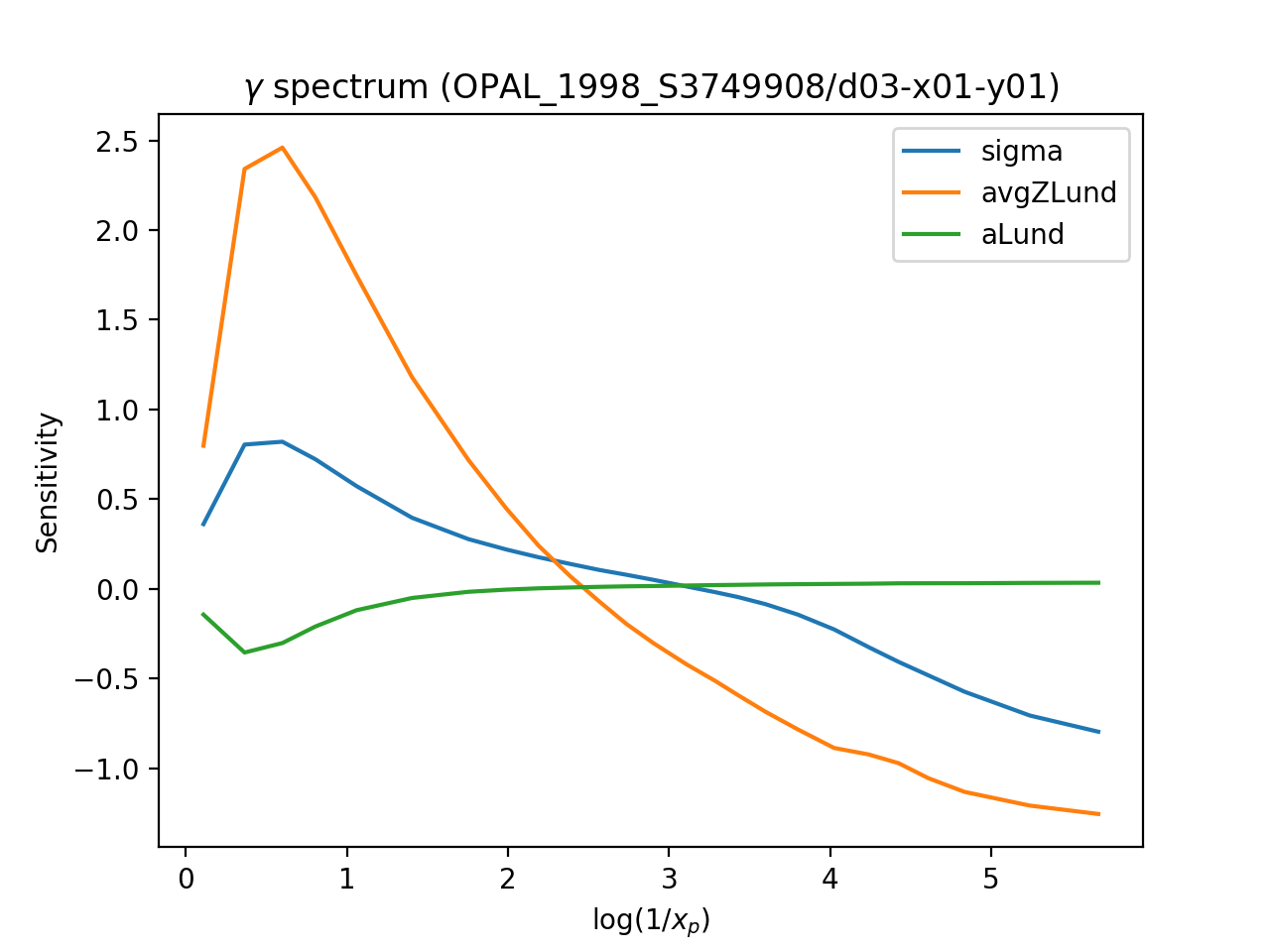}
  \includegraphics[width=0.44\textwidth]{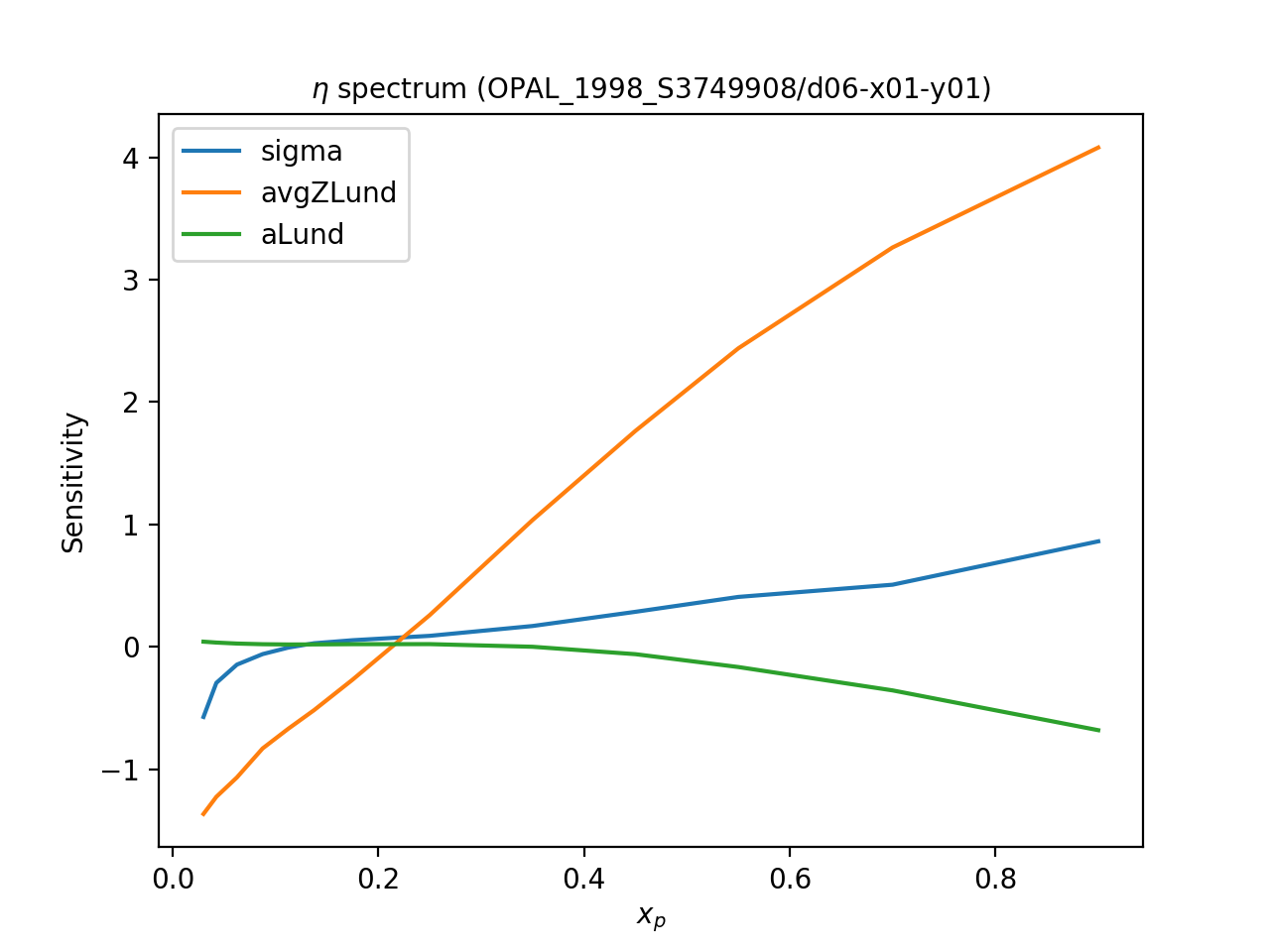}\\
  \includegraphics[width=0.44\textwidth]{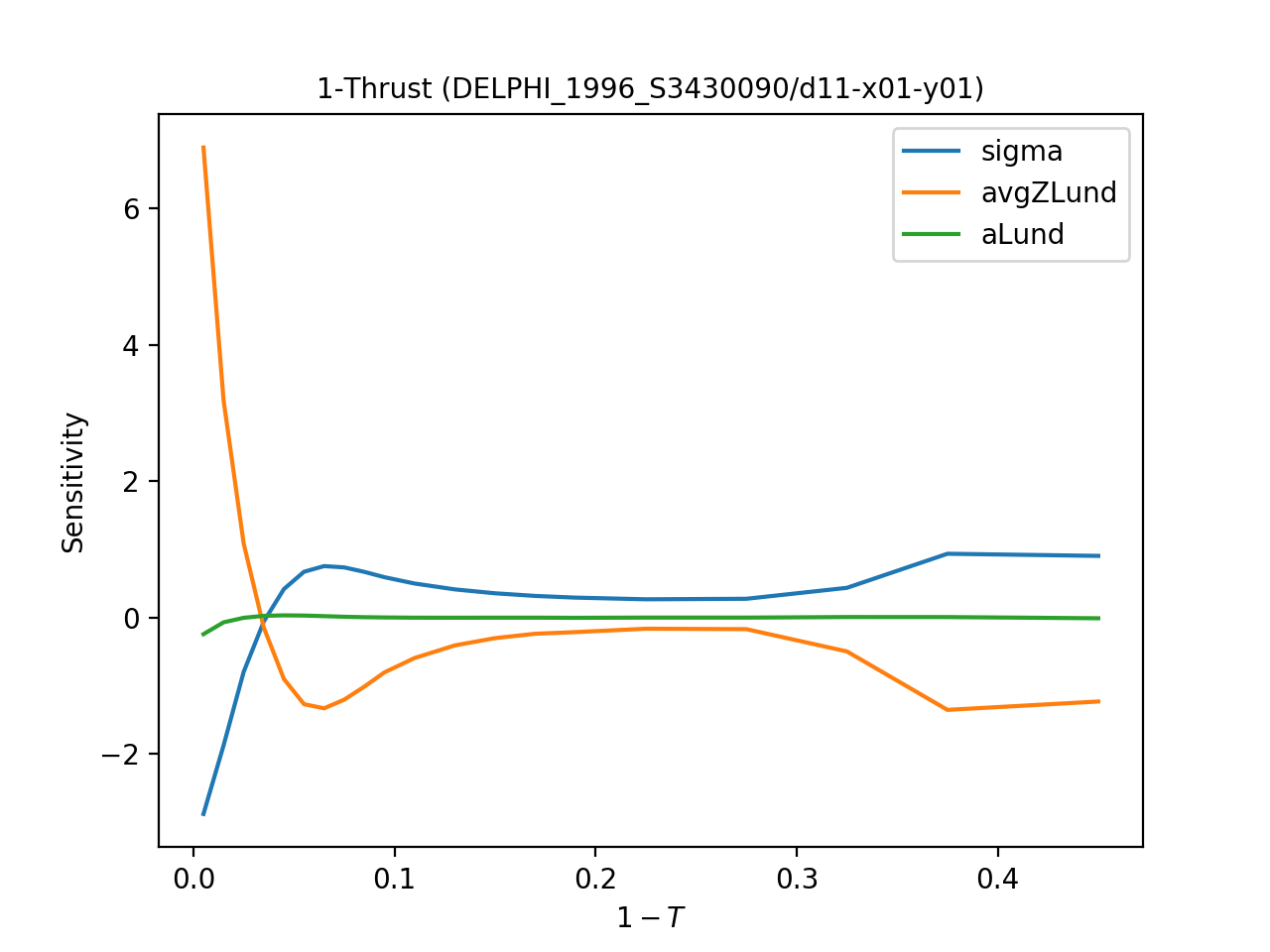}
  \includegraphics[width=0.44\textwidth]{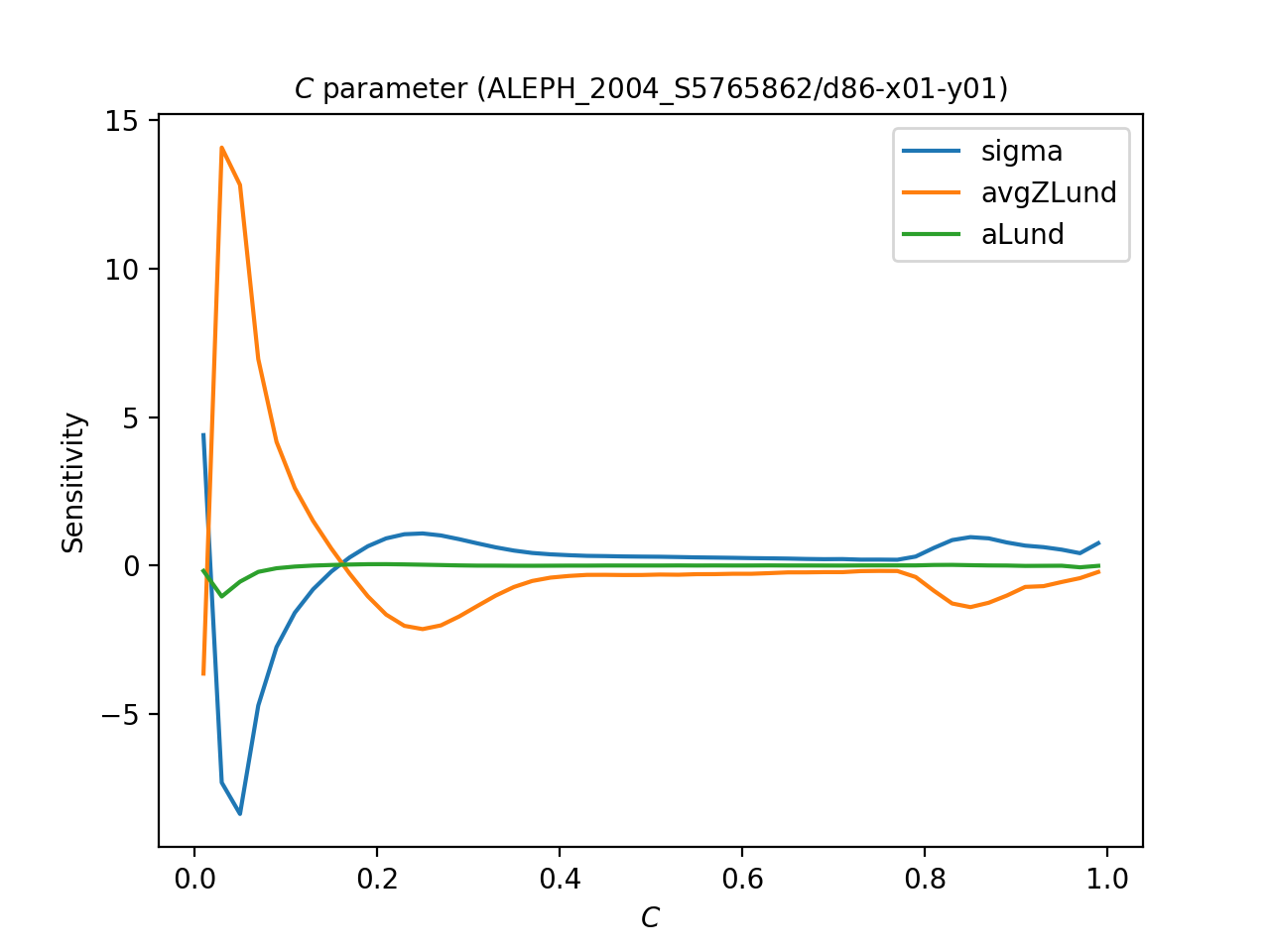}\\
\caption{\label{fig:sensitivity2} Same as in Fig. \ref{fig:sensitivity1} but for the $\gamma$ scaled momentum (a),
    the $\eta$ scaled momentum (b), the Thrust distribution (c)
    and the $C$-parameter distribution (d).}
\end{figure}

The sensitivity of different observables 
to the Lund fragmentation function parameters' is shown 
in Figs.~\ref{fig:sensitivity1} and \ref{fig:sensitivity2}. The observables are selected from different measurements 
and corresponding to the charged multiplicity, log of scaled momentum, spectra
of charged pions, neutral pions, photons and $\eta$-mesons, the Thrust and the $C$-parameter. As we can see
from Figs. \ref{fig:sensitivity1}-\ref{fig:sensitivity2}, all the observables are sensitive to the variations of the
parameters notably to \texttt{avgZLund}. \\

We tune the parameters of the string fragmentation function
in \textsc{Pythia} 8 to a set of sensitive \textsc{Lep} and \textsc{Slc} measurements at the $Z$-boson 
peak produced by 
\textsc{Aleph} \cite{Decamp:1991uz, Buskulic:1992hn, Buskulic:1995au, Barate:1996fi, Barate:1996uh,  Heister:2001kp, Heister:2003aj},
\textsc{Delphi} \cite{Abreu:1990cc, Adam:1995rf, Abreu:1996na}, 
\textsc{L3} \cite{Adeva:1991it, Adriani:1992hd, Acciarri:1994gza, Achard:2004sv}, 
\textsc{Opal} \cite{Akers:1994ez, Ackerstaff:1998ap, Ackerstaff:1998hz, Abbiendi:2004qz} and
\textsc{Sld} \cite{Abe:1996zi, Abe:1998zs, Abe:2003iy}. Several
qualitatively different tunes are made, by including or excluding
different data sets (and/or by modifying their weights). Our first
tune, labeled T1, only includes the spectra of $\pi^{0,\pm}, \gamma$
and $\eta$ particles. A second tune, labeled T2, also includes
event-shapes and jet-rate observables, and hence represents a more
global fit. The weights assigned to each measurement in these two
tunes are colled in Tables \ref{Tab1}-\ref{Tab6}. 
Further, to access to the compatibility between the different
experiments, we tune to each experiment individually including all
the aforementioned observables with unit weights. The resulting five
independent tunes are labeled by the names of the corresponding experiments.\\

At the technical level, we set up the event generation for an incoming
$e^+ e^-$ pair with QED initial-state effects switched off. (This
corresponds to the definition of the unfolded experimental
measurements used in the tuning.)  
We adopted the same definition of particle 
stability as it was used by LEP, i.e a given particle is stable if its mean 
lifetime satisfies $c\tau_0 > 100$ mm\footnote{Note: this criterion is
of course only applied during the tuning process and not when we later
simulate dark-matter fragmentation spectra, where all unstable
particles are decayed irrespective of lifetime.}. Finally, the strong coupling constant for Final State Radiation (FSR)
was set to be $\alpha_S(M_Z^2)=0.1365$ with a one-loop running as in the Monash tune. All the other parameters and
settings in \textsc{Pythia} are fixed to their default values.  The specific commands used in the tuning setup are shown in appendix \ref{sec:code}.

\section{Results}
\label{sec:results}
In this section, we present results of our tunes and related
uncertainties. We compare two different methods: eigentunes and manual
variations of parameters. We find that the eigentune method does not
provide acceptably conservative uncertainty estimates and therefore
advocate for a more elaborate method which we describe in section \ref{tune:uncertainties}.
\subsection{Tunes}
\begin{figure}[!t]
\centering
  \includegraphics[width=0.48\textwidth]{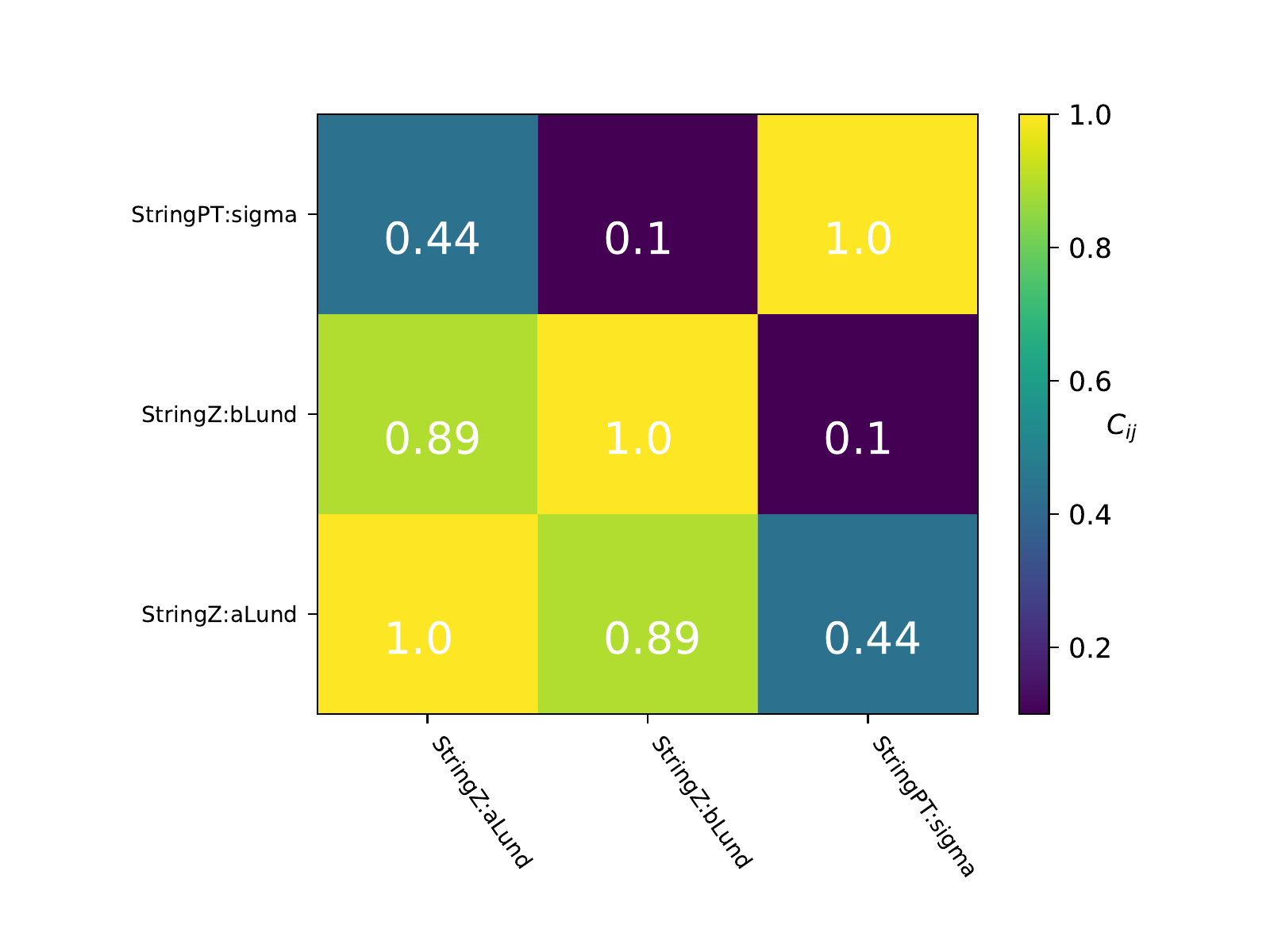} \hfill
  \includegraphics[width=0.48\textwidth]{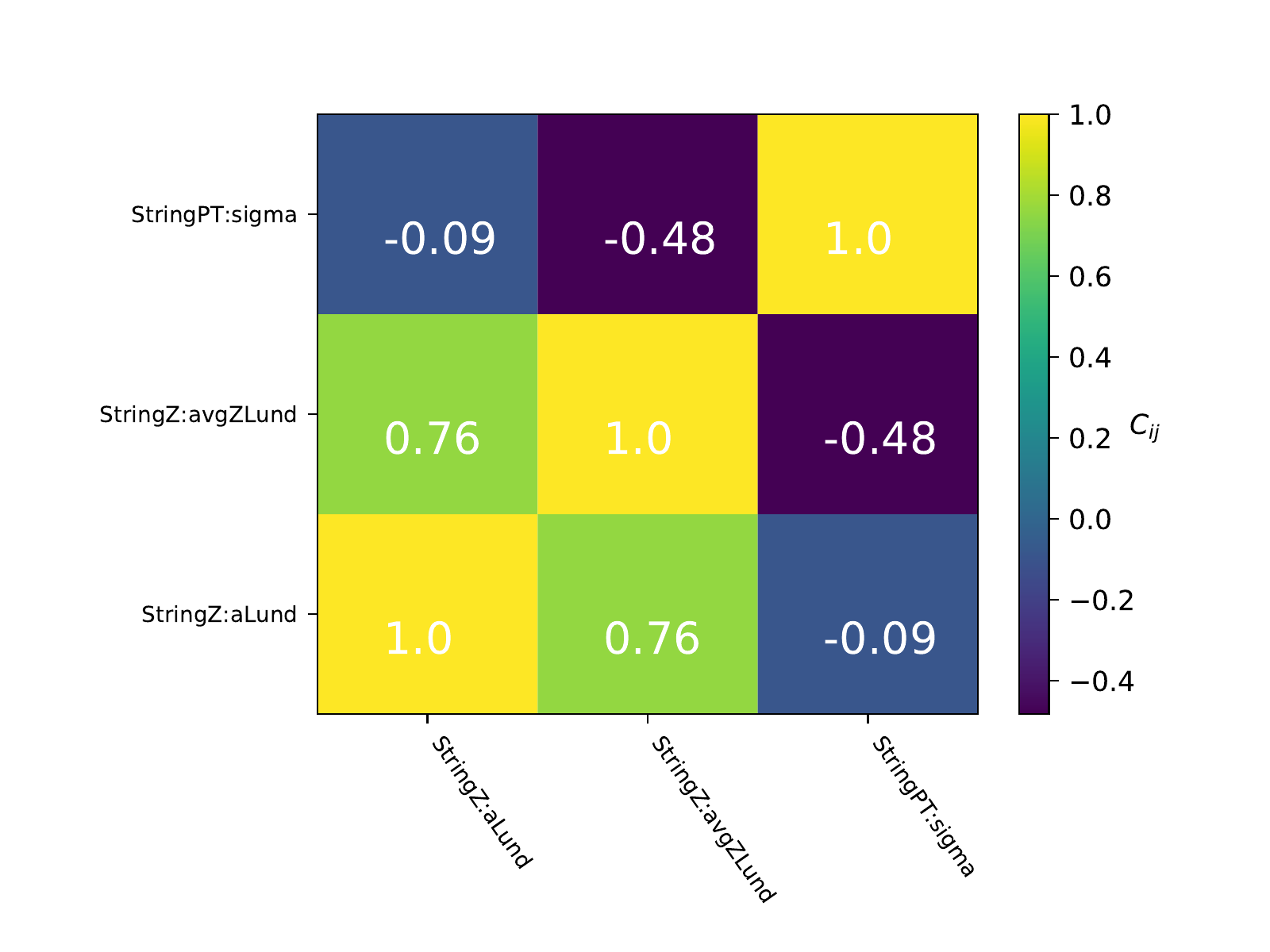}
  \caption{\label{fig:correlation} The parameters correlation ($C_{ij}$) as evaluated from the Hessian matrix at the  fit minimum.
    (a) shows result with  standard parametrization of the Lund string in terms of the $a$, $b$ parameters,
    (b) results with a new parametrization in terms of the $a$ and $<z_\rho>$ parameters.}
\end{figure}

We start by discussing the correlation among the parameters in the two parameterizations of the Lund fragmentation function. To illustrate this, we show in Fig. \ref{fig:correlation}, the correlation matrix $C_{ij}$ obtained 
from the T2 tune in the old (left) and the new (right) parameterizations of the Lund fragmentation function. Clearly 
\texttt{StringZ:aLund} and \texttt{StringZ:bLund} are highly correlated ($C_{12} = 89\%$). Using 
the new parametrization reduces the correlation by about $13\%$ where now the \texttt{StringZ:bLund}
is replaced by \texttt{StringZ:avgZLund}. Furthermore, we note that
the remaining correlation is almost entirely associated with the mean
multiplicities. If these are removed from the fit, the strong
correlation between \texttt{StringZ:aLund} and
\texttt{StringZ:avgZLund} is considerably reduced. Furthermore, in the old parametrization, \texttt{StringZ:aLund}
and \texttt{StringPT:sigma} have a correlation coefficient 
of $44\%$ which is reduced to $-9\%$ in the new parametrization of the Lund function. 

\begin{figure}[!t]
\centering
  \includegraphics[width=0.70\textwidth]{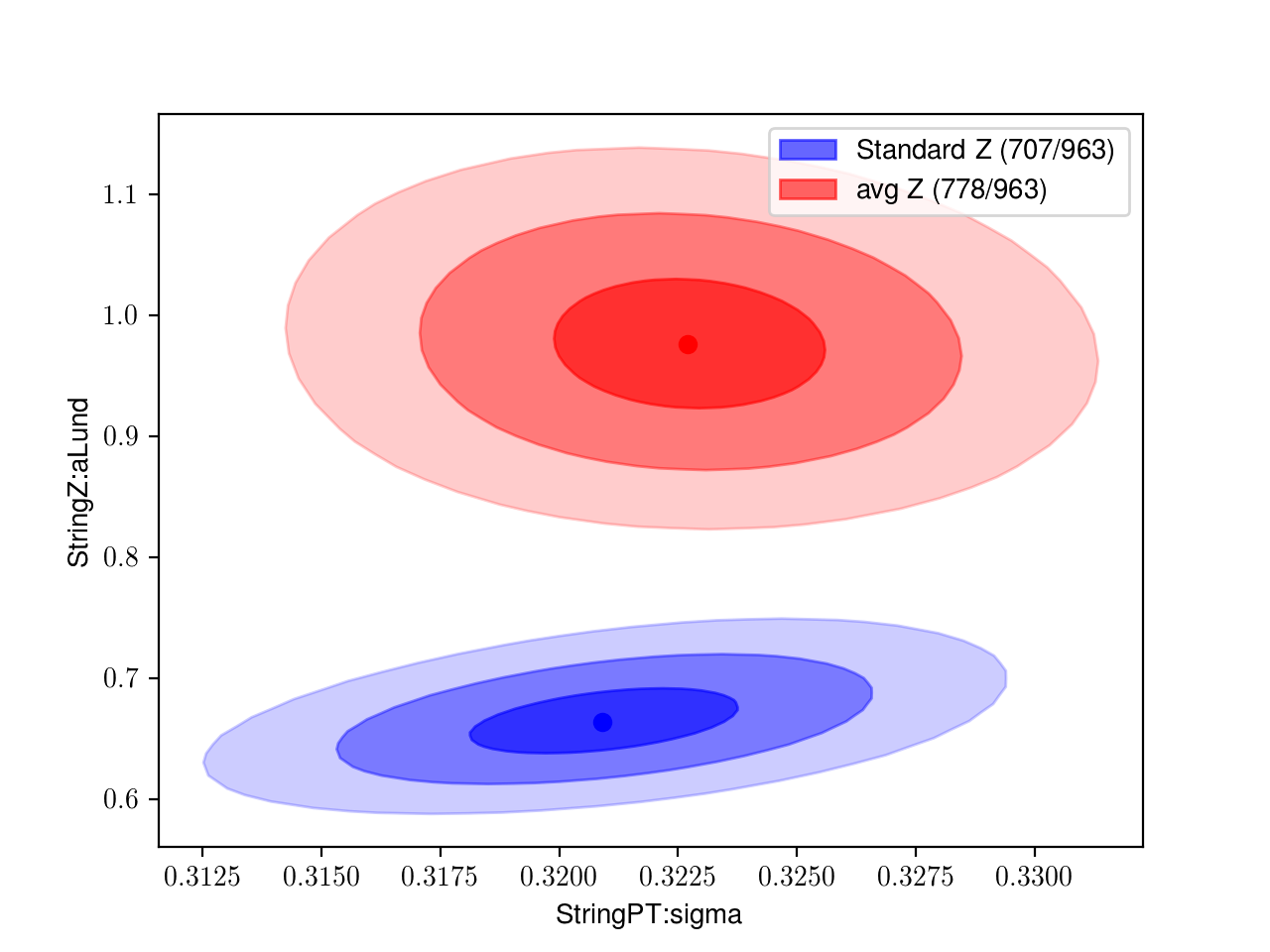}
  \caption{\label{fig:Zparametrization} Results of tunes using the standard parametrization 
  of the Lund string (blue) in terms of the $a$, $b$ parameters,
    and with a new parametrization (red) in terms of the $a$ and $\left<z_\rho\right>$ parameters.}
\end{figure}

\begin{figure}[!h]
 \centering
 \includegraphics[width=0.48\linewidth]{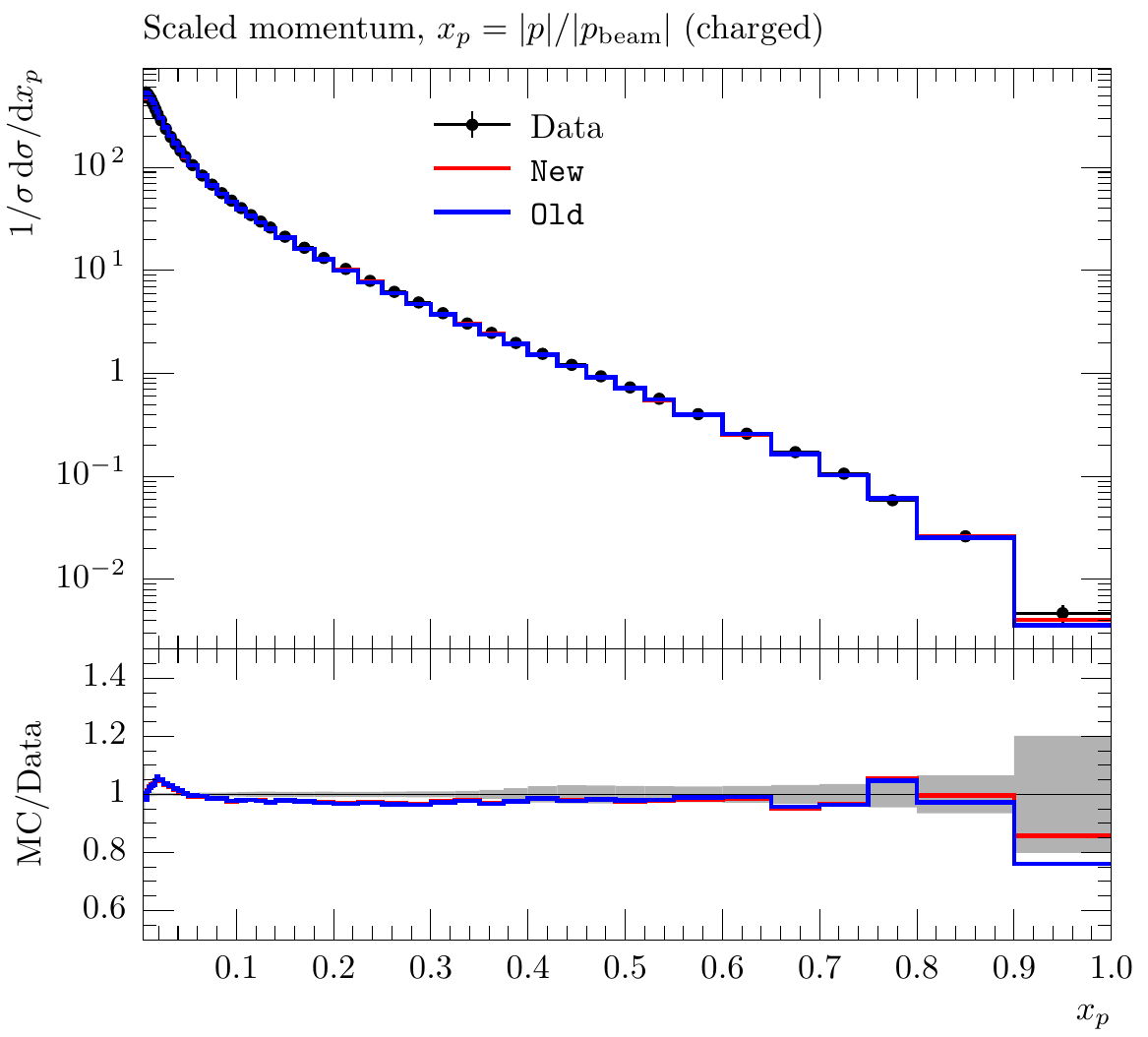}
 \hfill
 \includegraphics[width=0.48\linewidth]{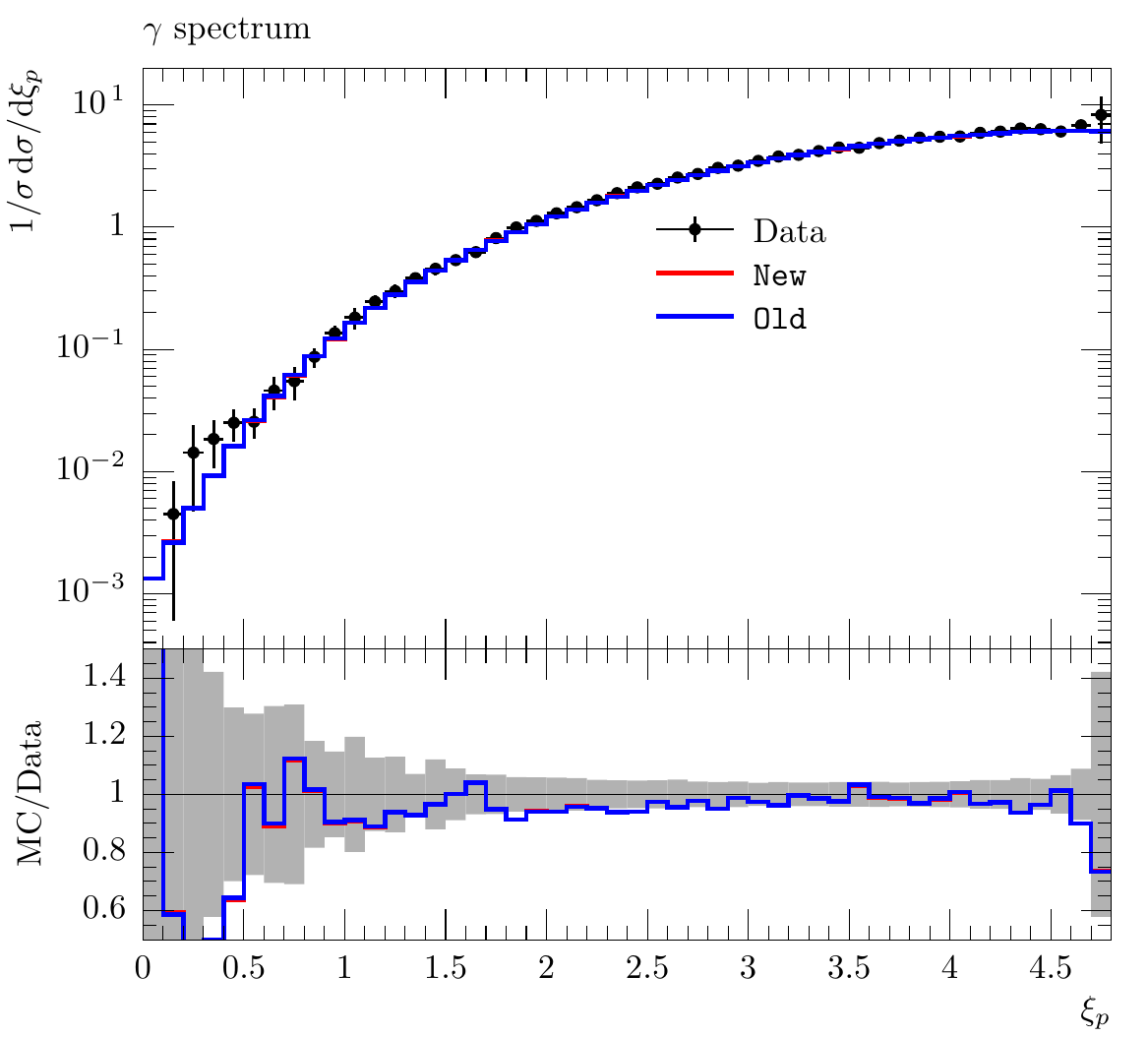}
 \caption{\label{fig:comparisonOldNew} Comparison between the results of the 
 tune in the old (\texttt{StandardZ}) and the new (\texttt{AvgZ}) parametrization of 
 the Lund fragmentation function for scaled momentum distribution 
 of charged particles (\emph{left}) and  
 photon scaled momentum (\emph{right}). Data is taken from
 \cite{Barate:1996fi}.}
\end{figure}

Since these are just two different parametrisations of the same
function, it is interesting to check whether one gets compatible
values for the tune parameters when fitting the two different forms to
the same data. For this purpose, we make a comparison between the two parameterizations showing, as an example, the T2 tune. 
We show the results of this comparison in the 
(\texttt{StringPT:sigma}, \texttt{StringZ:aLund}) plane in Fig. \ref{fig:Zparametrization}. 
While it looks obvious from Fig. \ref{fig:Zparametrization} that the two tunes give
inconsistent results even at the $3\sigma$ level, the two tunes are in
good agreement with each other on data as can be seen in
Fig. \ref{fig:comparisonOldNew} (the same conclusion applies for all
the other measurements which are not shown here). This can explained
via the correlation with the remaining parameter, $b$: while the 
two tunes gives very different central values for \texttt{StringZ:aLund} and \texttt{StringPT:sigma}, the obtained 
value for \texttt{StringZ:bLund} is also different\footnote{In the new parametrization of the string 
fragmentation function, \texttt{StringZ:bLund} can be obtained by solving 
numerically eqn. \ref{eq:zrho}.}. This is a clear reminder that
misleading conclusions can be reached if correlations are neglected. \\

As  pointed out in the previous section,
including the $5\%$ theory uncertainty 
affects significantly the quality of the tune.
This is can be seen  in Table~\ref{tab:T2tune},
where the results of the tune before and after including the 
$5\%$ uncertainty are compared. 
While the resulting parameters are  consistent in the two fits,
the goodness-of-fit per degree of freedom is improved by a factor of $~7$, bringing it close to unity for the second fit. 
The uncertainties on the parameter's determination  are also
affected, being significantly larger when the additional 5\%
uncertainty is added to the fit. We therefore consider that the added
$5\%$ uncertainty does provide a useful basic protection against
overfitting, and prefer the more conservative uncertainties this way,
being consistent with a $\Delta\chi^2 \pm 1$ around a central $\chi^2$
of order unity.\\

\begin{table}[!h]
  \begin{center}
    \begin{tabular}{lcc}
      \hline
      Parameter &  without $5\%$ & with $5\%$ \\
      \hline
      \verb|StringPT:Sigma| & $0.3151\substack{+0.0010\\ -0.00010}$ & $0.3227\substack{+0.0028\\ -0.0028}$\\
      \verb|StringZ:aLund|  & $1.028\substack{+0.031\\ -0.031}$ & $0.976\substack{+0.054\\-0.052}$\\
      \verb|StringZ:avgZLund| & $0.5534\substack{+0.0010\\ -0.0010}$ & $0.5496\substack{+0.0026\\ -0.0026}$ \\
      \hline
      $\chi^2$/ndf   & 5169/963 & 778/963 \\
      \hline
    \end{tabular}
  \end{center}
    \caption{\label{tab:T2tune} Results of tunes using the new
    parametrization 
    of the Lund fragmentation function in terms of the $a$ and 
    $\left<z_\rho\right>$ parameters. The second (third) column shows
    the result before (after) including a flat $5\%$ uncertainty to the theory prediction.}
\end{table}

\begin{table}[t!]
  \begin{center}
    \begin{tabular}{lcccc}
      \hline
      Tune     & \verb|StringZ:aLund| & \verb|StringZ:avgZLund| & \verb|StringPT:sigma| & $\chi^2$/ndf\\
      \hline
      \textsc{Aleph}  & $0.827\substack{+0.066\\ -0.062}$ & $0.5447\substack{+0.0044\\ -0.0044}$ & $0.3105\substack{+0.0045\\ -0.0045}$ & 284.7/382\\
      \textsc{Delphi} & $0.67\substack{+0.11\\ -0.09}$ & $0.5290\substack{+0.0062\\ -0.0063}$ & $0.3110\substack{+0.0062\\ -0.0061}$ & 82/113\\
      \textsc{L3}     & $1.186\substack{+0.093\\ -0.10}$ & $0.5708\substack{+0.0054\\ -0.0055}$ & $0.3303\substack{+0.0072\\ -0.0072}$ & 98/155 \\
      \textsc{Opal}   & $0.55\substack{+0.11\\ -0.095}$ & $0.511\substack{+0.011\\ -0.012}$ & $0.318\substack{+0.013\\ -0.013}$ & 82.4/184\\
      \textsc{Sld}    & $0.95\substack{+0.12\\ -0.11}$ & $0.5271\substack{+0.0097\\ -0.010}$ & $0.327\substack{+0.017\\ -0.017}$ & 34.4/116 \\
      \hline
      COMBINED & $0.976\substack{+0.054\\-0.052}$ & $0.5496\substack{+0.0026\\ -0.0026}$ & $0.3227\substack{+0.0028\\ -0.0028}$ & 778/963\\
      \hline
    \end{tabular}
  \end{center}
    \caption{\label{tab:experiments} Results of the tunes performed 
    separately to all the considered measurements from a given experiment. The COMBINED result corresponds to the 
    T2 tune given in Table \ref{tab:T2tune}. }
\end{table}

\begin{figure}\centering
  \includegraphics[width=0.48\textwidth]{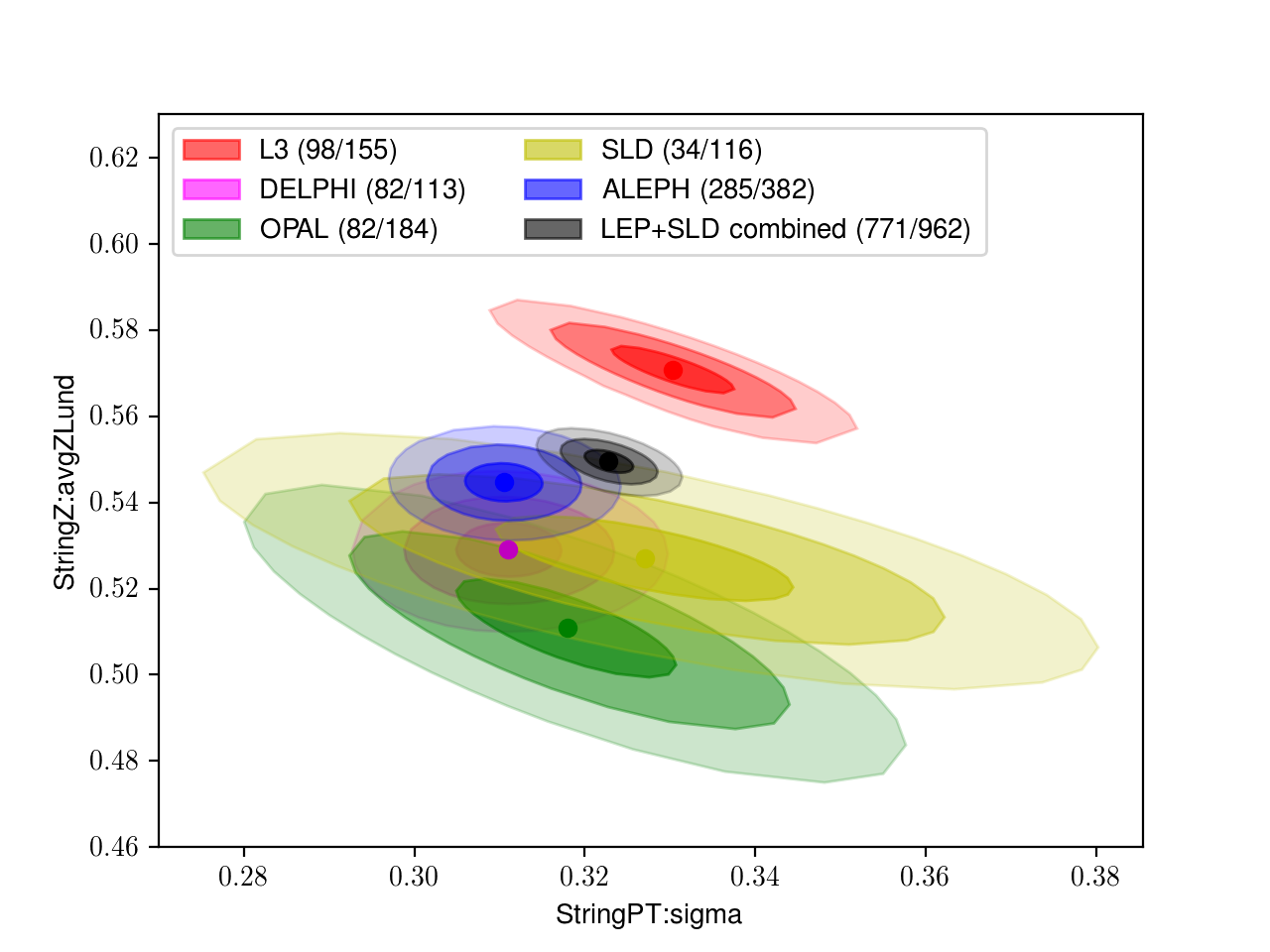}
  \includegraphics[width=0.48\textwidth]{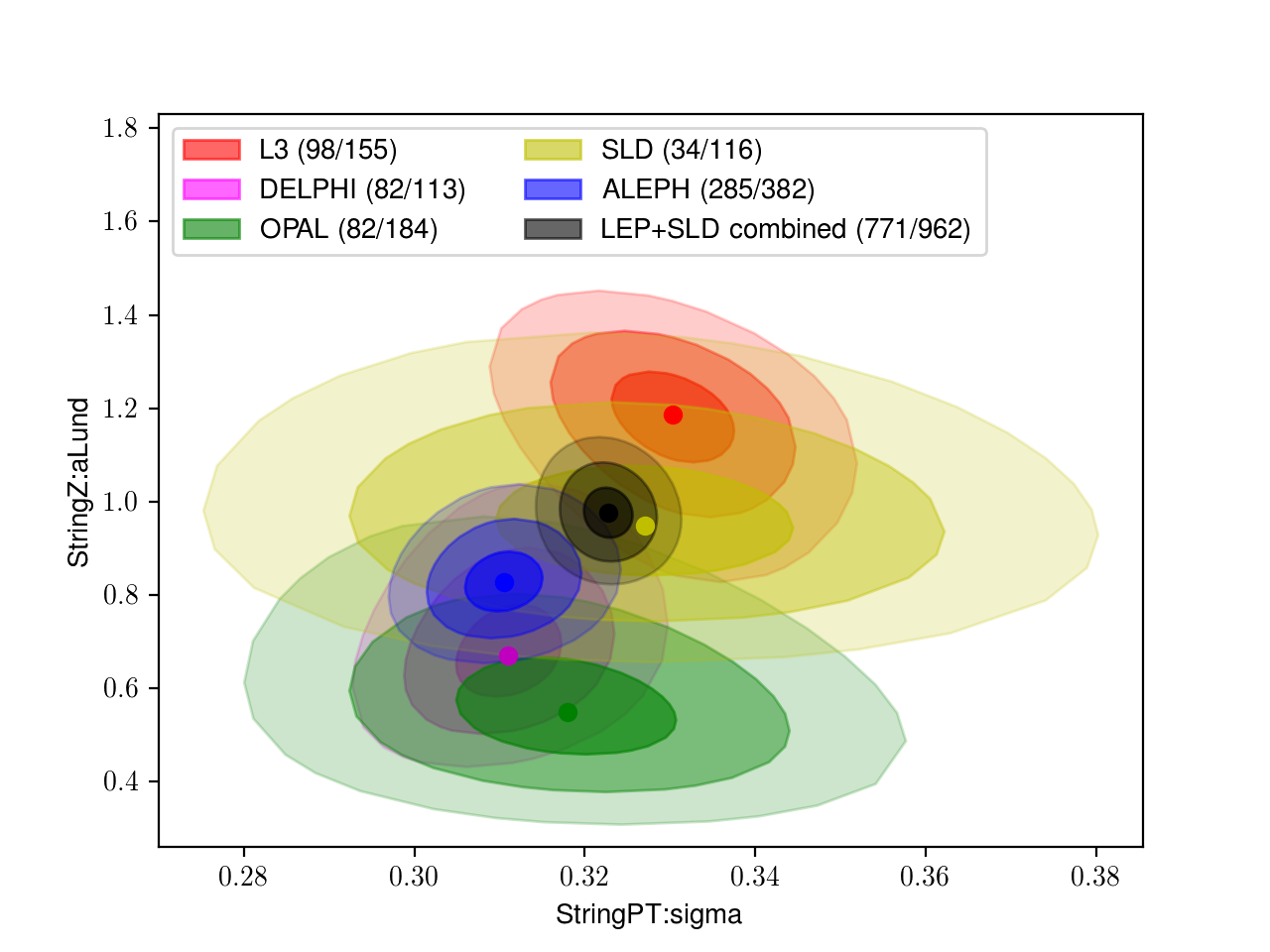}
  \caption{\label{fig:experiments} Results of tunes performed separately to all of the measurements 
  from a given experiment; \textsc{Aleph} (blue), \textsc{Delphi} (magenta), \textsc{L3} (red), \textsc{Opal} (green), \textsc{Sld} (yellow) and 
  COMBINED (gray). The contours corresponding to one, two and three sigma deviations are also shown.}
\end{figure}

We now turn to a study of possible tensions in the data measured  by the different experiments by making independent tunes 
including all of the sensitive measurements by each experiment. Five independent tunes are performed corresponding to the data measured by
\textsc{Aleph}, \textsc{Delphi}, \textsc{L3}, \textsc{Opal} and \textsc{Sld}. The results of these tunes are shown in figure \ref{fig:experiments}
and Table \ref{tab:experiments}. We can see that the tunes to \textsc{Aleph}, \textsc{Delphi}, \textsc{Opal} 
and \textsc{Sld} are in agreement regarding the obtained value of
\verb|StringZ:avgZLund| contrarily to \textsc{L3}. Again, due to the
correlations of $a$ with $b$ (or $<z_\rho$), we cannot conclude
anything from comparisons of each individual parameters and,
therefore, when compared on data, various tunes are expected to be in
good agreement with each other.

\begin{table}[t!]
  \begin{center}
    \begin{tabular}{lcccc}
      \hline
      Tune     & \verb|StringZ:aLund| & \verb|StringZ:avgZLund| & \verb|StringPT:sigma| & $\chi^2$/ndf\\
      \hline
      Charged multiplicity & $1.061\substack{+0.089\\-0.096}$ & $0.518\substack{+0.011\\-0.012}$ & $0.410\substack{+0.017\\-0.016}$ & 43.4/104\\
      Scaled momentum & $0.598\substack{+0.053\\ -0.049}$ & $0.5295\substack{+0.0070\\ -0.0072}$ & $0.324\substack{+0.012\\ -0.012}$ & 70.7/180\\
      $\gamma$ & $0.61\substack{+0.32\\-0.23}$ & $0.517\substack{+0.035\\ -0.039}$ & $0.344\substack{+0.067\\ -0.062}$ & 52.4/70\\
      $\pi^0$ & $1.22\substack{+0.18\\-0.16}$ & $0.566\substack{+0.014\\-0.014}$ & $0.340\substack{+0.020\\-0.020}$ & 31/117\\
      $\pi^{\pm}$ & $0.757\substack{+0.082\\ -0.073}$ & $0.5029\substack{0.0098\\ -0.0099}$ & $0.336\substack{+0.011\\ -0.011}$ & 72.5/205\\
      $T$ & $1.34\substack{+0.27\\ -0.20}$ & $0.498\substack{+0.018\\ -0.019}$ & $0.241\substack{+0.022\\ -0.023}$ & 124/194\\
      $C$-parameter & $1.65\substack{+0.35\\ -0.42}$ & $0.621\substack{+0.053\\ 0.038}$ & $0.390\substack{+0.067\\ -0.043}$ & 23.4/71\\
      \hline
      $\gamma, \pi^{0,\pm}$ (T1) & $0.821\substack{0.065\\ -0.060}$ & $0.5291\substack{+0.0057\\ -0.0057}$ & $0.3304\substack{+0.0060\\ -0.0060}$ & 321/514\\
      All (T2) & $0.976\substack{+0.054\\-0.052}$ & $0.5496\substack{+0.0026\\ -0.0026}$ & $0.3227\substack{+0.0028\\ -0.0028}$ & 778/963\\
      \hline
    \end{tabular}
  \end{center}
  \caption{\label{tab:observables} Results of tunes performed separately to measurements of
    charged multiplicity, charged scaled momentum, $\gamma$ spectra, $\pi^0$ spectra, $\pi^\pm$ 
    spectra, Thrust distribution and $C$-parameter. Results of tunes combining 
    measurements of $\gamma$,$\pi^{\pm}$ and $\pi^0$ (T1)
    or all measurements (T2) are also reported.}
\end{table}

\begin{figure}[!h] 
   \centering
  \includegraphics[width=0.48\textwidth]{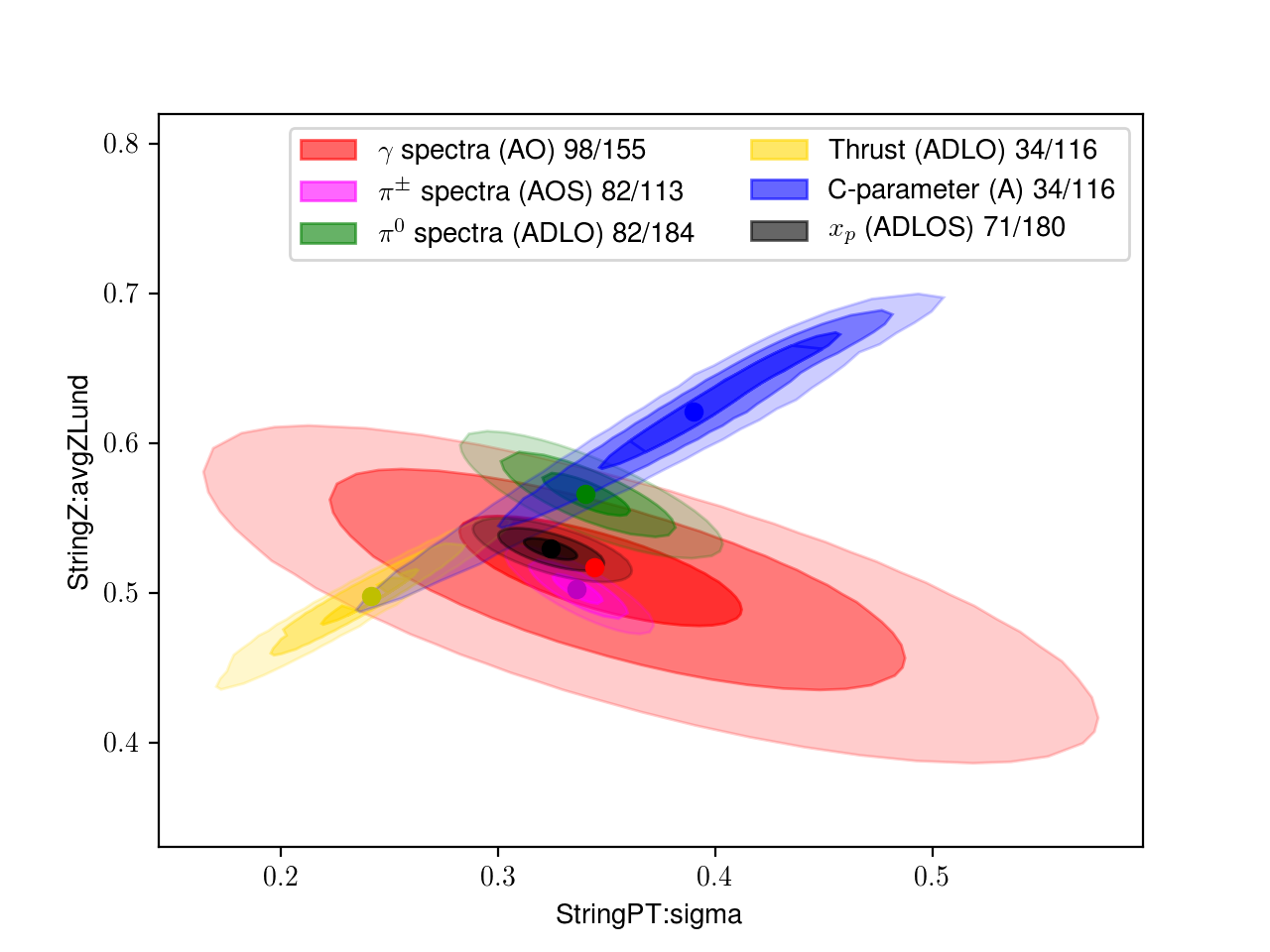}
  \includegraphics[width=0.48\textwidth]{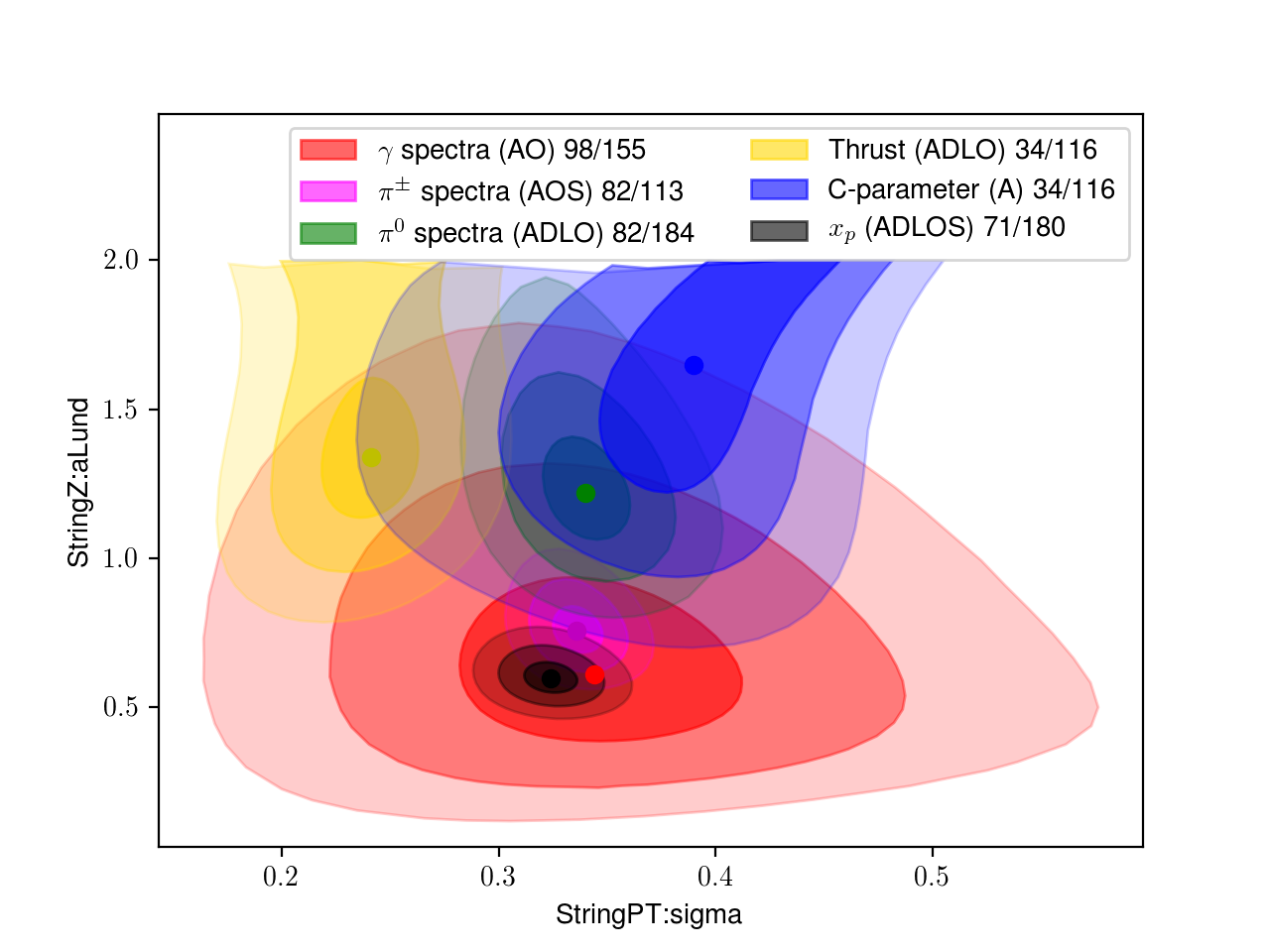}
  \caption{\label{fig:observables} Results of tunes performed separately to measurements of
    $\gamma$ spectra (red), $\pi^{\pm}$ spectra (magenta), $\pi^{\pm}$ spectra (green),
    Thrust distribution (yellow), $C$-parameter (blue) and charged particles scaled momentum (black).
    Measurements from \textsc{Aleph} (A), \textsc{Delphi} (D), \textsc{Opal} (O), \textsc{L3} (L) and \textsc{Sld} (S) are used.
    The contours corresponding to a one, two and three standard deviations are also shown.
  }
\end{figure}

Tuning the same observable from different experiments is very important to 
assess the constraining power of a given observable and to validate 
sensitivity studies which were carried in section \ref{sec:tune}. The results of these fits
are depicted in Fig. \ref{fig:observables} and Table \ref{tab:observables}. In this part of the tuning, we
focused on charged multiplicity, $\gamma$, $\pi^0$, $\pi^\pm$, and charged particles' spectra, 
$C$-parameter and the Thrust distribution. We, first, observe that the obtained value of \verb|StringZ:aLund|
from the tuning to the $C$ and $T$ parameters is not consistent at all either with the well established 
\textsc{Monash} tune or with the other results. This is an expected
result given the fact that the $C$ and $T$ parameters have less
sensitivity (expect in their first few bins) on the fragmentation
model and they are mainly
sensitive to the shower parameters, which are not varied in this study. Furthermore, for the same observables, 
the \verb|StringZ:avgZLund| and \verb|StringPT:sigma| parameters
are highly correlated as can be seen from Fig. \ref{fig:observables}.  \\

\begin{table}[t!]
  \begin{center}
    \begin{tabular}{lcccc}
      \hline
      Variation     & \verb|StringZ:aLund| & \verb|StringZ:avgZLund| & \verb|StringPT:sigma| \\
      \hline
      Central & 0.9757 & 0.5496 & 0.3227\\
      \hline
      OneUp & 0.9233 & 0.5476 & 0.3230\\
      OneDw & 1.0300 & 0.5516 & 0.3225\\
      TwoUp & 0.9757 & 0.5509 & 0.3200\\
      TwoDw & 0.9758 & 0.5483 & 0.3255\\
      ThreeUp & 0.9757 & 0.5507 & 0.3233\\
      ThreeDw & 0.9758 & 0.5485 & 0.3222\\
      \hline
    \end{tabular}
  \end{center}
\caption{\label{tab:eigentunes} The Hessian variations (eigentunes) for the nominal tune
including all observables corresponding to a one standard deviation (68\% CL) interval.}
\end{table}
 
\begin{table}[!h]
 \begin{center}
  \begin{tabular}{lc}
   \hline
   Parameter & Value \\
   \hline
   \verb|StringZ:aLund| & $0.5999\pm0.2$ \\
   \verb|StringZ:avgZLund| & $0.5278^{+0.027}_{-0.023}$ \\
   \verb|StringPT:sigma| & $0.3174^{+0.042}_{-0.037}$ \\
   \hline
  \end{tabular}
 \end{center}
  \caption{\label{tab:individual} Result of the single fit to all the measurements 
  as obtained from Fig. \ref{fig:fits_single}. The quoted errors correspond to the $68\%$ CL 
  uncertainty on the fit.}
\end{table}

\begin{figure}\centering
  \includegraphics[width=0.95\textwidth]{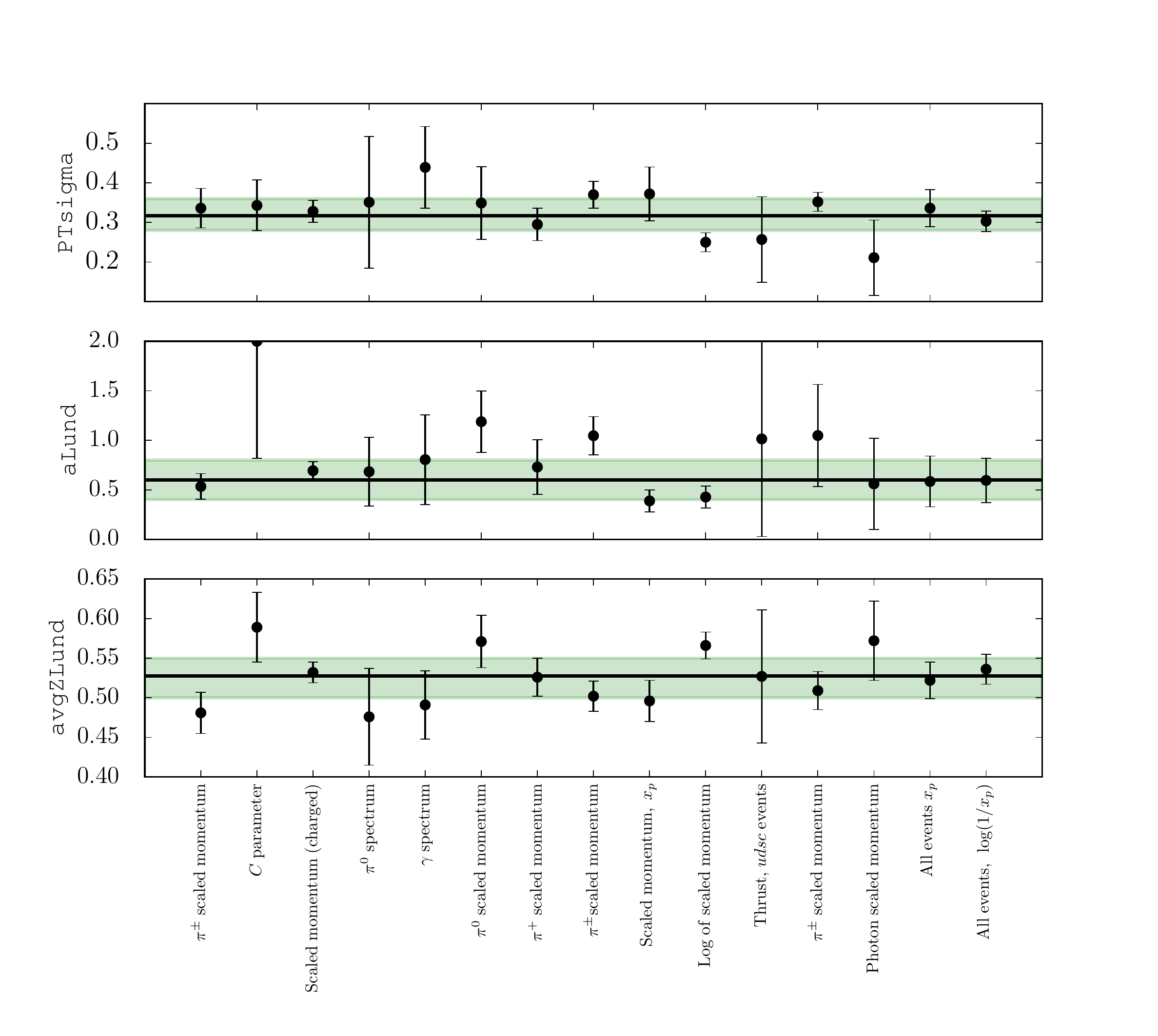}
  \caption{\label{fig:fits_single} Results of tunes performed separately to each of the observables.
    The weighted average of the  tunes to the individual measurements is shown with a black line.
    A green shaded area indicated  the 68\% CL interval on the parameters.}
\end{figure}

\subsection{Uncertainties}
\label{tune:uncertainties}
After discussing in details the results of the tuning and independent fits, we move to the question of QCD uncertainties. Those can be separated into the perturbative uncertainties, related to the parton showers evolution, and the non-perturbative ones, related to the determination of the parameters of the fragmentation model.  Uncertainties on the non-perturbative part, are specific to the chosen model and the data used to constrain them, leaving more ambiguities in the uncertainty estimate.\\

Uncertainties on parton showering in \textsc{Pythia}8 are estimated using the automatic setup developed in \citep{Mrenna:2016sih} which aims for a comprehensive
uncertainty bands by variation the central renormalization scale 
by a factor of $2$ in the two directions with a full NLO scale compensation.
Such perturbative uncertainties are investigated independently
from those on the parameters of the Lund fragmentation function.
Furthermore, the framework also allows for variations of non-singular terms in 
the Dokshitzer-Gribov-Lipatov-Altarelli-Parisi (DGLAP) splitting
kernels as an additional, complementary handle on the perturbative
uncertainties (the singular part of the DGLAP functions is 
universal while non-singular terms can be used to represent potential
effects of process dependence).  In most of the cases, variations of
the non-singular terms
give small uncertainties ($\simeq 1\%$) and can safely be neglected. \\

On the non-perturbative side, two methods are employed to obtain uncertainties on the parameters of the Lund fragmentation function. 
The \textsc{Professor} toolkit allows to estimate uncertainties on the 
fitted parameters through the eigentunes method.
This method diagonalises the $\chi^2$ covariance matrix around the best fit point, and uses variations along the principal directions (eigenvectors) in the space of the optimised parameters to build a set of $2\cdot N_{\textrm{params}}$ variations.
Variations are obtained moving along the eigenvectors for direction corresponding to a fixed change  in the goodness-of-fit (GoF) measure. 
If the GoF measure follows a $\chi^2$ statistics, one can define the $\Delta\chi^2$ corresponding to a given confidence level interval.
However, due to the intrinsic limitations of the phenomenological models used in event generators and the lack of correlation in systematic uncertainties between observables and bins in the tunes, 
this is usually not the case.
Heuristic choices of $\Delta\chi^2$ are typically chosen in tunes to obtain a reasonable coverage of the uncertainties in the experimental data.
In our study this is obtained by the addition of a 5\% uncertainty to the MC predictions, allowing for $\chi^2$/NDoF of order one in our tunes.
The resulting eigentunes  are however still found to provide small uncertainties which cannot be interpreted as conservative. The uncertainty on the parameters of the Lund fragmentation function are very small (below the one percent level) and inconsistent with the uncertainties of the data used in the tune\footnote{We also checked their impact on the gamma-ray spectra in different final states and for different DM masses including the ones corresponding to the pMSSM best fit points and have found that the bands obtained from the eigentunes are negligibly small.}. In Table \ref{tab:impact} we also show the  uncertainties from QCD on the photon spectra in the peak region for $\chi\chi \to g g$ for $m_\chi=25$ GeV where the nominal values of the parameters correspond to the result of T2 tune and the corresponding eigentunes are shown in Table \ref{tab:eigentunes}.

Therefore, we use an alternative method to estimate 
the uncertainty on the Lund fragmentation function's parameters. We,
first, make a fit each measurement. Thus, for $N$ measurements, we get
$N$ best-fit points for each parameter. We then take the $68\%$ CL
errors on the parameters to be our estimate of the uncertainty taking
care to exclude observables with little or no sensitivity to the given
parameters. The results of these fits along with their $68\%$ CL errors are shown in both Fig. \ref{fig:fits_single} and Table \ref{tab:individual}. We checked, that the predictions of this single fit agree fairly well with the T2 tune. We discuss how to obtain comprehensive uncertainty bands from the  $68\%~\textrm{CL}$ errors on the parameter. We denote a variation on the parameters $\verb|(StringPT:sigma, StringZ:avgZLund, StringZ:aLund)|\equiv(\sigma, z_\rho, a)$ by $(X_1,X_2,X_3)$ with $X_i=+,-,0$. The $+$($-$) corresponds to a variation of the parameter $X_i$ in the positive (negative) direction with respect to the nominal value denoted by $0$. Besides the nominal tune corresponding to $(0,0,0)$, there are $3^3 - 1 = 26$ possible variations. However, there are some variations which don't give observable effects on the spectra and, therefore, can be neglected. To understand this, we discuss the effects of different parameters on the most significant observables, i.e the event shapes and scaled momenta. If one increases $\sigma$, we get few hard hadrons because if a particle takes more mean transverse momentum $<p_T>$ then there will be no enough phase space for the other hadrons. On the other hand, decreasing $\sigma$ will result in many low $p_T$ hadrons. The effect of $z_\rho$ is similar to $\sigma$ but on the total momentum. The $a$ parameter has a similar effect on the particle momenta but with an reversed behavior in a sense that if $z_\rho$ and $a$ are varied in the same direction, their effect will almost compensate and therefore one gets no bands in the high energy bins. The same $a$ parameter has no effect on the event shapes as $\sigma$ and $z_\rho$ do. We stress finally that varying $\sigma$, and $z_\rho$ in opposite directions has large effect on the lowest bins of the event shapes (Thrust for example). Hence, the $(+,-,X)$ and $(-,+,X)$ have the most significant effects on the event shapes. In conclusion, the envelope that we consider as a faithful representation of the QCD uncertainty is obtained from $10$ variations; $(+,+,0), (-,-,0), (+,+,+), (-,-,-), (+,-,+), (+,-,-), (-,+,+), (-,+,-), (-,+,0)$ and $(+,-,0)$ in addition to the nominal one. Their effects are shown in Figs. \ref{comparison-1} and \ref{comparison-2} and used in the impact on DM fits. Furthermore, the updated tables for particle spectra will be provided with their uncertainties (obtained from these variations and from shower uncertainties as well).

\begin{table}[!h]
\centering
 \begin{tabular}{lc}
  \hline
  $x_\gamma$   &   $(\textrm{d}N/\textrm{d}x_\gamma)_\texttt{T2}  \pm \delta_\textrm{had.} \pm \delta_\textrm{shower}$  \\
  \hline
  $0.00125$ & $7.59\substack{+0.05\%\\-0.0\%}\substack{+8.1\%\\-4.8\%}$\\
  $0.002$ & $13.79\substack{+0.18\%\\-0.26\%}\substack{+8.3\%\\-4.9\%}$ \\
  $0.003$ & $22.29\substack{+0.13\%\\-0.0\%}\substack{+8.2\%\\-4.9\%}$ \\
  $0.005$ & $31.95\substack{+0.2\%\\-0.04\%}\substack{+8.1\%\\-4.8\%}$ \\
  $0.008$ &   $40.74\substack{+0.12\%\\-0.05\%}\substack{+7.7\%\\-4.6\%}$ \\
  $0.0125$ & $45.83\substack{+0.08\%\\-0.09}\substack{+7.1\%\\-4.3\%}$ \\
  $0.02$ &   $45.01\substack{+0.13\%\\-0.02}\substack{+6.5\%\\-4.0\%}$ \\
  $0.03$ & $39.43\substack{+0.13\%\\-0.0\%}\substack{+5.2\%\\-3.3\%}$ \\
  $0.05$ & $30.73\substack{+0.0\%\\-0.15\%}\substack{+3.1\%\\-2.1\%}$\\
  $0.08$ & $21.36\substack{+0.0\%\\-0.06\%}\substack{+0.4\%\\-0.5\%}$ \\
  $0.125$ & $12.98\substack{+0.13\%\\-0.23\%}\substack{+1.6\%\\-3.0\%}$\\
\hline
 \end{tabular}
\caption{\label{tab:impact} Scaled momentum of photons in the process $\chi\chi \to gg$ for $m_\chi=25$ GeV where 
only the peak region of the spectra is shown. In this table, we show the 
predictions from the weighted 
tune denoted by \texttt{T2} (the central values of the parameters and their eigentunes are shown in Tables \ref{tab:T2tune} and \ref{tab:eigentunes}). The $68\%$ CL on hadronisation parameters are shown as
first errors for each bin while uncertainties due to shower variations are the second errors.}
\end{table}
 
\begin{table}[!t]
\begin{center}
\begin{tabular}{ c | c | c | c | c  }
\hline
\cline{1-5}
$E_\gamma$ [GeV] &  \textsc{Hadronization}  & \textsc{Shower} & \textsc{Herwig} & \textsc{Total} \\ \hline
$0.4$ & $^{+6.17\%}_{-5.57\%}$ & $^{+8.24\%}_{-4.91}$ & $-21.31\%$  & $^{+10.29\%}_{-22.56\%}$ \\ \hline
$1.2$ & $^{+2.57\%}_{-2.47\%}$  & $^{+5.94\%}_{-3.68}$ & $-6.57\%$ & $^{+6.47\%}_{-7.92\%}$\\ \hline
$2.0$ &.$^{+1.23\%}_{-1.28\%}$  & $^{+3.51\%}_{-2.24\%}$ & $-1.51\%$ & $^{+3.71\%}_{-2.98\%}$\\ \hline
$2.8$ & $^{+0.70\%}_{-0.69\%}$ & $^{+1.78\%}_{-1.18\%}$ & $-0.39\%$ & $^{+1.91\%}_{-2.34\%}$\\ \hline
$3.6$ & $^{+0.42\%}_{-1.03\%}$ & $^{+0.10\%}_{-0.15\%}$ & $-0.31\%$  & $^{+0.43\%}_{-1.08\%}$\\ \hline
$4.4$ & $^{+1.08\%}_{-1.21\%}$ & $^{+0.68\%}_{-1.24\%}$ & $-0.70\%$ & $^{+1.27\%}_{-2.14\%}$\\ \hline
$5.2$ & $^{+1.43\%}_{-1.63\%}$ & $^{+1.40\%}_{-2.46\%}$ & $-0.97\%$ & $^{+2.00\%}_{-3.10\%}$\\ \hline
$6.0$ & $^{+1.84\%}_{-1.97\%}$ & $^{+2.09\%}_{-3.88\%}$ & $-1.60\%$ & $^{+2.78\%}_{-4.63\%}$\\ \hline
$6.8$ & $^{+2.14\%}_{-2.33\%}$ & $^{+2.60\%}_{-4.56\%}$ & $-1.61\%$ & $^{+3.36\%}_{-5.36\%}$\\ \hline
$7.6$ & $^{+2.86\%}_{-2.56\%}$ & $^{+3.02\%}_{-4.96\%}$ & $-1.51\%$ & $^{+4.15\%}_{-5.78\%}$\\ \hline
$8.4$ & $^{+3.28\%}_{-3.04\%}$ & $^{+3.72\%}_{-5.97\%}$ & $-1.40\%$ & $^{+4.95\%}_{-6.84\%}$\\ \hline
$9.2$ & $^{+2.72\%}_{-4.05\%}$ & $^{+4.10\%}_{-6.54\%}$ & $-2.20\%$ & $^{+4.92\%}_{-8.00\%}$\\ \hline
$10$ & $^{+3.55\%}_{-4.12\%}$ & $^{+4.13\%}_{-6.73\%}$ & $-2.04\%$ & $^{+5.44\%}_{-8.15\%}$\\ \hline
$14$ & $^{+4.34\%}_{-4.35\%}$ & $^{+6.30\%}_{-9.58\%}$ & $-1.44\%$ & $^{+7.65\%}_{-10.62\%}$ \\ \hline
$18$ & $^{+5.90\%}_{-4.99\%}$ & $^{+6.88\%}_{-10.58\%}$ & $+0.76\%$ & $^{+9.09\%}_{-11.69\%}$ \\ \hline
$22$ & $^{+4.15\%}_{-4.27\%}$ & $^{+7.58\%}_{-11.99\%}$ & $+0.88\%$ & $^{+8.68\%}_{-12.72\%}$ \\ \hline
$30$ & $^{+5.50\%}_{-5.18\%}$ & $^{+9.76\%}_{-8.91\%}$  & $+2.41\%$ & $^{+11.45\%}_{-10.30\%}$ \\ \hline
$38$ & $^{+6.77\%}_{-11.04\%}$ & $^{+8.97\%}_{-14.17\%}$ & $-5.49\%$ & $^{+11.23\%}_{-18.78\%}$ \\ \hline
$42$ & $^{+0.54\%}_{-10.74\%}$ & $^{+11.82\%}_{-8.57\%}$ & $-15.15\%$ & $^{+11.83\%}_{-20.45\%}$ \\ \hline
$52$ & $^{+25.19\%}_{-22.96\%}$ & $^{+5.75\%}_{-12.29\%}$ & $-8.71\%$ & $^{+25.83\%}_{-27.46\%}$ \\ \hline
$64$ & $^{+13.33\%}_{-53.33\%}$ & $^{+0.0\%}_{-9.25\%}$ & $-46.81\%$ & $^{+13.33\%}_{-71.56\%}$ \\ \hline
\cline{1-5}
\end{tabular}
\end{center}
\caption{Uncertainties on the photon spectra in the region ($0.4~$GeV $\leq E_\gamma \leq 64$~GeV for $W^+W^-$ annihilation 
channel of DM corresponding to $m_\chi=90.6$ GeV. We show hadronization uncertainties, shower uncertainties and uncertainties from the relative difference 
between \textsc{Herwig} and \textsc{Pythia}. Total uncertainties are shown by summing in quadrature.}
\label{tab:uncertaintiesWW}
\end{table}    

\subsection{Impact on Dark Matter Spectra and Fits}
\begin{figure}[!t]
\centering
\includegraphics[width=0.48\linewidth]{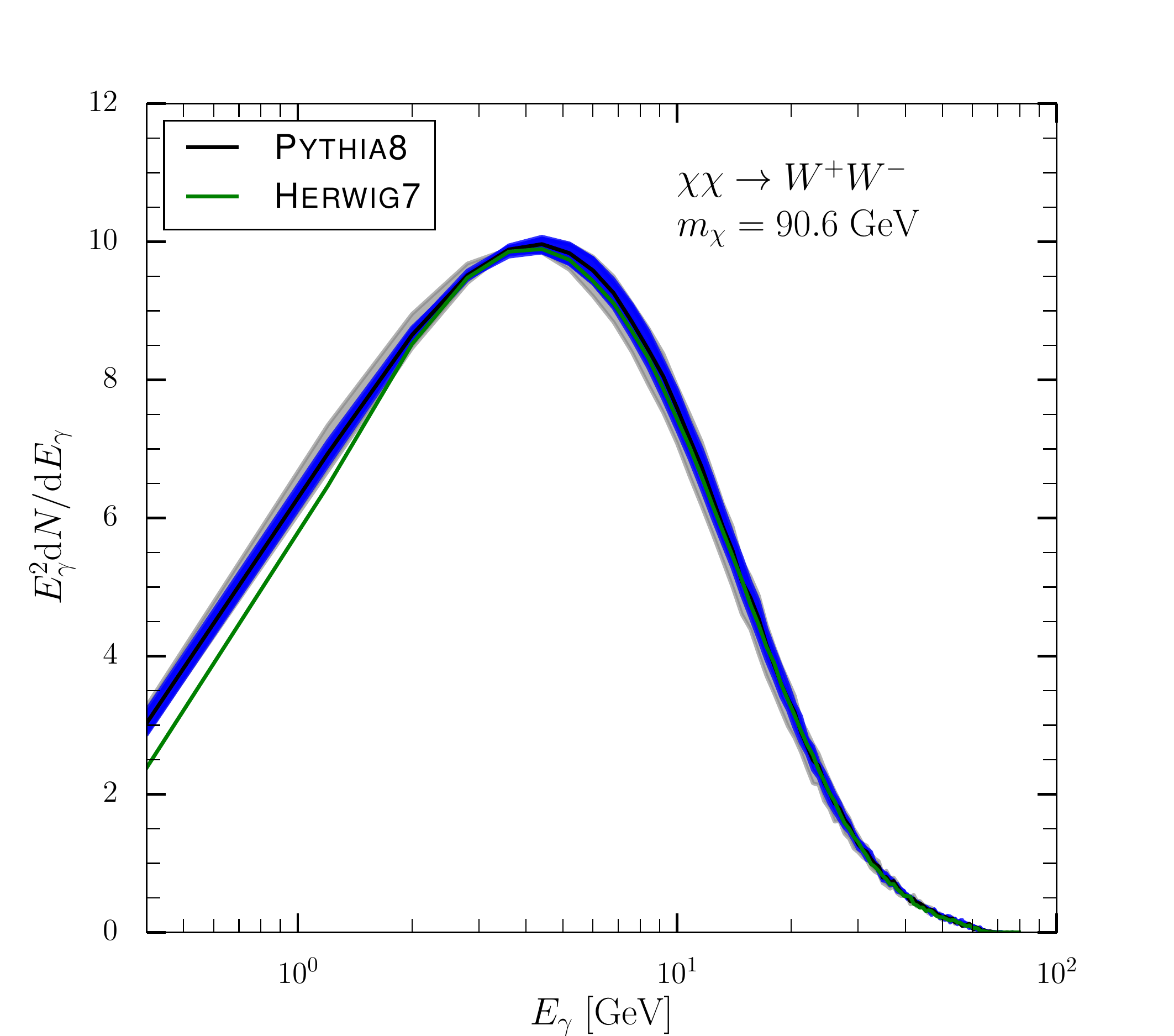}
\hfill
\includegraphics[width=0.48\linewidth]{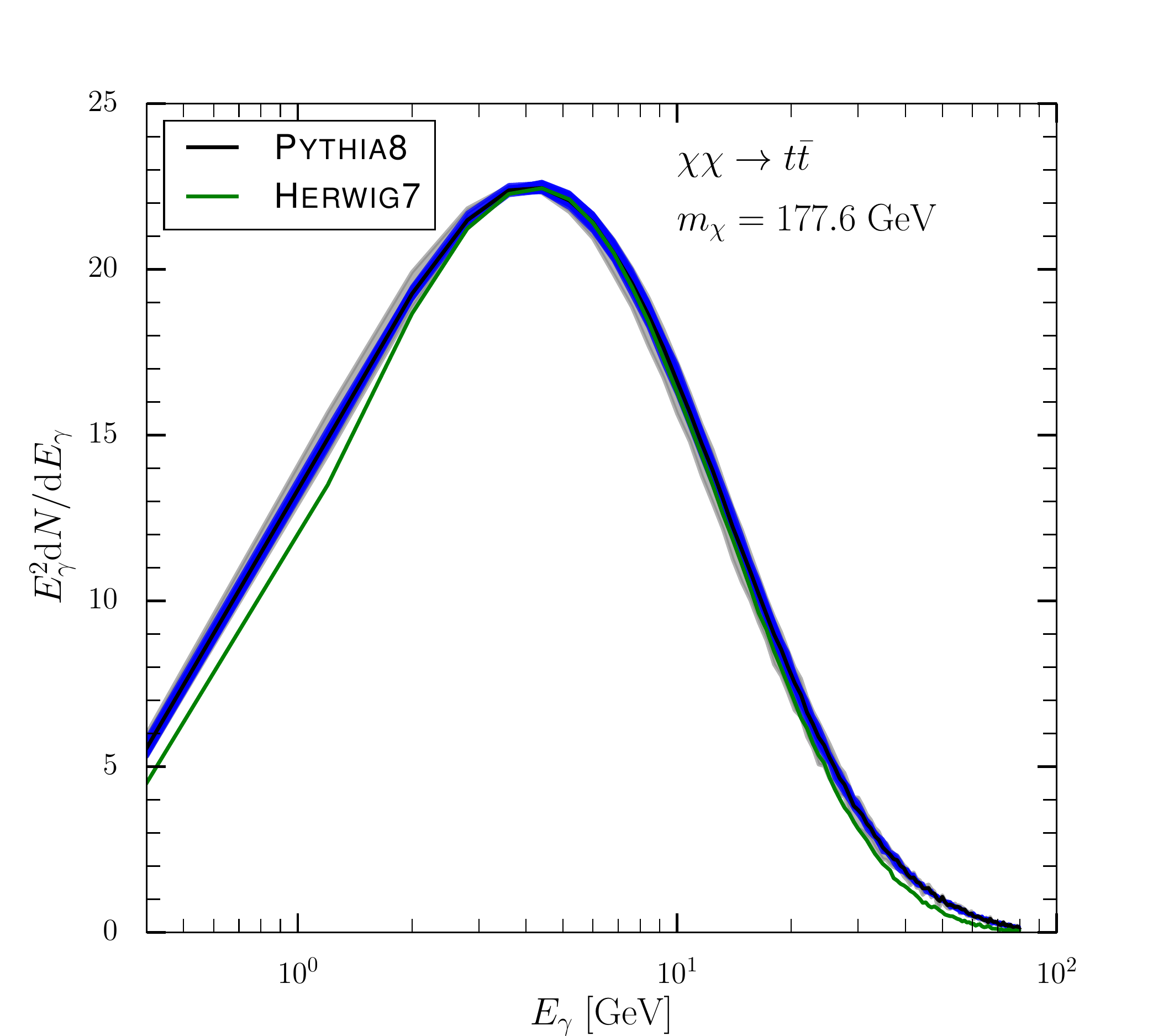}
\caption{Photon energy distribution for dark matter annihilation into $W^+ W^-$ with $m_\chi=90.6$ GeV (\emph{left}) 
and into $t\bar{t}$ with $m_\chi = 177.6$ GeV (\emph{right}). In the two cases, the result corresponding to the new tune is shown in black line. Both the uncertainties from parton showering (gray bands) and from hadronisation (blue bands) are shown. Predictions from \textsc{Herwig7} are shown as a gray solid line.}
\label{fig:DMimpact}
\end{figure}

In this subsection, we study the impact of QCD fragmentation function, parton shower and MC event generators \footnote{Estimated from differences produced in the spectrum between \textsc{Pythia8} and \textsc{Herwig7}.} uncertainties on the photon spectra of two representative DM annihilation channels: $W^+W^-$ and $t\bar{t}$. Our motivation is to see how the best fit of the GCE, using PASS8 data performed in the pMSSM \cite{Achterberg:2017emt}, can be affected upon including realistic uncertainties. In that analysis, the best-fit was found for two neutralino masses, i.e $m_\chi=90.6$ GeV and $m_\chi=177.6$ GeV corresponding to the $W^+W^-$ and $t\bar{t}$ DM annihilation channels respectively. \\

In Fig. \ref{fig:DMimpact} we show the photon spectra for
$m_\chi=90.6$ GeV in the $W^+W^-$ channel (left panel) and for
$m_\chi=177.6$ GeV in the $t\bar{t}$ channel (right panel) with the
new tune (black line) and the \textsc{Herwig} prediction (green
line). The bands show the \textsc{Pythia} parton-shower (gray bands) and hadronisation (blue bands) uncertainties. We can see that the predictions from \textsc{Pythia} and \textsc{Herwig} agree very well except for $E_\gamma \leqslant 2~$GeV where differences can reach about $21\%$ for $E_\gamma \sim 0.4$ GeV. Furthermore, one can see that uncertainties can be important for both channels. Particularly, in the peak region which corresponds to energies where the photon excess is observed in the galactic center region. Indeed combining them in quadrature assuming the different type of uncertainties are uncorrelated, they can go from few percents where the GCE lies to about 70\% in the high energy bins. Furthermore hadronisation uncertainties are the dominant ones around the peak of the photon spectrum whereas the ones from parton showering are the main source of uncertainties while moving away toward the edges of the spectra. 

In Table \ref{tab:uncertaintiesWW}, we show the uncertainties in photon spectra for $0.4~\textrm{GeV} \leqslant E_\gamma \leqslant 64~\textrm{GeV}$ taking the example of the $WW$ final state where we can see that, for $1~\mathrm{GeV} \leqslant E_\gamma \leqslant 5~\mathrm{GeV}$, hadronization uncertainties are below $\simeq 2\%$. After including the other components they can reach up to about $\sim 4\%$-$8\%$. One possible reason for the smallness of hadronization uncertainties in this region is that \textsc{Lep} measurements of photon spectra at $\sqrt{s}=91.2$ GeV don't allow for large variations in the peak region. On the other hand, for high energy bins, notably for $E_\gamma \geq 42$ GeV, differences between \textsc{Herwig} and \textsc{Pythia} becomes very important ($\sim -46.81\%$) which is expected due to differences in the algorithms used for QED bremmstrahlung off e.g. quarks. Besides, we notice that the region of high energy bins has low statistics (about $0.02$ permille of the total photons have energies between $50$ and $65$ GeV). However, constraining and including the large differences in the photon spectra at large energies as seen in Fig. \ref{fig:DMimpact} will become
important to continue searches for relatively low mass Dark Matter particles with upcoming experiments (e.g. CTA \cite{Ong:2017ihp}).
This is especially important if such experiments are mainly 
sensitive to high energetic photons at energies of around $100$-$1000$ GeV. \\

As demonstrated in this section, the set of QCD uncertainties we
present in this study do provide a realistic and resonably
conservative estimate of the uncertainties allowed by data. These
uncertainties can have sizeable impacts  on fits like the GCE in
models like the pMSSM, hence we believe it will be relevant to include
them in future phenomenological analyses of gamma-ray dark matter searches.

\section{Conclusions}
\label{sec:conclusions}

In this work, we presented for the first time a dedicated study of QCD uncertainties on photon spectra from DM annihilation processes in the context of \textsc{Pythia~8}. First, we showed predictions of several different modern MC event generators (\textsc{Herwig} 7.1.3, \textsc{Pythia} 8.235 and \textsc{Sherpa} 2.2.5) and demonstrated that their relative differences do not, in our opinion, give a reliable picture of the allowed uncertainty on the modeling of particle spectra; in some regions of the constraining observables, notably near the peaks of the spectra, their differences can be very small and do not exhaustively span the range allowed by the data, while in other regions, notably in the tails of distributions, their differences can vastly overestimate the uncertainties allowed by data. \\

The problem is that, while the generators do use qualitatively different physics models and this does lead to differences between them, their default parameter sets have largely been optimised to give ``central'' fits to  the same data. There is no explicit intent behind the central fits to explore the allowed ranges of variation in a statistical sense. 
We therefore studied the complementary approach of defining a set of parametric variations within a single modeling paradigm, taking the current default \textsc{Monash} tune of \textsc{Pythia}~8.2 as our baseline. We performed several retunings of the light-quark fragmentation functions using selected measurements from LEP, which we report on in this paper. We then discussed the different ways QCD uncertainties can be estimated (manual and eigentunes) and use a combination of the two to obtain a fairly conservative estimate of the uncertainty. \\

Next, we have studied the impact of QCD uncertainties on 
two benchmark points of the pMSSM were the best fit of the Fermi-LAT Pass 8 GCE were found. Respectively, a neutralino mass of $m_\chi=90.6$ GeV annihilating to $W^+W^-$, and $m_\chi=177.6$ GeV annihilating to $t\bar{t}$. We have found 
that the variation of the photon spectrum, combining the  uncertainties, can go from a few percent to about 50 \%, which has a large impact on the fit of the GCE. Therefore QCD uncertainties should be taken into account in DM phenomenological studies when indirect detection searches with gamma-ray data are considered. \\

We have validated our findings against the standard reference in the field, the PPPC4\-DMID~\cite{Cirelli:2010xx}, and generally find good agreement, except for a few discrepancies in tails of distributions; these are remarked upon in appendix \ref{sec:pppc4dmid} in which we also summarize the salient changes that have happened during the roughly eight years between the release dates for the original \textsc{Pythia} version used for the PPPC4DMID study (8.135) and that used for our study (8.235). Full data tables which can be used to update those in the PPPC4DMID will be published online as a follow-up to this work\footnote{Until then, please contact the authors of this work to obtain the tables.}. \\

Our findings motivate new directions for the study of QCD uncertainties and their applications. The list of applications includes but is not restricted to 
\begin{itemize}
\item Studies of Higgs boson decays to hadrons at the LHC and future colliders. In particular, the conventional method of estimating the QCD uncertainties on the heavy quark fragmentation functions by comparing the central predictions of several different MC generators, may not span the full envelope. This can affect measurements related to both $H\to c\bar{c}$ and $H\to b\bar{b}$ decays. While the QCD uncertainties presented in this paper cannot be used directly for such studies, we remark that \textsc{Pythia8} has two parameters that control the heavy-quark-to-hadron fragmentation function. These parameters are introduced in \cite{Bowler:1981sb} and are correlated to \verb|StringZ:bLund|. Therefore, a dedicated and more systematic study of the full fragmentation function (including the heavy component) may be in order for Higgs studies. We note that, in principle, the underlying event and the associated topic of colour reconnections (CR) could represent another source of QCD uncertainty, specific to $pp$ collisions, though due to the relatively long lifetime of the (SM) Higgs boson compared with the typical timescale of hadronisation ($\Gamma_H \sim 4~\mathrm{MeV}$ compared with $\Lambda_{\mathrm{QCD}}\sim 200~\mathrm{MeV}$) we would not expect Higgs decays to be sensitive to CR.
\item Top quark mass measurements. The reconstruction of the top quark mass from final-state observables is sensitive to several processes that occur during and after top quark production, and also here $b$-quark fragmentation plays an important role. Moreover, since the top quark width is larger than the hadronisation scale and the produced top quarks are colour-connected to the initial states, CR effects are expected to be relevant \cite{Argyropoulos:2014zoa}, with current mass determinations from the LHC citing a CR uncertainty of $+0.31\pm 0.08~$GeV \cite{Sirunyan:2018gqx}. An estimate of QCD uncertainties on the fragmentation function (including the heavy part) and any dependence it exhibits on global and local event properties, could be relevant to achieve improved accuracy for top quark mass determinations.
\item Other stable final-state products of DM annihilation. We plan to extend this study to include the spectra of antiprotons, positrons, and neutrinos (and in principle of electrons and protons as well), in future work. 
\item A final aspect not touched on in this work is QCD fragmentation uncertainties on the spectra of secondary particles produced from cosmic-ray interactions.
\end{itemize}

\paragraph*{Acknowledgements}
The reparameterisation of the Lund symmetric fragmentation function benefited from a pilot project carried out by Ms.~Sophie~Li, a student at Monash University. 
The authors would like to thank Marco Cirelli and Gennaro Corcella for useful discussions. 
SA acknowledges support from the Helmholtz Gemeinschaft.
The work of AJ is sponsored by CEPC theory program and by the National Natural Science Foundation of China 
under the Grants No. 11875189 and No.11835005. R. RdA, has been supported by the Ram\'on y Cajal program of the Spanish
MICINN and also thanks the support of the Spanish MICINN's
Consolider-Ingenio 2010 Programme  under the grant MULTIDARK
CSD2209-00064, the Invisibles European ITN project
(FP7-PEOPLE-2011-ITN, PITN-GA-2011-289442-INVISIBLES, the  
``SOM Sabor y origen de la Materia" (FPA2014-57816-P) and the Spanish 
MINECO Centro de Excelencia Severo Ochoa del IFIC 
program under grant SEV-2014-0398. PS is supported in part by the 
Australian Research Council, contract FT130100744.

\appendix
\section{Observables and their Weights \label{app:observables}}
\label{sec:weights}
 \begin{table}[!h]
  \begin{center}
   \begin{tabular}{l|l}
    \hline 
    \hline
    Observable  \hspace{3cm} &  \hspace{1cm} Associated Weight \\ \hline
    $\pi^0$ spectrum         & \hspace{3cm} $34.0$ \\ \hline
    $\pi^\pm$ spectrum       & \hspace{3cm} $70.0$ \\ \hline
    $\gamma$ spectrum        & \hspace{3cm} $70.0$ \\ \hline
    $\eta$ spectrum          & \hspace{3cm} $1.0$ \\ \hline \hline
   \end{tabular}
  \end{center}
  \caption{Identified photon and meson spectra and their weights. Data from
  \cite{Buskulic:1994ft, Barate:1996fi, Heister:2001kp, Adriani:1992hd, Akers:1994ez, Ackerstaff:1998ap, Abe:1998zs}.}
  \label{Tab1}
 \end{table}
 
  \begin{table}[!h]
  \begin{center}
   \begin{tabular}{l|l}
    \hline 
    \hline
    Observable  \hspace{3cm}                             & \hspace{1cm} Associated Weight \\ \hline
    Mean charged multiplicity                            & \hspace{3cm} $10.0$ \\ \hline
    Mean charged multiplicity for rapidity $|Y| < 0.5$   & \hspace{3cm} $10.0$ \\ \hline
    Mean charged multiplicity for rapidity $|Y| < 1.0$   & \hspace{3cm} $10.0$ \\ \hline
    Mean charged multiplicity for rapidity $|Y| < 1.5$   & \hspace{3cm} $10.0$ \\ \hline
    Mean charged multiplicity for rapidity $|Y| < 2.0$   & \hspace{3cm} $10.0$ \\ \hline
    Mean $\pi^0$ multiplicity                            & \hspace{3cm} $10.0$ \\ \hline
    Mean $\pi^\pm$ multiplicity                          & \hspace{3cm} $10.0$ \\ \hline  \hline 
    \end{tabular}
  \end{center}
  \caption{Mean particle multiplicities and their weights. 
  Data from \cite{Barate:1996fi, Abreu:1996na, Ackerstaff:1998hz, Amsler:2008zzb}.}
  \label{Tab2}
 \end{table}
 
 \begin{table}[!h]
\begin{center}
\begin{tabular}{l|l}
\hline \hline 
Observable \hspace{1cm}                                        &  Associated Weight \\ \hline
In(out-)-plane $p_\bot$ in GeV w.r.t. (thrust) sphericity axes & \hspace{1cm} $2.0$ \\ \hline 
Mean out-of-plane $p_\bot$ in GeV w.r.t. thrust axis vs. $x_p$ & \hspace{1cm} $2.0$ \\ \hline
Scaled momentum $x_p=|p|/|p_\text{beam}|$                      & \hspace{1cm} $20.0$ \\ \hline
Log of scaled momentum, $\log(1/x_p)$                          & \hspace{1cm} $20.0$ \\ \hline
Energy-energy correlation, EEC                                 & \hspace{1cm} $2.0$ \\ \hline
Sphericity, $S$                                                & \hspace{1cm} $2.0$ \\ \hline
Aplanarity, $A$                                                & \hspace{1cm} $2.0$ \\ \hline
Planarity, $P$                                                 & \hspace{1cm} $2.0$ \\ \hline 
 $D$ parameter                                                  & \hspace{1cm} $2.0$ \\ \hline
$C$ parameter                                                  & \hspace{1cm} $2.0$ \\ \hline  \hline
\end{tabular}
\end{center}
\caption{Event shapes and the associated weights. Data from 
\cite{Decamp:1991uz, Barate:1996fi, Heister:2003aj, Abreu:1996na, Achard:2004sv, Abbiendi:2004qz}. }
\label{Tab3}
 \end{table}
 
  \begin{table}[tbp]
 \begin{center}
 \begin{tabular}{l | l}
 \hline \hline
 1-Thrust                                                       & \hspace{1cm} $2.0$ \\ \hline
 Thrust major, $M$                                              & \hspace{1cm} $2.0$ \\ \hline
Thrust minor, $m$                                              & \hspace{1cm} $2.0$\\ \hline
Oblateness, $O=M-m$                                            & \hspace{1cm} $2.0$ \\ \hline
Charged multiplicity distribution                              & \hspace{1cm} $20.0$ \\ \hline
Two-jet resolution variable, $Y3$ (charged)                    & \hspace{1cm} $2.0$ \\ \hline
Rapidity w.r.t. thrust axes, $y_T$ (charged)                   & \hspace{1cm} $2.0$ \\ \hline
Heavy jet mass $(E_{\text{CMS}} = 91.2 \text{ GeV})$           & \hspace{1cm} $2.0$ \\ \hline
Total jet broadening $(E_{\text{CMS}} = 91.2 \text{ GeV})$     & \hspace{1cm} $2.0$ \\ \hline
Wide jet broadening $(E_{\text{CMS}} = 91.2 \text{ GeV})$      & \hspace{1cm} $2.0$ \\ \hline
Jet mass difference $(E_{\text{CMS}} = 91.2 \text{ GeV})$      & \hspace{1cm} $2.0$ \\ \hline
Rapidity w.r.t. sphericity axes, $y_S$                         & \hspace{1cm} $2.0$ \\ \hline
Mean $p_\perp$ in GeV vs. $x_p$                                & \hspace{1cm} $2.0$ \\ \hline
Planarity, $P$                                                 & \hspace{1cm} $2.0$ \\ \hline
Heavy hemisphere masses, $M_h^2/E_\text{vis}$                  & \hspace{1cm} $2.0$ \\ \hline \hline
 \end{tabular}
 \end{center}
 \caption{Event shapes and the associated weights (\emph{contd}). Data from 
\cite{Decamp:1991uz, Barate:1996fi, Heister:2003aj, Abreu:1996na, Achard:2004sv, Abbiendi:2004qz}. }
\label{Tab4}
 \end{table}

 \begin{table}[tbp]
  \begin{center}
   \begin{tabular}{l | l}
\hline \hline
Observable \hspace{1cm}                                        &  Associated Weight \\ \hline
Light hemisphere masses, $M_l^2/E_\text{vis}$                  & \hspace{1cm} $2.0$ \\ \hline
Difference in hemisphere masses, $M_d^2/E_\text{vis}$          & \hspace{1cm} $2.0$ \\ \hline
Wide hemisphere broadening, $B_\text{max}$                     & \hspace{1cm} $2.0$ \\ \hline
Narrow hemisphere broadening, $B_\text{min}$                   & \hspace{1cm} $2.0$ \\ \hline
Total hemisphere broadening, $B_\text{sum}$                    & \hspace{1cm} $2.0$ \\ \hline
Difference in hemisphere broadening, $B_\text{diff}$           & \hspace{1cm} $2.0$ \\ \hline
Moments of event shapes at $91$ GeV                                 & \hspace{1cm} $2.0$ \\ \hline \hline    
   \end{tabular}
  \end{center}
\caption{Event shapes and the associated weights (\emph{contd}). Data from 
\cite{Decamp:1991uz, Barate:1996fi, Heister:2003aj, Abreu:1996na, Achard:2004sv, Abbiendi:2004qz}. }
\label{Tab5}
 \end{table}
 
    \begin{table}[tbp]
  \begin{center}
   \begin{tabular}{l|l}
    \hline 
    \hline
    Observable  \hspace{3cm}                                           &  \hspace{1cm} Associated Weight \\ \hline
 Differential $3$-jet rate in the Durham algorithm & \hspace{3cm} $2.0$ \\ \hline
 Differential $3$-jet rate in the Jade algorithm & \hspace{3cm} $2.0$ \\ \hline
  Differential $4$-jet rate in the Durham algorithm & \hspace{3cm} $2.0$ \\ \hline
 Differential $4$-jet rate in the Jade algorithm & \hspace{3cm} $2.0$ \\ \hline
  Differential $5$-jet rate in the Durham algorithm & \hspace{3cm} $1.0$ \\ \hline
 Differential $5$-jet rate in the Jade algorithm & \hspace{3cm} $1.0$ \\ \hline
Durham jet resolution $2\to1$ and $3\to2$    &  \hspace{3cm} $2.0$ \\ \hline
Durham jet resolution $4\to3$, $5\to 4$ and $6 \to 5$    &  \hspace{3cm} $1.0$ \\ \hline
    $2$-jet fraction $(E_{\text{CMS}}=91.2 \text{ GeV})$               &  \hspace{3cm} $2.0$ \\ \hline
    $3$-jet fraction $(E_{\text{CMS}}=91.2 \text{ GeV})$               &  \hspace{3cm} $2.0$ \\ \hline
$4$-jet fraction $(E_{\text{CMS}}=91.2 \text{ GeV})$               &  \hspace{3cm} $2.0$ \\ \hline
    $5$-jet fraction $(E_{\text{CMS}}=91.2 \text{ GeV})$               &  \hspace{3cm} $1.0$ \\ \hline
        $n \geq 6$-jet fraction $(E_{\text{CMS}}=91.2 \text{ GeV})$               &  \hspace{3cm} $1.0$ \\ \hline\hline
   \end{tabular}
  \end{center}
  \caption{Jet rates and their weights. Data from \cite{Abreu:1996na, Heister:2003aj}.}
  \label{Tab6}
 \end{table}
\clearpage
\section{Photon and Pion Spectra for $m_\chi = 25$ GeV and $m_\chi = 250$ GeV}
\label{sec:photonSpectra}
\begin{figure}[!h]
\centering
\includegraphics[width=0.41\linewidth]{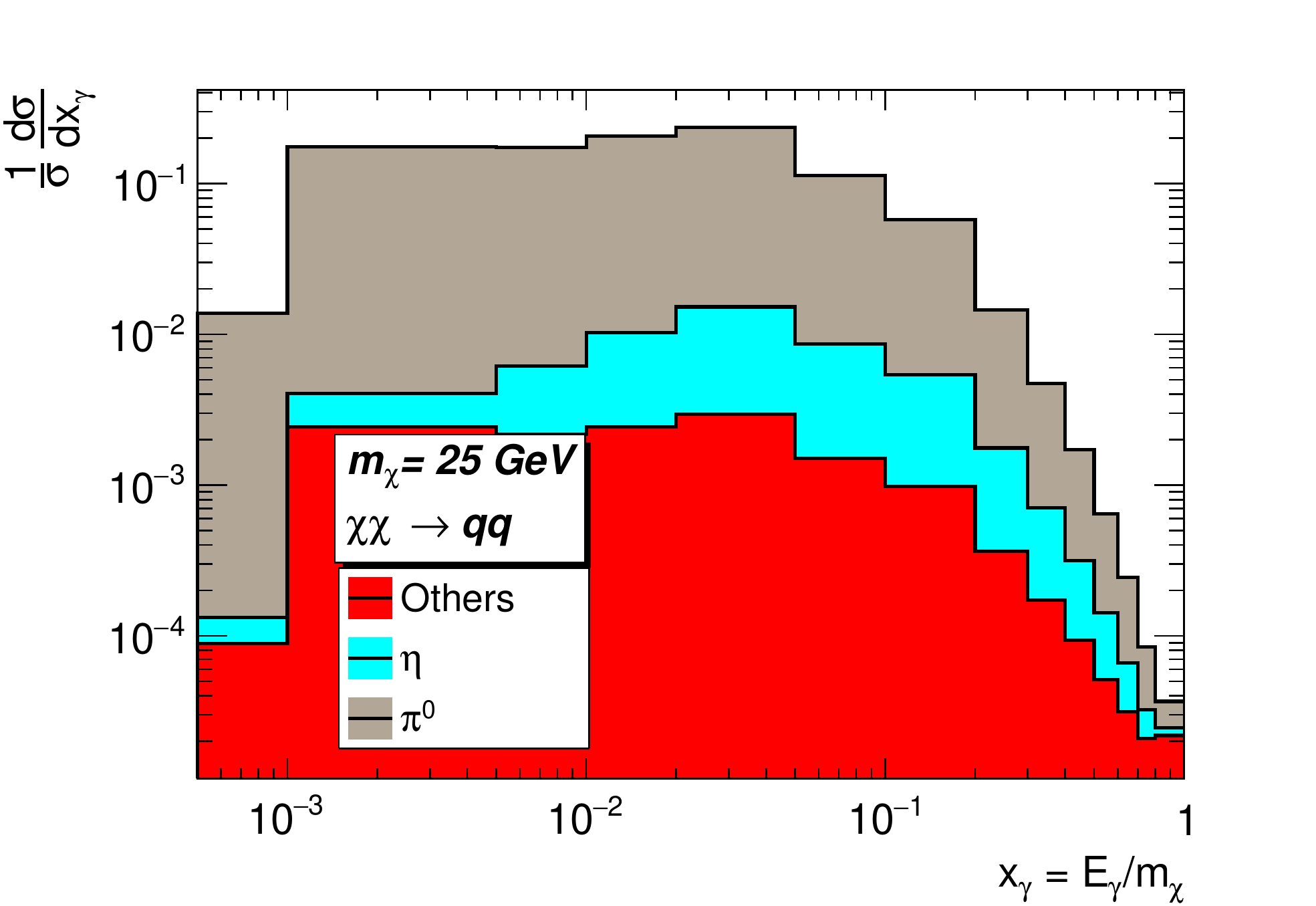}
\includegraphics[width=0.41\linewidth]{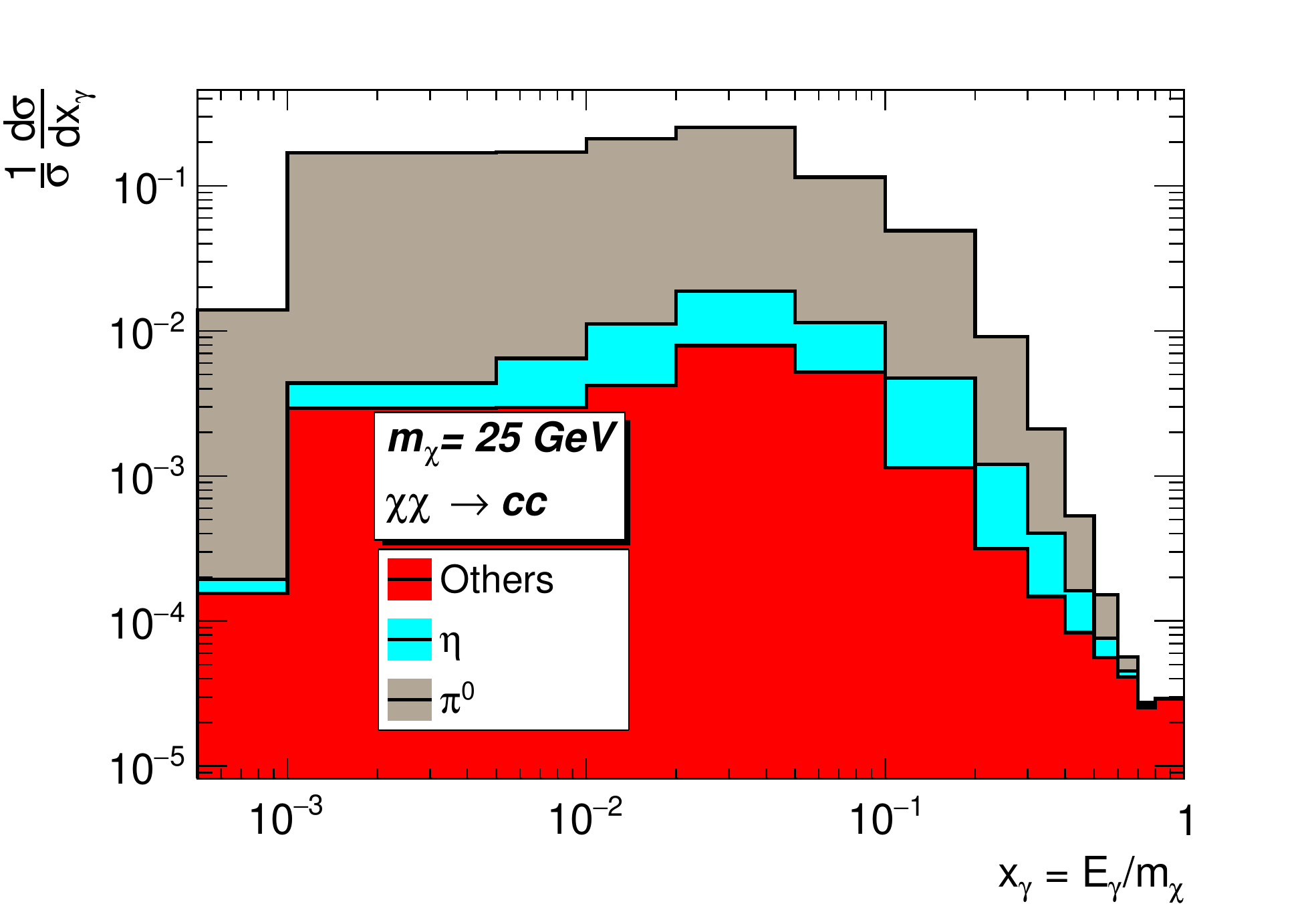}\\
\includegraphics[width=0.41\linewidth]{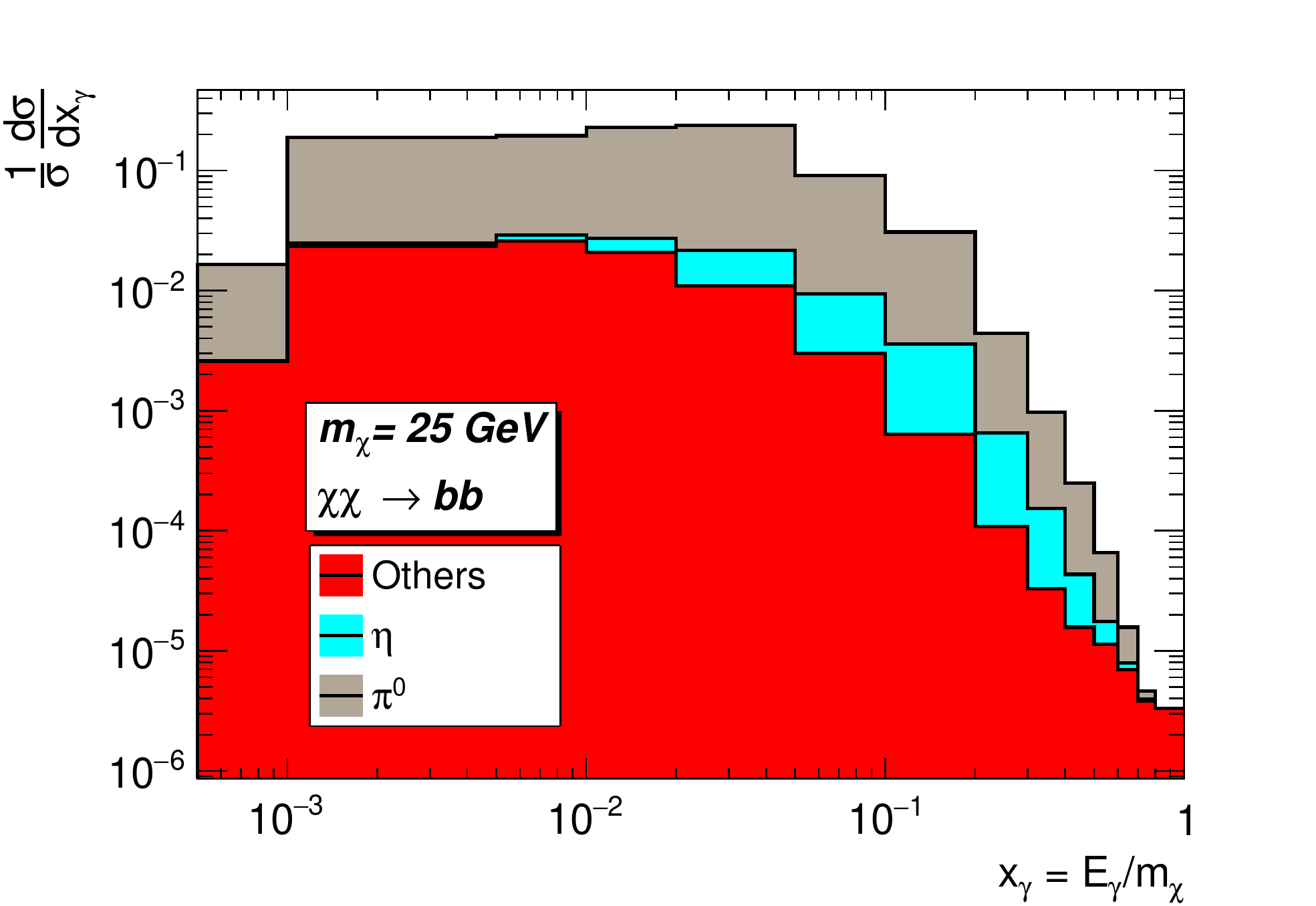}
\includegraphics[width=0.41\linewidth]{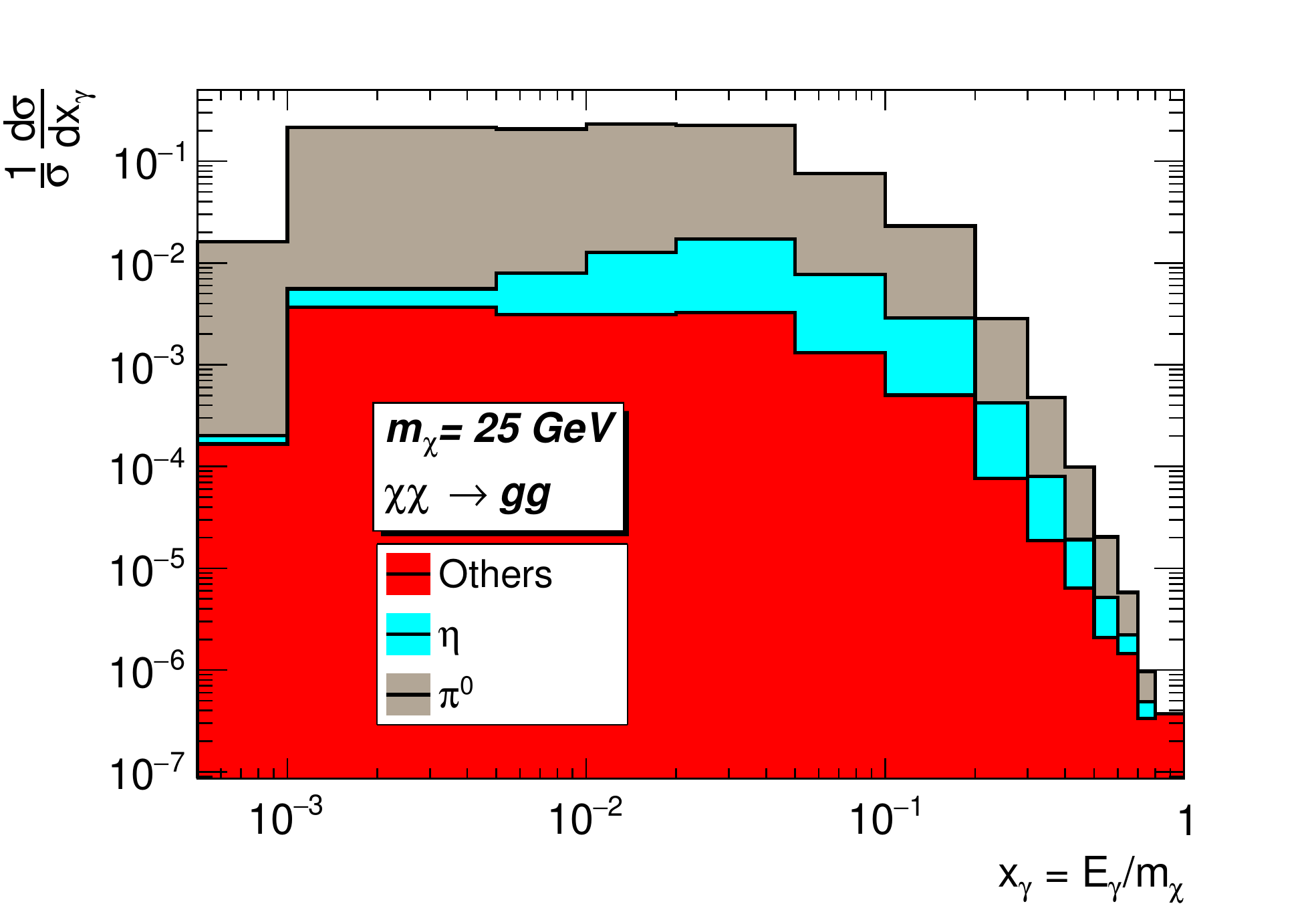}
\caption{$\gamma$ momentum for 
DM annihilation into $q\bar{q}$, $c\bar{c}$, $b\bar{b}$ and $gg$ for $m_\chi=25$ GeV.}
\label{photons-sources-25}
\end{figure}
\begin{figure}[!b]
\centering
 \includegraphics[width=0.41\linewidth]{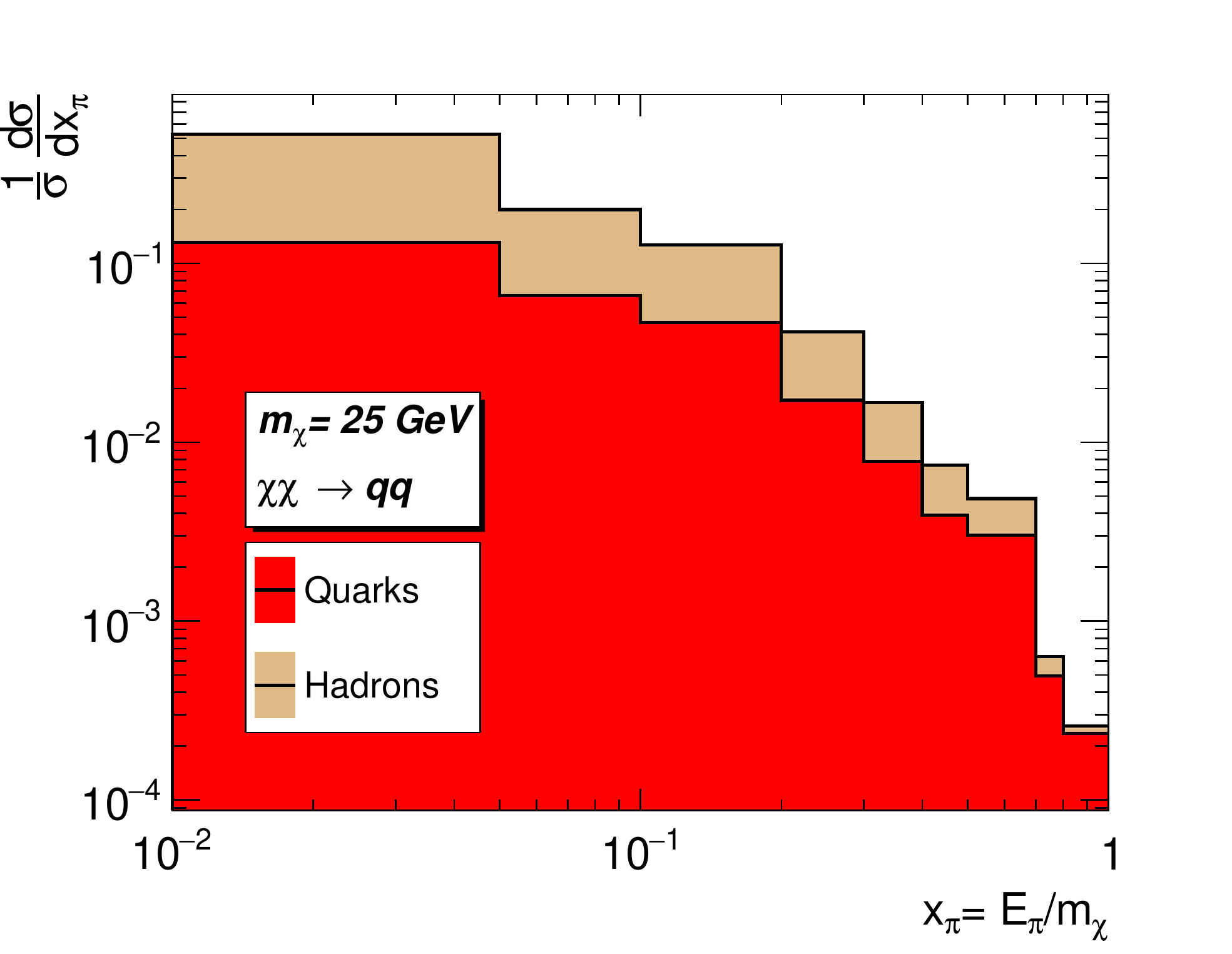}
\includegraphics[width=0.41\linewidth]{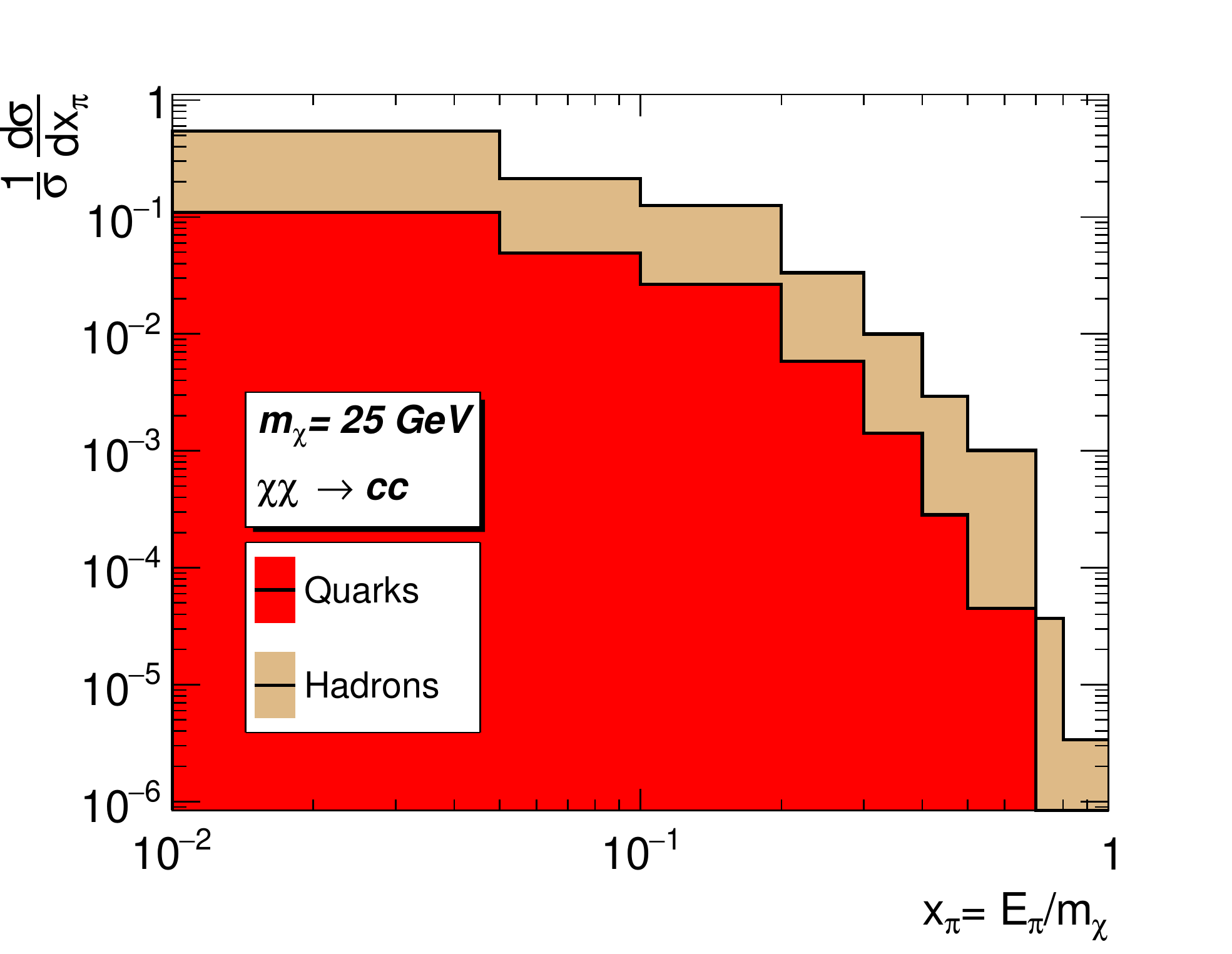}\\
\includegraphics[width=0.41\linewidth]{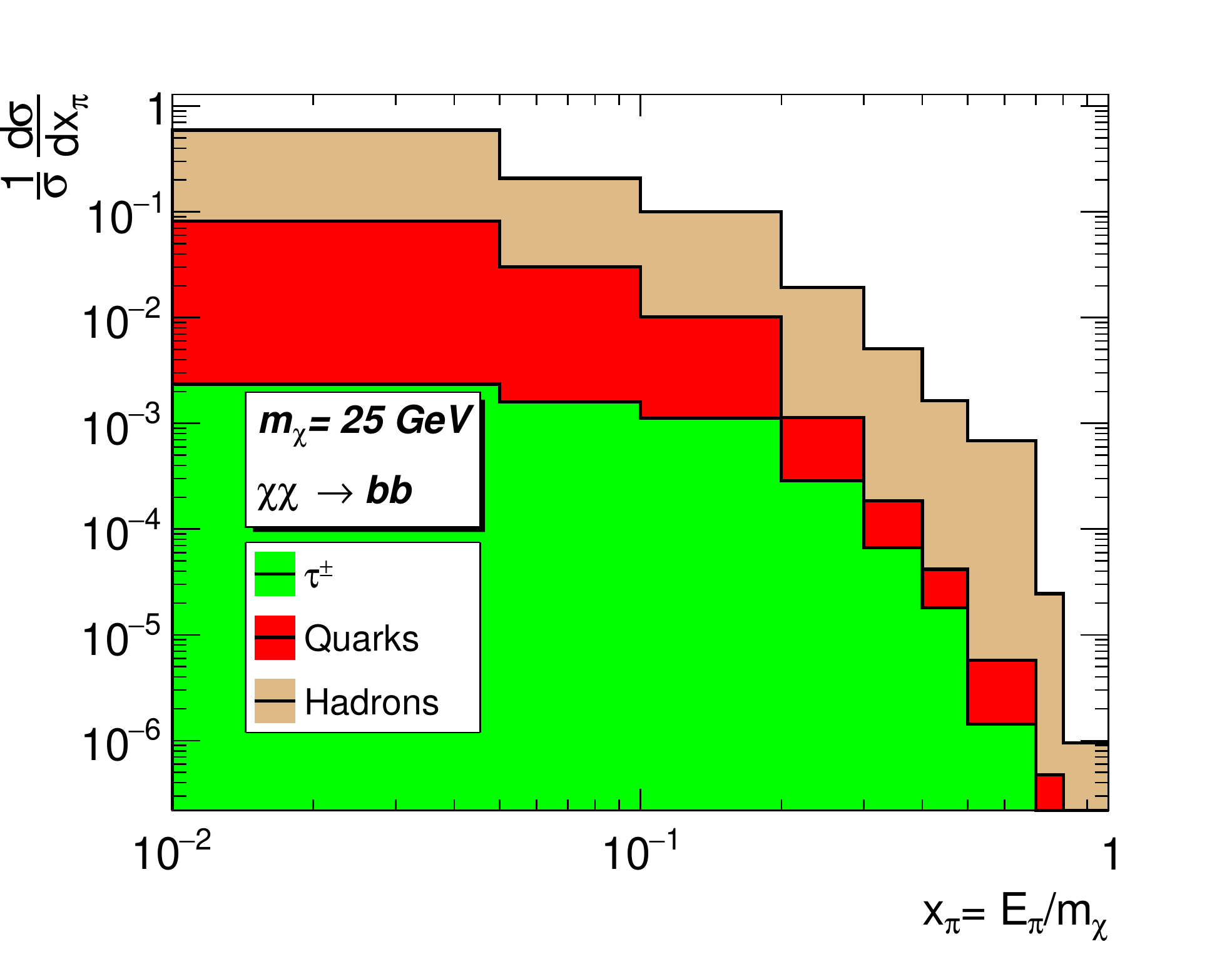}
\includegraphics[width=0.41\linewidth]{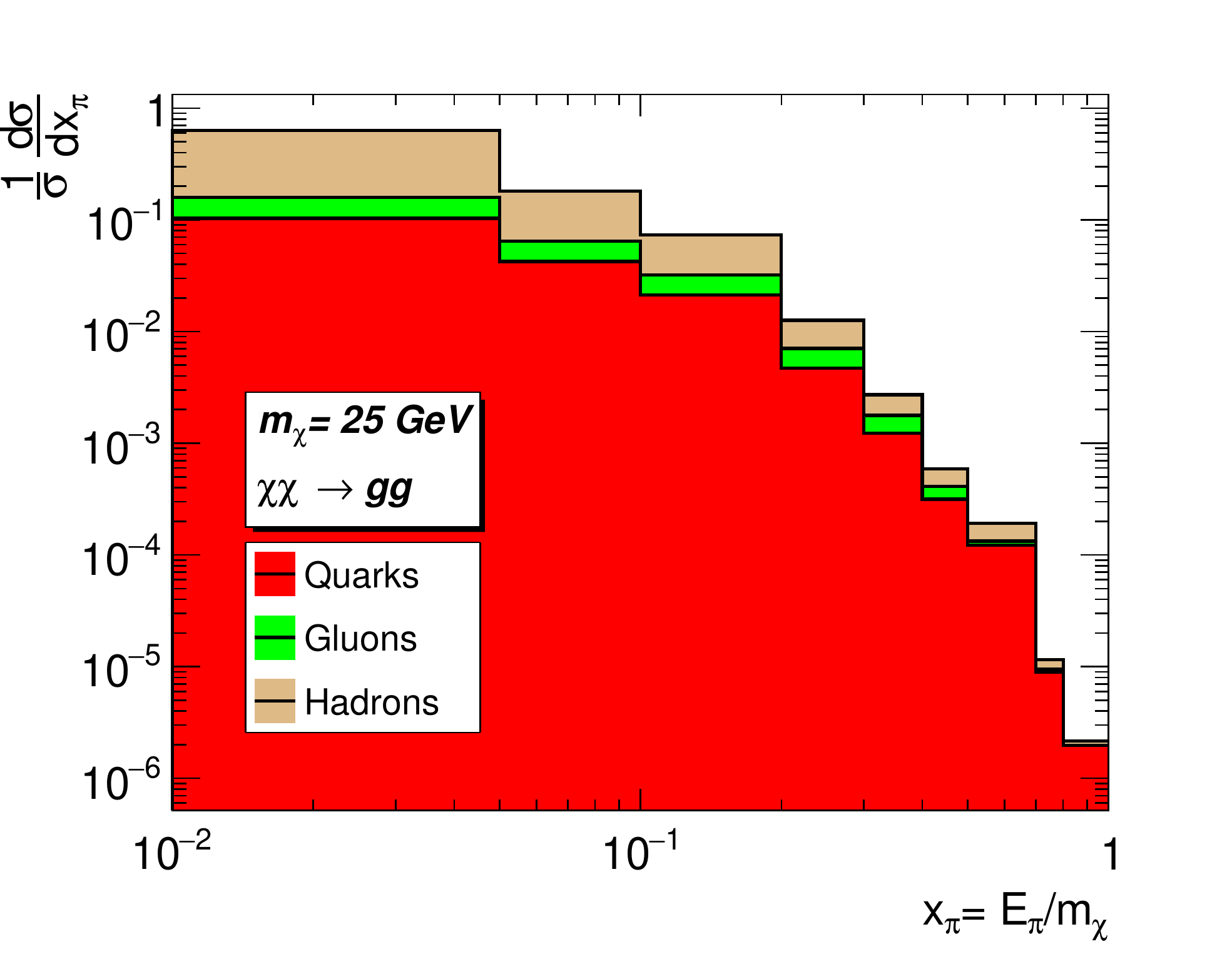}
\caption{$\pi^0$ momentum for 
DM annihilation into $q\bar{q}$, $c\bar{c}$, 
$b\bar{b}$ and $gg$ for $m_\chi=25$ GeV.}
\label{pions-sources-25}
\end{figure}

\begin{figure}[!t]
\centering
 \includegraphics[width=0.41\linewidth]{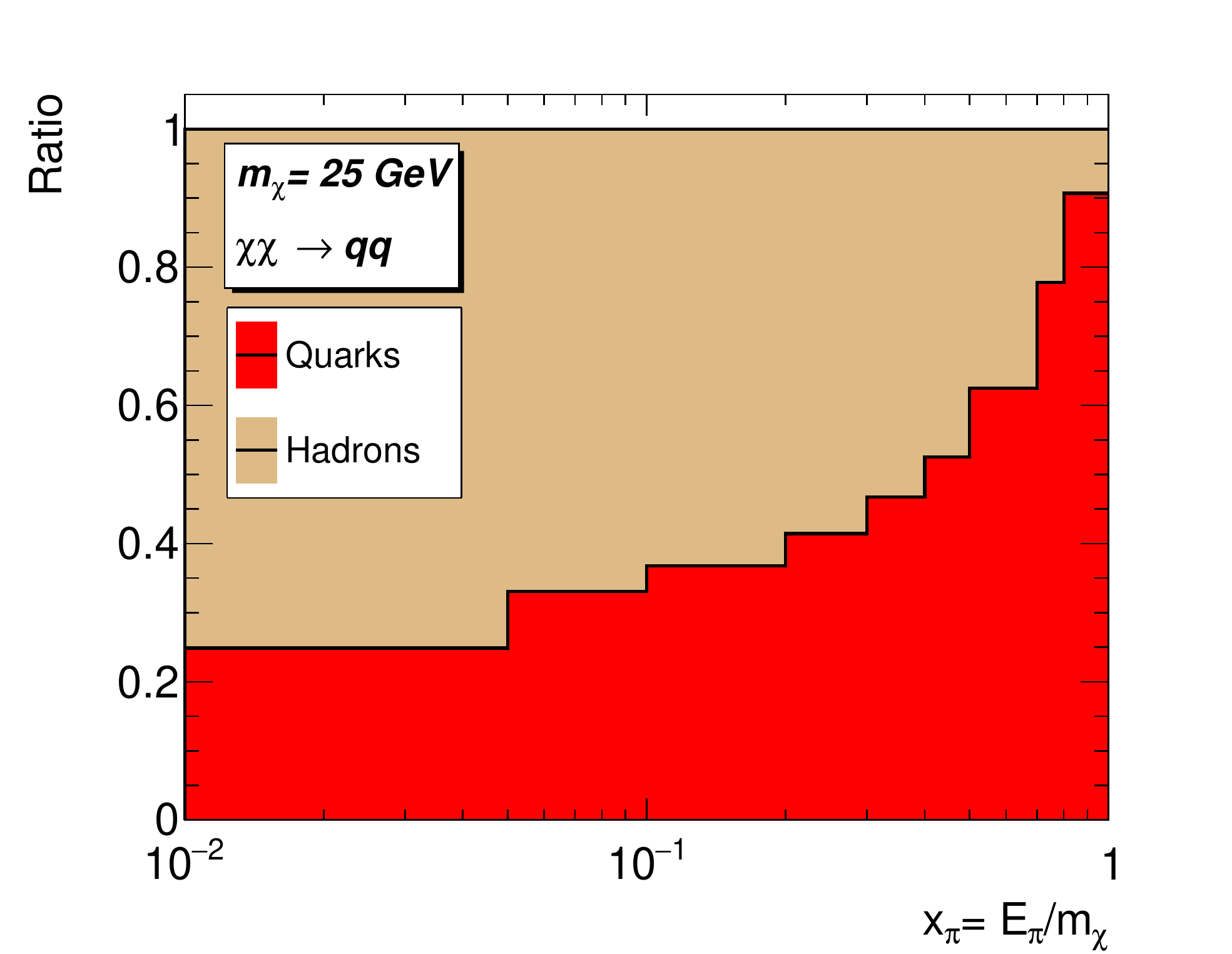}
\includegraphics[width=0.41\linewidth]{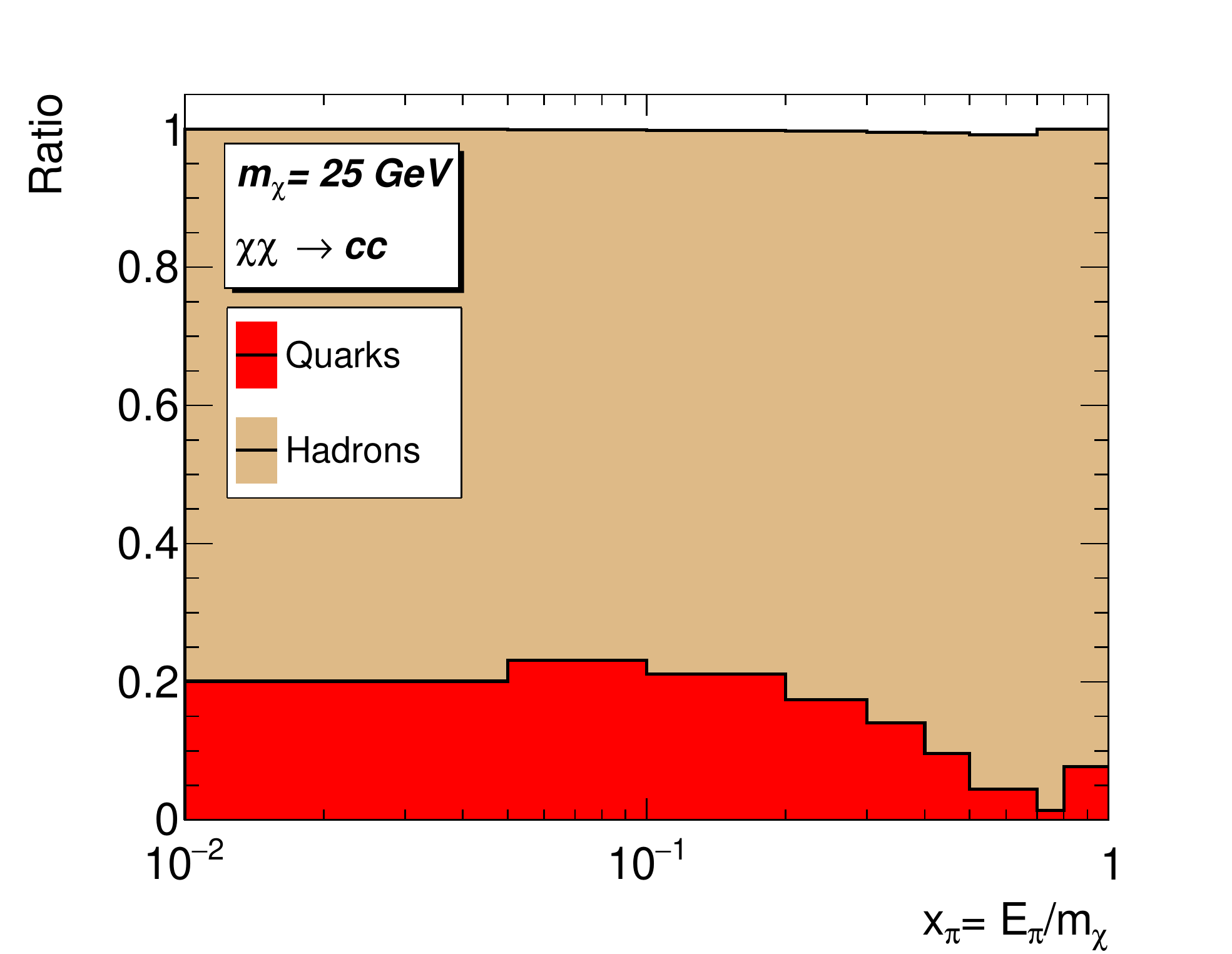}]\\
\includegraphics[width=0.41\linewidth]{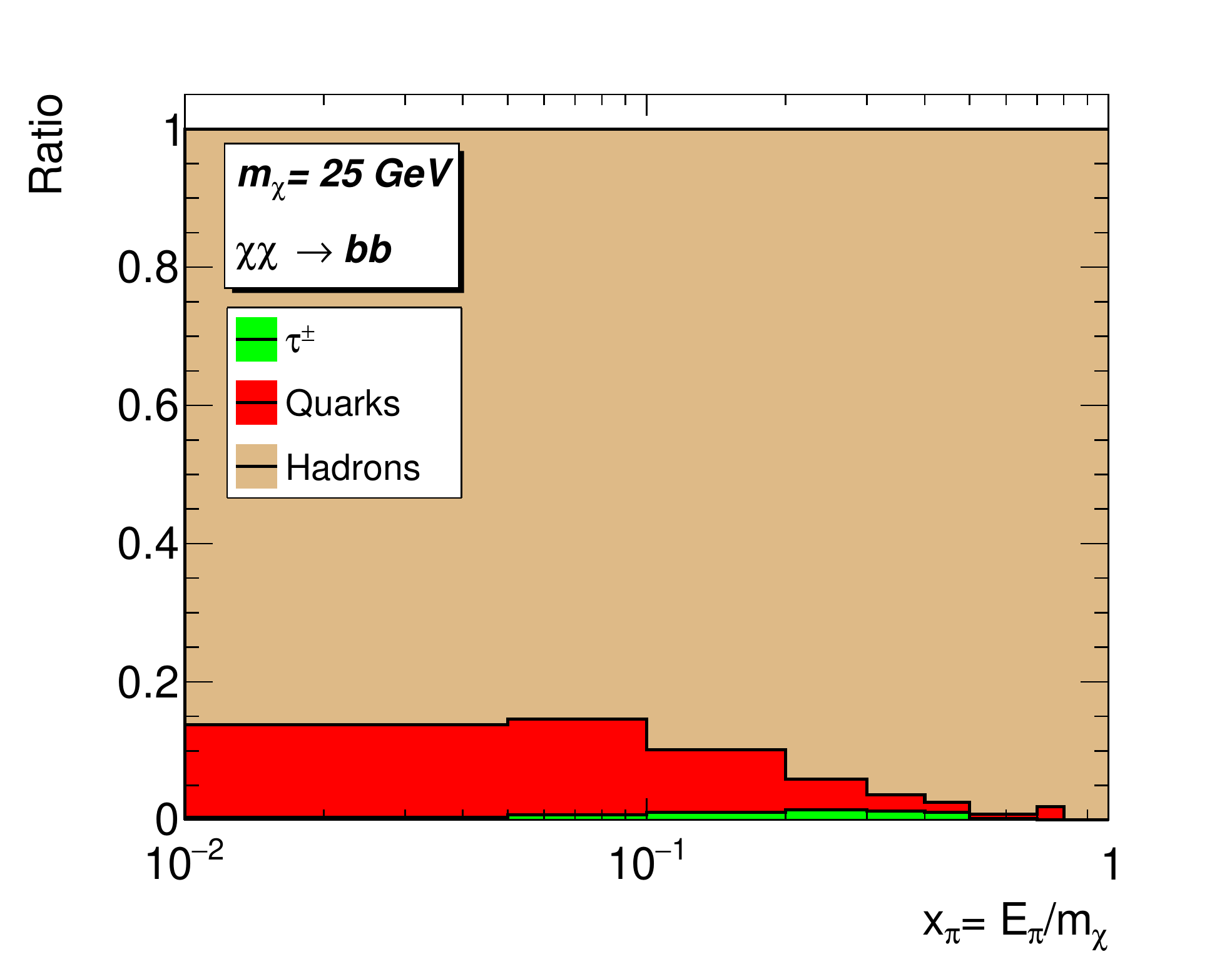}
\includegraphics[width=0.41\linewidth]{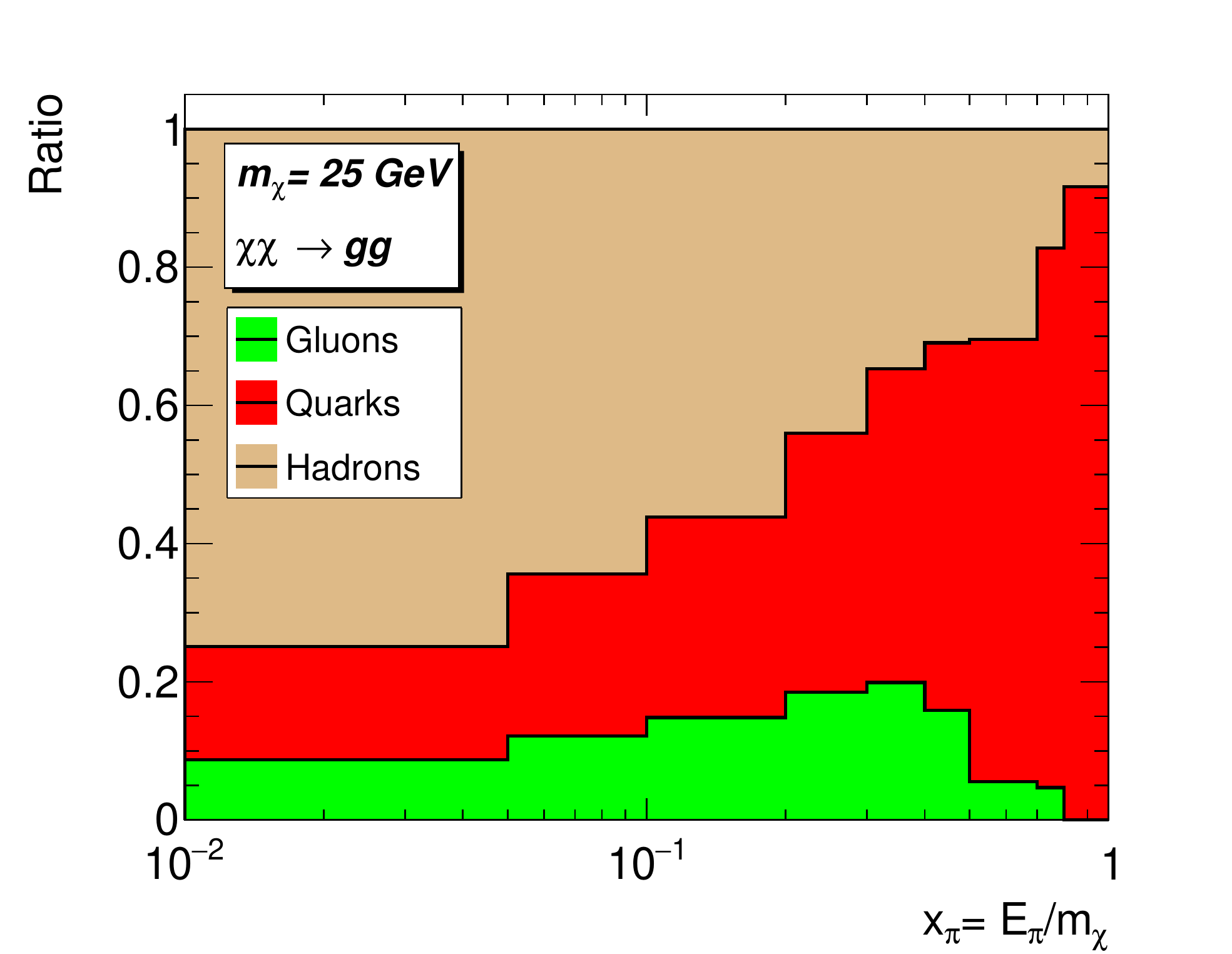}
\caption{Different contributions to the $\pi^0$ scaled momentum for 
dark matter annihilation into $q\bar{q}$ (top left), $c\bar{c}$ (top right), 
$b\bar{b}$ (bottom left) and $gg$ (bottom right) for $m_\chi=25$ GeV.}
\label{pions-sources-25-ratio}
\end{figure}

\begin{figure}[!h]
\centering
 \includegraphics[width=0.41\linewidth]{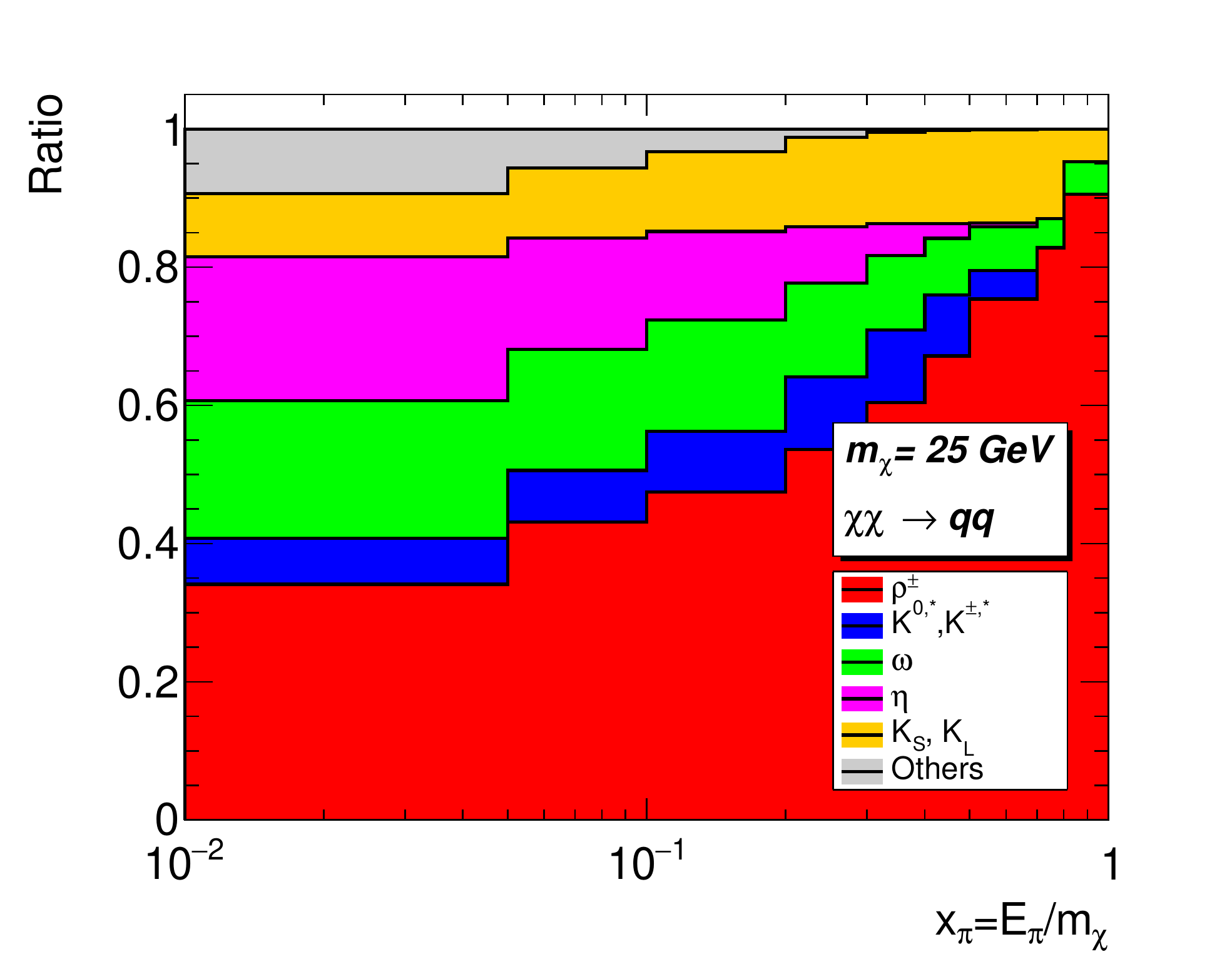}
\includegraphics[width=0.41\linewidth]{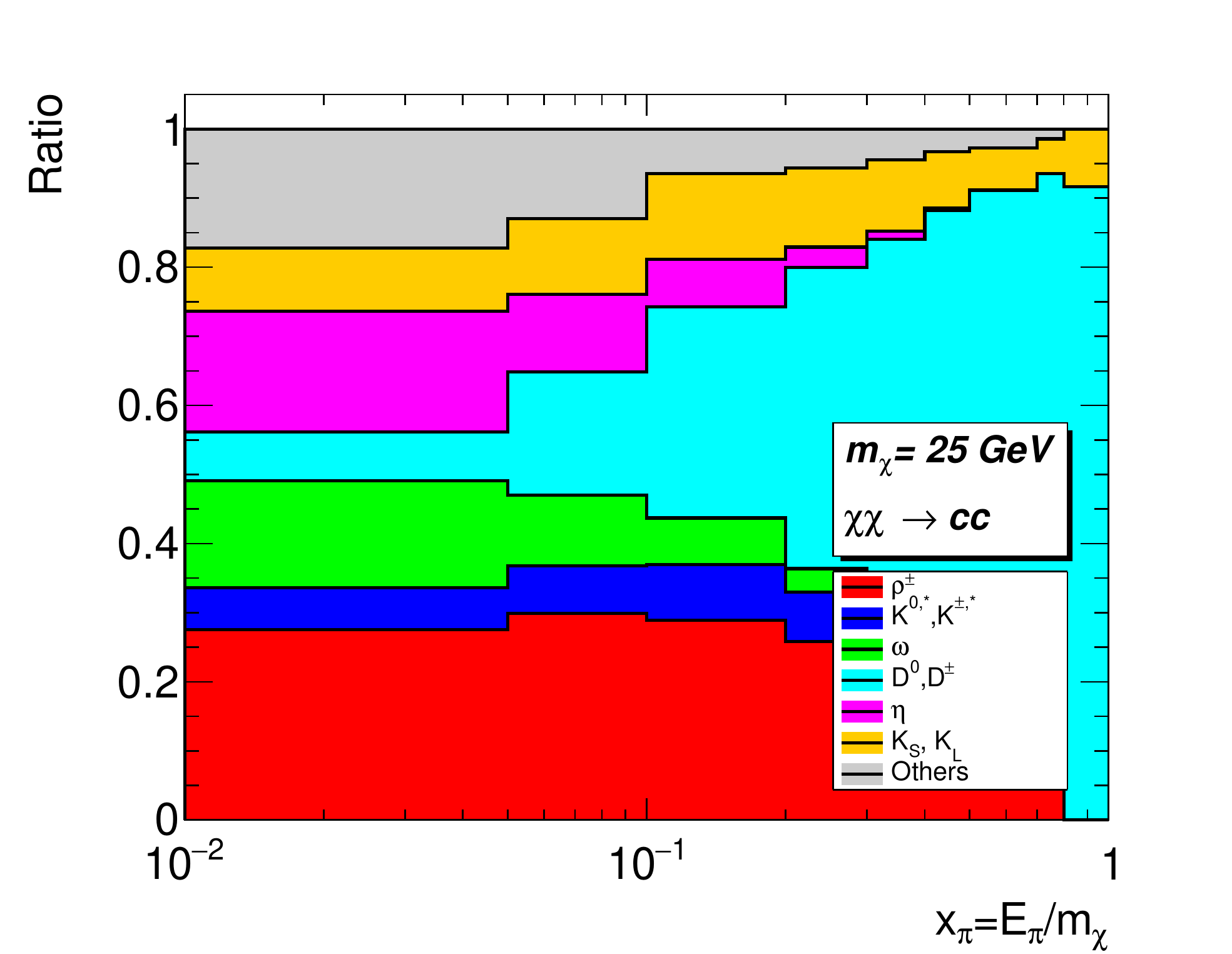}\\
\includegraphics[width=0.41\linewidth]{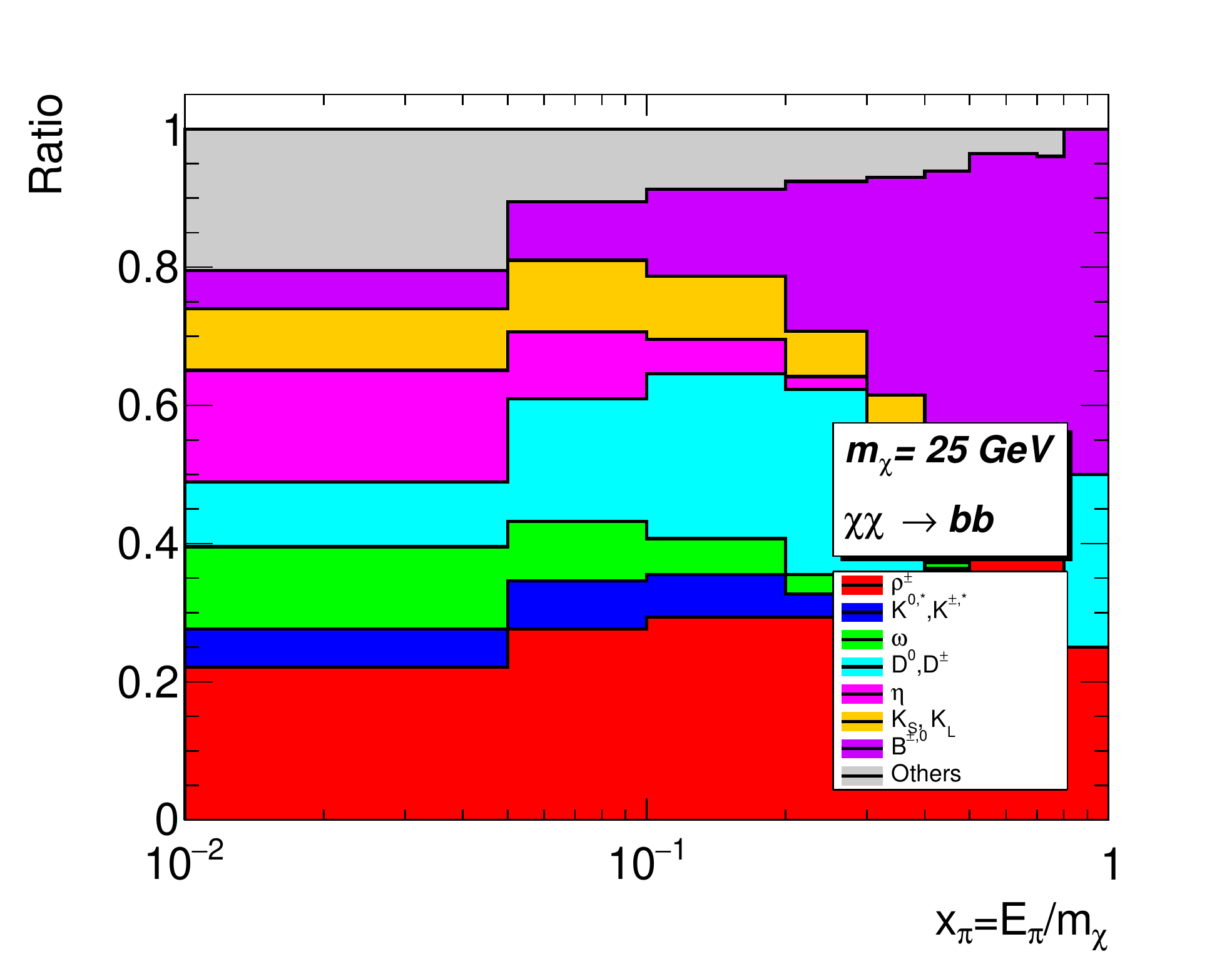}
\includegraphics[width=0.41\linewidth]{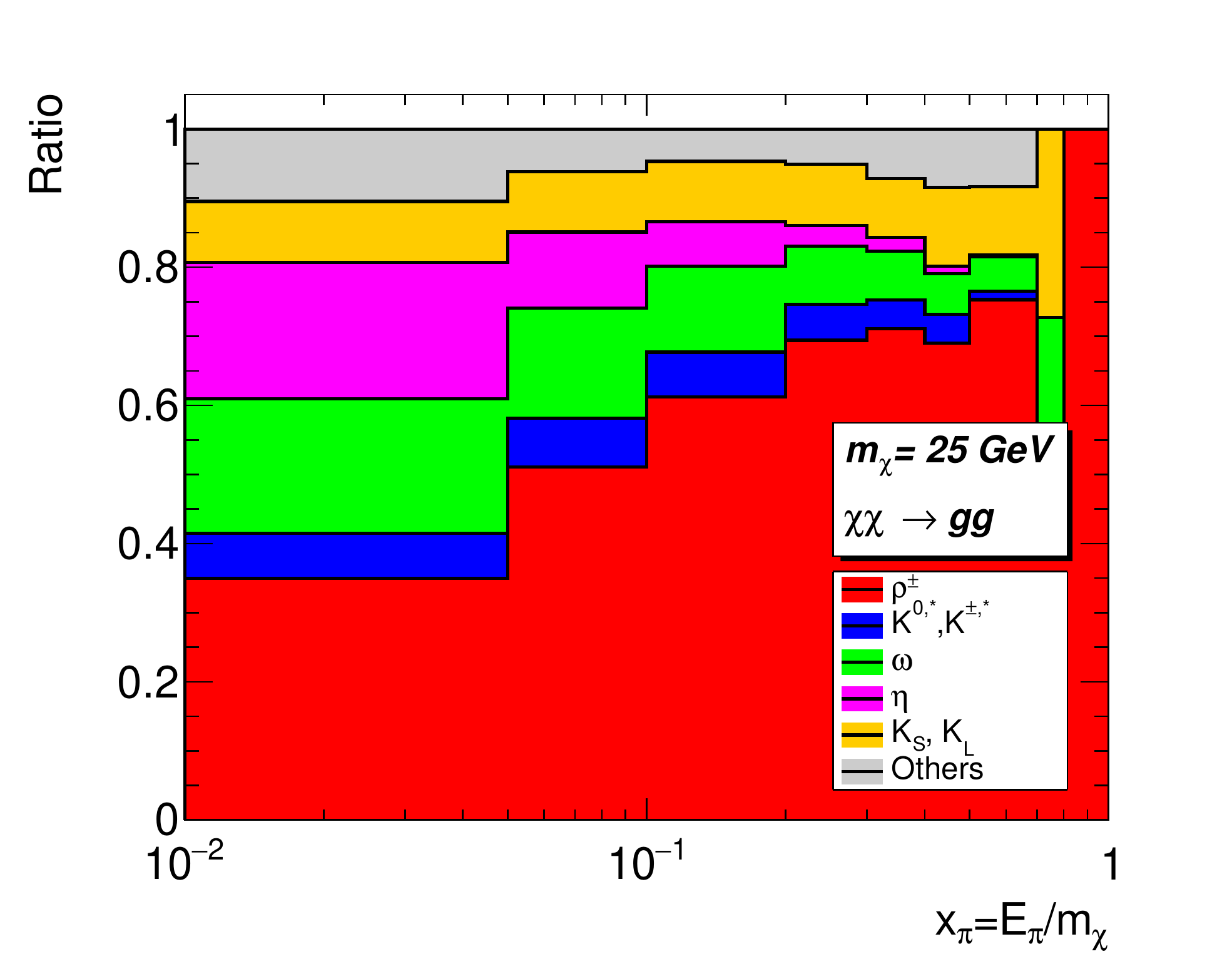}
\caption{Same as Fig. \ref{pions-sources-25-ratio} but for pions coming from hadrons}
\label{pions-sources-25-ratio-hadrons}
\end{figure}

\begin{figure}[!t]
\centering
\includegraphics[width=0.41\linewidth]{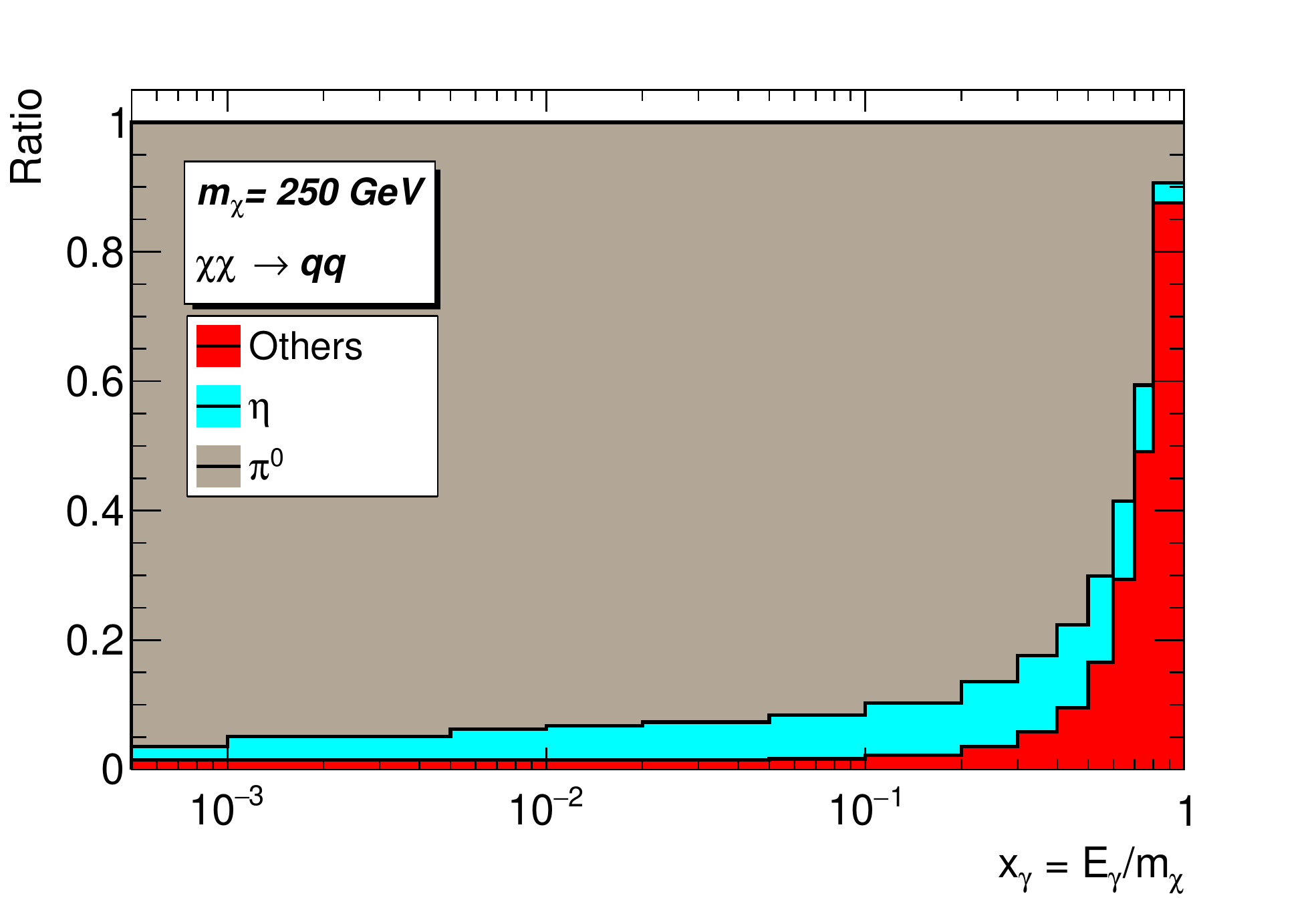}
\includegraphics[width=0.41\linewidth]{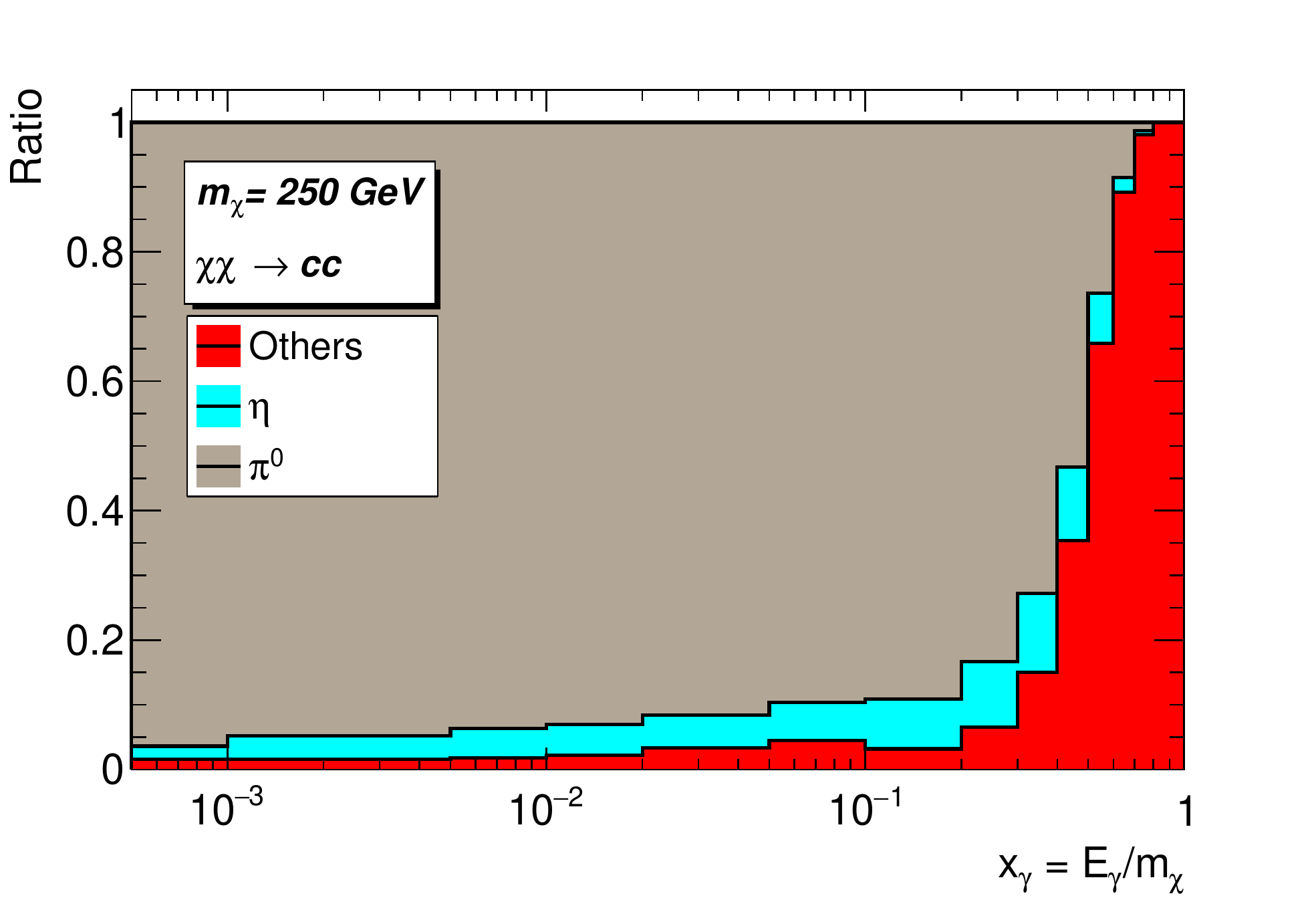}\\
\includegraphics[width=0.41\linewidth]{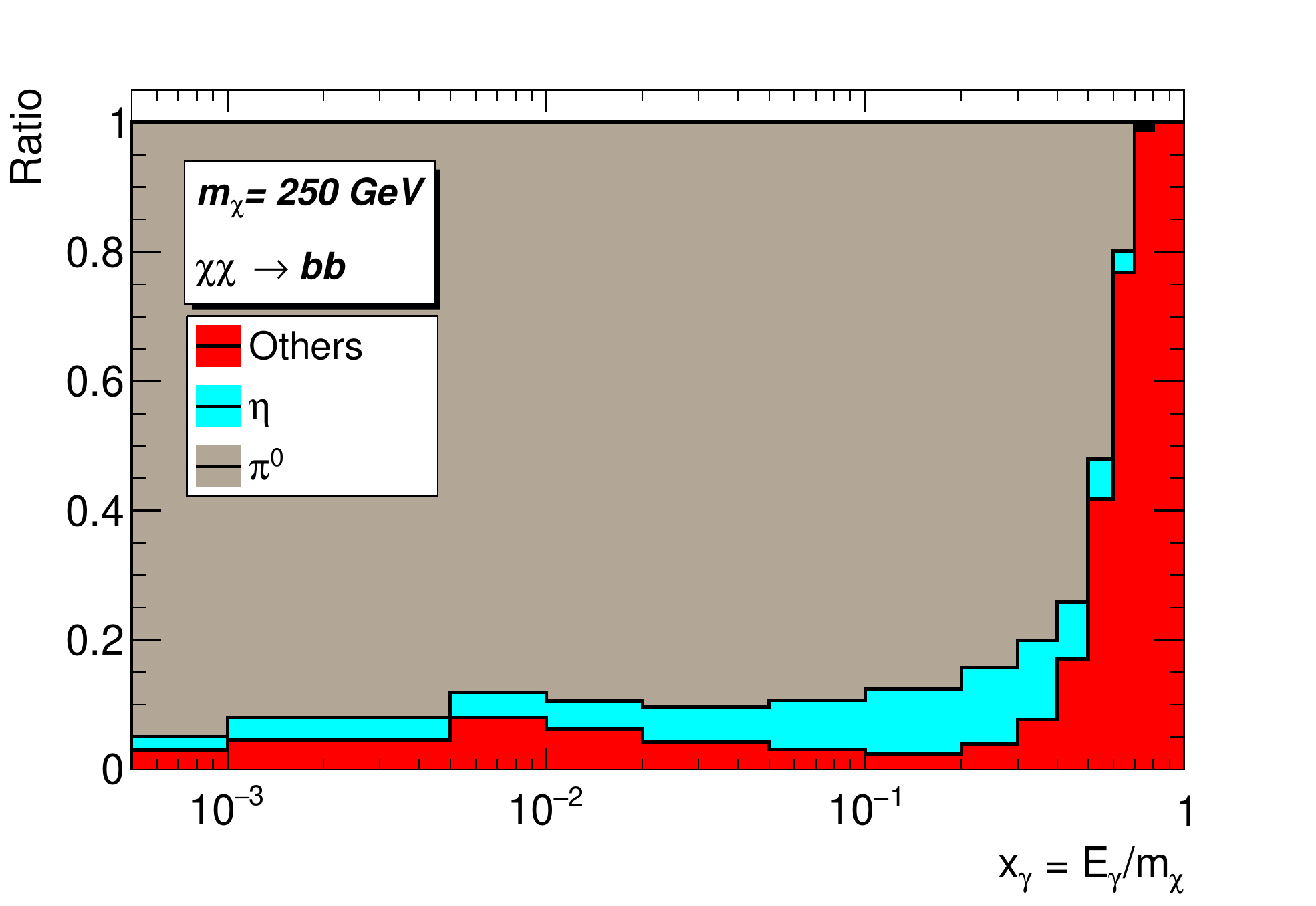}
\includegraphics[width=0.41\linewidth]{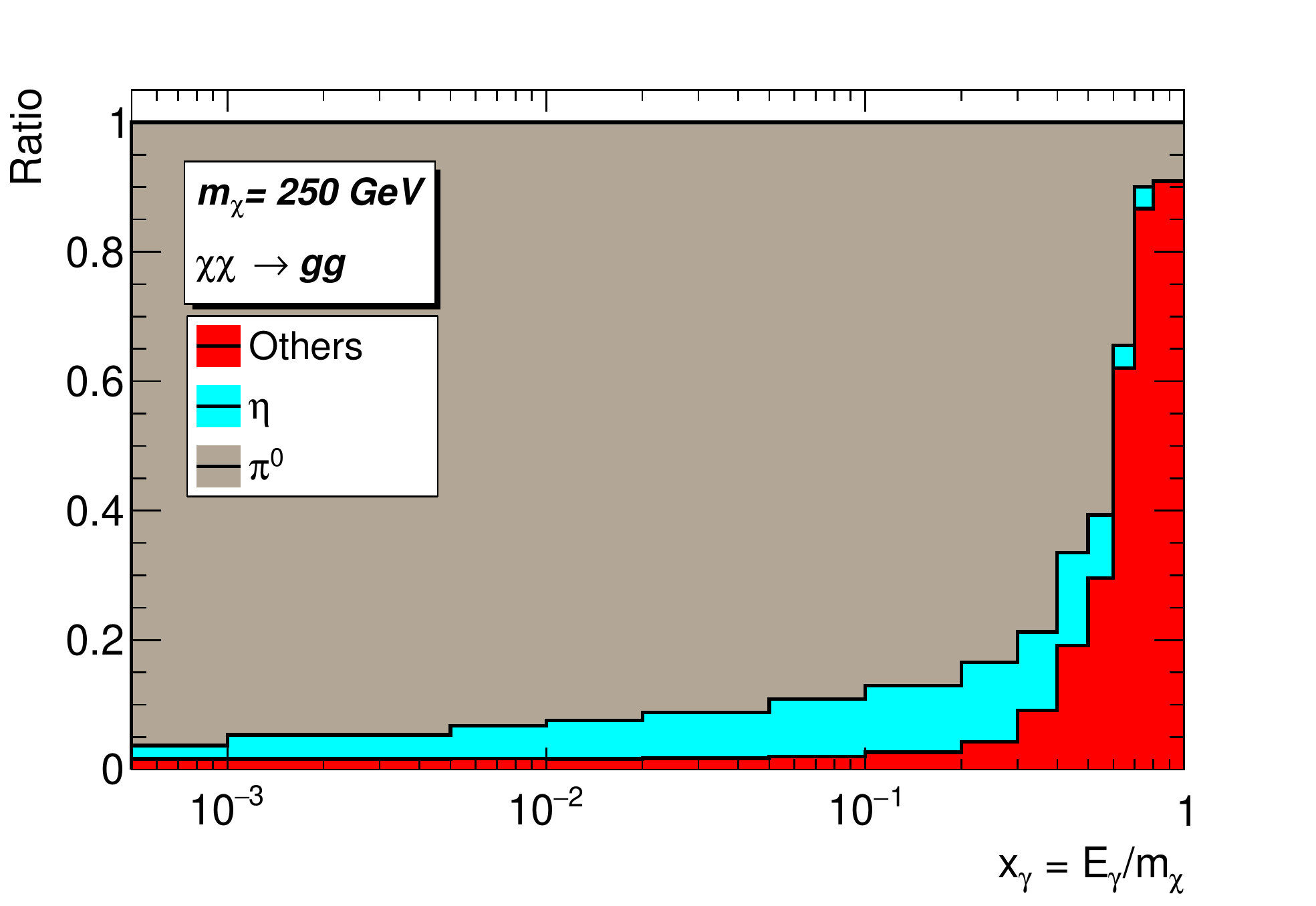}
\caption{Contribution to $\gamma$-spectrum for  
dark matter annihilation into $q\bar{q}$ (top left), $c\bar{c}$ (top right), 
$b\bar{b}$ (bottom left) and $gg$ (bottom right) with $m_\chi=250$ GeV. The distribution
is normalized to the total number of photons.}
\label{photons-sources-250-light-ratio}
\end{figure}

\begin{figure}[!h]
\centering
\includegraphics[width=0.41\linewidth]{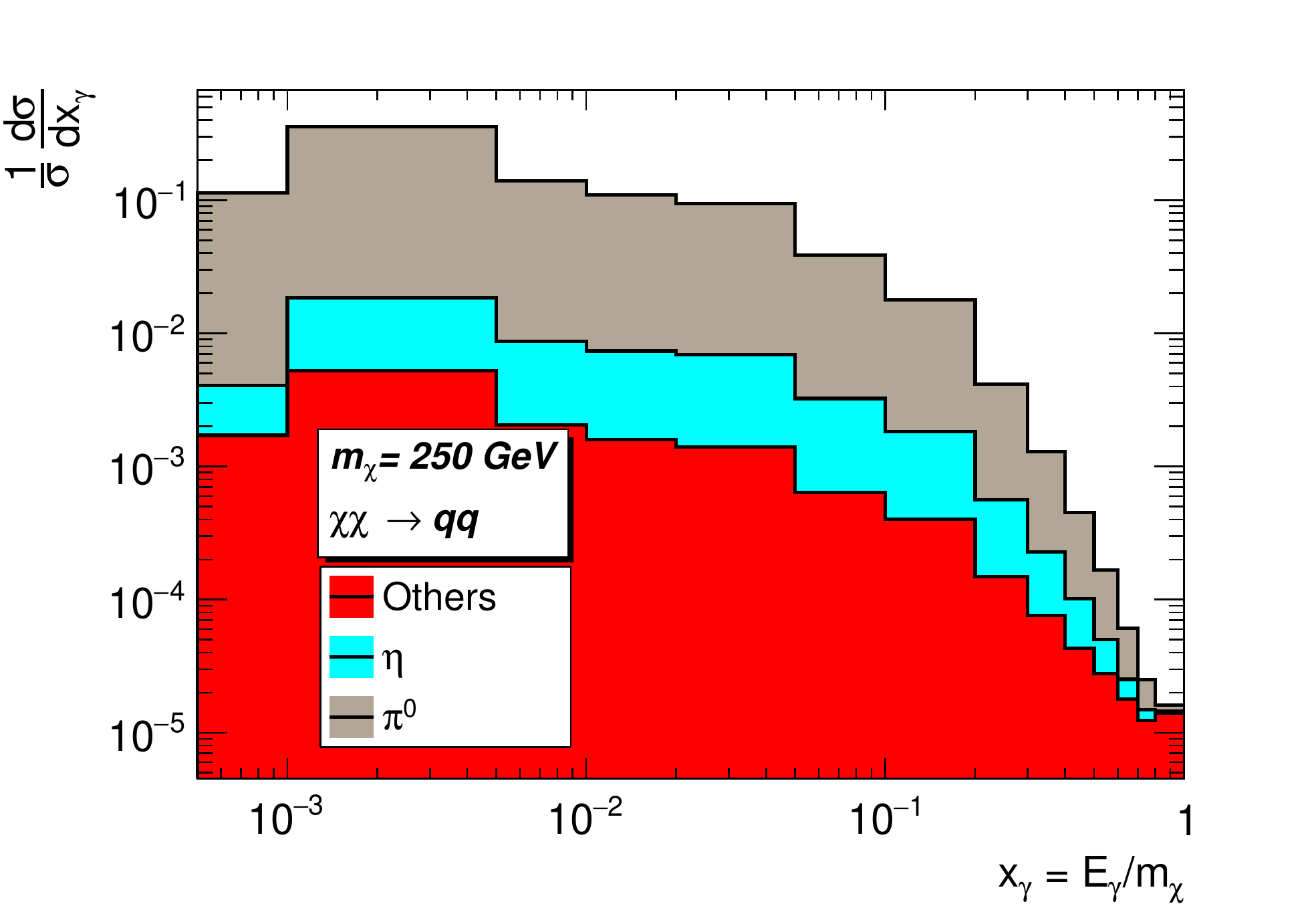}
\includegraphics[width=0.41\linewidth]{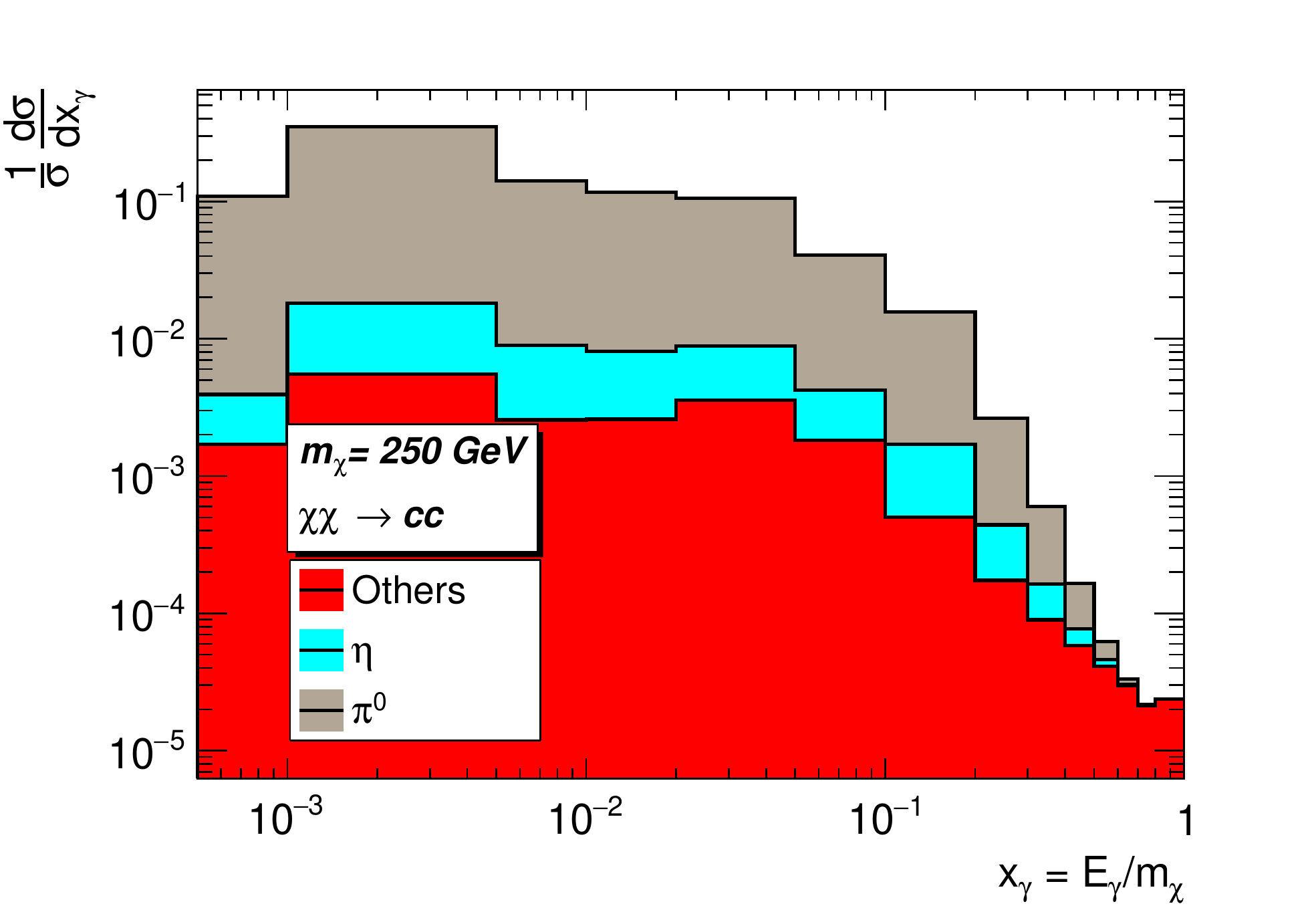}\\
\includegraphics[width=0.41\linewidth]{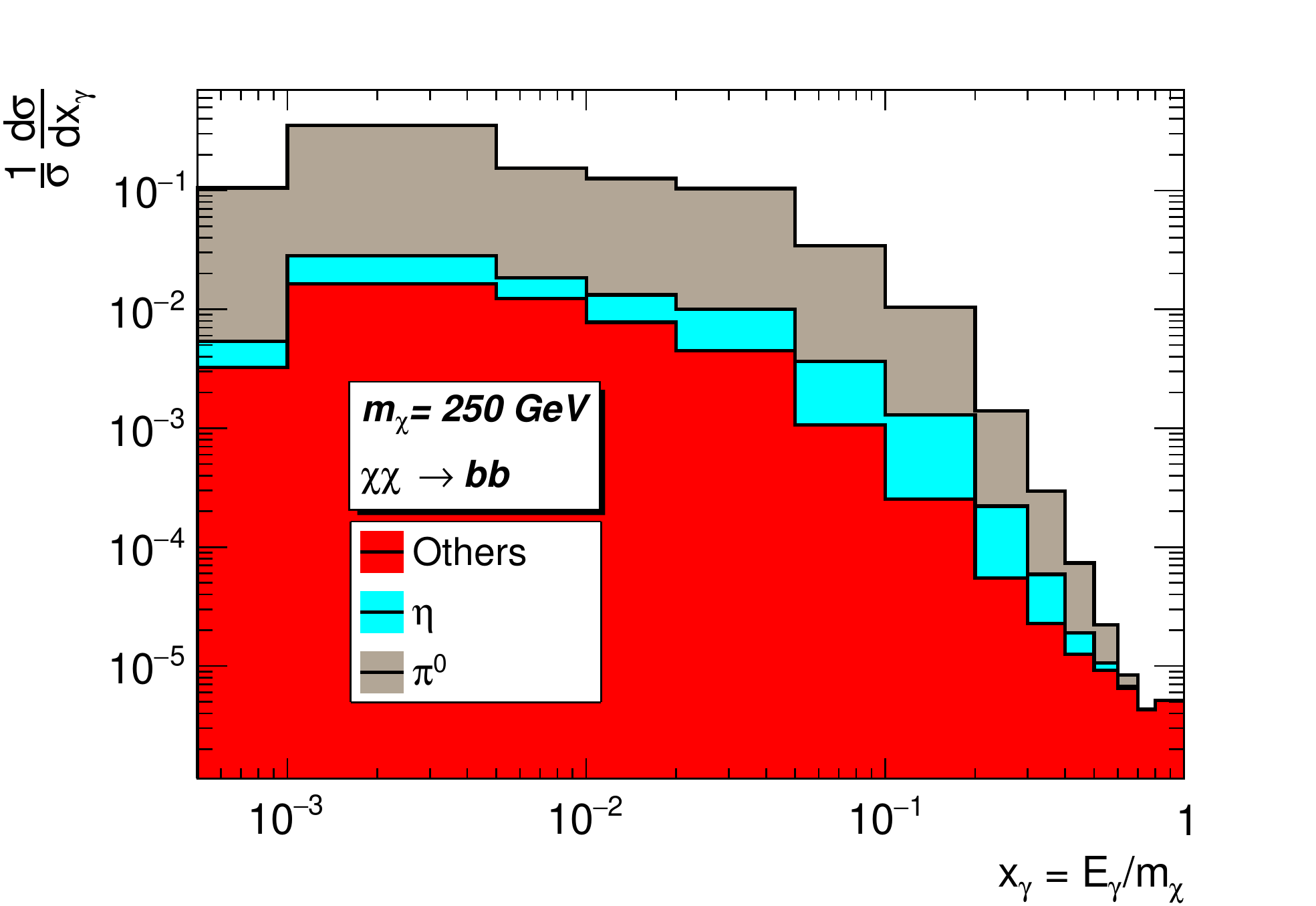}
\includegraphics[width=0.41\linewidth]{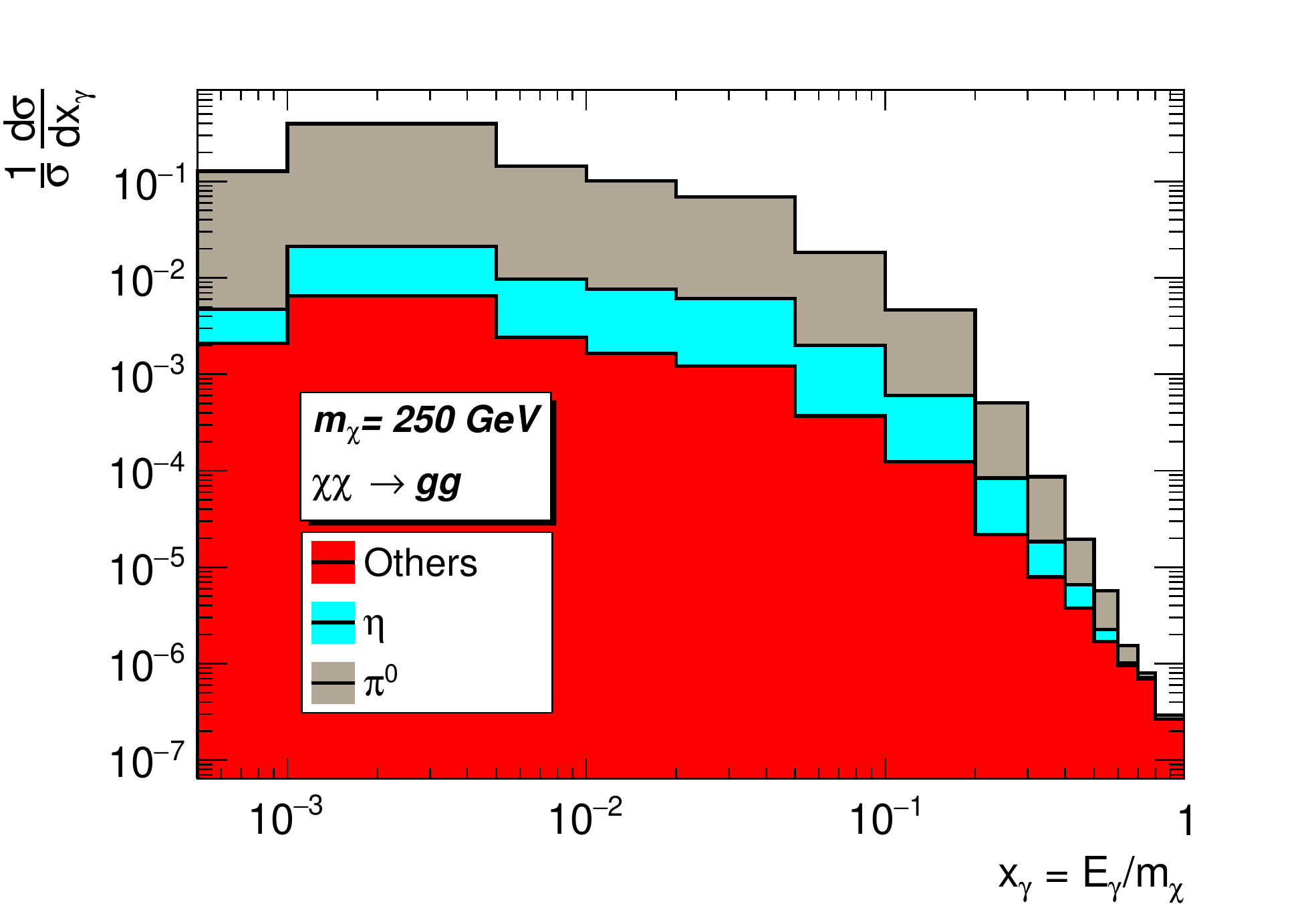}
\caption{Same as Fig. \ref{photons-sources-250-light-ratio} but with 
the differential distribution is shown.}
\label{photons-sources-250-light}
\end{figure}

\begin{figure}[!t]
\centering
\includegraphics[width=0.41\linewidth]{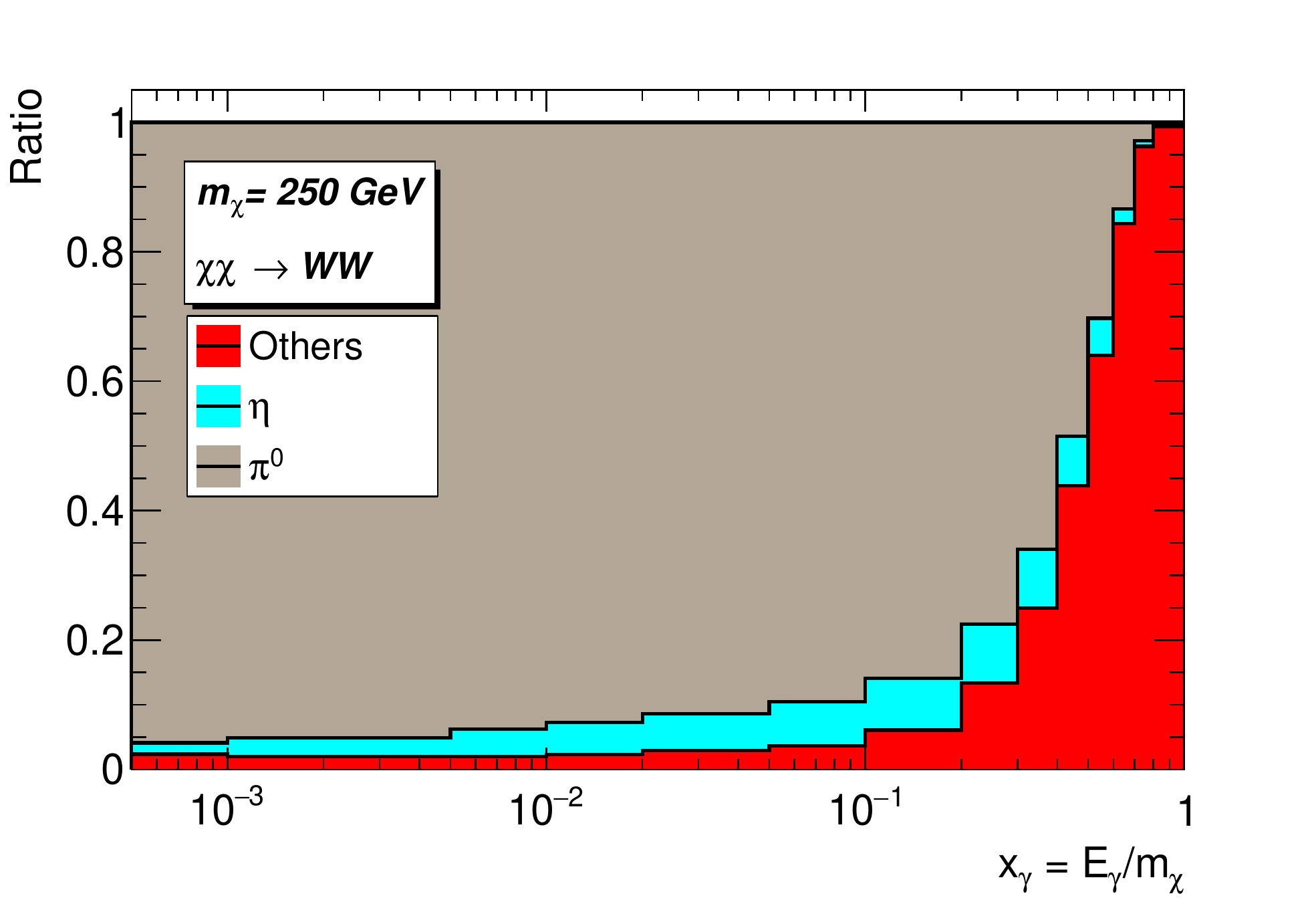}
\includegraphics[width=0.41\linewidth]{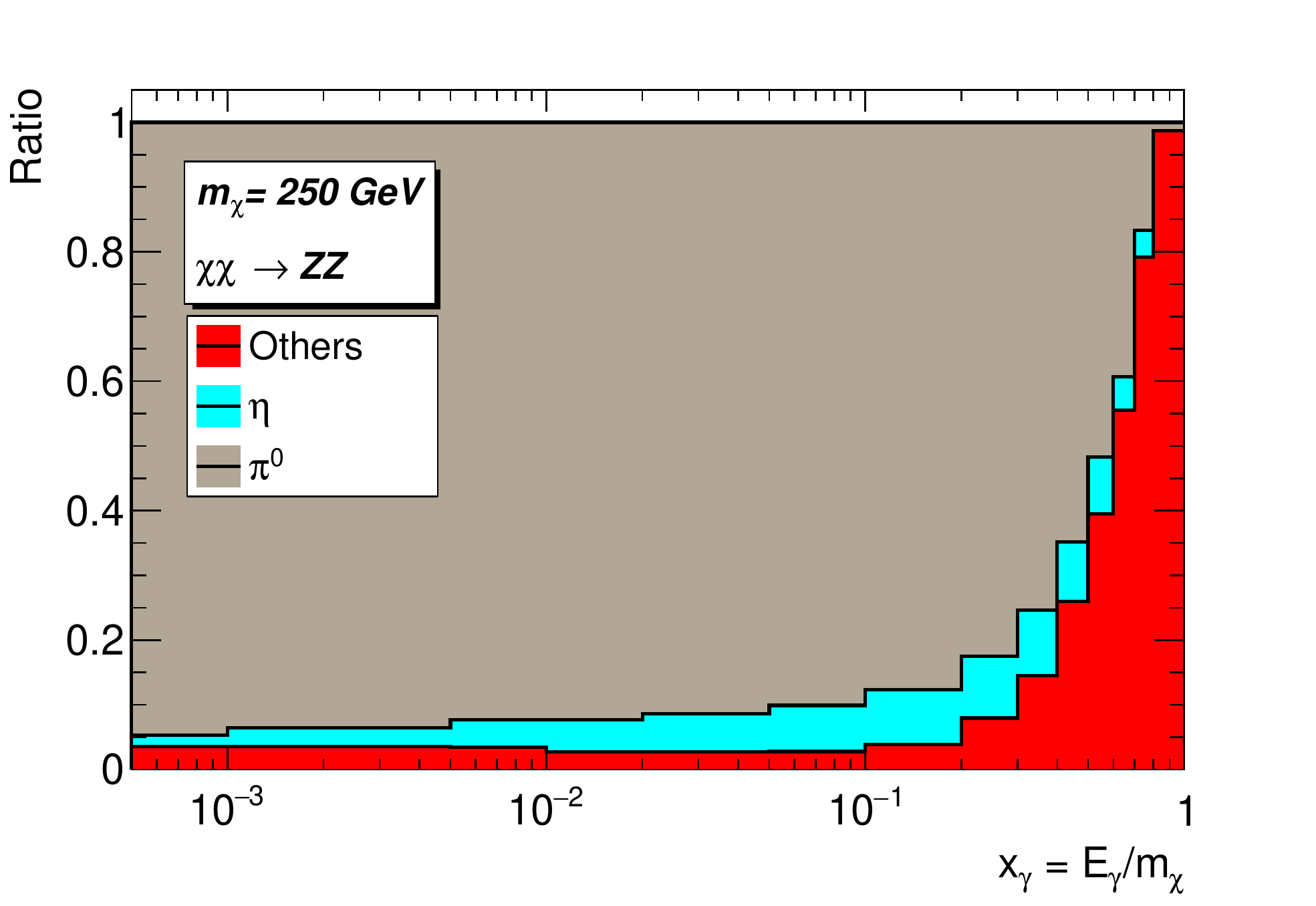}\\
\includegraphics[width=0.41\linewidth]{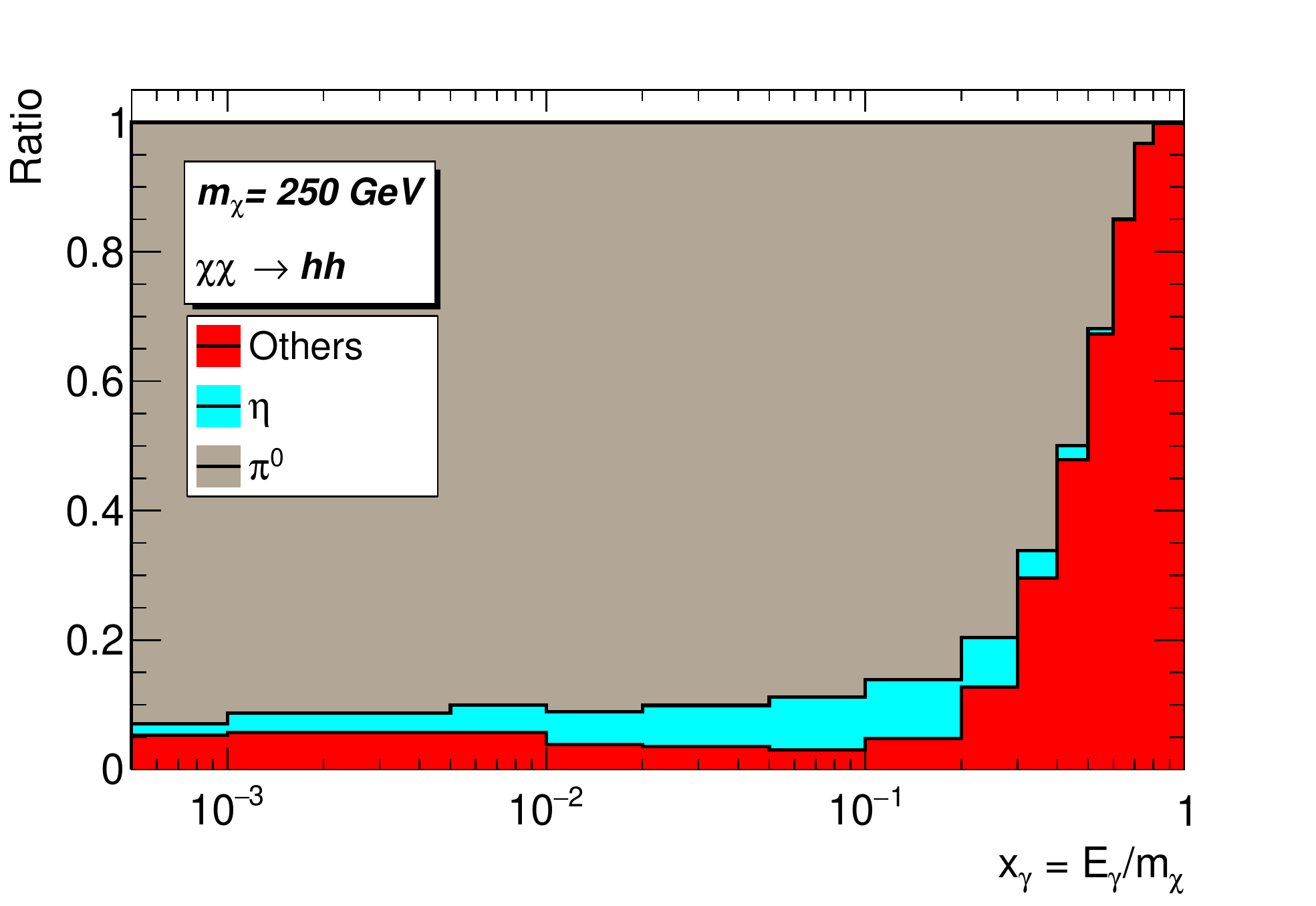}
\includegraphics[width=0.41\linewidth]{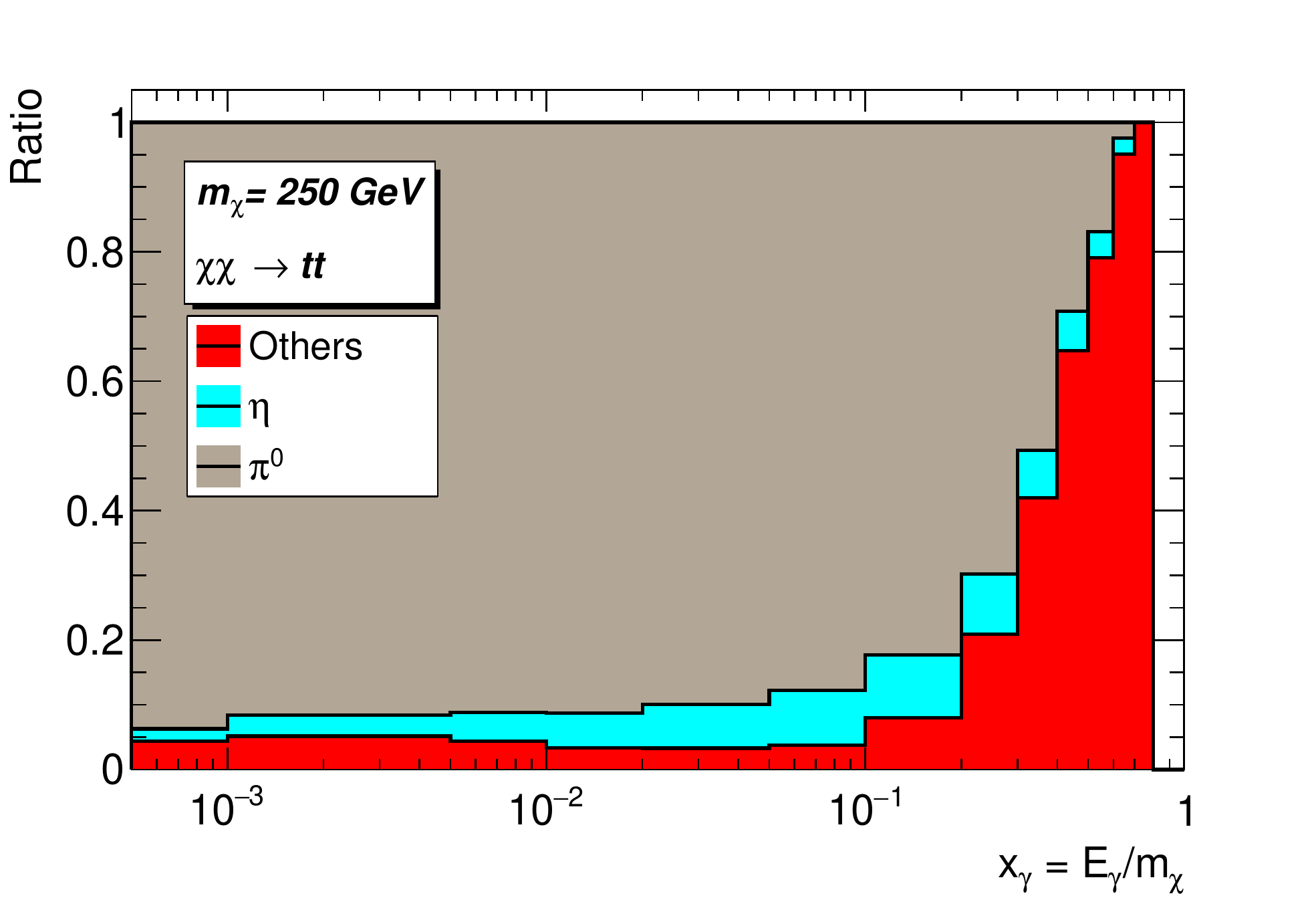}
\caption{Same as Fig. \ref{photons-sources-250-light-ratio} but for 
$WW$ (top left), $ZZ$ (top right), $hh$ (bottom left) and $t\bar{t}$ (bottom right)
final states.}
\label{photons-sources-250-ratio}
\end{figure}

\begin{figure}[!h]
\centering
\includegraphics[width=0.41\linewidth]{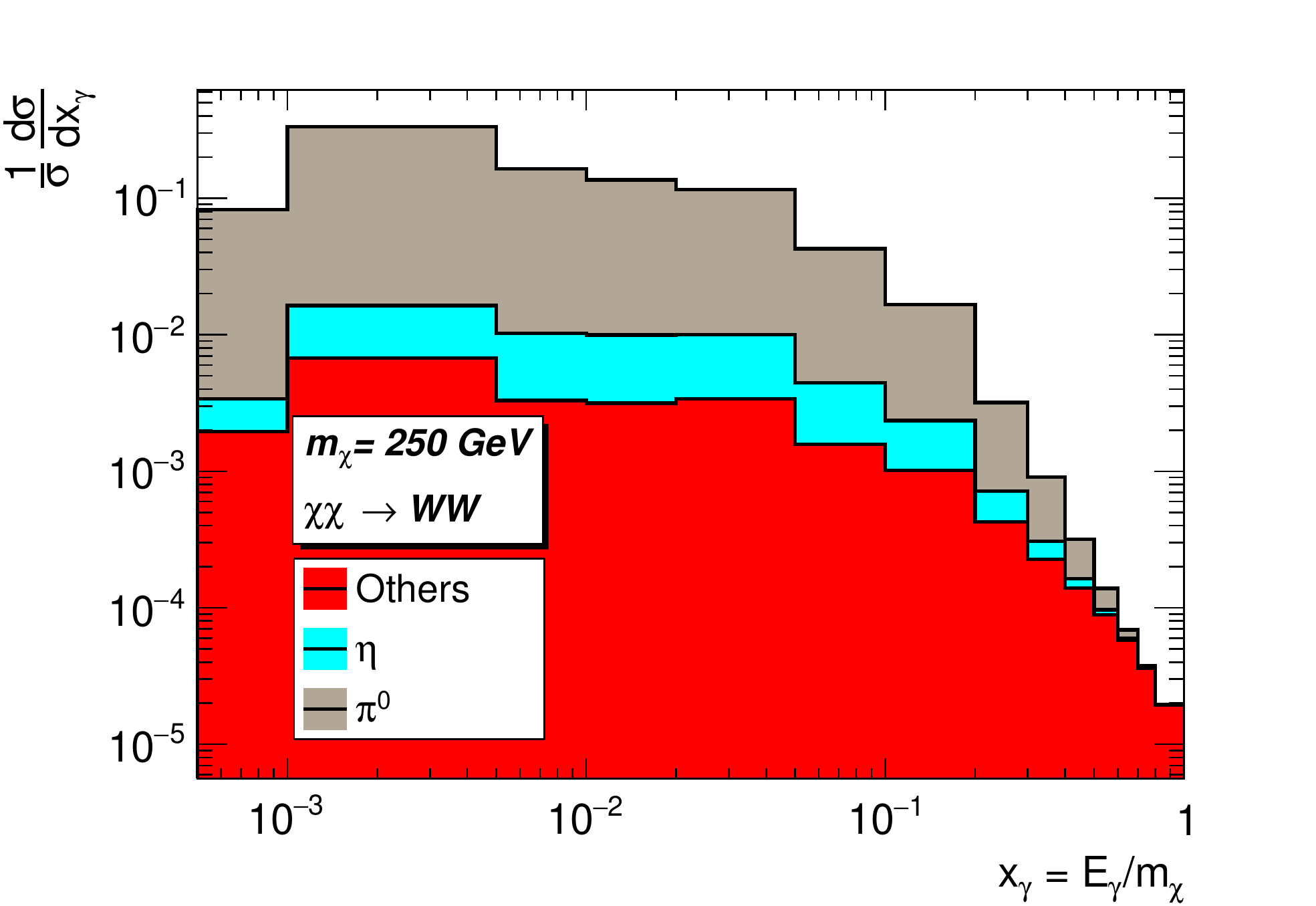}
\includegraphics[width=0.41\linewidth]{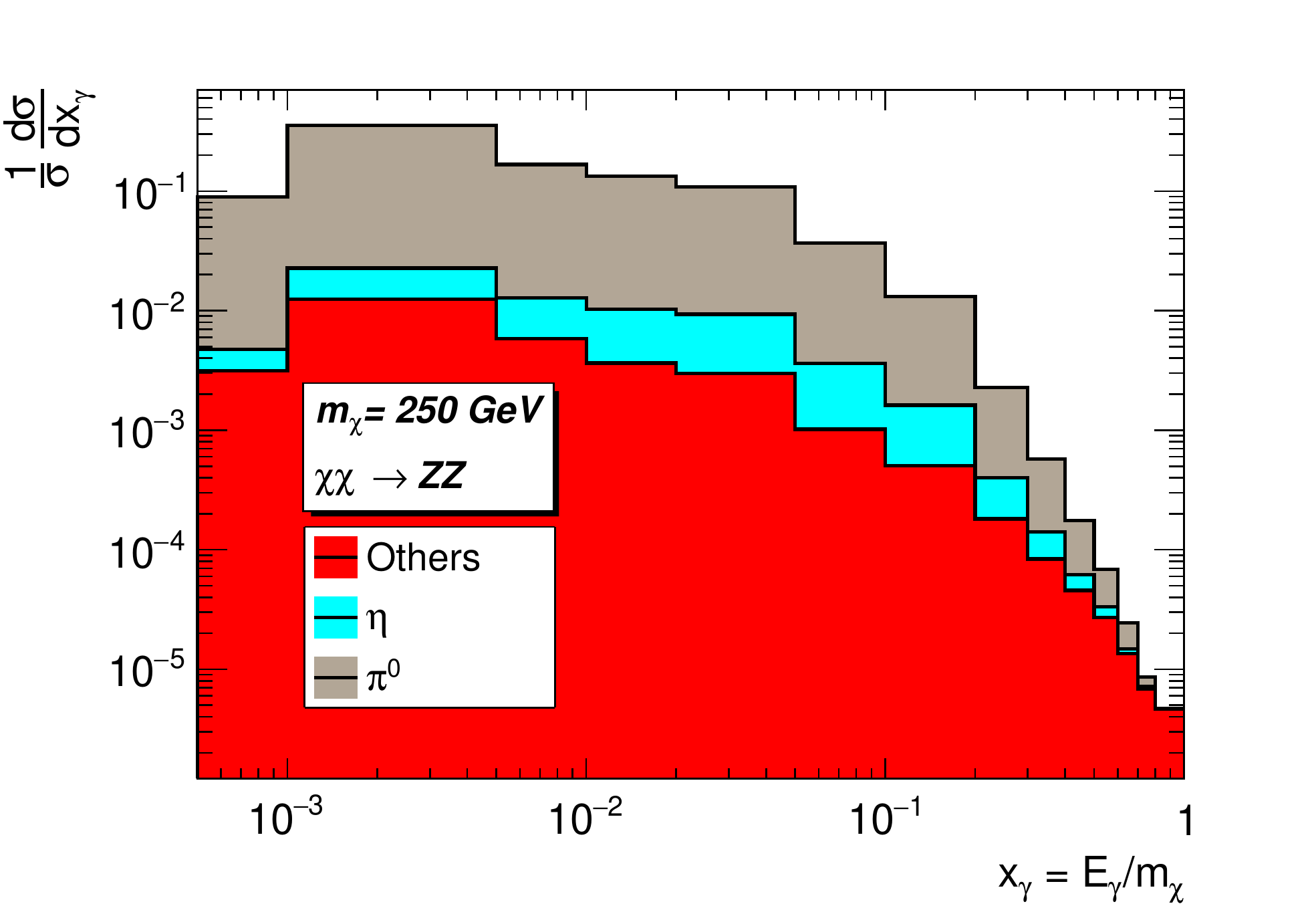}\\
\includegraphics[width=0.41\linewidth]{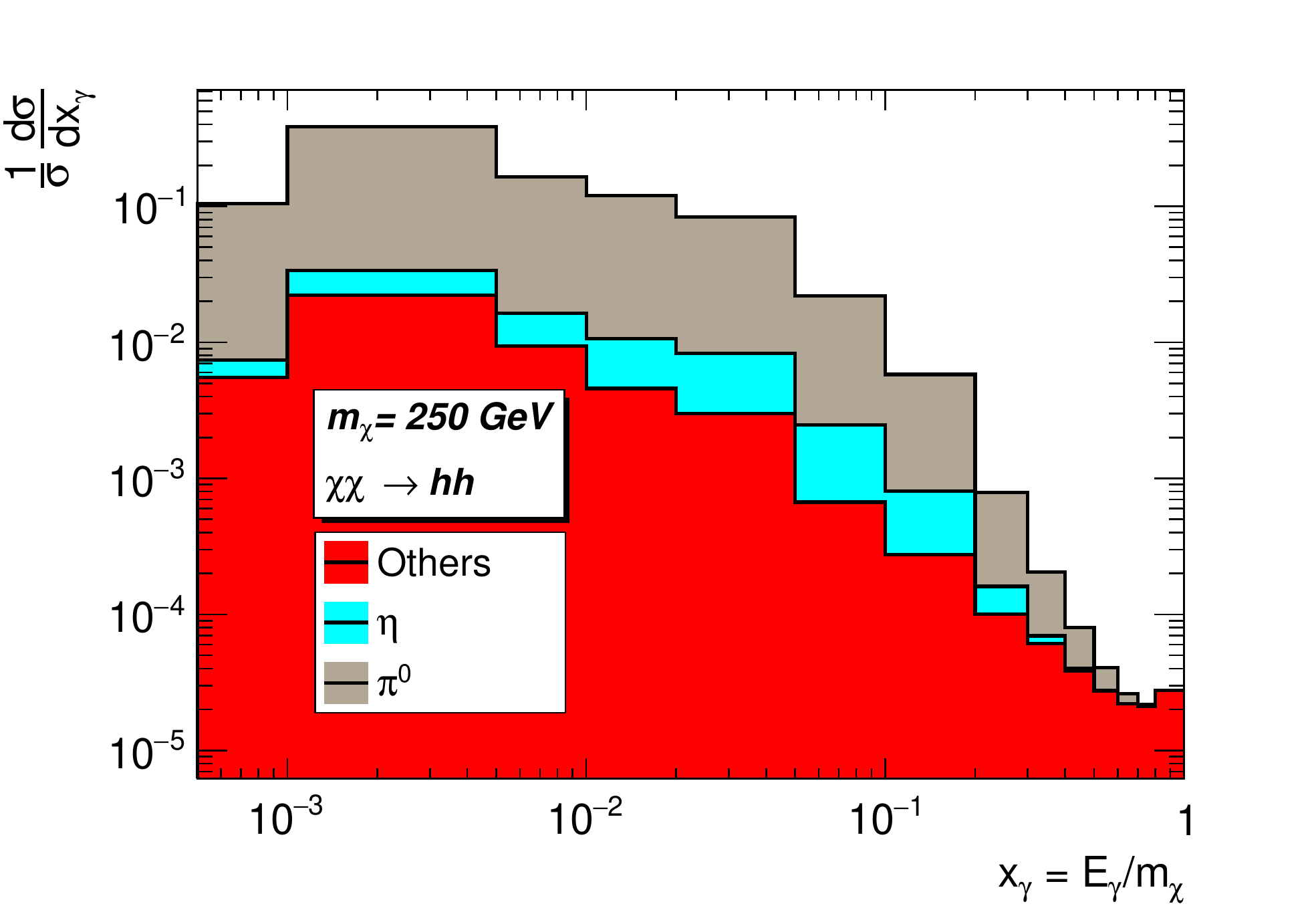}
\includegraphics[width=0.41\linewidth]{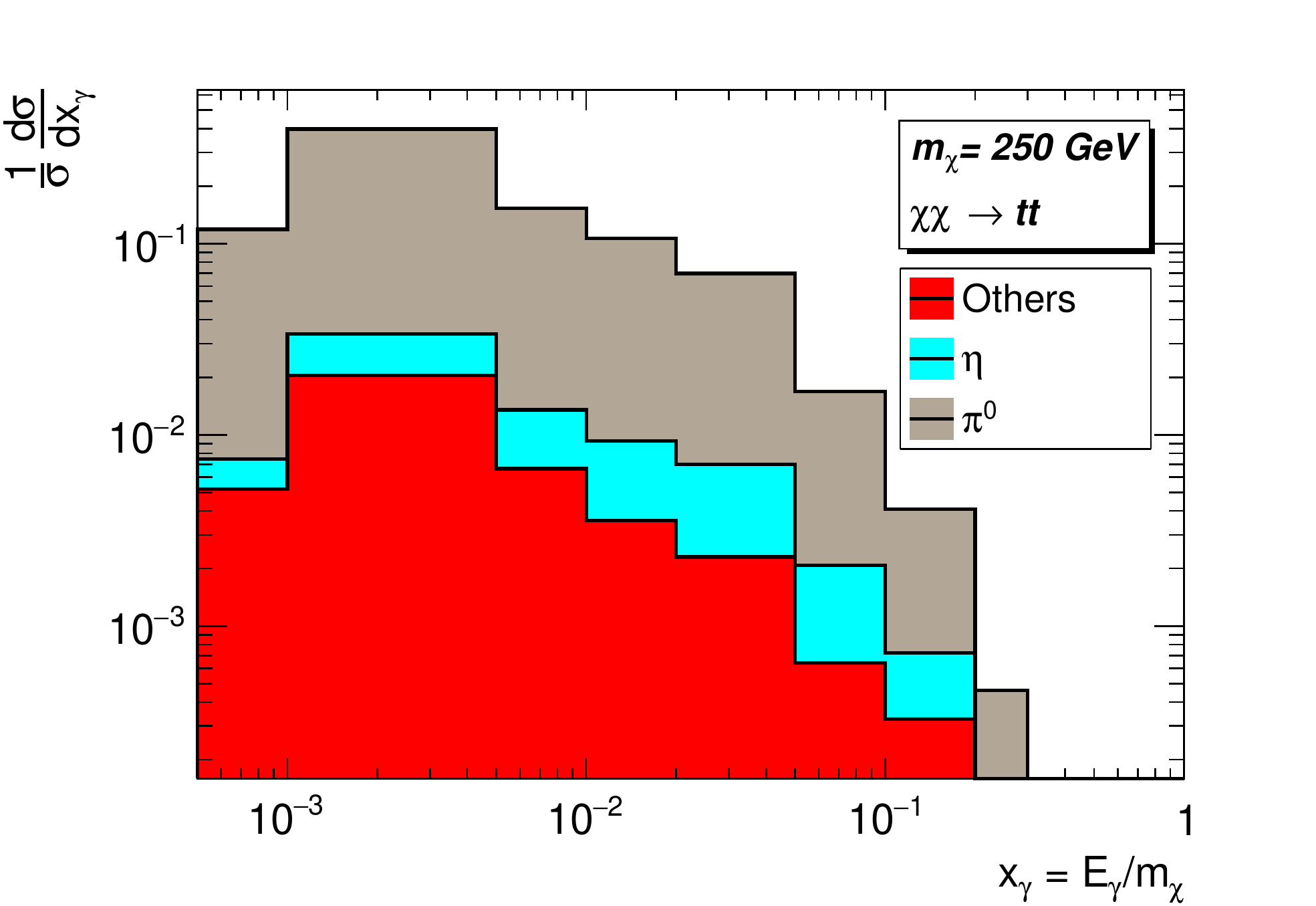}
\caption{Same as Fig. \ref{photons-sources-250-light} but for 
$WW$ (top left), $ZZ$ (top right), $hh$ (bottom left) and $t\bar{t}$ (bottom right)
final states.}
\label{photons-sources-250}
\end{figure}

\begin{figure}[!t]
\centering
 \includegraphics[width=0.41\linewidth]{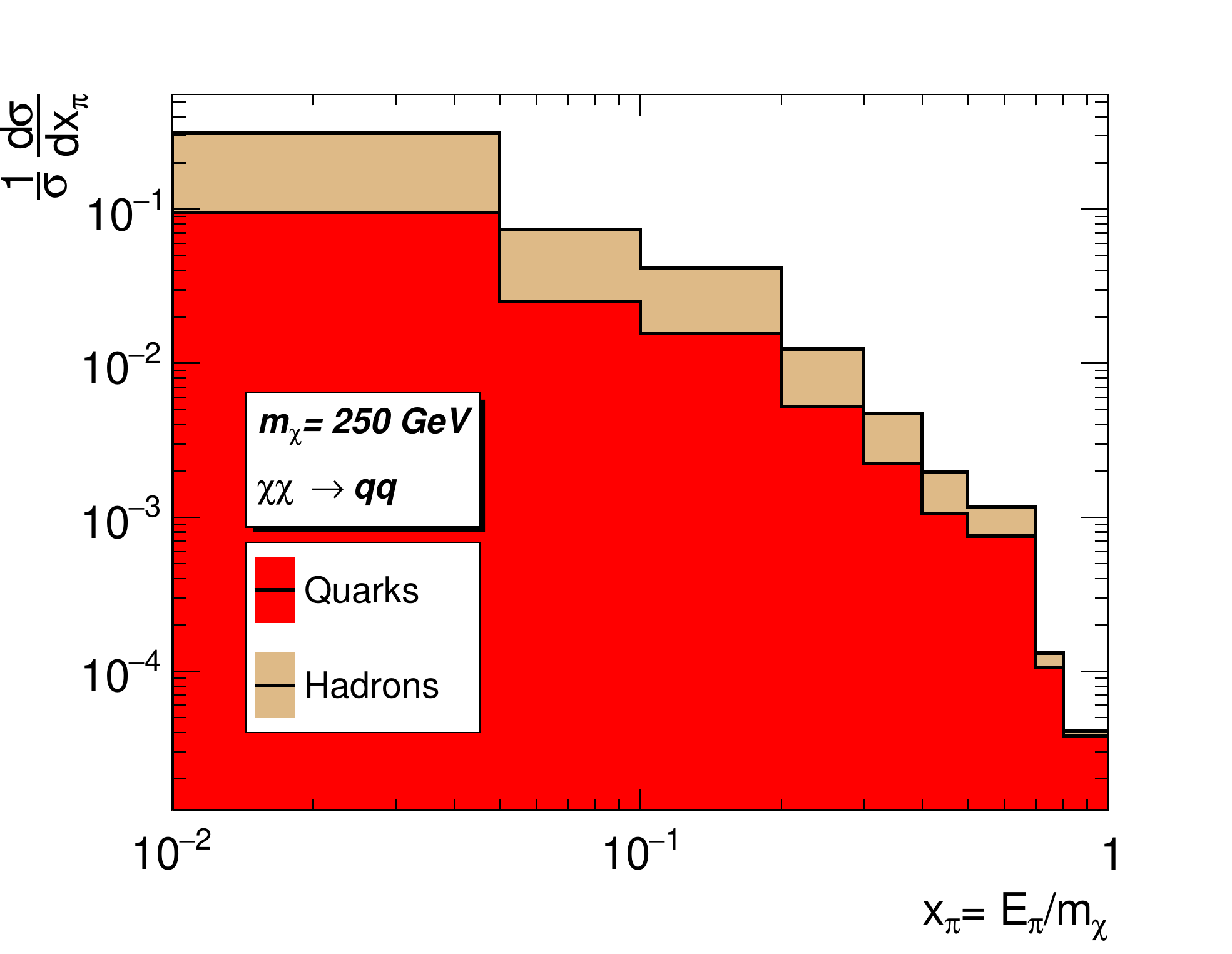}
\includegraphics[width=0.41\linewidth]{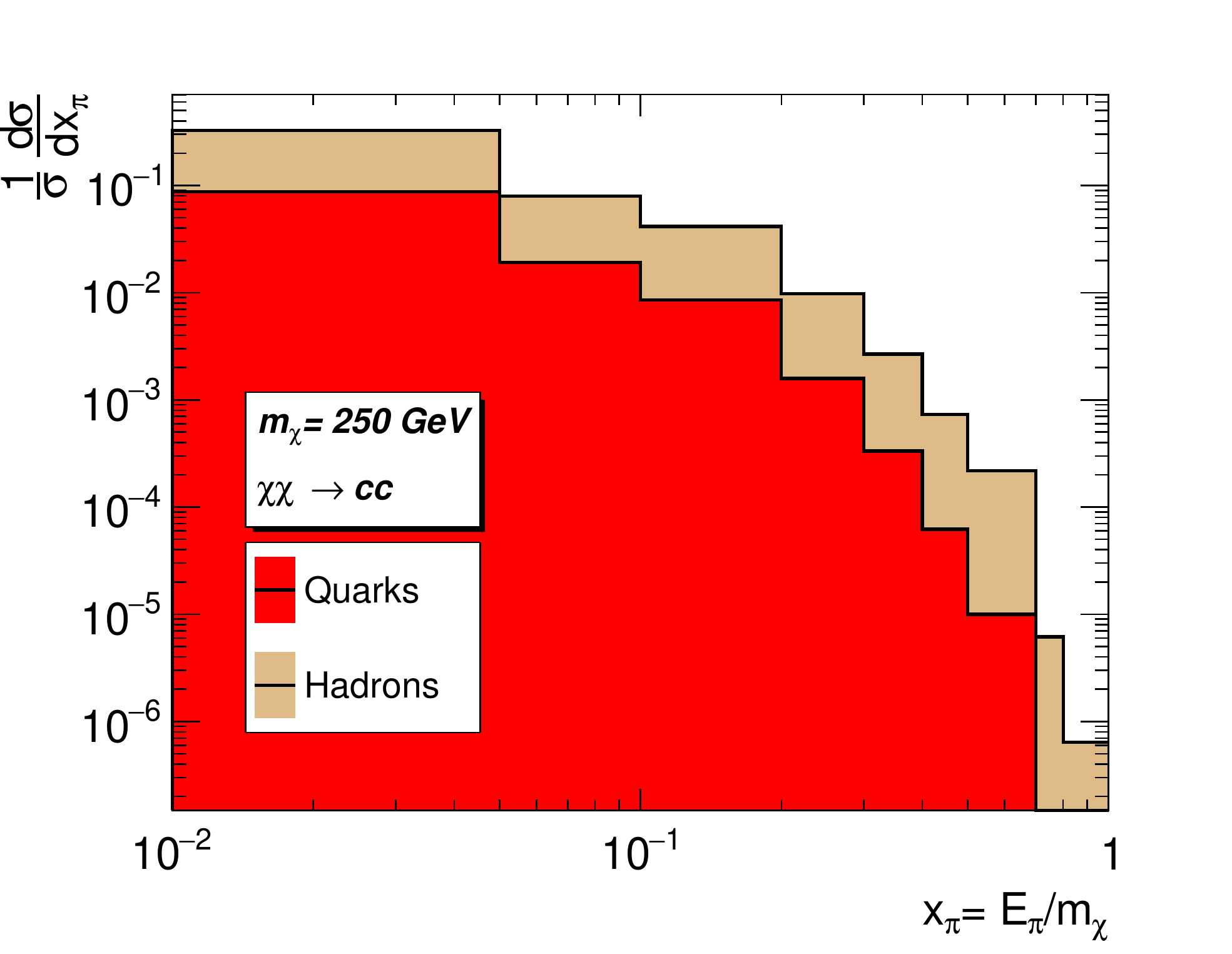}\\
\includegraphics[width=0.41\linewidth]{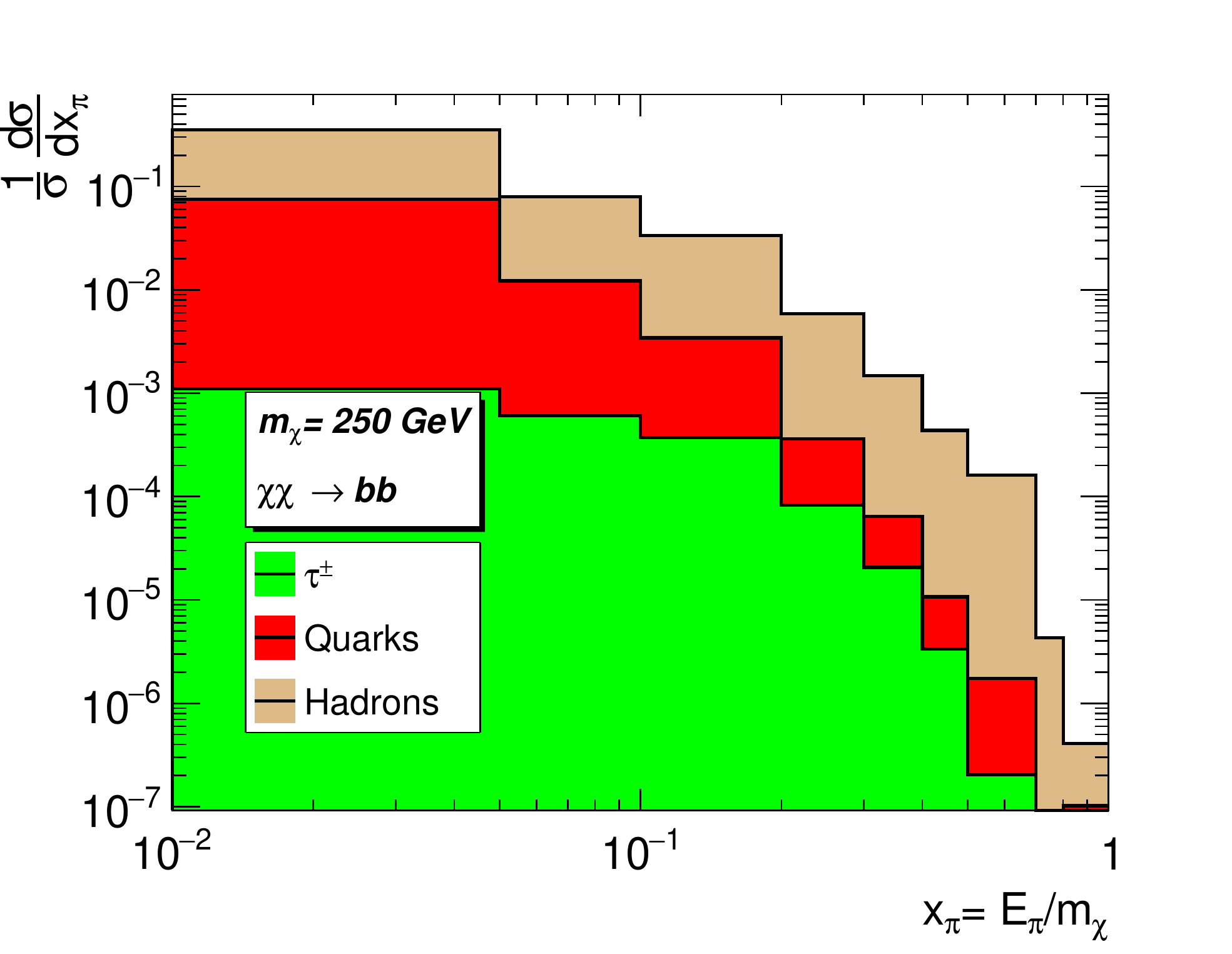}
\includegraphics[width=0.41\linewidth]{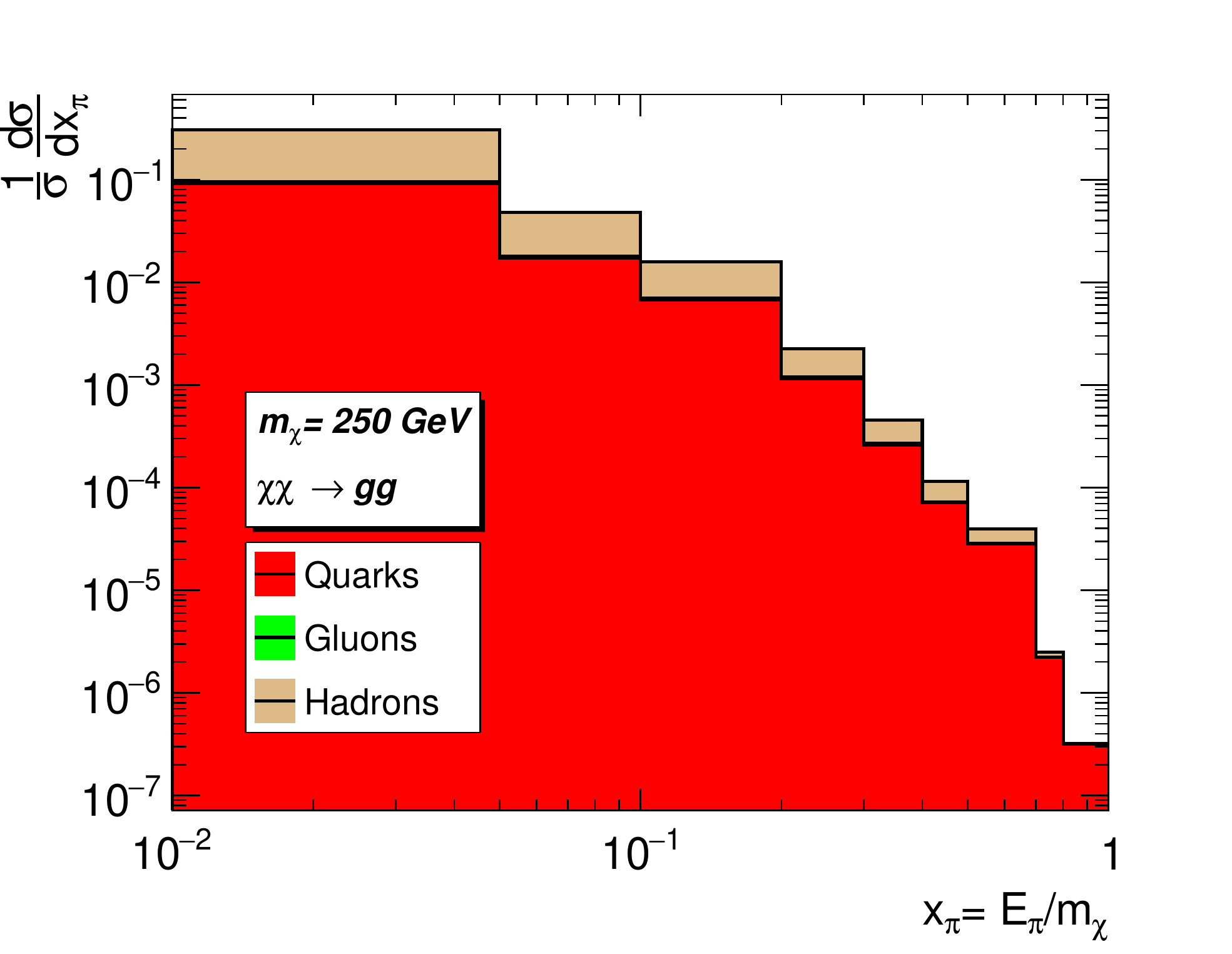}
\caption{Same as Fig. \ref{pions-sources-25} but for $m_\chi=250$ GeV.}
\label{pions-sources-250-light}
\end{figure}

\begin{figure}[!h]
\centering
 \includegraphics[width=0.41\linewidth]{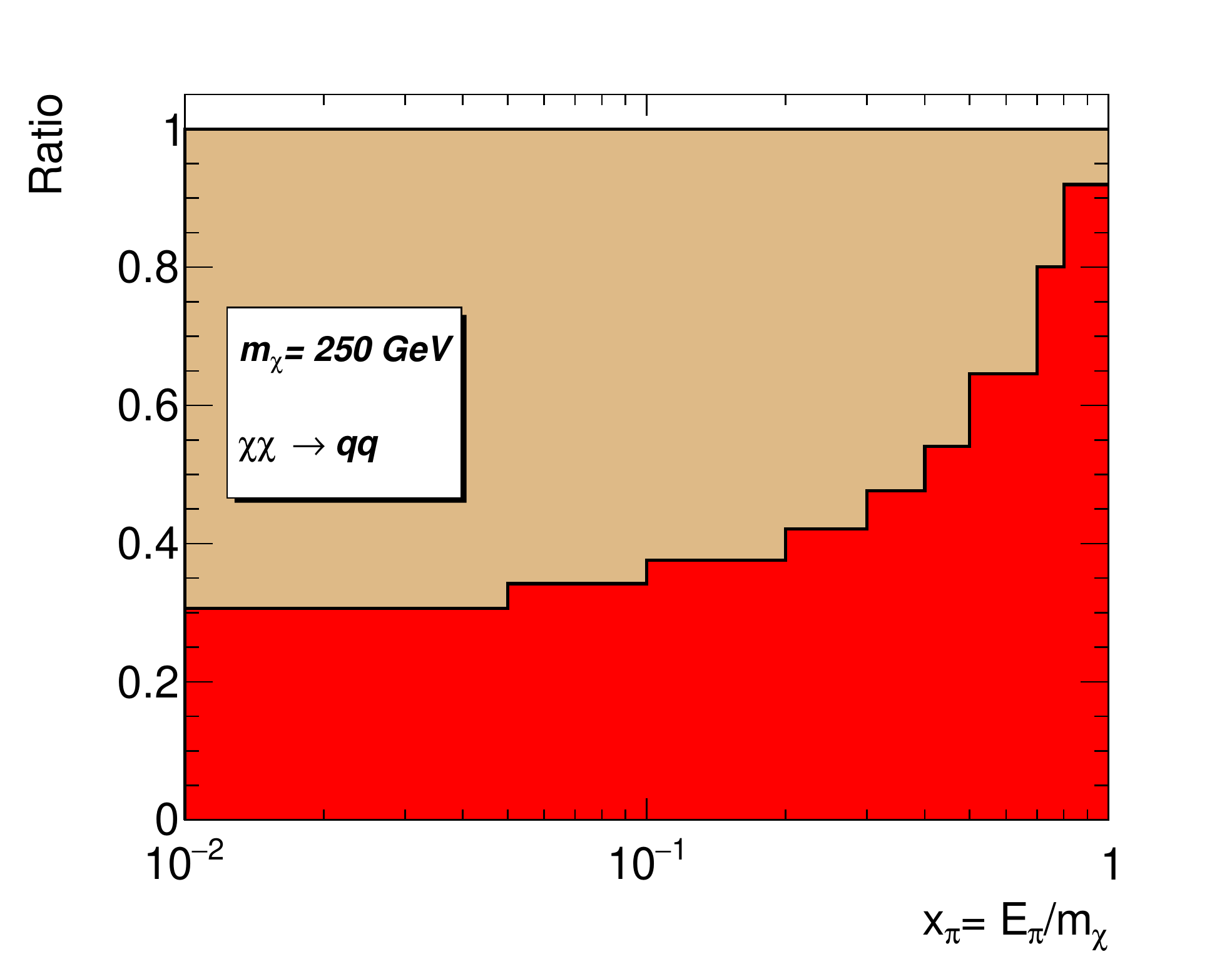}
\includegraphics[width=0.41\linewidth]{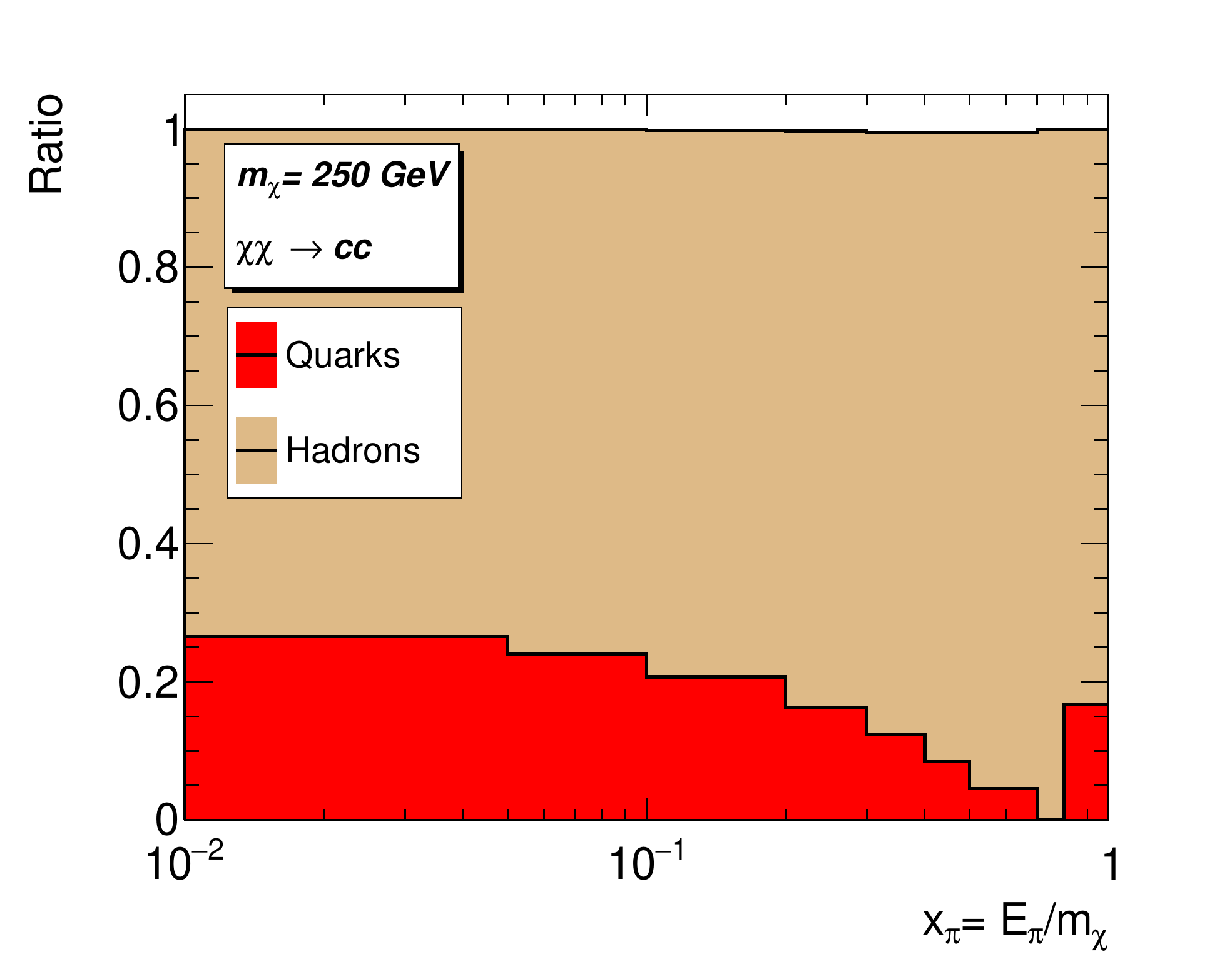}\\
\includegraphics[width=0.41\linewidth]{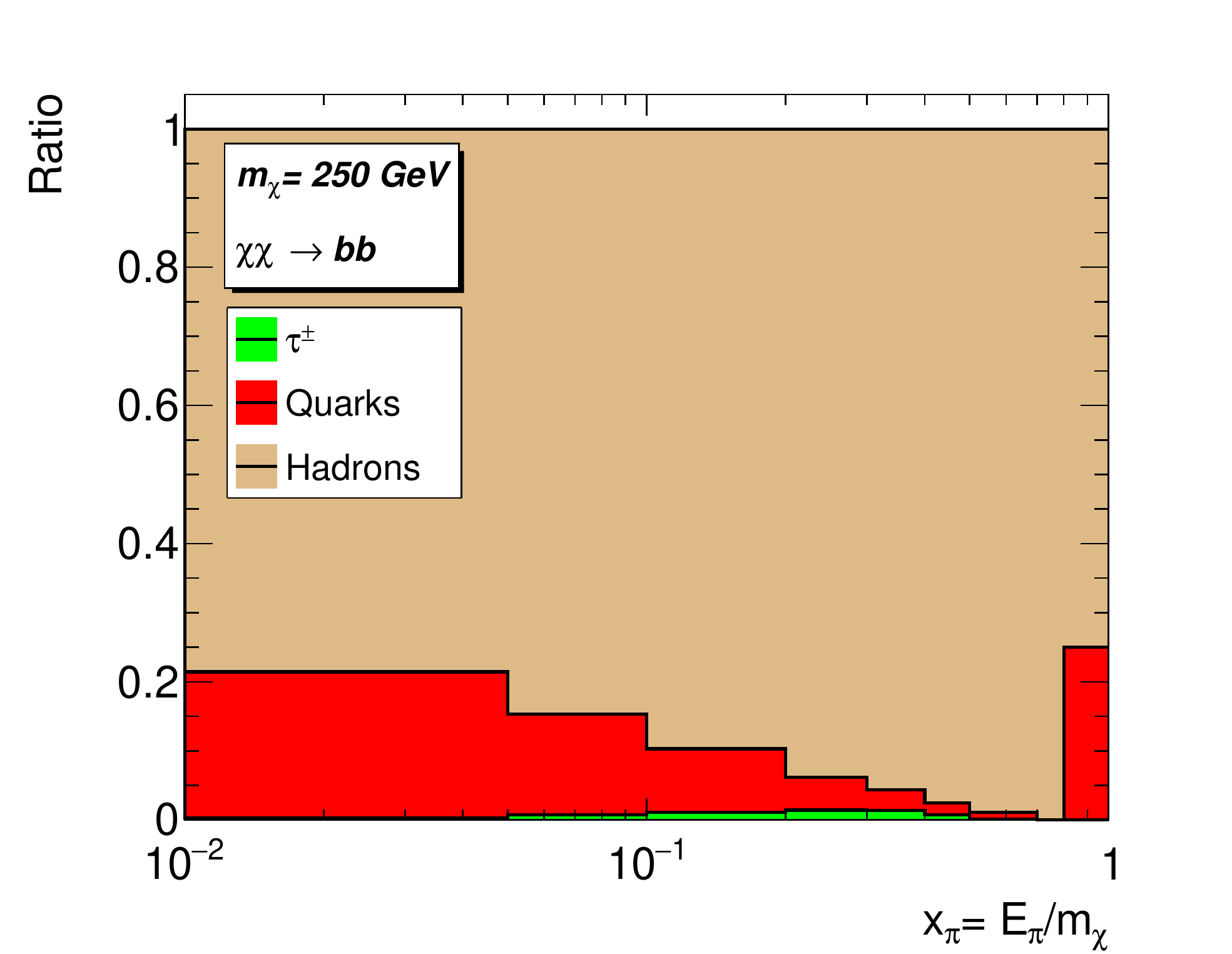}
\includegraphics[width=0.41\linewidth]{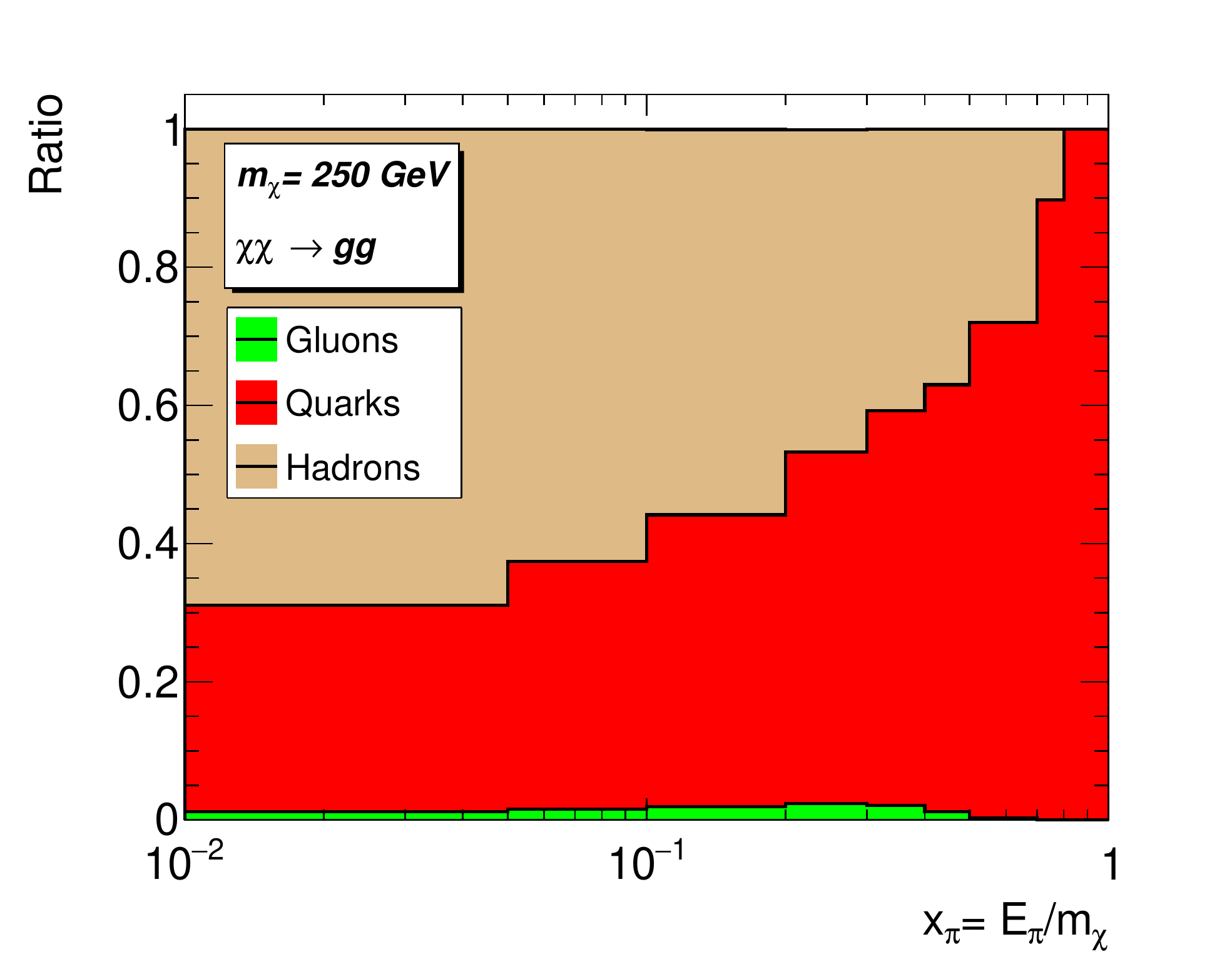}
\caption{Same as Fig. \ref{pions-sources-25-ratio} but for $m_\chi=250$ GeV.}
\label{pions-sources-250-light-ratio}
\end{figure}

\begin{figure}[!t]
\centering
 \includegraphics[width=0.41\linewidth]{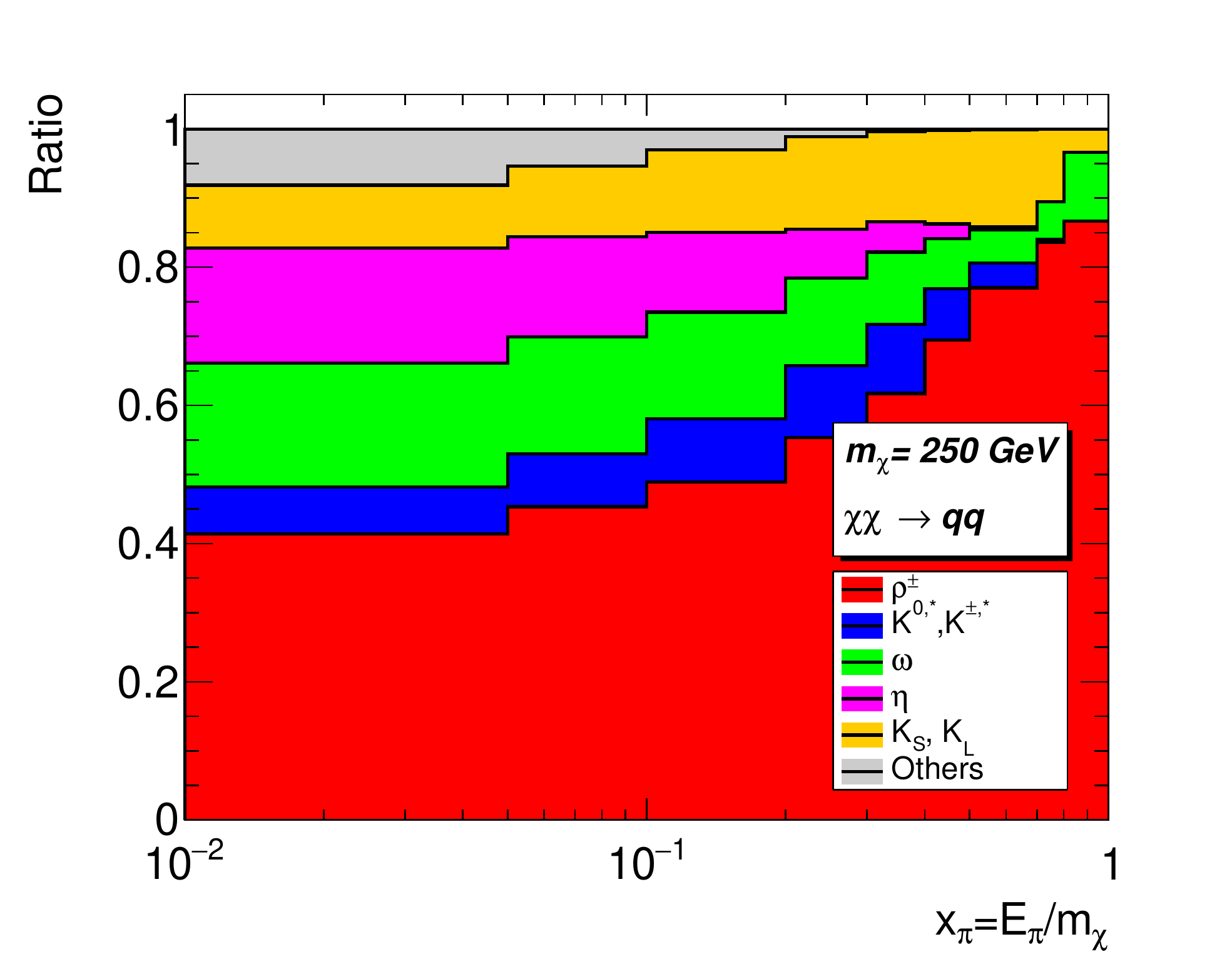}
\includegraphics[width=0.41\linewidth]{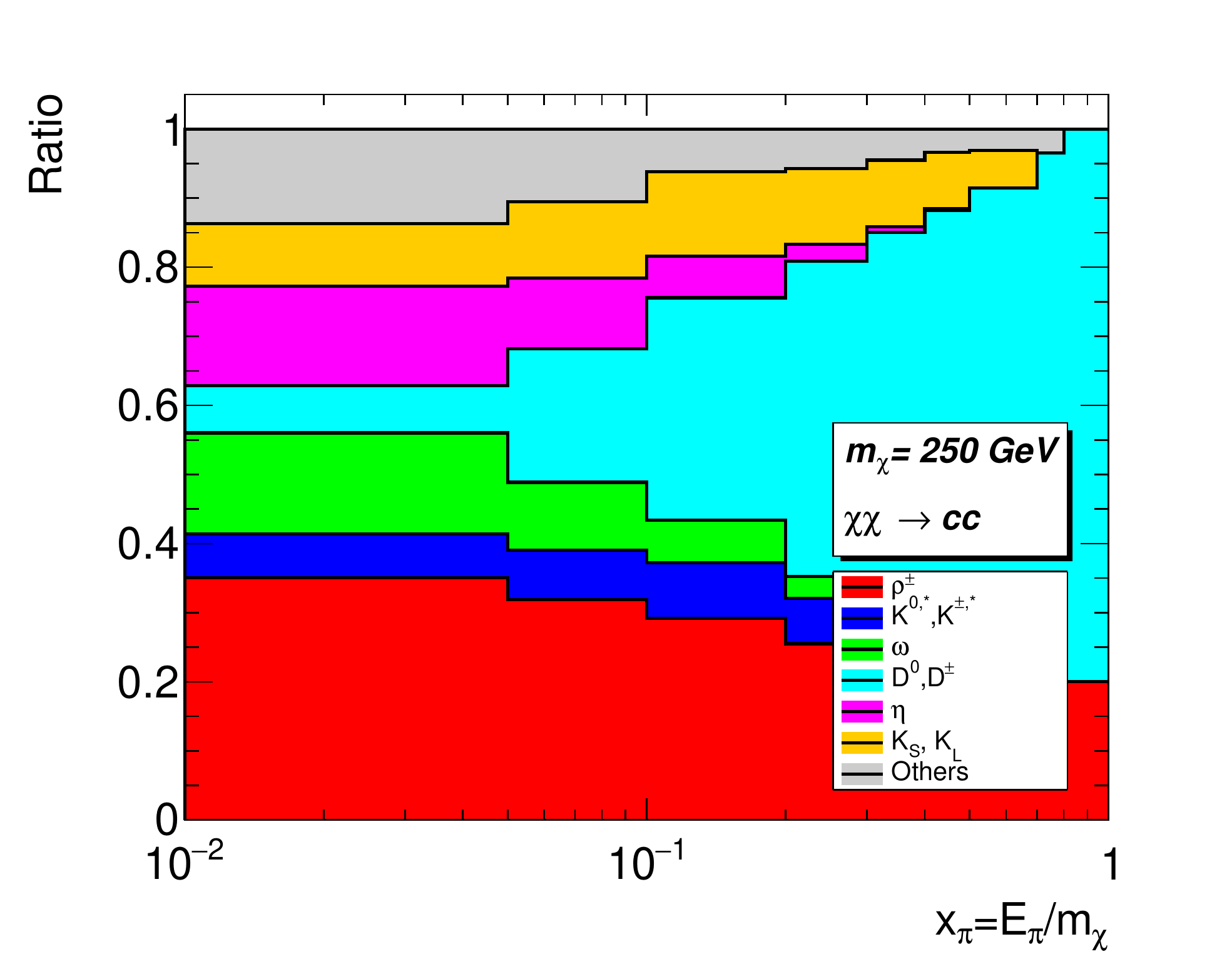}\\
\includegraphics[width=0.41\linewidth]{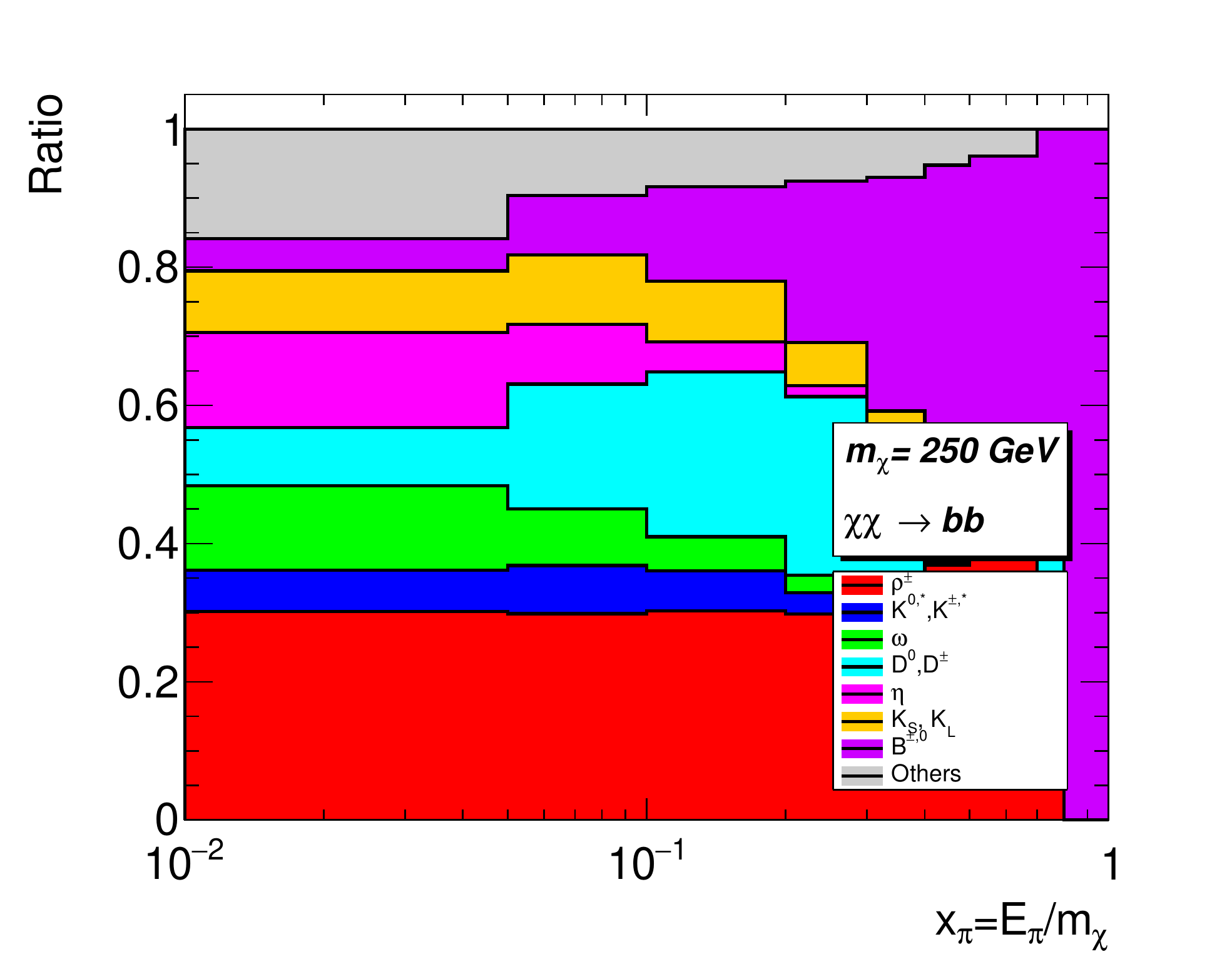}
\includegraphics[width=0.41\linewidth]{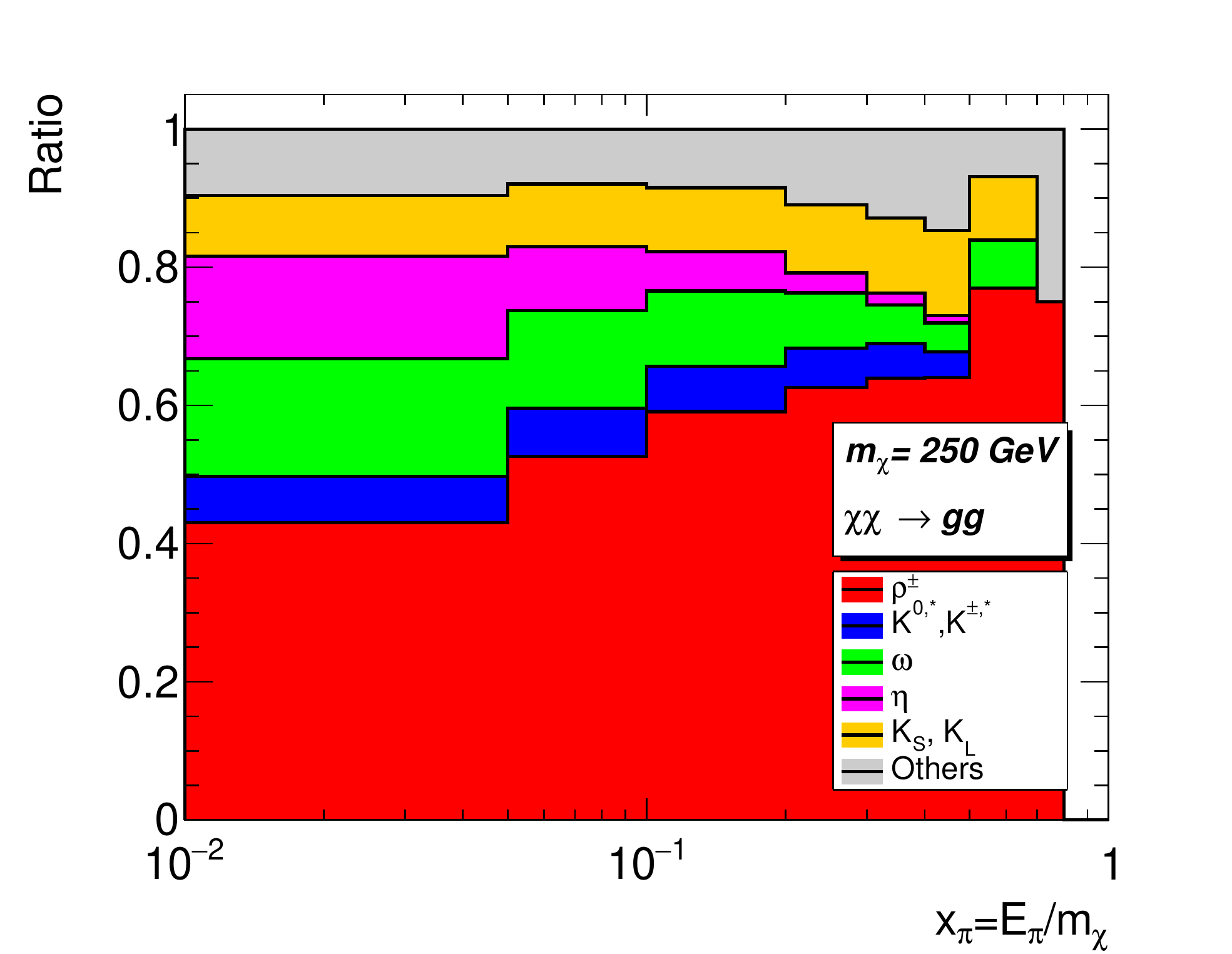}
\caption{Same as Fig. \ref{pions-sources-25-ratio-hadrons} but for $m_\chi=250$ GeV.}
\label{pions-sources-250-light-ratio-hadrons}
\end{figure}

\begin{figure}[!h]
\centering
 \includegraphics[width=0.41\linewidth]{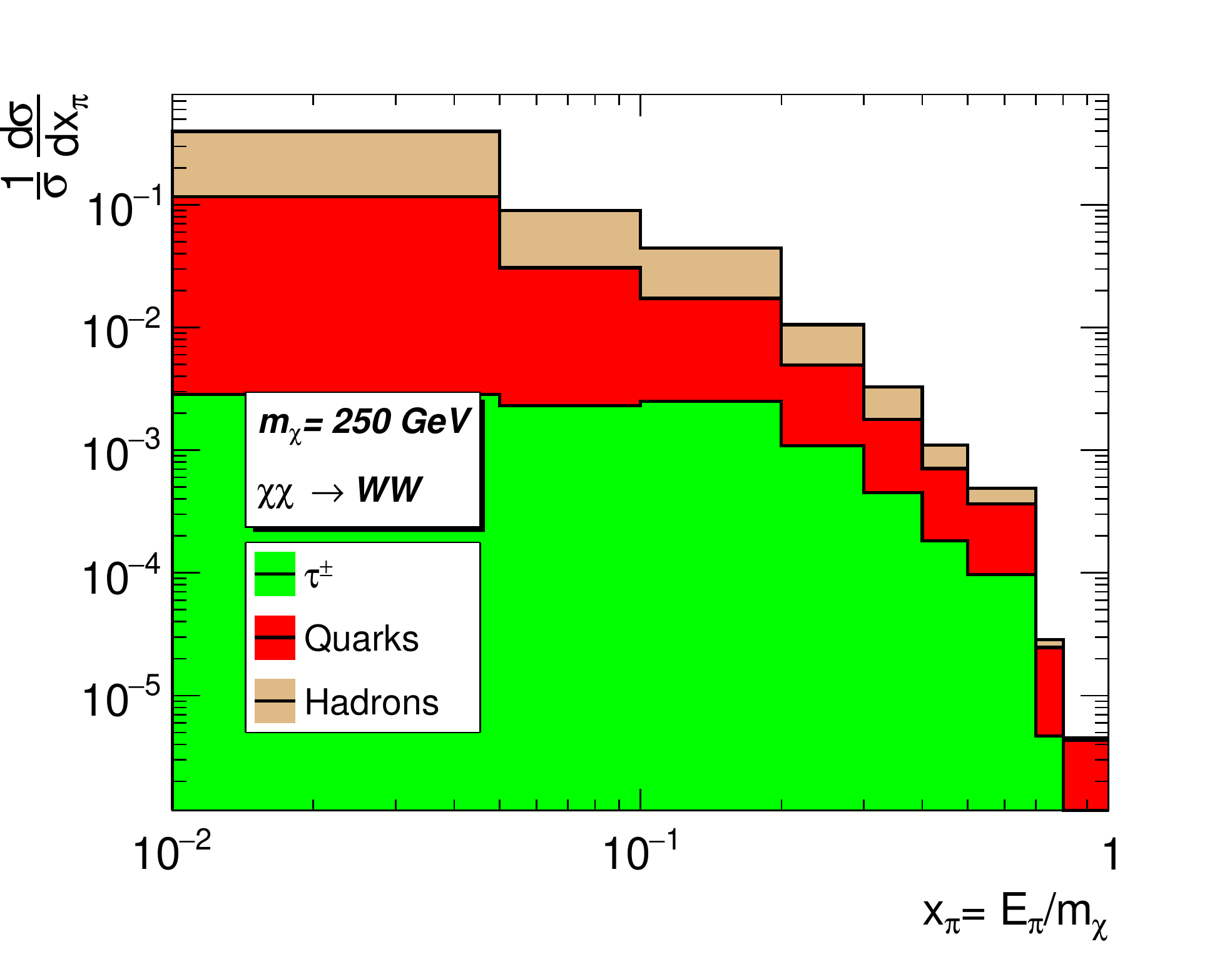}
\includegraphics[width=0.41\linewidth]{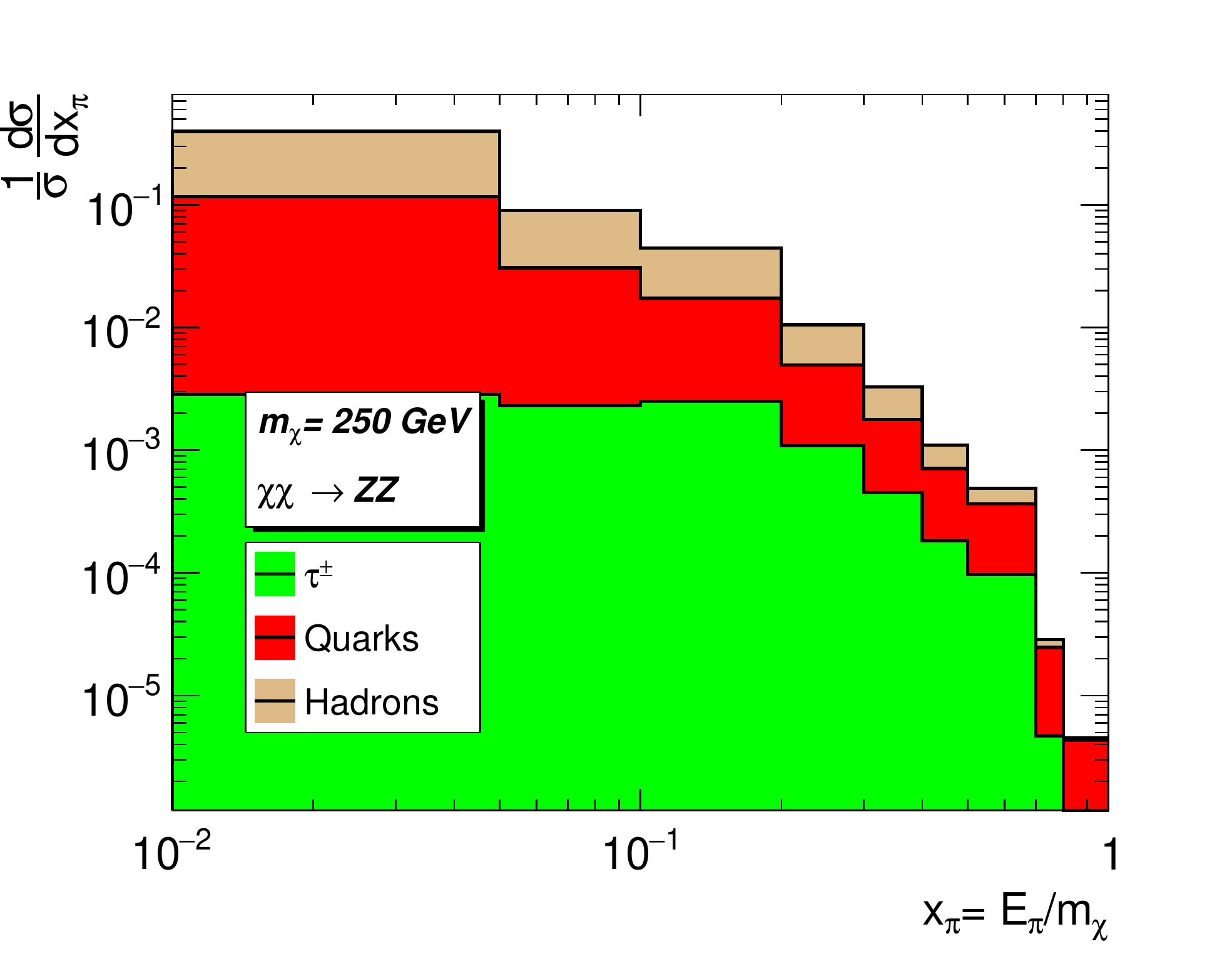}\\
\includegraphics[width=0.41\linewidth]{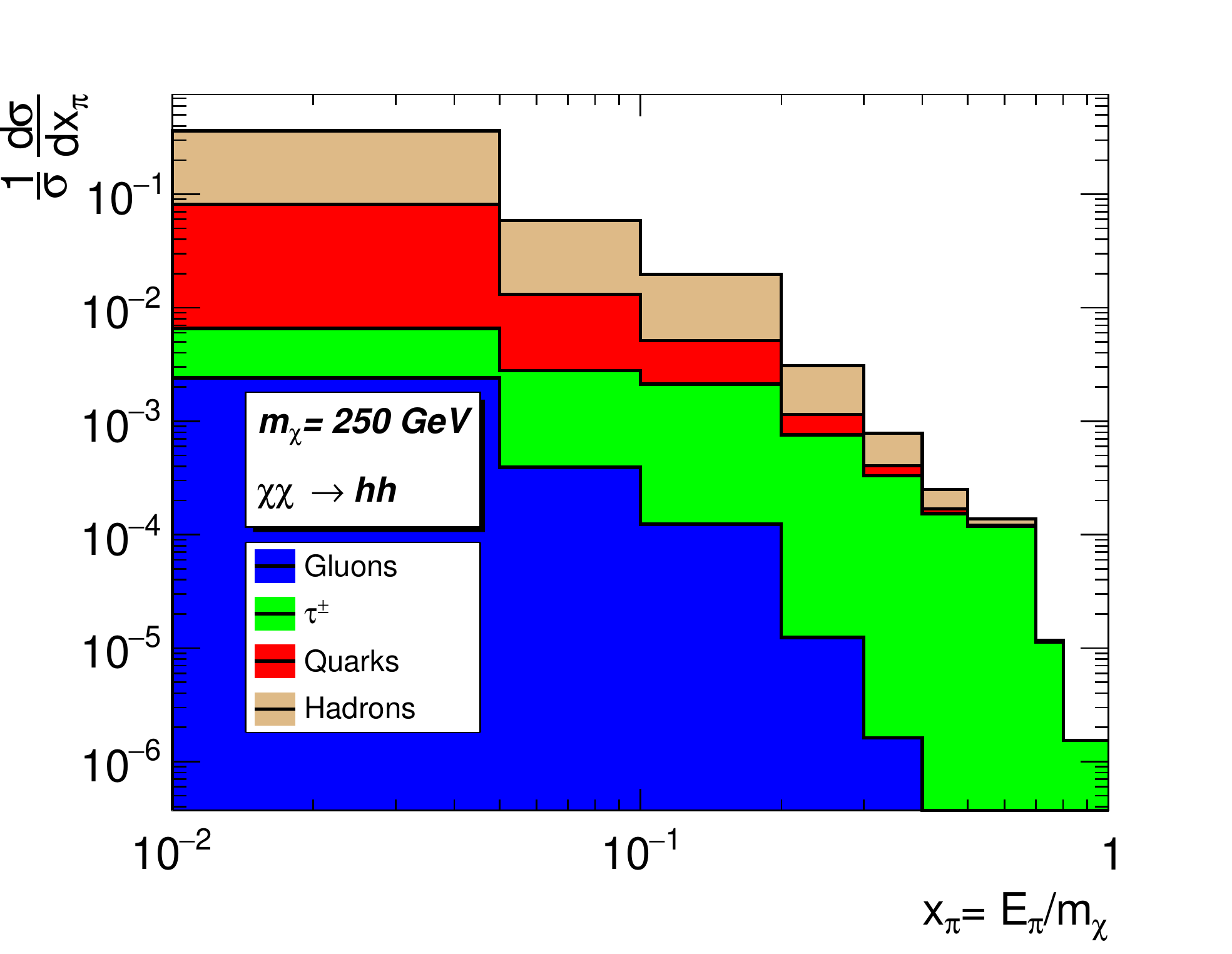}
\includegraphics[width=0.41\linewidth]{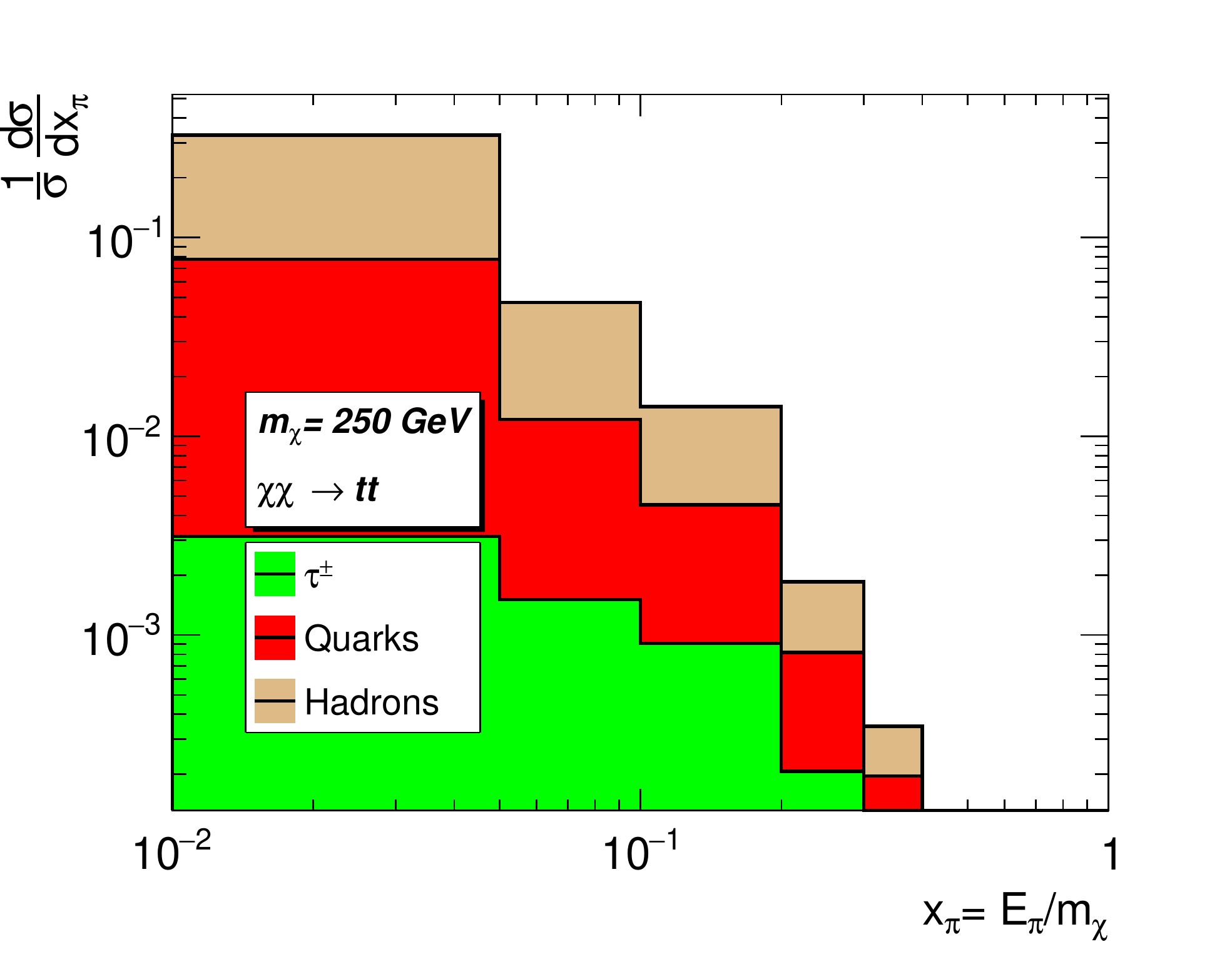}
\caption{Same as Fig. \ref{pions-sources-250-light} but for $WW$ (top left), 
$ZZ$ (top right), $hh$ (bottom left) and $t\bar{t}$ (bottom right) final states.}
\label{pions-sources-250}
\end{figure}

\begin{figure}[!t]
\centering
 \includegraphics[width=0.41\linewidth]{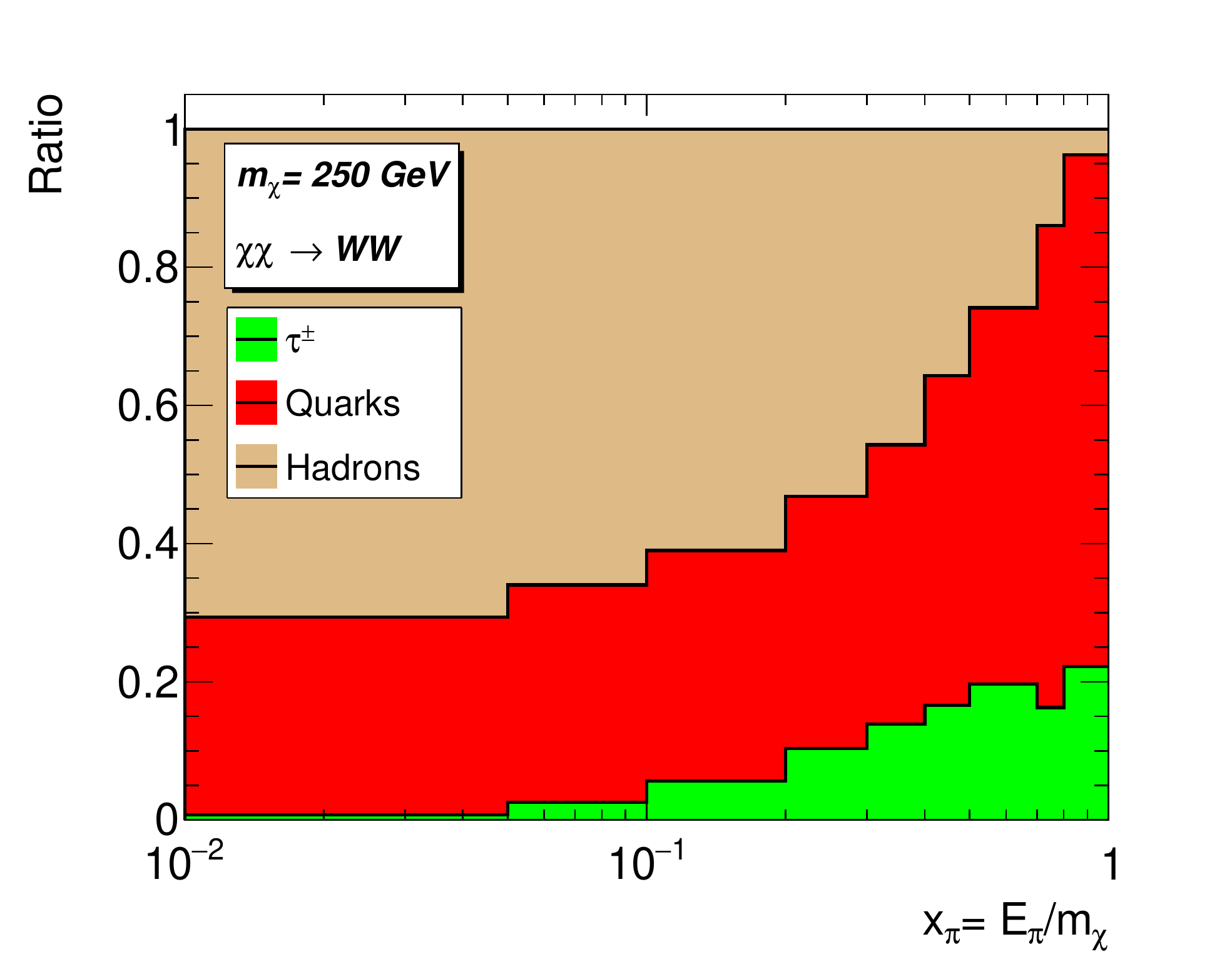}
\includegraphics[width=0.41\linewidth]{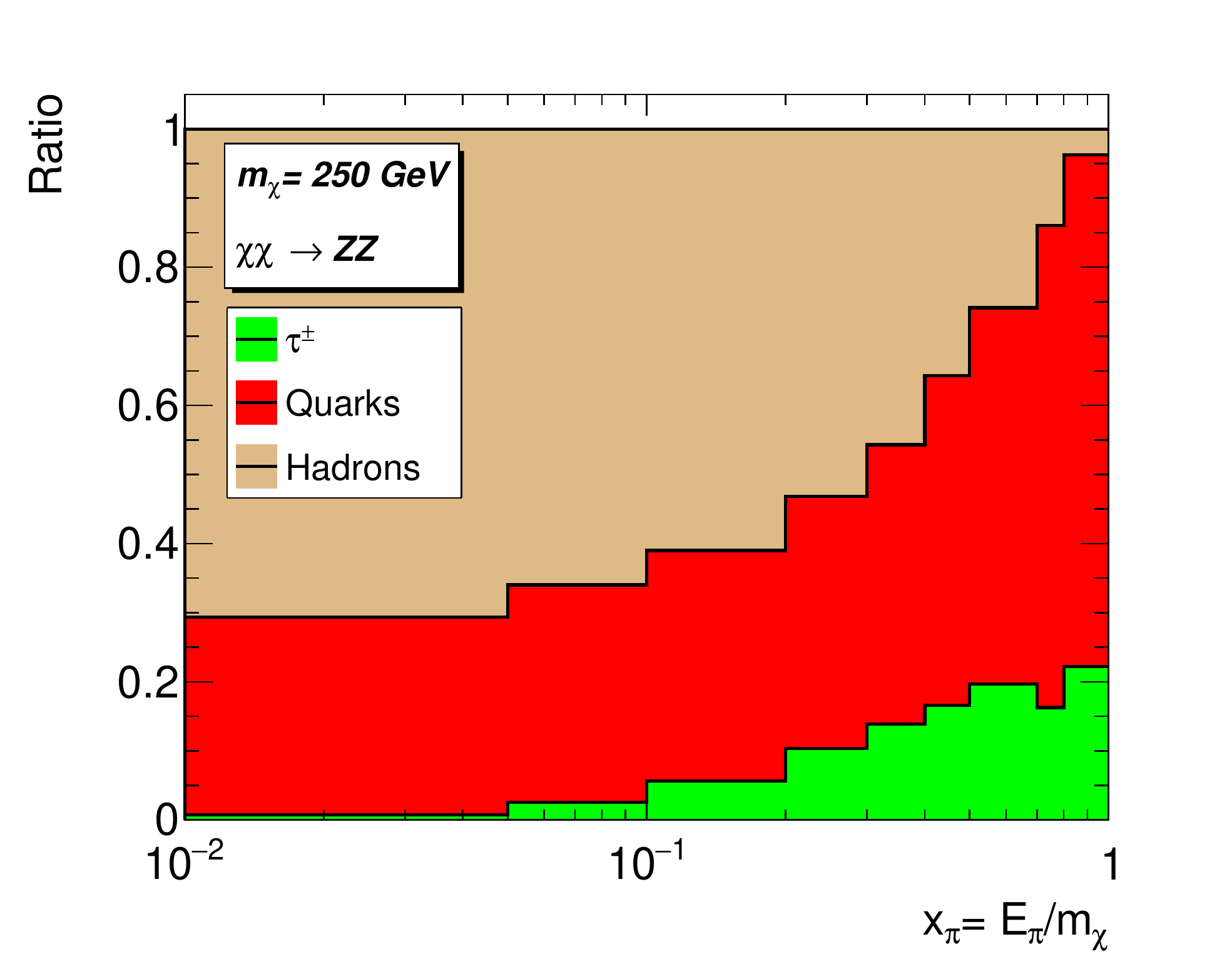}\\
\includegraphics[width=0.41\linewidth]{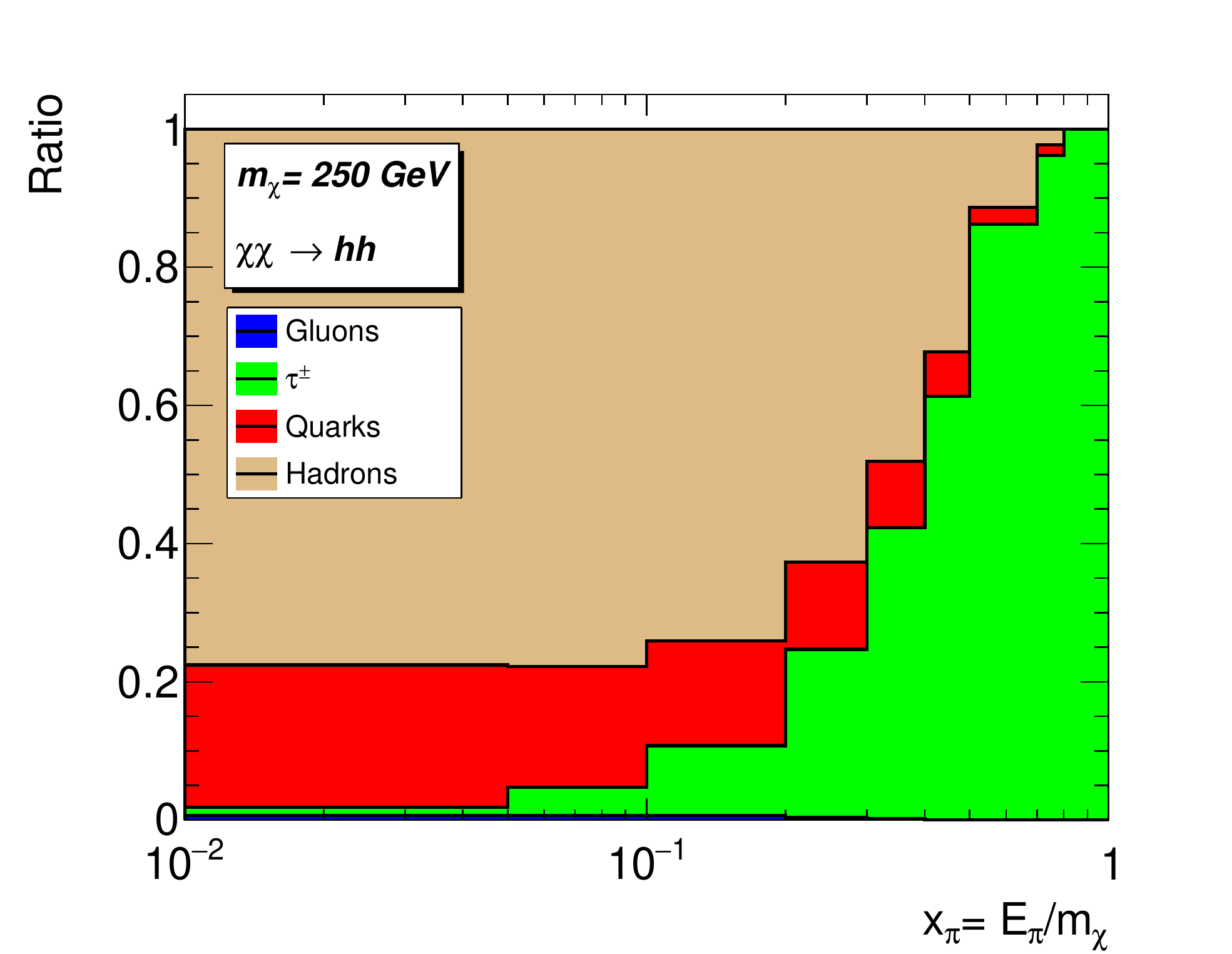}
\includegraphics[width=0.41\linewidth]{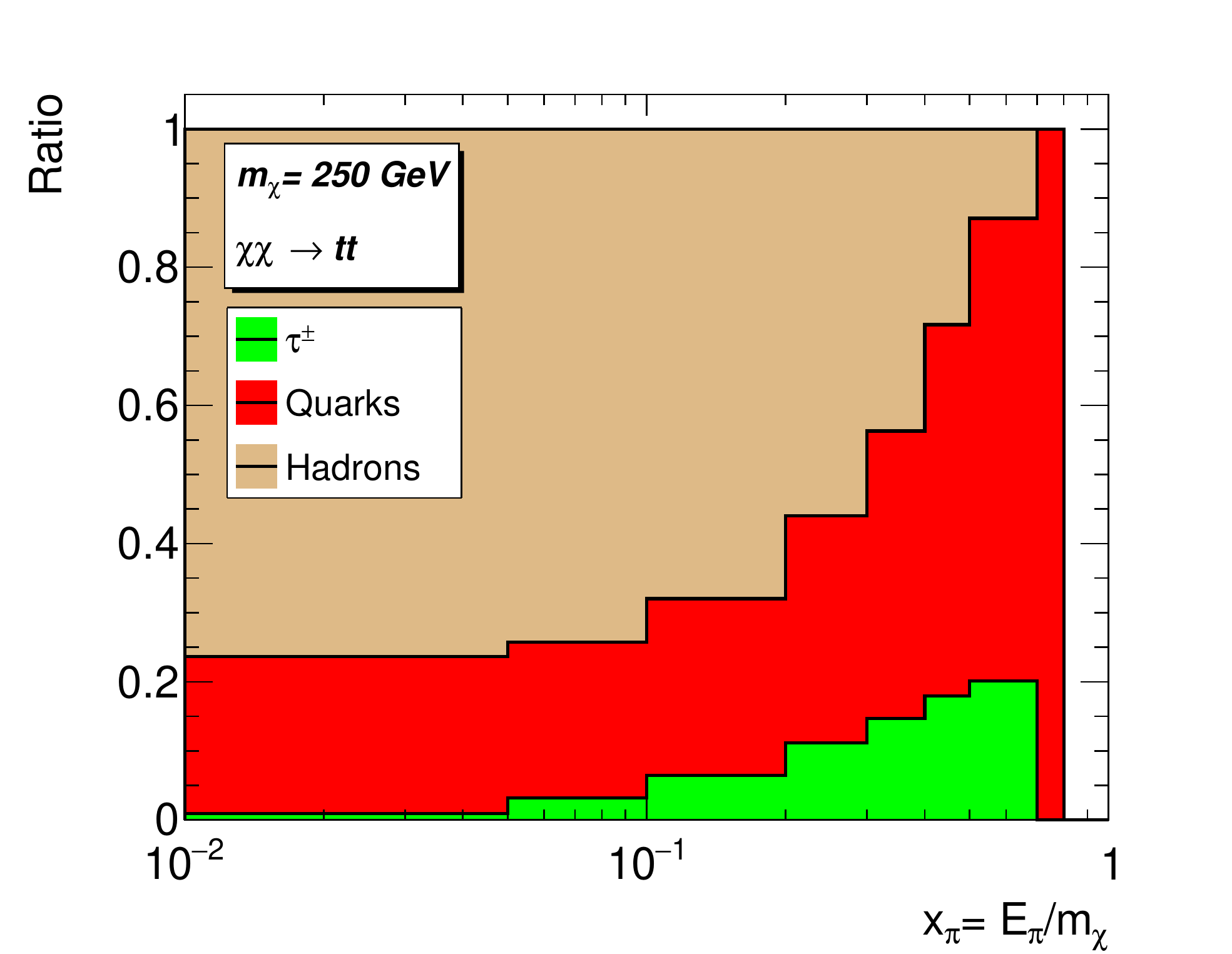}
\caption{Same as Fig. \ref{pions-sources-250-light-ratio} but for $WW$ (top left), 
$ZZ$ (top right), $hh$ (bottom left) and $t\bar{t}$ (bottom right) final states.}
\label{pions-sources-250-ratio}
\end{figure}

\begin{figure}[!h]
\centering
 \includegraphics[width=0.41\linewidth]{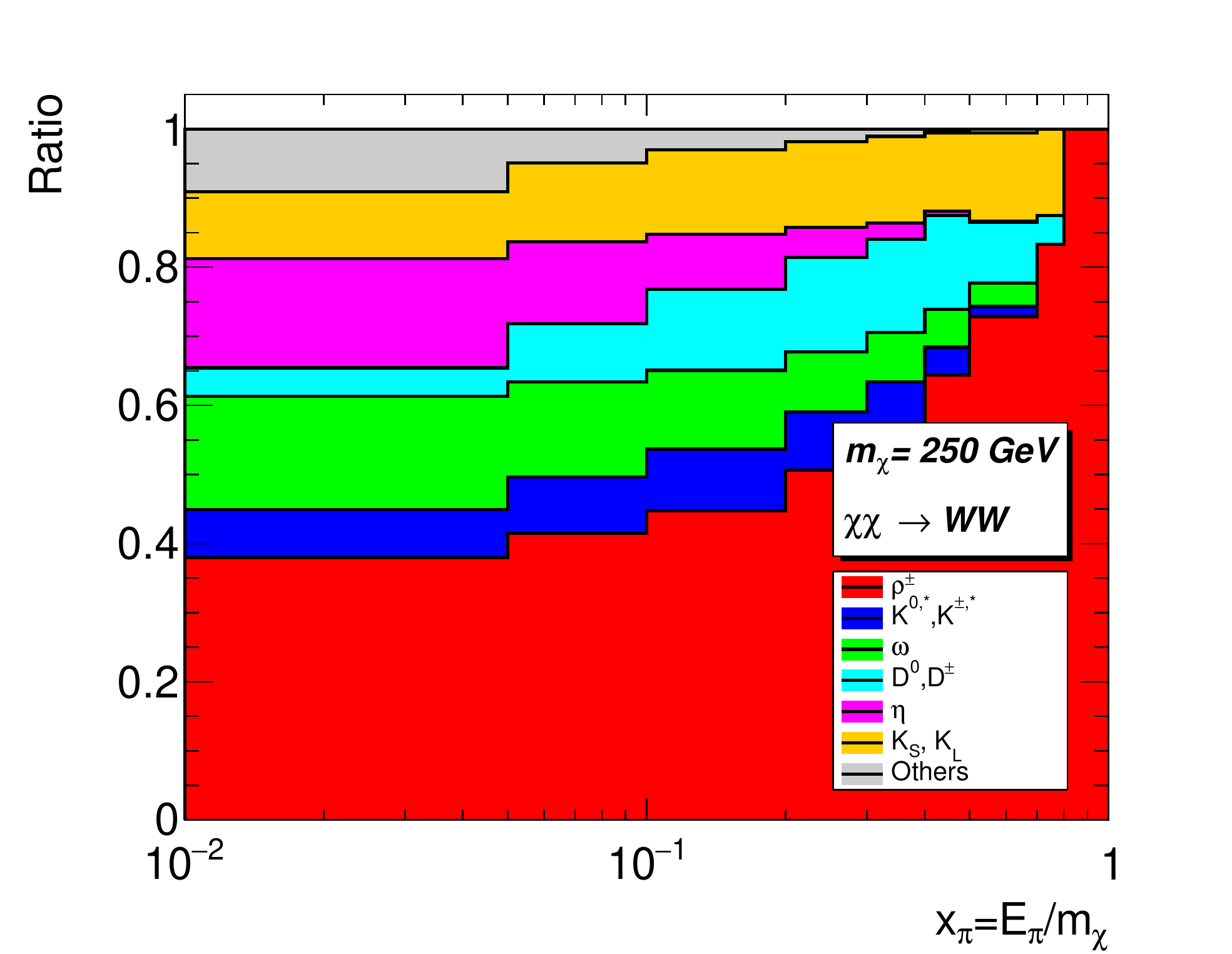}
\includegraphics[width=0.41\linewidth]{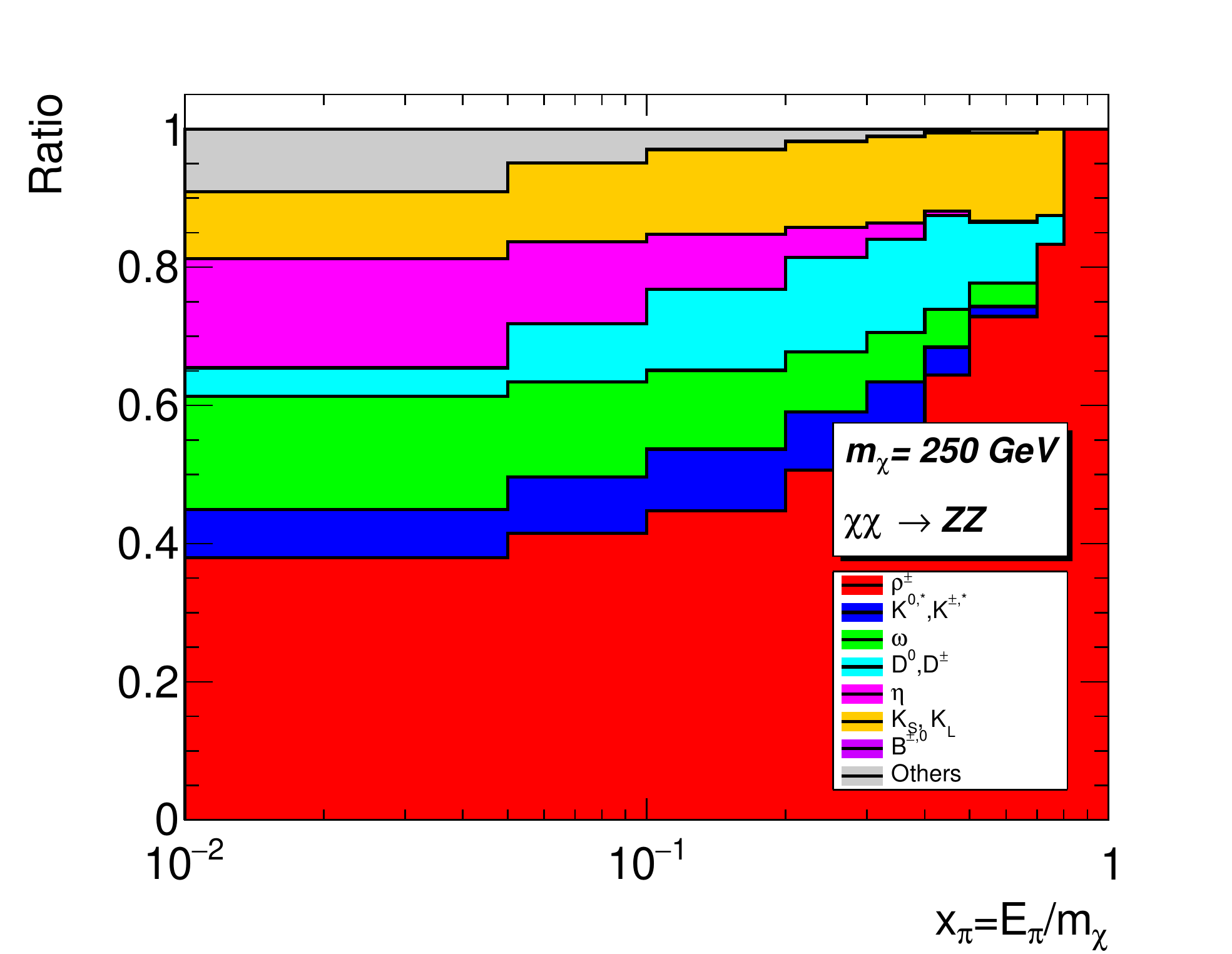}\\
\includegraphics[width=0.41\linewidth]{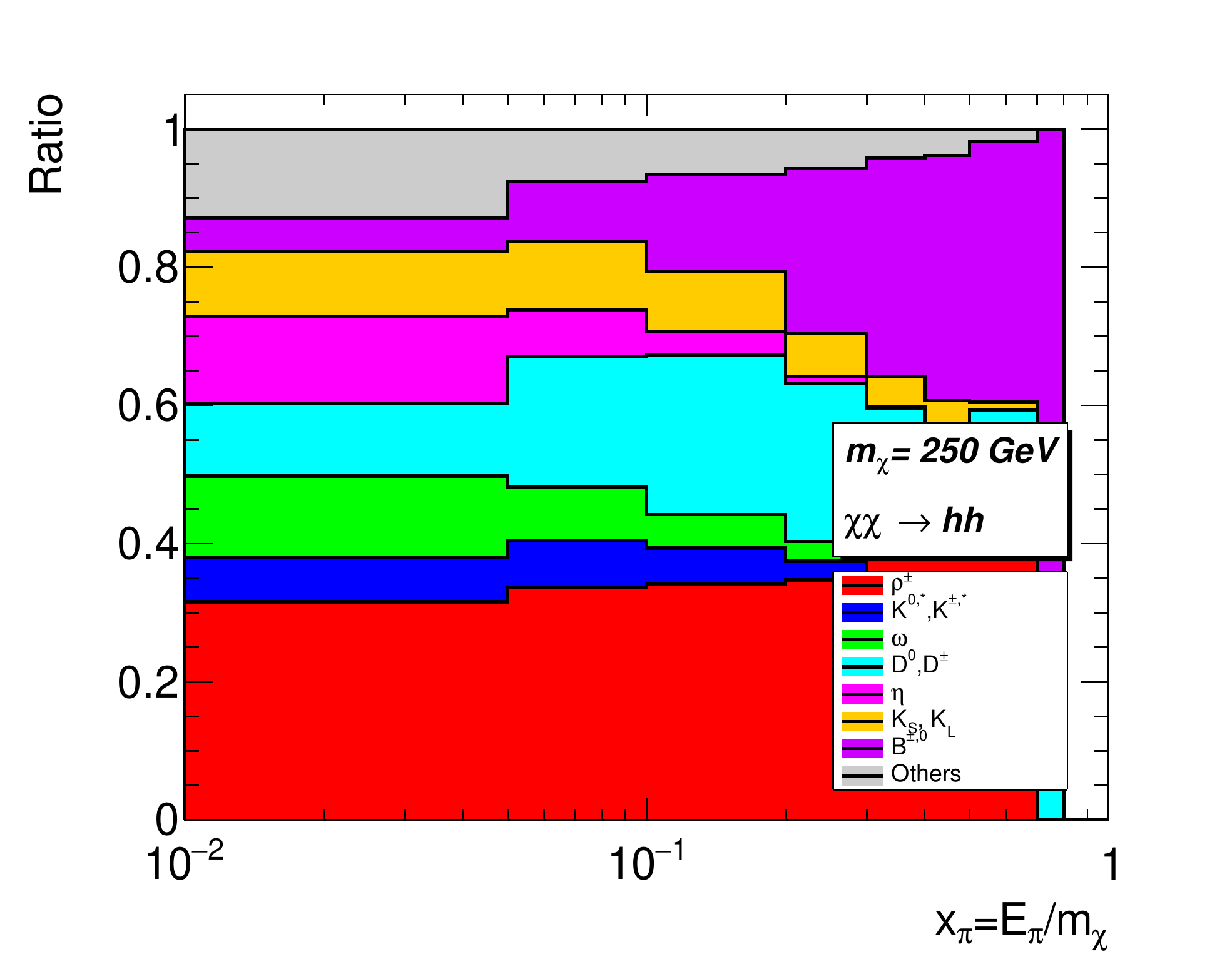}
\includegraphics[width=0.41\linewidth]{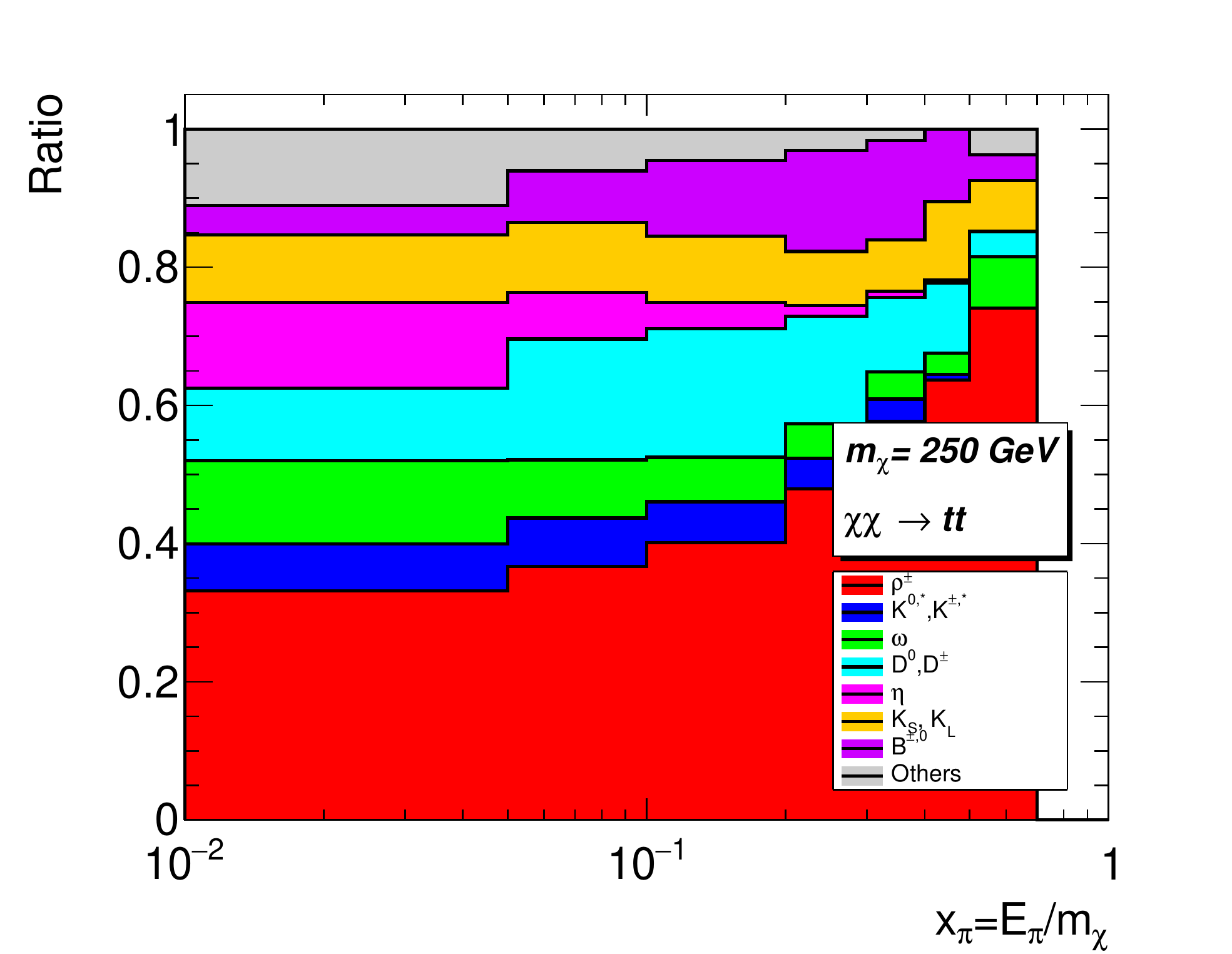}
\caption{Same as Fig. \ref{pions-sources-250-light-ratio-hadrons} but for $WW$ (top left), 
$ZZ$ (top right), $hh$ (bottom left) and $t\bar{t}$ (bottom right) final states.}
\label{pions-sources-250-ratio-hadrons}
\end{figure}
\clearpage
\section{Code}
\label{sec:code}
In this section, we show the flags that can used by a \textsc{Pythia} program to generate the spectra with the uncertainties. In \textsc{Pythia~8}, there is an example program (called \texttt{main07.cc}) which can be used to generate a generic (model-independent) resonance that decays into SM particles. Another option is to generate parton level events using an external tool and read the output (usually in the form of LHEF files, see~\cite{Alwall:2006yp}) by \textsc{Pythia} to add (cascades of) resonance decays, showering, and hadronisation. However, final-state particle spectra depend more on the kinematics of the process from which they originate (either a resonant or non-resonant production of e.g. jets), and the dark matter mass. Therefore, photon spectra in DM annihilation typically do not depend sensitively on the new physics model that predict it. The main difference is a normalization factor which is equal to the annihilation cross section for a given benchmark point in a given new physics scenario.\footnote{We have tested this using several benchmark points in the MSSM which was compared to the predictions of a generic process in \textsc{Pythia} and found that the shape of particle spectra is indeed the same in the two cases. However, to get the correct predictions for fluxes, the particle spectra in the generic picture should be scaled by the corresponding cross section (which can computed independently using an external tool).} Below, we provide a list of flags that were used to produce our photon spectra. 
\begin{verbatim}
Beams:idA  = -11 
Beams:idB  =  11 
Beams:eCM  = 1000.0 
PDF:lepton = off
SpaceShower:QEDShowerByL = off
\end{verbatim}
The first two flags set up the initial state beams as an electron and positron. The third flag sets the center-of-mass energy of the collision which is twice the DM mass (here we have $E_{cm}=2 m_\chi = 1000$~GeV). The fourth and the fifth flags switch off QED radiation from the incoming leptons (these flags are switched on by default). This is important since the incoming DM particles have no electric charge. 

The \texttt{main07.cc} example program sets up a generic production process for a fictitious resonance that decays to a pair of SM particles. Below, we show the necessary options needed to generate the spectra
\begin{verbatim}
! id:all = name antiName spinType chargeType colType m0 mWidth mMin mMax tau0
999999:all= GeneralResonance void 1 0 0 1000. 1. 0. 0. 0.
999999:isResonance = true
\end{verbatim}

\begin{verbatim}
! id:addChannel = onMode bRatio meMode product1 product2
!999999:addChannel = 1 0.15 101 1 -1 # for \chi \chi \to d\bar{d} 
!999999:addChannel = 1 0.15 101 2 -2 # for \chi \chi \to u\bar{u} 
!999999:addChannel = 1 0.15 101 3 -3 # for \chi \chi \to s\bar{s} 
!999999:addChannel = 1 0.15 101 4 -4 # for \chi \chi \to c\bar{c} 
!999999:addChannel = 1 0.15 101 5 -5 # for \chi \chi \to b\bar{b} 
999999:addChannel = 1 0.15 101 6 -6 # for \chi \chi \to t\bar{t} 
!999999:addChannel = 1 0.15 101 11 -11 # for \chi \chi \to e^+ e^- 
!999999:addChannel = 1 0.15 101 13 -13 # for \chi \chi \to \mu^+ \mu^- 
!999999:addChannel = 1 0.15 101 15 -15 # for \chi \chi \to \tau^+ \tau^- 
!999999:addChannel = 1 0.15 101 21 21  # for \chi \chi \to g g
!999999:addChannel = 1 0.15 101 22 22  # for \chi \chi \to \gamma \gamma 
!999999:addChannel = 1 0.15 101 23 23  # for \chi \chi \to Z^0 Z^0 
!999999:addChannel = 1 0.15 101 24 -24 # for \chi \chi \to W^+ W^- 
!999999:addChannel = 1 0.15 101 25 25  # for \chi \chi \to h^0 h^0
\end{verbatim}
In this setup a fictitious spin-$0$ resonance (with a PDG code $9999999$) is produced in $e^+ e^-$ collisions and then decays into a pair of SM particles. The flag \verb|999999:isResonance = true| is important especially for low dark matter masses because if it is set to \verb|false| then there will be no shower added after the decay of the resonance. Note that in the above example, all channels except the $t\bar{t}$ final state are shown commented out; this represents a simple way to focus on one specific channel at a time. (\textsc{Pythia} automatically rescales the total branching fraction to unity.)

The following two commands define particle stability. As shown below, a particle is treated to be stable if its proper lifetime larger than $100$ $\textrm{mm}/c$.  At the LHC and LEP, $\texttt{tau0Max}$ should be setup to $10$ and $100$ respectively\footnote{These two flags were used in the setup of our tuning and should not be used in DM simulations. Another application of these two flags is that they can be used for comparison between LEP and LHC. There are, however, seven hadrons which are treated as stable at the LHC but decay before reaching the detector at LEP; $K_S^0$ (26.8), $\Sigma^-$ (44.3), $\Lambda^0$ (78.9), $\Sigma^+$ (24.0), $\Xi^-$ (49.1), $\Xi^0$ (87.1), $\Omega^-$ (24.6) where the numbers inside brackets refer to their proper lifetime in $\textrm{mm}/c$.}
\begin{verbatim}
ParticleDecays:limitTau0 = on
ParticleDecays:tau0Max   = 100.0
\end{verbatim}
However, particles produced from DM annihilation travel for very long distances and therefore some particles (treated as stable in e.g. LEP) will decay before reaching the detector.
 There are five particles that are treated as stable in LEP but they decay in astrophysical processes; $\pi^\pm, \mu^\pm, K^\pm, K_0^L, $ and the neutron. There are two ways to set these particles as unstable; either to change $\texttt{ParticleDecays:tau0Max}$ to a very large value $\simeq 10^{15}$ or to force them to decay (which is more safe) using the following commands
\begin{verbatim}
13:mayDecay   = true
211:mayDecay  = true
321:mayDecay  = true
130:mayDecay  = true 
2112:mayDecay = true
\end{verbatim}

Which will force $\mu^\pm, \pi^\pm, K^\pm, K_L^0$ and the neutron to decay. The parameters of the Lund fragmentation function can be changed using the following commands

\begin{verbatim}
StringZ:deriveBLund = on
StringZ:aLund        = 0.5999 #0.80 #0.40
StringZ:avgZLund     = 0.5278 #0.50 #0.55
StringPT:Sigma       = 0.3174 #0.28 #0.36
\end{verbatim}
Where the first number for each parameter corresponds to the result of the central tune while the the last two numbers are obtained as $68\%$ CL error (we refer the reader to section \ref{sec:results} for more details about the uncertainties). Finally, uncertainty on the parton showering can be obtained by using the following commands;

\begin{verbatim}
UncertaintyBands:doVariations = on
        UncertaintyBands:List = { 
                alphaShi fsr:muRfac = 0.5,
                alphaSlo fsr:muRfac = 2.0,
                hardHi fsr:cNS      = 2.0, 
                hardLo fsr:cNS      = -2.0} 
\end{verbatim}

 Where the first command is needed to switch on the shower variations and the last four commands to select the variations with required amount, e.g. $\texttt{alphaShi fsr:muRfac} = \texttt{0.5}$ $(\texttt{alphaSlo fsr:muRfac} = 2)$ refer to the variation of the renormalization scale by a factor of 2 in the negative (positive) direction. $\texttt{hardHi fsr:cNS}$ corresponds to variation of the non-singular term in the DGLAP splitting function. \\

Finally, we recommend that the other parameters and changes which occurred in different \textsc{Pythia} versions (see next section) to be kept with their default value to guarantee a correct modeling of particle spectra. Most importantly the flag \verb|TimeShower:QEDshowerByOther| should not be set off especially for heavy charged SM particles far from their threshold.

\begin{figure}[!t]
\centering
\includegraphics[width=0.32\linewidth]{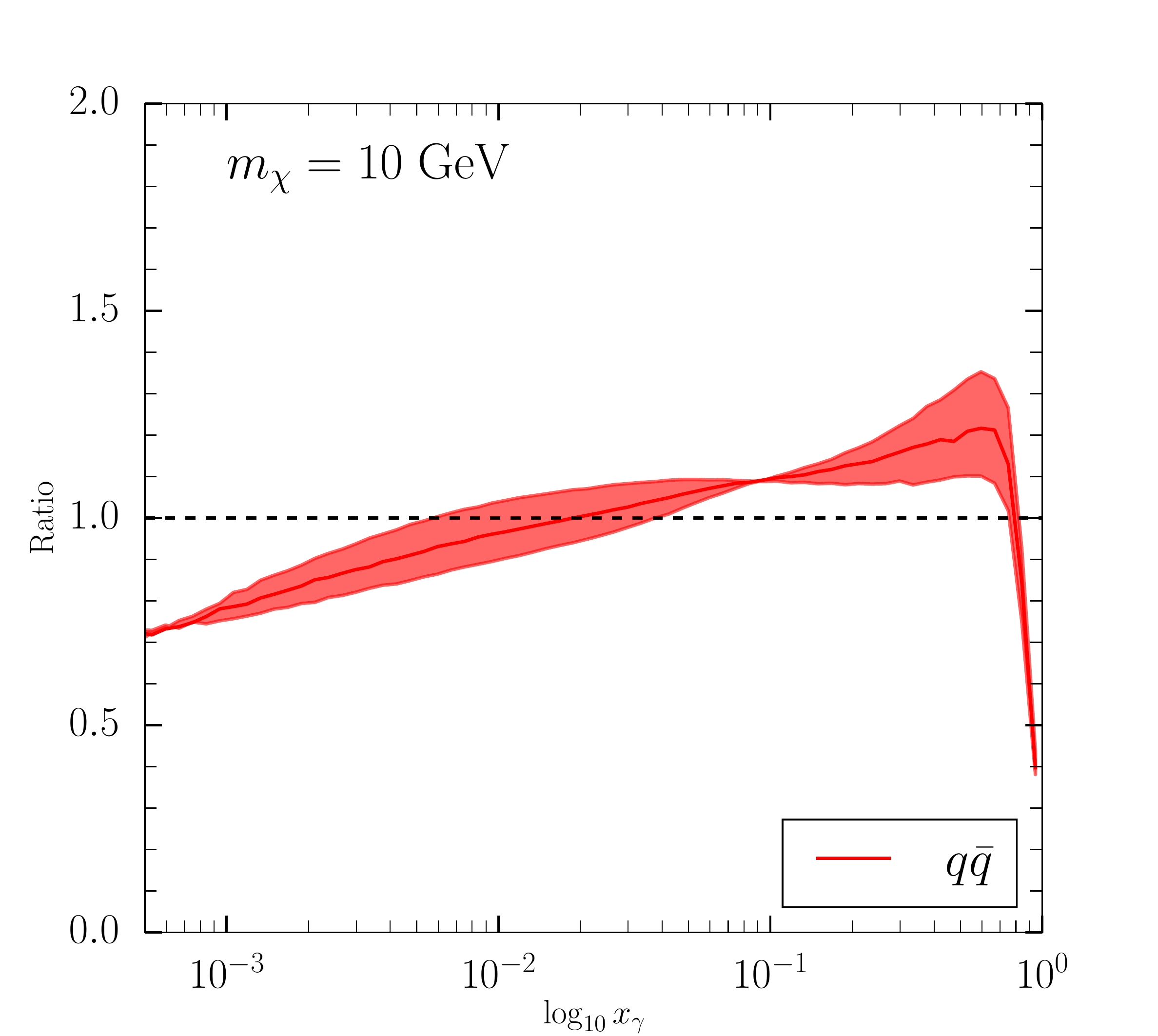}
\hfill
\includegraphics[width=0.32\linewidth]{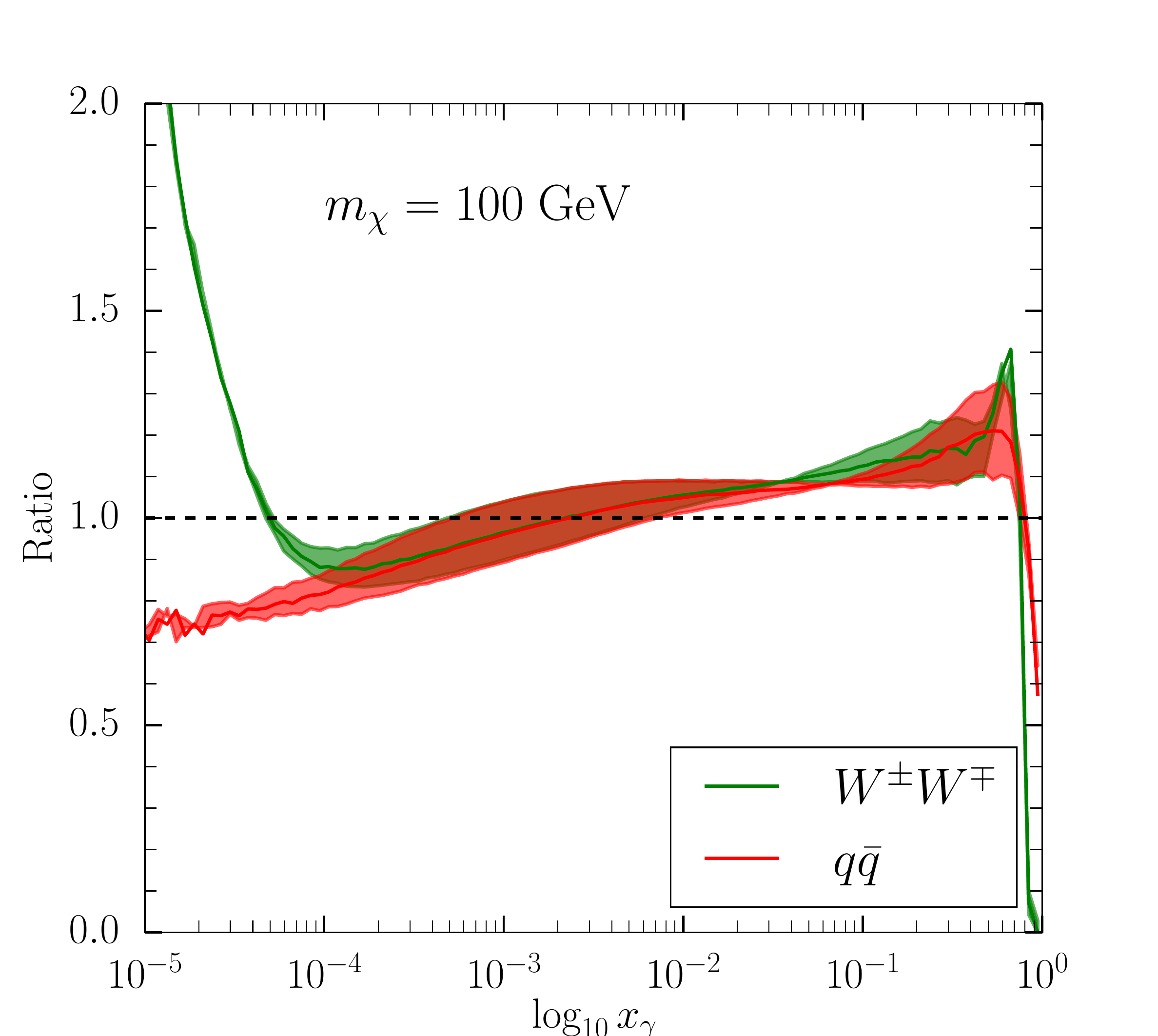}
\hfill
\includegraphics[width=0.32\linewidth]{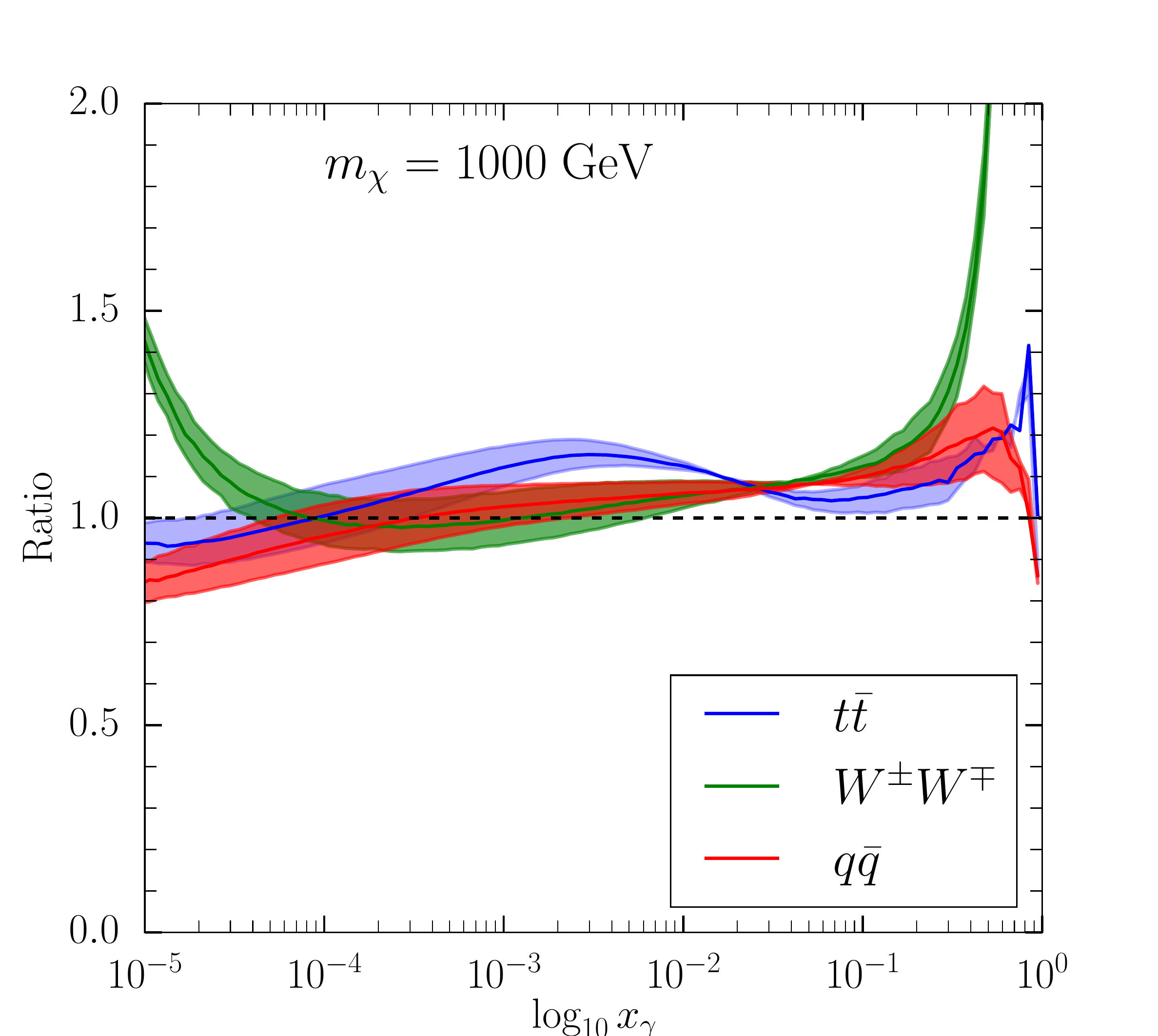}
\caption{Photon spectra obtained using our tune normalized to the results of \cite{Cirelli:2010xx} for $m_\chi = 10$ GeV (\emph{left pane}), $m_\chi = 100$ GeV (\emph{center pane}) and $m_\chi = 1000$ GeV (right pane). The spectra are shown for DM annihilation into $q\bar{q}$ (red), $W^\pm W^\mp$ (green) and $t\bar{t}$ (blue). The dashed bands show the QCD uncertainties on the parameters of the Lund fragmentation function.}
\label{fig:comparison}
\end{figure}

\section{Our results relative to the PPPC4DMID}
\label{sec:pppc4dmid}

The PPPC4DMID~\cite{Cirelli:2010xx} is widely used for DM studies. The authors of \cite{Cirelli:2010xx} made a detailed MC study of particle spectra in DM annihilation including Electro-Weak (EW) corrections \cite{Ciafaloni:2010ti}. The output of this study was a complete recipe, in the form of interpolating grids, for particle spectra ($\gamma, e^+, \bar{p}, \bar{\nu} \ldots$) in DM annihilation covering a wide range of DM masses (from $5$ GeV to $100$ TeV) and annihilation channels. 

The spectra currently available in the PPPC4DMID were obtained using \textsc{Pythia} version 8.135, which was published in January 2010. Since then, a number of salient changes and updates have been made to the \textsc{Pythia} code. Firstly, the original default tune parameters (which were poorly documented but dated from around 2009) were replaced, in \textsc{Pythia}~8.2, by the \textsc{Monash} 2013 tune \cite{Skands:2014pea}, which included a complete overhaul of the final-state fragmentation parameters. Secondly, additional capabilities are available in the newer version we use for this study, such as QED showering off heavy charged particles (e.g., $W$ bosons and top quarks) which can have important contributions when they are produced very far from their threshold as was discussed in section \ref{sec:physics-modeling}. A more complete list of changes (relative to when the original PPPC4DMID study was done) 
relevant to the modeling of photon spectra in final-state fragmentation processes is as follows: 
\begin{itemize}
\item  \textsc{Pythia} 8.135: version used for the cookbook.
\item \textsc{Pythia} 8.170: particle masses, widths, and decay branching fractions updated using the PDG 2012 values. 
\item \textsc{Pythia} 8.175: new option included (off by default) to allow photon radiation in leptonic two-body decays of hadrons, via \verb|ParticleDecays:allowPhotonRadiation = on|. 
\item \textsc{Pythia} 8.175: The lower shower cutoff \verb|TimeShower:pTminChgL| for photon radiation off charged leptons was reduced from  $0.0005$ to $10^{-6}$. 
\item \textsc{Pythia} 8.176: new option for weak showers introduced (off by default). 
\item \textsc{Pythia} 8.200: default tune parameters changed to those of the Monash 2013 tune. The default tune parameter values from 8.135 are still available as an option using \verb|Tune:ee = 3|.
 \item \textsc{Pythia} 8.219: new flag \verb|TimeShower:QEDshowerByOther| allows charged resonances like $W^\pm$ to radiate photons.
\end{itemize}

We therefore believe that it would be useful to provide our results as updates to the PPPC4DMID tables. These are available from the authors of this study, and will also be released publicly in a future update. (We also note that the EW corrections considered in the PPPC4DMID study factorize off
the non-perturbative QCD modeling and they can be added to our predictions without any problem).

To illustrate the numerical size of the differences, we display in Fig.~\ref{fig:comparison} the ratio of our predictions to the results of \cite{Cirelli:2010xx} in the photon spectra for three DM masses; $m_\chi=10, 100$ and $1000$ GeV. We have chosen three final states, i.e $q\bar{q}, q=u,d,s$, $W^\pm W^\mp$ and $t\bar{t}$. We can see that the relative differences between our tuning and the predictions of the Cookbook can be quite important, particularly in the edges of the distributions (small $x_\gamma$ and large $x_\gamma$). As these differences cannot be accounted for by QCD uncertainties (shown as dashed bands in Fig. \ref{fig:comparison}), we urge to use the updated predictions from this study. \\

\begin{figure}
    \centering
    \includegraphics[width=0.65\linewidth]{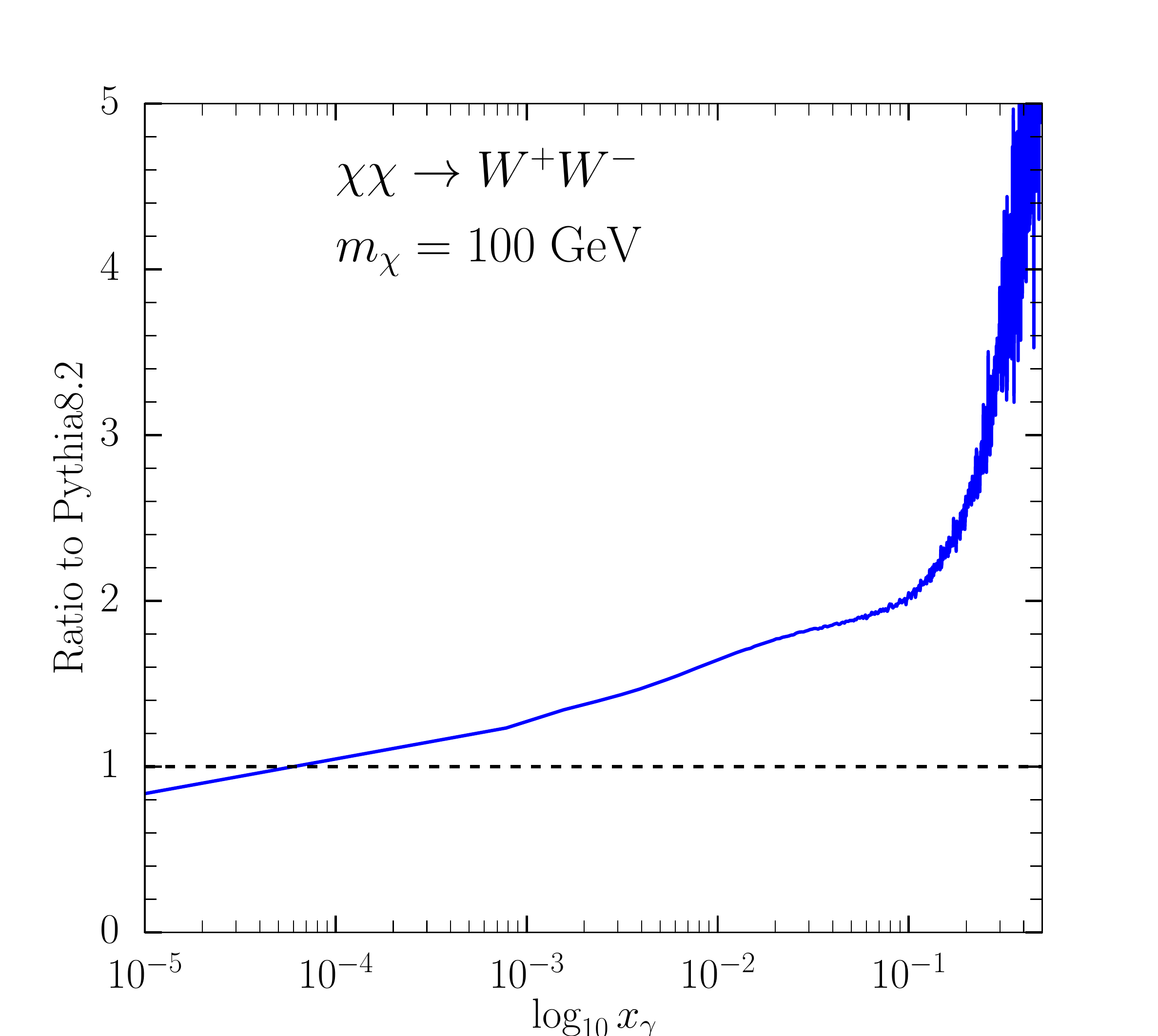}
    \caption{Photon spectra obtained using the fit functions given in \cite{Cembranos:2010dm} normalized to the results of our tune for $\chi\chi\to W^+ W^-$ and $m_\chi=100$ GeV.} 
    \label{fig:PY8PY6}
\end{figure}

Before closing this section, we would like to shortly discuss the comparison with the predictions of \textsc{Pythia6-418} used in an analysis done by the authors of \cite{Cembranos:2010dm}. In \cite{Cembranos:2010dm}, a complete analysis of the photon spectra in WIMP DM annihilation has been performed and fitting functions have been provided for different DM masses, and annihilation channels.  As an example, we compare our predictions with those obtained in Eqn. 8 of \cite{Cembranos:2010dm} for $\chi\chi \to W^+ W^-$ with $m_\chi = 100$ GeV. We display the results of the comparisons in Fig. \ref{fig:PY8PY6}. We find that the agreement is relatively not so good as compared to the PPP 4 DMID (medium panel of Fig. \ref{fig:comparison}) especially in the peak region. This suggests that the improvements in \textsc{Pythia} MC event generator are converging to solid picture thanks to the developments of the used models as well as the wealth of data that were used to tune the parameters.


\bibliographystyle{JHEP}
\bibliography{bibliography.bib}


\end{document}